\documentclass[12pt]{article}
\usepackage{latexsym,graphicx,a4,epsfig,psfrag,here,color,citesort}
\usepackage{amsmath,amssymb}

\newcommand{\mathbold}[1]{\mbox{\boldmath $\bf#1$}}
\newcommand{\eq}{\begin{eqnarray}}
\newcommand{\en}{\end{eqnarray}}
\renewcommand{\theequation}{\arabic{section}.\arabic{equation}}

\newcommand{\nnnl}{\nonumber\\}
\newcommand{\fs}{\, .}
\newcommand{\scs}{\, , \,}
\newcommand{\bea}{\begin{eqnarray}}
\newcommand{\eea}{\end{eqnarray}}
\newcommand{\ed}{\end{document}}

\newcommand{\pid}{$\pi D$\,\,}
\newcommand{\kad}{$\bar K D$\,\,}
\newcommand{\piH}{$\pi H$\,\,}
\newcommand{\kaH}{$\bar K H$\,\,}
\newcommand{\dafne}{DA$\Phi$NE\,\,}
\newcommand{\co}{\,,\,}
\newcommand{\msigma}{\Sigma_+}
\begin{document}

\begin{titlepage}

\begin{flushright}
Preprint HISKP--TH--07/20
\end{flushright}

\vspace{1cm}
\begin{center}{\Large{\bf Hadronic
 atoms in QCD + QED}}

\vspace{0.5cm}
November 16, 2007

\vspace{0.5cm}
{\bf J.~Gasser$\,^a$, V.E.~Lyubovitskij$\,^{b,\dagger}$ and 
A. Rusetsky$\,^{c,\ddagger}$}

\vspace{2em}
\footnotesize{\begin{tabular}{c}
$^a\,$
Institut f\"ur theoretische Physik, Universit\"at Bern\\
$\hspace{2mm}$Sidlerstr. 5, CH--3012 Bern, Switzerland\\[2mm]
$^b\,$
Institut f\"ur Theoretische Physik, Universit\"at T\"ubingen\\
$\hspace{2mm}$Auf der Morgenstelle 14, D--72076 T\"ubingen, Germany\\[2mm]
$^c\,$ 
Helmholtz--Institut f\"ur Strahlen-- und Kernphysik,\\
 $\hspace{2mm}$   Universit\"at Bonn,
Nussallee 14--16, D--53115 Bonn, Germany\\[2mm]
$^\dagger\,$
On leave of absence from: Department of Physics, \\
$\hspace{2mm}$ Tomsk State University, 634050 Tomsk, Russia\\[2mm]
$^\ddagger\,$ 
On leave of absence from: High Energy Physics Institute,\\
$\hspace{2mm}$Tbilisi State University,
University St.~9, 380086 Tbilisi, Georgia\\
\end{tabular}  }

\vspace{1cm}

\begin{abstract}
We  review the theory of hadronic atoms
in QCD + QED, based on a 
 non--relativistic effective Lagrangian framework.
We first provide an introduction to the theory, and
 then describe several applications: meson--meson, meson--nucleon 
 atoms and meson--deuteron compounds.
Finally, we compare the quantum field theory framework used here
with the traditional approach, which is  based on  
quantum--mechanical potential scattering.
\end{abstract}

\vspace{1cm}
\footnotesize{\begin{tabular}{ll}
{\bf{Pacs:}}$\!\!\!\!$& 36.10.Gv, 12.39.Fe, 11.10.St
\\
{\bf{Keywords:}}$\!\!\!\!$& Hadronic atoms, QCD, QED, 
Chiral perturbation theory, \\
& Non--relativistic effective Lagrangians\\
&\\
&\\
&\\
{\em \underline{ Corresponding Author: }}$\!\!\!\!$& Akaki Rusetsky\\
&Tel: +49 228 732372, Fax: +49 228 733728\\
& Email: rusetsky@itkp.uni-bonn.de \\
\end{tabular}} 
\end{center}
\end{titlepage}

\setcounter{page}{2}

\tableofcontents

\newpage
%%%%%%%%%%%%%%%%%%%%%%%%%%%%%%%%%%%%%%%%%%%%%%%%%%%%%%%%%%%%%%
\section{Introduction}
\label{sec:intro}

Consider a hydrogen atom, and replace the electron with a 
negatively charged pion. This compound  is called {\it pionic hydrogen} and
 is the simplest example of a
 {\it hadronic atom}.
 In many aspects, the properties of 
hadronic atoms
are similar to those of the hydrogen atom, because both bound states are
predominantly formed by the static Coulomb force. The
typical size of  pionic hydrogen e.g. is characterized by its Bohr radius
\eq
r_B=(\alpha\mu_c)^{-1}\, ,\quad\quad \mu_c=m_pM_\pi(m_p+M_\pi)^{-1}\scs
\en 
where $\alpha\simeq 1/137$ denotes the fine--structure constant, and 
where $\mu_c$
 is the reduced mass of the system. The mass of the proton and of the 
charged pion
 are denoted by $m_p$ and $M_\pi$, respectively. 
The distance $r_B\simeq 220~{\rm fm}$   is much smaller than the 
hydrogen radius, but still much larger than
the range of strong interactions, which is typically of the order of a 
few fm.  It is for this reason that strong interactions
do not change the structure of the bound--state spectrum in a 
profound manner. At leading order
in an expansion in $\alpha$, the energy of   $S$-wave states of
pionic hydrogen is still given
by the standard quantum--mechanical formula
\eq\label{eq:Ecoul}
E_{n}=m_p+M_\pi-\frac{\mu_c\alpha^2}{2n^2}\, ,\quad\quad
n=1,2,\cdots\fs
\en
However, in distinction to ordinary hydrogen,  the ground state of 
pionic hydrogen 
is unstable. It decays mainly via the charge--exchange
 reaction $\pi^-p\to\pi^0n$, and via the electromagnetic 
channel $\pi^-p\to\gamma n$, with width  $\Gamma_1\sim 1~{\rm eV}$~\cite{Sigg:1995}.
 This  corresponds to a lifetime   $\tau_{1} \sim  10^{-15}~{\rm s}$,
 which is  much smaller than the lifetime
of the charged pion,  $\tau_{\pi}\sim  10^{-8}~{\rm s}$, so
that the pion in the atom can be considered  a practically stable particle.
 Despite of its
short lifetime, pionic hydrogen can be considered  a  quasi-stable bound
 state, since the pion travels many times around the proton 
before decaying, as the ratio  $\frac{1}{2}\,\mu_c\alpha^2/\Gamma_{1}\sim
 10^3$ indicates. 

To understand the significance of experiments performed with hadronic atoms, 
we imagine a simplified picture,
where the interaction between the constituents of a hadronic atom consists
of only two terms: the long--range Coulomb part and the short--range
strong interaction. Equation~(\ref{eq:Ecoul}) describes the 
bound--state spectrum only approximately, at leading order in an expansion in 
$\alpha$. The leading correction  
to Eq.~(\ref{eq:Ecoul}), which emerges
at order $\alpha^3$, is due to strong interactions only -- there is no
interference between Coulomb and strong interactions at this order.
 Consequently,  
measuring very accurately the difference between the true 
energy levels of  pionic
hydrogen and their pure Coulomb values Eq.~(\ref{eq:Ecoul}),
one can extract information about the strong interactions between the pion
and the proton. Because the size of
the atom is much larger than the characteristic radius of strong 
interactions, the bound--state observables can feel only the low--momentum
behavior of the strong pion--nucleon $S$--matrix -- in other words, the energy
shift must be expressed in terms of the threshold parameters of the 
pion--nucleon scattering amplitude: the scattering lengths, effective ranges, 
etc.  Because the characteristic 3-momenta of the hadrons
within the atom are of size 
$|{\bf p}_{\rm av}|\simeq r_B^{-1}=\alpha\mu_c\propto\alpha$,
the relative strength of the corrections which contain scattering lengths,
effective ranges, $\cdots$ should be ordered according to
$1:\alpha^2:\cdots$. As a result of this,  the leading--order 
energy shift depends only
on the pion--nucleon scattering lengths. The situation closely 
resembles a well--known example in classical electrodynamics, which tells 
us that an arbitrarily complicated charge distribution can be characterized 
by a few parameters of the multipole expansion, if the distance to the system 
is much larger than the size of the system itself.
 It is also clear, why the
observables of the atom cannot depend on anything else but on the threshold
parameters of the pion--nucleon amplitude: the distances characteristic for the
atom are already asymptotic for the pion--nucleon system, where nothing but
the $S$-matrix elements can be observed. In other words, the measurement of
the observables of the pionic hydrogen does not probe the inner region of the
pion--nucleon interaction.

The insensitivity of 
the pionic hydrogen observables to the short--range details of the
pion--nucleon interaction is very fortunate: it
provides us with the possibility to directly extract the values of
the pion--nucleon scattering lengths from  atomic experiments. A different
 method
for determining the same quantities is to measure the scattering cross sections
at different energies, and to extrapolate the result to
threshold~\cite{Hoehler,KochPi,Koch1,Koch2,VPI,Rasche-phase}. 
The former method, however, is free from the
 difficulties which are related to this extrapolation procedure.
This property is even more important in other hadronic systems, where
the scattering amplitude near threshold is hardly accessible by  other
experimental technique.

Deser, Goldberger, Baumann and Thirring (DGBT) were the first who derived
the model--independent relation between the (complex) 
energy shift of the ground state  in  pionic hydrogen
and the strong pion--nucleon scattering amplitude at leading order in 
$\alpha$~\cite{Deser}. The result reads\footnote{
We use throughout the Landau symbols $O(x)$ [$o(x)$] for quantities that
vanish like $x$ [faster than $x$] when $x$ tends to zero. Furthermore,
it is understood that this holds modulo logarithmic terms, i.e. we write also
$O(x)$ for $x\ln x$.
}
\eq\label{eq:Deser_hist}
\Delta E_{1}^{\rm str}-\frac{i}{2}\,\Gamma_1=-\frac{2\pi}{\mu_c}\,
|\tilde\Psi_{10}(0)|^2\, A(\pi^-p\to\pi^-p)+O(\alpha^4)\, ,
\en
where $|\tilde\Psi_{10}(0)|^2=\alpha^3\mu_c^3/\pi$ denotes the square of the
Coulomb wave function of the atom at the origin and is, therefore,
   a measure of
 the probability that the charged pion and the proton in 
  the atom  come very close
to each other. 
  Further, $A(\pi^-p\to\pi^-p)$ is the $\pi^-p$ elastic scattering amplitude at 
threshold, which describes strong interactions between these particles after
  they come close. The scattering amplitude 
is normalized so that
  \eq
A(\pi^-p\to\pi^-p)=a_{0+}^++a_{0+}^-+\cdots\, .
  \en
Here, the quantities on the r.h.s. denote the $S$-wave $\pi N$ isospin even/odd
  scattering lengths, which are defined according to Ref.~\cite{Hoehler},
and the ellipsis stands for  isospin breaking terms.

The imaginary part of the quantity $A(\pi^-p\to\pi^-p)$ is given by
unitarity. For simplicity, let us assume for a moment that the atom decays
only into the strong channel $\pi^-p\to\pi^0 n$. Then, the imaginary part 
is\footnote{Note that, as we shall see below, 
adding the second
decay channel $\pi^-p\to\gamma n$ merely leads to the replacement
${\rm Im}\, A(\pi^-p\to\pi^-p)\to (1+P^{-1})\,{\rm Im}\, A(\pi^-p\to\pi^-p)$, 
where $P=\sigma(\pi^-p\to\pi^0 n)/\sigma(\pi^-p\to\gamma n)$ is the
so--called Panofsky ratio $P\simeq 1.546$, which
describes the relative probability
for the ``strong'' and ``electromagnetic'' decay channels. Since the value
of the Panofsky ratio is taken from other 
measurements and is thus considered to be an
input in the analysis of the pionic hydrogen data, this analysis proceeds
exactly similar to the 1-channel case.}
\eq\label{eq:Im1channel}
{\rm Im}\, A(\pi^-p\to\pi^-p)=2p^\star|A(\pi^-p\to \pi^0 n)|^2\, ,
\en
where $p^\star$ is the CM momentum of the outgoing $\pi^0$ in the scattering
process $\pi^- p\to \pi^0 n$ at threshold.
 The normalization is chosen such that, in the isospin limit, 
the threshold amplitude is $A(\pi^-p\to \pi^0 n)=- a_{0+}^-$.
Finally, the decay width $\Gamma_{1}$
of the ground state  is calculated from Eqs.~(\ref{eq:Deser_hist},
\ref{eq:Im1channel}). Therefore, measuring the spectrum of pionic atoms
provides information on the pion--nucleon scattering lengths.

Hadronic scattering lengths are important low--energy
characteristics of quantum chromodynamics (QCD) for the following reason. 
At  small momenta of order of the pion mass, 
QCD can be described by 
a  low--energy effective field theory (EFT), which contains only
hadronic degrees of freedom -- the approach goes under the name chiral
perturbation theory (ChPT)~\cite{Weinberg_ChPT}.
The crucial point is
that, in certain cases, 
 (e.g., for  $\pi\pi$ scattering~\cite{Bijnens1,Bijnens2,Anant,Colangelo:2000jc,Colangelo:2001df,Caprini}),
it is possible to obtain a very accurate prediction for the hadronic 
scattering lengths in ChPT and, consequently, comparing the theoretical values
with the results obtained in the experiment, one can perform a
direct study of the properties of the low--energy QCD. 
For a method which does not make use of chiral symmetry to determine
 the scattering lengths, see Refs.~\cite{nonchiral}.
 Scattering lengths in the framework of pure QCD can now also 
be determined from numerical simulations of QCD  on a 
lattice~\cite{scatt_lattice}, see also the reviews
by Leutwyler \cite{leutlatt} and Colangelo \cite{colangelo_kaon07}.
 Last but not least, the hadronic scattering lengths often serve as an input 
in the determination of other important low--energy characteristics of strong 
 interactions (like the pion--nucleon $\sigma$-term \cite{sigma}
or the $\pi NN$ coupling 
constant \cite{GMO,Ericson:2000md,Sainio-new} 
in the case of the pion--nucleon scattering 
lengths). Below, we
discuss  the fundamental physics content 
of various hadronic scattering lengths. Here we  merely note that 
the knowledge of the exact values of these quantities
would greatly advance our knowledge of the fundamental features of strong 
interactions, allowing for a variety of high--precision tests of QCD at 
low energy.

In order to carry out the comparison of the experimentally 
determined had\-ro\-nic
scattering lengths with the theoretical predictions, the accuracy of the
lowest--order formulae, analogous to Eq.~(\ref{eq:Deser_hist}), is 
not sufficient
in most cases. For this reason, 
there have been numerous 
attempts to improve the accuracy in these relations by 
including electromagnetic corrections 
and strong isospin breaking effects. These investigations
had first been
carried out within the framework of  potential scattering theory:
 strong interactions are described by  potentials, 
 assumed to be isospin symmetric.
 Isospin violation then stems from Coulomb corrections, from the
mass differences in the free Green functions (masses of all particles 
involved are put equal to their physical masses), from the coupling to the
electromagnetic channels, etc. Unfortunately, the results of  
 potential--model--based calculations turned out to be  not unique -- 
 in some cases, they  even differ  to an extent that matters 
at the level of the experimental precision. 

The following two comments are in order at this point.
\begin{itemize}

\item[i)]
The scattering lengths which one attempts to extract 
from the investigation of hadronic atoms 
are the ones in pure QCD, in the isospin symmetry limit $m_u=m_d$.
In particular, the electromagnetic effects are switched off.
In potential scattering theory, the Coulomb potential can be easily
turned on or off. It is however well known that  splitting off 
electromagnetic effects in  quantum field theory (QFT)
is ambiguous and leads in general to scale dependent
results.
\item[ii)]
Whereas the relation of hadronic atom observables to the
threshold parameters in the hadron--hadron scattering amplitude is universal,
the relation of these threshold 
parameters to the  purely hadronic scattering lengths 
is not universal and depends
on a complicated interplay of strong and electromagnetic interactions.
  Potential models do not include the full content of  isospin 
breaking effects {\it ab initio} and therefore do not, in general, suffice to work out this relation.
For example, the effects due to 
the short--range electromagnetic interactions and the quark mass dependence
of the hadronic scattering amplitudes -- which naturally emerge in ChPT --
  are omitted in  simple hadronic 
potentials. On the other hand, in many cases
 the bulk of the total correction is provided exactly by these omitted terms.

\end{itemize}

These two points illustrate the major challenge to  any theoretical
investigation:
in order to be able to fully exploit the high--precision data from present
and future experiments on hadronic atoms, one needs
 a systematic framework to relate hadronic atom characteristics to observables
of the underlying theory. Chiral perturbation theory is an obvious candidate for such a framework.
However, despite remarkable progress 
in describing low--energy scattering processes with {\em elementary hadrons} 
in ChPT, the study of {\em bound states} within the same theory remained 
{ terra incognita} for a long time.
 In certain cases, the problem 
could be solved by  brute force with the use of methods of QFT, namely,
 Bethe--Salpeter equation or  quasipotential 
reductions thereof \cite{Silagadze,Sazdjian1,Sazdjian2,Jallouli:2006an,Dubna1,Dubna2,Dubna3}.
On the other hand, the matter  turned out to be very involved in general -- it 
became obvious that new ideas are needed to overcome the difficulties on 
the way towards a general theory of hadronic atoms in QCD+QED.

A crucial observation was that
the calculation of the spectrum of hadronic atoms
can be treated separately from the chiral expansion 
\cite{Labelle,Soto1,Holstein,Ravndal1,Ravndal2,Bern1}.
The reason for this consists in 
the large difference of the pertinent momentum
scales: whereas the characteristic momenta in ChPT are of order of the pion
mass, the typical bound--state momentum is the inverse 
of the Bohr radius. This difference in scales prompts using a 
non--relativistic effective Lagrangian approach to bound states, which has
first been  introduced by Caswell and Lepage in the framework of 
QED~\cite{Caswell}. 

Let us shortly describe  the pertinent low--energy EFT,  which is valid at 
typical momenta of size $|{\bf p}|\ll M_\pi$. 
The structure of this EFT 
is much simpler than the structure of ChPT, for the
following reasons. i) As  just mentioned, the system can be described 
by means of an effective non--relativistic  
Lagrangian.
 Relativistic corrections are included in a perturbative manner.
  In bound--states, these corrections will obviously 
contribute at higher orders in $\alpha$, because ${\bf p}$ is a quantity
of order $\alpha$. ii) The bound 
states in the 
non--relativistic theory are described by a Schr\"{o}dinger equation, 
instead of the Bethe--Salpeter equation. This leads to dramatic 
simplifications
and allows one to design a systematic perturbative approach to the calculation
of the bound--state observables on the basis of the Feshbach
formalism~\cite{Feshbach1,Feshbach2}. Finally, the parameters of the
effective Lagrangian are determined 
by matching this theory to ChPT in the scattering sector. At the end of the
calculation, the reference to the (auxiliary) non--relativistic EFT
completely disappears from the final results, which express the 
observables of hadronic atoms it terms of the parameters 
of ChPT.

The application of the above simple idea for the description of hadronic atoms
turned out to be very fruitful. Indeed,   within a few years, most of 
the (theoretically) relevant  hadronic bound states
have been revisited using this  approach (see, e.g.,
 Refs.~\cite{Bern1,Bern2,Bern3,Bern4,Mojzis,SchweizerHA,Zemp,Raha1,Raha2,Raha3,Raha4}.
 For  calculations using  field theoretical methods different 
from the ones just described, see 
 Refs.~\cite{Efimov,Belkov,Volkov,Kuraev:1997ed,pioniumgg,Ivanov1,Ivanov2,Ivanov3,Ivanov4,Ivanov7,Ivanov8,Irgaziev,Krewald}).
Data analysis now rests 
on the basis of  low--energy effective field theories of QCD+QED, and
 is in general free of any model assumptions\footnote{Non--relativistic Lagrangians in the 
spirit of Ref.~\cite{Caswell} are heavily used for the description of heavy quark bound 
states in QCD, see Ref.~ \cite{lepagethacker} for the introduction of the method. 
Two recent reviews on the subject are Refs.~\cite{Brambilla:2004jw,quarkonia2}.}.

The article is organized as follows.
In section~\ref{sec:fundamental} 
we start with a detailed discussion of the fundamental physics background
behind hadronic atom experiments.
In sections~\ref{sec:nonrel}, \ref{sec:including_gamma}, \ref{sec:boundstates}
and \ref{sec:splitting},
 we construct the general theory of hadronic atoms in QCD+QED.
We keep this part rather self--contained,
 avoiding extensive references to existing literature. 
For this reason,
these sections also contain the essentials of  non--relativistic
effective field theories, to an extent that is needed for understanding
the material related to hadronic atoms.
 In section~\ref{subsec:widthtable} we collect useful general information on the relation
between spectra and scattering lengths.
We then describe
in sections~\ref{sec:pionium}--\ref{sec:deuteron}
several hadronic atoms, which have been 
treated so far in the framework of QFT, see 
Table \ref{tab:atoms}.
\begin{table}[t]
\begin{center}
\begin{tabular}{|r c c|}\hline
&&\\[-3.5mm]
Atom & Name & Symbol\\[2mm] \hline
$\pi^+\pi^-$&pionium&$A_{2\pi}$\\[2mm]
$\pi^\mp K^\pm$&$\pi K$ atom&$A_{\pi K}$\\[2mm]
$\pi^-p$&pionic hydrogen&\piH\\[2mm]
$K^-p$&kaonic hydrogen&\kaH\\[2mm]
$\pi^-d$&pionic deuterium&\pid\\[2mm]
$K^-d$&kaonic deuterium&\kad\\[2mm]\hline
\end{tabular}
\end{center}
\caption{Hadronic atoms investigated in this report. The corresponding names
 and  symbols used in the text 
are also displayed.
\label{tab:atoms}}
\end{table}

 In section~\ref{sec:potential} we briefly discuss the relation
of our approach to conventional potential models.
Finally, section~\ref{sec:concl} contains a summary and concluding remarks.
 Appendix \ref{app:notations} collects some notation, appendix 
\ref{app:generalized} discusses the unitarity condition for non
 hermitian Hamiltonians, and in appendix \ref{app:unitarity} we discuss 
matching for  non--relativistic coupling constants in \piH$\!\!\!$, using unitarity.

We note that it was not our aim
to cover the large number of articles concerned with the rich 
and well--developed phenomenology of hadronic atoms within the traditional
quantum--mechanical setting, based on  hadronic potentials, or to provide 
a complete bibliography on this issue. There are excellent 
textbooks (see, e.g., Refs.~\cite{Ericson-Weise,Scheck,Deloff:2003ns}) or review 
articles~(e.g. Ref.~\cite{Gal}), which should be considered as  complementary to
 the present work. In addition, the issue of  deeply bound exotic atoms
lies  beyond the scope of the present review and will not be 
discussed  at all. In order to follow the developments in this field,
we refer the interested reader e.g. 
 to the EXA05 proceedings~\cite{EXA05}. 
For a short review, see, e.g., Ref.~\cite{Weise:2005ss} and references therein.

%%%%%%%%%%%%%%%%%%%%%%%%%%%%%%%%%%%%%%%%%%%%%%%%%%%%%%%%%%%%%%%%%%%%%%%%%%
\setcounter{equation}{0}
\section{Physics background}
\label{sec:fundamental} 

As already mentioned,  experiments on hadronic atoms can
provide stringent information on  fundamental properties of QCD.
In this section we  consider in some detail
the potential of individual experiments in this respect. 

\subsection{Pionium}

  Pionium is presently investigated
 by the DIRAC collaboration 
 at CERN~\cite{DIRAC,Adeva:2003up,Gomez,DIRAC_PRELIMINARY,Goldin:2005ce,DIRAC_RESULT,DIRAC:addendum}. 
 The experiment is a very challenging one: the lifetime of  pionium in the
  ground state is of the order $\tau_1\simeq 3\times 10^{-15}~{\rm s}$. 
 Because the ratio of the binding energy with the width 
 is of order  $10^4$, one concludes that
 pionium is a quasi-stable bound state as well.
At present, 
 it is not possible to measure this lifetime directly.
  Instead,
 one first measures the ionization probability of pionium into a charged
pion pair in different targets. From the calculated
 relation~\cite{Afanasev,Trautmann} 
between this probability and the lifetime of  pionium (which
 in vacuum almost exclusively decays into a $\pi^0\pi^0$ pair), 
one obtains the lifetime.
 The DIRAC collaboration has reported~\cite{DIRAC_RESULT}
the result 
 \eq\label{eq:tau_pionium}
\tau_1=  {2.91~}^{+0.45}_{-0.38}\,\mbox{(stat)}
 {~}^{+0.19}_{-0.49}\,\mbox{(syst)}  \times
10^{-15} ~\mathrm{s}= {2.91~}^{+0.49}_{-0.62} \times
 10^{-15} ~\mathrm{s}.
\en
 Efforts to reduce the experimental uncertainty are ongoing~\cite{tauscher_kaon2007}.

The decay width of pionium in the ground state into a $2\pi^0$
pair is related to the $S$-wave $\pi\pi$
scattering lengths\footnote{We use a convention where $\pi\pi$ 
scattering lengths 
$a_I$ are dimensionless.} $a_I$ with total isospin $I=0,2$, 
\eq\label{eq:DeserGamma}
\Gamma_{1,2\pi^0}=\frac{2}{9}\,\alpha^3p_1^\star(a_0-a_2)^2+\cdots\, ,
\en
where $p_1^\star=(M_\pi^2-M_{\pi^0}^2-\frac{1}{4}\, M_\pi^2\alpha^2)^{1/2}$
denotes the magnitude of the  3-momentum
of the $\pi^0\pi^0$ pair in the final state (with higher order terms in
$\alpha$ ignored) , and the ellipses
stand for the isospin breaking corrections. 
 In the derivation of Eq.~(\ref{eq:DeserGamma}) one has not used 
chiral symmetry: this relation is universal and holds -- at leading
order in isospin breaking -- as long as the scale of strong interactions is
much smaller than characteristic atomic distances.  
The DIRAC collaboration plans to measure  $\Gamma_{1,2\pi^0}$
with an accuracy of 10\%. Once a reliable evaluation of the isospin breaking 
corrections in the relation Eq.~(\ref{eq:DeserGamma}) is carried out, this
measurement enables one to 
determine the value of $|a_0-a_2|$ with 5\% accuracy.
 As we shall see later, these isospin breaking corrections
 are of the order of  $6\%$ and therefore not at all negligible at the 
accuracy one is interested here.

What makes the above enterprise particularly 
interesting is the fact that the difference $a_0-a_2$ is very sensitive
to the value of the quark condensate in QCD~\cite{small}, 
and thus to the manner in which chiral symmetry is spontaneously broken.
 The so--called
``standard'' scenario is based on the 
assumption that the condensate is large. This, for example, signifies that
in the expansion of the pion mass in terms of the quark mass,
\eq\label{eq:l3}
M_\pi^2=M^2-\frac{\bar l_3}{32\pi^2F^2}\, M^4+O(M^6)\, ,\,
M^2=2\hat m B\,,\, \hat m=\frac{1}{2}(m_u+m_d)\,,
\en
the first term is dominant~\cite{Colangelo:2001sp}. 
Here, $F$ stands for the pion decay constant 
$F_\pi$
 in the chiral 
limit\footnote{We use $F_\pi$=92.4 MeV.},
 and $\bar l_3$ denotes a particular low--energy constant
(LEC) that appears in the $O(p^4)$ Lagrangian of ChPT, whereas
 $B$ is related to the quark condensate in the chiral limit~\cite{GL_ChPT,Gasser:1984gg}.
 Assuming the standard scenario of chiral symmetry breaking,
a very accurate prediction of scattering lengths was achieved by merging 
two--loop ChPT with Roy 
equations~\cite{Colangelo:2000jc,Colangelo:2001df},
\eq\label{eq:a0a2}
a_0=0.220\pm 0.005\,,\quad a_2=-0.0444\pm 0.0010\,,\quad a_0-a_2=0.265\pm
0.004\, .
\en
If the experimental value of $a_0-a_2$ does not agree with this 
 prediction, this would unambiguously indicate that chiral symmetry
breaking in QCD proceeds differently
from the standard picture~\cite{small}. As a result of this, the first term in 
the expansion Eq.~(\ref{eq:l3}) would be non--leading, and the chiral 
 expansion must be reordered~\cite{small}.

Measuring the energy levels of 
pionium~\cite{Ovsyannikov1,Ovsyannikov2} enables one to extract
a different combination of the $S$-wave scattering lengths. Indeed,
 the strong  energy
  shift of $S$--states is given by~\cite{SchweizerHA}
\eq\label{eq:DE}
\Delta E_{n}^{\rm str}=-\frac{\alpha^3M_\pi}{6n^3}\,(2a_0+a_2)+\cdots\, ,
\en
where $n$ denotes the principal quantum number, and where the ellipsis
 stands for isospin breaking corrections. Hence, measuring both,
$\Gamma_{1,2\pi^0}$ and $\Delta E_{n}$, and assuming $a_0-a_2>0$, 
one may extract the 
values of $a_0$ and $a_2$ separately
and compare them with the prediction  Eq.~(\ref{eq:a0a2}).

The present situation concerning the verification of  
the prediction Eq.~(\ref{eq:a0a2}) is the following. 
Lattice results~\cite{scatt_lattice,leutlatt,colangelo_kaon07}, the data 
from DIRAC~\cite{DIRAC_RESULT} on pionium lifetime and from NA48 
on the cusp in $K\to 3\pi$ 
 decays~\cite{Batley:2005ax,Cabibbo1,Cabibbo2,Prades,cusp1,cusp2} 
 neatly confirm the prediction, 
although partly with considerably larger error bars.
On the other hand,  preliminary NA48/2 data on $K_{e4}$
 apparently preferred larger values~\cite{na48earlier} 
of $a_0$ than the one displayed in Eq.~(\ref{eq:a0a2}).
 This puzzle was very recently resolved by 
the following observation~\cite{gasser_kaon07,isospinCGR}. 
The electromagnetic corrections  applied in the  NA48/2 $K_{e4}$ 
data  were calculated using the PHOTOS Monte Carlo code~\cite{photos}, 
 and by applying in addition the Sommerfeld factor~\cite{blochprivate}. 
However, this does not yet take care of all 
isospin breaking effects in this decay~\cite{marseillekl4}. Indeed, 
the kaon can first decay 
into a $\pi^0\pi^0$ or $\pi^0\eta$ pair,
 that then re--scatter into the outgoing charged pions,
 through $\pi^0\pi^0\to \pi^+\pi^-$ and $\pi^0\eta\to 
\pi^0\pi^0\to\pi^+\pi^-$.   
Since the charged pion is heavier  than the neutral one by about 
4.6 MeV, these intermediate states generate a cusp in the {\it phase} 
of the relevant form factor. As a result of this, the
 phase is pushed upwards by about half a degree, and does not vanish 
at the threshold for $\pi^+\pi^-$ production. Once this is taken 
into account, the $K_{e4}$ data from NA48/2 are in nice agreement 
with Eq.~(\ref{eq:a0a2}). There is, however, a discrepancy with the E865 
data~\cite{Pislak1,Pislak2}: the isospin corrections just mentioned spoil the 
good agreement with the prediction Eq.~(\ref{eq:a0a2}). 
We refer to
 Refs.~\cite{gasser_kaon07,bloch_kaon07,colangelo_kaon07,leutlatt} for more
details.

\subsection{$\pi K$ atom}
 \label{subsec:pikatom}
The DIRAC collaboration at CERN 
 plans~\cite{DIRAC:addendum} to measure the $\pi K$ atoms in 
continuation of the already running experiment on pionium.
 From the measurements of the  energy shift and  decay width
of the ground state, one would in principle extract the two independent
 isospin combinations 
  $a_0^+$ and $a_0^- $ of the  $S$-wave $\pi K$ scattering lengths, because
 the strong energy shifts and decay widths of $\pi K$ atoms in their $S-$
 states are related to these scattering lengths in a manner 
very analogous to pionium,
 \eq\label{eq:GapiK}
  \Delta E_n^{\rm str}&=&  - \frac{2\alpha^3\mu_c^2}{n^3} 
 (a_0^++a_0^-)+\cdots\scs\nnnl
  \Gamma_n &=& \frac{8\alpha^3\mu_c^2}{n^3}\, p_n^\star(a_0^-)^2+\cdots\fs
 \en
 Here the ellipses stand for isospin breaking effects.

The theoretical interest in these scattering lengths  
 is twofold at least. First,
 we mention that there exists a low--energy theorem~\cite{Roessl}, 
which states that the
 expansion of the isospin odd scattering length $a_0^-$ is of the form
\bea\label{eq:roessl}
 a_0^- =\frac{M_\pi M_K}{8\pi F_\pi^2(M_\pi+M_K)}(1+O(M_\pi^2)).
\eea
 Here, $M_\pi, M_K$ and $F_\pi$ denote the physical meson masses and the pion
decay constant in the isospin symmetry limit.
 The theorem Eq.~(\ref{eq:roessl}) 
 states that  the corrections to the leading current algebra result
  are proportional to powers of the pion mass -- they vanish 
in the chiral limit $m_u=m_d=0$.\footnote{Implicitly, this result is contained in the
explicit expressions of the $\pi K$ amplitudes given in Ref.~\cite{BKM-Kpi}.}
Since there is no strong final state 
interaction in this channel, one expects  that 
these corrections are modest.
 On the other hand, the two--loop calculation performed in Ref.~\cite{piKbijnens} suggests
 that the  correction at order $p^6$  is substantial, and moreover larger than the  one at order $p^4$.
The numerical result obtained in Ref.~\cite{piKbijnens} for the scattering length $a_0^-$
agrees with an analysis~\cite{piKBuettiker} based on
Roy equations\footnote{Note, however, that the available 
experimental data below $1~\mbox{GeV}$ are, in general,
inconsistent with the solution of Roy equations, where the 
data above $1~\mbox{GeV}$ are used as phenomenological  input.}.

The situation is puzzling -- how can it be that higher orders are 
larger than low--order contributions? As was shown by Schweizer 
~\cite{piKschweizer}, in
the present case, the large two--loop correction stems mostly from 
counterterms at order $p^6$, estimated with resonance 
saturation in Ref.~\cite{piKbijnens}, see also Ref.~\cite{Ananthanarayan:2000cp} in this connection.
 Is this resonance estimate correct? If yes, ChPT is
turned upside down. If no, ChPT does not agree with the result from the Roy
equation analysis.

Second, the scattering length do also depend on certain LECs at order $p^4$,
whose values would be interesting to know in light of the large/small
condensate question~\cite{small}. So, 
in case one does have a precise
value of the scattering lengths, the first point could be clarified in the
sense that one knows what is right and what is wrong. Concerning the 
second point, it remains to be seen whether this would 
indeed allow one to pin down those LECs which play a dominant role
in connection with the large/small condensate issue. 
 A detailed analysis of this point remains to be performed.

\subsection{Pionic hydrogen and pionic deuterium}
\label{subsec:piHpiD}
 Measurements of the energy shift and 
decay width of the ground state of  \piH and 
 \pid  were carried out by the 
Pionic Hydrogen Collaboration at 
 PSI~\cite{Sigg:1995,Chatellard:1995,Sigg:1996qd,Chatellard:1997nw,Hauser:1998,Schroder:1999,Schroder:2001}. 
 The pionic hydrogen experiment 
was  upgraded
 in 1998~\cite{R98}, followed by an 
increased accuracy in particular 
 of the width measurement~\cite{Gotta:2005}
 (for a review, see, e.g., Ref.~\cite{Gotta:2004rq}). 
 The preliminary values   of the strong shift 
and  width of the ground state of \piH obtained in the upgrade~\cite{R98}  are~\cite{Gotta:2006}
\eq\label{eq:hyd_energywidth}
\Delta E_{1}^{\rm str}&=&-7.120\pm 0.011\,~{\rm eV}\, ,
\nonumber\\[2mm]
\Gamma_{1}&=&0.823\pm 0.019~{\rm eV}\fs
\en
The announced  aim~\cite{Gottaprecision} of the collaboration is
to extract $S$-wave $\pi N$ scattering lengths from this experiment with
1\% precision. This will be a unique experimental result for hadron physics.
Of course, in order to achieve this goal by 
measuring  \piH alone, 
one should remove  possible sources of theoretical uncertainties related
to isospin breaking effects, at an accuracy that matches the
experimental precision. At present, this seems to be a very difficult 
task, as we will show later in this report.

The measured shift and width of \pid are~\cite{Hauser:1998}
\eq \label{eq:deuteron}
\Delta E^{\rm str,d}_{1} = 2.46 \pm 0.048 \ {\rm eV}\, ,\quad\quad 
\Gamma_{1}^{\rm d} = 1.194 \pm 0.105 \ {\rm eV}\, . 
\en 
This accuracy is expected to improve in the near future~\cite{R06,strauchmenu2007}.

 We will show  in later sections that, using
multiple--scattering 
theory~\cite{Ericson-Weise,Afnan:1974ye,Mizutani:1977xw,ThomasLandau,Kolybasov:1972bn,Deloff:2001zp,Baru,Ericson:2000md,Baru:1997xf,Baru:1996pd,Tarasov:2000yi,Deloff:2003ns} or
chiral EFT in the two--nucleon
sector~\cite{Bernard1,Bernard2,Hanhart_Ko,Hanhart:2007ym,Valderrama:2006np,Platter:2006pt,Baru:2004kw,Hanhartnew,Weinberg-deuteron,Doring:2004kt,Epelbaum:2005pn,Borasoy:2003gf,Beane:2002aw,Raha2},
  one can relate the value of the $\pi d$ scattering length, which can be extracted from the 
experimental data, to the 
$\pi N$ scattering lengths. 
This results in  additional constraints
on the values of $a_{0+}^+$ and $a_{0+}^-$.

Let us  assume that one indeed is able to extract the exact values of
the $\pi N$ scattering lengths from the experiment with high precision.
 What  information about the fundamental properties of QCD are
contained in these  precise values? Of course, $\pi N$ scattering lengths
are quantities of fundamental importance in low--energy hadronic physics by themselves,
since they test the QCD symmetries and the exact pattern of the chiral 
symmetry breaking (see, e.g., Refs.~\cite{Weinberg66,Bernard:1995dp,BLI}). 
Moreover, since the knowledge of these 
scattering lengths
places a constraint on the  $\pi N$ interactions at low energy, it also affects
our understanding of more complicated systems where $\pi N$ interaction serves
as an input, e.g. $NN$ interaction, $\pi$-nucleus scattering, three--nucleon
forces, etc.

In addition to this, the high--precision values of the $\pi N$ scattering 
lengths are used as an input for the determination of different basic
parameters of QCD at low energies more accurately. One example 
is the $\pi NN$ coupling constant $g_{\pi NN}$, which is 
obtained from the Goldberger--Myazawa--Oehme (GMO) sum rule~\cite{GMO}, where
a particular combination of scattering lengths enters as a subtraction
constant,
\eq\label{eq:gmo}
\frac{g_{\pi NN}^2}{4\pi}&=&\biggl(\biggl(\frac{2m_N}{M_\pi}\biggr)^2-1\biggr)
\biggl\{\biggl(1+\frac{M_\pi}{m_N}\biggr)\frac{M_\pi}{4}\,(a_{\pi^-p}-a_{\pi^+p})
\nonumber\\[2mm]
&-&\frac{M_\pi^2}{8\pi^2}\int_0^\infty\frac{dk}{\sqrt{M_\pi^2+k^2}}\,
(\sigma_{\pi^-p}(k)-\sigma_{\pi^+p}(k))\biggr\}\, ,
\en
where $m_N$ is the nucleon mass in the isospin symmetry limit, 
$\sigma_{\pi^\pm p}(k)$ denotes the total elastic cross section for
the scattering of $\pi^+$ ($\pi^-$) on the proton, and $a_{\pi^\pm p}$ stand
for  isospin combinations of  $S$-wave $\pi N$ scattering
lengths,  $a_{\pi^\mp p}=a_{0+}^+ \pm a_{0+}^- $.
 For  recent investigations of the GMO sum rule, see
 Refs.~\cite{Ericson:2000md,Sainio-new}.

Other important quantities, which can be obtained by using 
the $S$-wave $\pi N$ scattering lengths as an input, are the so--called $\pi N$
$\sigma$-term and the strangeness content of the nucleon~\cite{sigma}. 
The $\sigma$-term,
which measures the explicit breaking of chiral symmetry in the one--nucleon
sector due to the $u$-- and $d$--quark masses, is given by the matrix element
of the chiral symmetry--breaking Hamiltonian $\hat m (\bar u u+\bar d d)$ 
between the one--nucleon states. The latter quantity is expressed through 
the value of the
scalar form factor of the nucleon at $t=(p'-p)^2=0$,
\eq\label{eq:scalar}
\bar u(p',s')\sigma_{\pi N}(t)u(p,s)=\langle p's'|\hat m(\bar uu+\bar dd)|ps\rangle
\, , \quad\quad
\sigma_{\pi N}\doteq \sigma_{\pi N}(0)\, ,
\en
where  the  Dirac spinors are normalized to $\bar uu=2m_N$.
The $\sigma$-term is related to the strangeness content
of the nucleon $y$ and the $SU(3)$ symmetry breaking piece of the strong
Hamiltonian,
\eq\label{eq:y}
&&\frac{m_s-\hat m}{2m_N}\,\langle ps|\bar uu+\bar dd-2\bar ss|ps\rangle=
\biggl(\frac{m_s}{\hat m}-1\biggr)(1-y)\sigma_{\pi N}\, ,
\nonumber\\[2mm]
&&y=\frac{2\langle ps|\bar ss|ps\rangle}{\langle ps|\bar uu+\bar dd|ps\rangle}\, .
\en
An analysis of the experimental data, carried out some 
time ago~\cite{sigma45}, gives
\eq\label{eq:GLS}
\sigma_{\pi N}\simeq 45~{\rm MeV}\, ,\quad\quad y\simeq 0.2\, .
\en 
Note that in this analysis, one has used $S$-wave $\pi N$ scattering 
lengths as 
input in the dispersion relations which provide the extrapolation of the
isospin even pion--nucleon scattering amplitude from threshold down to the
Cheng--Dashen point. The $\sigma$-term, which is obtained as a
result of the above analysis, is rather sensitive to these scattering lengths,
see, e.g., Ref.~\cite[Fig. 1]{sensitivity}. 
 Consequently, an accurate measurement of the scattering
lengths will have a large impact on the values of $\sigma_{\pi N}$ and $y$.
For an analysis that finds a substantially larger value of 
 $\sigma_{\pi N}$ than the one quoted above, see Pavan et al.~\cite{sigma}.

\subsection{Kaonic hydrogen and kaonic deuterium}
\label{subsec:kaH}

\begin{sloppypar}
The DEAR collaboration at 
LNF--INFN~\cite{Baldini,Bianco,Beer,Zmeskal} 
has performed a measurement
of the energy level shift and width of the \kaH  ground 
 state with a considerably better accuracy than the earlier
KpX experiment at KEK~\cite{KEK}. The present experimental values of these quantities 
are~\cite{SIDDHARTANEW}
 \end{sloppypar}
\eq\label{eq:dear}
 \Delta E_{1}^{\rm str}&=&193\pm 37~\mbox{(stat)}\pm 6~\mbox{(syst) eV}\, ,
\nonumber\\[2mm]
 \Gamma_{1}&=&249\pm 111~\mbox{(stat)}\pm 30~\mbox{(syst) eV}\, .
\en
 As can be  seen from the above result, the uncertainty is still 
tens of eV in the
 energy shift and more than 100 eV in the width. Now DEAR is being followed
by the SIDDHARTA experiment that features new silicon drift 
detectors~\cite{Lucherini:2007ki,SIDDHARTANEW}.
 The plans of the SIDDHARTA collaboration include the measurement
of both, the energy shift and 
 width of \kaH, with a precision of several
eV, i.e. at the few percent level, by 2008. Moreover, 
 SIDDHARTA will attempt the
first ever measurement of the energy shift of  \kad 
 with a comparable accuracy and possibly, kaonic helium and sigmonic atoms.
 Eventually, these experiments will allow one to extract the values of the
$K^-p$ and $K^-d$ threshold scattering amplitudes by using the pertinent
  DGBT--type formulae. Because these amplitudes depend on the other hand on the 
$S$-wave $\bar KN$ scattering lengths $a_0$ and 
 $a_1$, one can in principle determine these.

 The necessity to perform  measurements of \kad 
 energy shift and width of the ground state is also due to 
the fact that, unlike in the case of  pionic
 atoms, the measurement of the \kaH spectrum alone does not allow one
 -- even in principle -- 
 to extract independently $a_0$ and $a_1$.
This happens because the imaginary parts of the threshold amplitudes
 do not vanish in the isospin limit, being determined by the decays 
into inelastic strong channels $\pi\Sigma,\pi^0\Lambda,\cdots$ -- in other
 words, the separation of thresholds is governed by the breaking of
 $SU(3)$ symmetry. Consequently, one attempts here to
 determine four independent quantities (real and imaginary parts of $a_0$
and $a_1$) that requires performing four independent measurements -- e.g., the 
  energy level shifts and widths of  \kaH {\em and} 
\kad. Note however that,
 even though it is clear that $a_0$ and $a_1$ cannot be determined 
separately without measuring kaonic deuterium, it is still not evident
whether it is possible to do so if one performs such a measurement.
The reason is that the (complex) kaon--deuteron amplitude at threshold, which
is directly determined from the experiment and which is expressed in terms of
$a_0$ and $a_1$ through a multiple--scattering series, is generally
plagued by systematic uncertainties due to a
poor knowledge of the low--energy kaon--nucleon dynamics.
Thus, one needs to know whether these uncertainties are small 
enough not to hinder a determination of $a_0$ and $a_1$
from the  SIDDHARTA experiment.

In order to answer this question, a detailed analysis of the problem has been
performed in Ref.~\cite{Raha4}. This investigation revealed that
 -- at least, within the lowest--order approximation -- a combined 
analysis of DEAR/SIDDHARTA data on \kaH and \kad 
turns out to be more restrictive than one would {\it a priori} expect,
 see also the discussion in section~\ref{sec:deuteron} of this report. 
The combined analysis imposes stringent constraints on the 
theoretical description of the kaon--deuteron interactions at low energies and
provides  a tool for determining $a_0$ and $a_1$ with reasonable
accuracy from the forthcoming 
SIDDHARTA data. It remains to be seen, 
whether this conclusion stays intact, if higher--order corrections 
are systematically included.

What fundamental physical information 
can be gained from the values of $a_0$ and $a_1$? We believe that
it could be very useful to carry out a comparison of the 
$\bar KN$ scattering lengths measured in the DEAR/SIDDHARTA experiment
with different theoretical predictions based on the unitarization
of the lowest order ChPT amplitude
\cite{Siegel,OsetRamos,Krippa,MO,Nissler1,Nissler2,Nissler3,Verbeni1,Verbeni2,Oller:2006jw,BMN,OllerOsetRamos,Jido,Lutz}.
 It turns out that even the data from  \kaH alone
 impose rather stringent 
constraints on the values of the $\bar KN$ scattering lengths -- namely,
in certain cases 
there emerge difficulties to make DEAR data compatible with the
scattering sector~\cite{Raha1,Nissler1,Nissler2,Nissler3,Verbeni1,Verbeni2,Oller:2006jw,BMN,OllerOsetRamos,Jido,Lutz}.
It is clear that imposing additional constraints from \kad data
makes the issue of compatibility even more pronounced. 
In our opinion, it will be  important to check 
whether the unitarization approach passes this test.

 Finally we wish to note that
the original physics program of the DEAR/SID\-DHAR\-TA experiment was lar\-ge\-ly
 a direct extension of the \piH experiment to the strange quark 
sector, including, in addition, the measurement of the kaon--nucleon 
$\sigma$-term and the strangeness content of the nucleon~\cite{Bianco,Beer}. 
On the other hand, there are
 significant differences between these systems~\cite{Sainiosigma,Gasser_hep}:
\begin{itemize}
 \item[i)]
There are much less data on $\bar KN$ scattering than in the $\pi N$ case.
\item[ii)]
There are open strong channels below threshold.
\item[iii)] 
There is a sub-threshold resonance $\Lambda(1405)$.
\item[iv)]
The distance between the threshold and the Cheng--Dashen point is much larger 
than in the $\pi N$ case.
\end{itemize}
It will therefore be very difficult to extract the
 experimental value of the $\bar KN$ $\sigma$-term~\cite{Bianco,Beer}.

%%%%%%%%%%%%%%%%%%%%%%%%%%%%%%%%%%%%%%%%%%%%%%%%%%%%%%%%%%%%%

\setcounter{equation}{0}
\section{Non--relativistic effective theories: strong sector}
\label{sec:nonrel}

\subsection{Introductory remarks}

In relativistic quantum field theories, the number of particles 
is not a conserved quantity. As a result of this, $S$-matrix 
elements have a complicated analytic structure. In particular, 
they contain branch points at each energy that corresponds to 
a threshold for a newly allowed physical process~\cite{elop}.

This feature renders the description of bound states very 
complicated. On the other hand, recall that we are concerned 
with loosely bound states which generate poles in the complex 
energy plane. The poles are  located close to the elastic threshold 
and thus far away from any inelastic and crossed channel singularities. 
Further, the magnitudes of the characteristic momenta are much 
smaller than the particle masses, and the characteristic kinetic 
energies lie far below the inelastic thresholds generated by the 
presence of additional  massive particles in the final state. 
This suggests that a framework where  all singularities are 
treated on an equal footing is counterproductive and superfluous 
in this case: the physics at low energies can  equally well be 
described by a simpler effective non--relativistic field theory, 
in which the $S$-matrix elements possess only the elastic cut 
and the poles that are located close to this cut, whereas the 
contributions from distant singularities can be Taylor expanded 
in external momenta~\cite{Caswell}. 
This (truncated) Taylor series is generated  
by a finite number of operators in the non--relativistic Lagrangian,  
multiplied with unknown coupling constants. For properly chosen 
couplings, the non--relativistic and the relativistic theories are 
physically equivalent at low energy, up to a certain power in the 
momentum expansion. This procedure to fix the couplings in the 
non--relativistic Lagrangian is called {\it matching}.

In this non--relativistic theory, bound--state dynamics is described 
by the Schr\"{o}\-din\-ger equation, and the perturbation expansion 
of the bound--state energies is given by the Rayleigh--Schr\"{o}dinger 
series~\cite{Caswell}.  The crucial property of the non--relativistic 
theory which makes this calculation simple, is the fact that the 
production and annihilation of  massive particles can be forbidden 
by construction: the interaction Hamiltonian has vanishing matrix 
elements between states with different number of massive particles. 
 In the presence of photons, only matrix elements between 
 states that contain the same 
number of massive particles and an arbitrary number of photons
 can be non--vanishing. 

For the Coulomb bound state, momenta are of order $\alpha$. 
As a result of this, the contribution to the bound--state energy 
from  an operator in the interaction Lagrangian which contains 
$n$ space derivatives is suppressed by $\alpha^n$ as compared to 
the contribution from an operator with the same field content and 
without derivatives. Stated differently, 
one counts each space derivative  $\partial_i=O(v)=O(\alpha)$
(here $v$ stands for the velocity of a massive particle). 
If one wishes to evaluate the binding energy at a given order 
in $\alpha$, one may truncate the non--relativistic Lagrangian 
at a certain order in $v$ and carry out calculations with the 
truncated Lagrangian -- the error in the bound state occurs at 
a higher order in $\alpha$. The method works, provided that 
the power counting stays intact in higher--order calculations 
involving loops. This can be achieved by applying threshold  
expansions, in the scattering sector as well as in the  
Rayleigh--Schr\"{o}dinger perturbation series
(see, e.g.,
Refs.~\cite{Kinoshita,Labelle_QED,Pineda_ultrasoft,Pineda_QCD,Brambilla:1999xf,Brambilla:2004jw,Beneke,Pineda,Pineda_lamb,Antonelli,Bern4,Manohar}).
For a review on electromagnetic bound states
 see, e.g., Refs.~\cite{Czarnecki,Karshenboim:2005iy}.    

We illustrate in this and in the following two sections
the procedure in a simple theory: we consider a self--interacting
scalar field which is coupled to photons -- in the absence of the
self interaction, the Lagrangian reduces to the one of scalar 
electrodynamics. We determine the widths and energy levels of the 
bound states at next--to--leading order in $\alpha$, following closely 
the procedure outlined in Refs.~\cite{Antonelli,Gall}. As we shall see 
later, the evaluation of the energy levels and widths in hadronic 
atoms in the framework of ChPT amounts to a rather straightforward 
generalization of the techniques presented here.

\subsection{The Lagrangian and the reduction formulae}

We start from a situation where only one hard scale is present,
and where the external momenta involved are much smaller than this scale.
 This situation is described e.g. by  a theory that contains solely
massive charged scalar particles
$\phi^\pm$ of physical mass $M$, 
and where the external momenta are considered to be small
 in comparison to $M$.
In the relativistic case, we introduce a complex massive self--interacting 
scalar field  $\phi$.
 Photons will be included in the following section.
 The Lagrangian is given by
\eq\label{eq:L4}
{\cal L}=\partial_\mu\phi (\partial^\mu\phi)^\dagger 
- M_r^2 \phi\phi^\dagger+\frac{\lambda_r}{4!}\,(\phi\phi^\dagger)^2
+{\rm counterterms}\,, 
\en
where 
$\phi=\frac{1}{\sqrt{2}}\,(\phi_1+i\phi_2)$ is the charged 
scalar field, $M_r=M_r(\mu_0)$ and $\lambda_r=\lambda_r(\mu_0)$ 
stand for the renormalized mass and the quartic coupling, respectively.  
We use the modified minimal subtraction ($\overline{\rm MS}$) 
prescription throughout this paper.  
The scale of dimensional renormalization is denoted by $\mu_0$.

In the following, we restrict the discussion to the sector with  vanishing 
total charge and, in particular, consider 
the scattering of two oppositely charged particles 
$\phi^+\phi^-\to\phi^+\phi^-$. 
The pertinent Green function is given by  
\eq\label{eq:GR}
G(p_1,p_2;q_1,q_2)&\!=\!&\int d^4x_1 d^4x_2 d^4y_1 d^4y_2 
\,{\rm e}^{ip_1x_1+ip_2x_2-iq_1y_1-iq_2y_2} \nonumber\\[2mm]
&\!\times\!& i^{4} \,\langle 0|T\phi(x_1)\phi^\dagger(x_2)
\phi^\dagger(y_1)\phi(y_2)|0\rangle_c\, ,
\en 
where the subscript $c$ refers to the connected part. 

The scattering amplitude 
\eq\label{eq:TR}
\langle p_1,p_2\,\,\mbox{out}|q_1,q_2\,\,\mbox{in}\rangle&=&
\langle p_1,p_2\,\,\mbox{in}|q_1,q_2\,\,\mbox{in}\rangle\nnnl
&+&i(2\pi)^4\delta^4(p_1+p_2-q_1-q_2)
T({\bf p}_1, {\bf p}_2;{\bf q}_1,{\bf q}_2) 
\en
is obtained from the Green function in a standard manner, 
through amputating external legs and multiplying with
appropriate renormalization factors.

Let us now construct the non--relativistic theory, which reproduces  
this scattering amplitude  at small 3-momenta 
${\bf p}_i^2,{\bf q}_j^2\ll M^2$. 
We first introduce the non--relativistic field operators
$\Phi_\pm(x)$ ($\Phi^\dagger_\pm(x)$) 
that annihilate (create) a non--relativistic (positive/negative) 
charged particle from the vacuum.  
The non--relativistic Lagrangian ${\cal L}_{NR}(x)$ consists of an infinite
tower of operators which are constructed from the fields $\Phi_\pm(x)$, its
 conjugated fields $\Phi^\dagger_\pm(x)$ and 
space derivatives thereof. 
Time derivatives at order $\geq 2$ can be
 eliminated by using the equations of motion and field
redefinitions. The guiding principles for constructing
the Lagrangian are the following.

\begin{itemize}

\item[i)]
The effective theory must
be rotationally
invariant and obey $P,T$ discrete symmetries, under which
the non--relativistic field transforms as follows, 
\eq\label{eq:PT}
P\Phi_\pm(x^0,{\bf x})P^\dagger=\Phi_\pm(x^0,-{\bf x})\, ,\quad\quad
T\Phi_\pm(x^0,{\bf x})T^\dagger=\Phi_\pm(-x^0,{\bf x})\, ,
\en 
where $T$ is an anti-unitary operator.
Lorentz invariance is realized
in form of relations between different low--energy couplings. 

\item[ii)]
Since the number of the heavy particles is conserved,
all vertices in the non--relativistic Lagrangian must contain an 
equal number of $\Phi_\pm(x)$ and $\Phi^\dagger_\pm(x)$. 
Particle creation and annihilation, which occurs
in the relativistic theory, is implicitly included in the couplings
of the non--relativistic Lagrangian. In the case  considered in this section,
these couplings are real and the Lagrangian is a hermitian operator.

\item[iii)]
The $T$-invariance implies
\eq\label{eq:T-hermit}
T{\cal L}_{NR}(x^0,{\bf x})T^\dagger={\cal L}_{NR}(-x^0,{\bf x})\, .
\en

\item[iv)]
It is necessary to specify counting rules, 
which allow one to order different operators in the Lagrangian. 
We introduce a formal small parameter $v$ -- a velocity of the massive
particle in units of the speed of light -- and count each 3-momentum 
${\bf p}$ and  space derivatives ${\bf \nabla}$ as $O(v)$.
Further, the kinetic energy of the massive particles $p^0-M$ 
(or $i\partial_t-M$) is counted at $O(v^2)$. This rule amounts to 
an expansion in  inverse powers of the heavy mass $M$. All terms in the 
Lagrangian can  be ordered according to the counting in the parameter $v$. 
This ordering can be performed separately in each $n$-particle 
sector -- the sectors with different numbers of massive particles do not mix. 

\end{itemize}

According to the above rules, the non--relativistic 
effective Lagrangian in the
two--particle sector with $Q=0$ is given by 
\eq\label{eq:LNR}
{\cal L}_{NR}&=&\sum_{\pm} \Phi_\pm^\dagger
\biggl(i\partial_t-M+\frac{\triangle}{2M}
+\frac{\triangle^2}{8M^3}+\cdots\biggr)\Phi_\pm
+\frac{g_1}{M^2}\,(\Phi_+^\dagger\Phi_-^\dagger) (\Phi_+\Phi_-)   
\nonumber\\[2mm]
&+&\frac{g_2}{4M^4}\,\biggl\{(\Phi^\dagger_+ 
\stackrel{\leftrightarrow}{\triangle}\Phi^\dagger_-) 
(\Phi_+\Phi_-)+h.c.\biggr\}+\frac{g_3}{2M^4}\,(\Phi^\dagger_+\Phi_+)
\stackrel{\leftrightarrow}{\triangle}(\Phi^\dagger_-\Phi_-)
+ \cdots \, , \nonumber\\
& &
\en
where $u\stackrel{\leftrightarrow}{\triangle}v=u(\triangle v)+(\triangle u) v$
and $g_i$ denote non-relativistic couplings, 
which are functions of the parameters $\lambda_r(\mu_0)$, 
$M_r(\mu_0)$ and $\mu_0$. Ellipses stand for higher derivatives. 

The calculation in perturbation theory with the Lagrangian 
${\cal L}_{NR}$ proceeds as follows. One splits the Lagrangian 
into a free and an interacting part, 
${\cal L}_{NR}={\cal L}_{NR}^0+{\cal L}_{NR}^{\rm int}$ 
with ${\cal L}_{NR}^0= \sum_{\pm} 
\Phi_\pm^\dagger(i\partial_t-M+\triangle/2M)\Phi_\pm$. 
The free propagator is given by  
\eq\label{eq:prop} 
i\langle 0|T\Phi_\pm(x)\Phi^\dagger_\pm(0)|0\rangle
=\int\frac{d^4p}{(2\pi)^4}\,
\frac{{\rm e}^{-ipx}}{M+\frac{{\bf p}^2}{2M} - p^0-i0}\, .
\en
It vanishes for negative times, which shows that 
the field $\Phi_\pm(x)$ indeed annihilates the vacuum.
As expected, the free particle has
a non--relativistic dispersion law $p^0=M+{\bf p}^2/2M$. In order to arrive
at a relativistic dispersion law, one has to sum up all 
higher--order relativistic corrections proportional 
to $\triangle^2/8M^3\, \cdots$.  Taking into account that $\triangle$ is of
order $v^2$, and counting the propagator in momentum space as $O(v^{-2})$, 
the corresponding series  for the propagator is of 
 the form $a/v^2+b+cv^2+\cdots$ and thus indeed corresponds 
to an expansion which makes sense.

For later use, we note that  the one--particle 
states in the non--relativistic theory -- with the propagator 
given by Eq.~(\ref{eq:prop}) --
are normalized according to
\eq\label{eq:alpha2}
\langle{\bf k}|{\bf q}\rangle=(2\pi)^3\,\delta^3({\bf k}-{\bf q})
\qquad [\mbox{non--relativistic}]\fs
\en
Calculation of the scattering amplitude of two oppositely charged particles 
in the non--relativistic theory proceeds analogously to the relativistic 
framework: one evaluates the connected Green functions 
\eq\label{eq:G} 
\!\! G_{NR}(p_1,p_2 \!\!\!\! &;& \!\!\!\! 
q_1,q_2)\!=\!\int d^4x_1 d^4x_2 d^4y_1 d^4y_2 
\,{\rm e}^{ip_1x_1+ip_2x_2-iq_1y_1-iq_2y_2} \nonumber\\[2mm]
&\!\!\times\!\!& i^{4} \langle 0|T \Phi_+(x_1) \Phi_-(x_2) 
             \Phi^\dagger_+(y_1) \Phi^\dagger_-(y_2) |0\rangle_c\, . 
\en 
Here, the non--relativistic self--energy 
corrections in the external legs are summed up, so that 
the correct relativistic dispersion law $p^0=w({\bf p})=(M^2+{\bf p}^2)^{1/2}$
is recovered. Because light (massless) particles are absent here, 
there are no loop corrections to the self--energy, and the pertinent 
wave function renormalization factor $Z_{NR}=1$. 
Amputating external legs on the mass shell, we finally obtain
\eq\label{eq:residue_TNR}
\hspace*{-1.cm}&\!\!&i(2\pi)^4\delta^4(p_1+p_2-q_1-q_2)
T_{NR}({\bf p}_1, {\bf p}_2;{\bf q}_1,{\bf q}_2)
\nonumber\\[2mm]
\hspace*{-1.cm}&\!\!\!=\!\!\!&
\prod_i^2\lim_{p_i^0\to w({\bf p}_i)}(w({\bf p}_i)-p_i^0)
\prod_j^2\lim_{q_j^0\to w({\bf q}_j)}(w({\bf q}_j)-q_j^0)
G_{NR}(p_1, p_2;q_1,q_2)  .
\en

\subsection{Matching condition}

The relativistic and non--relativistic theories must be physically equivalent
at small momenta. This equivalence is enforced by the matching condition,
which enables one to express the non--relativistic couplings 
$g_i$ in Eq.~(\ref{eq:LNR}) through the parameters $\lambda_r$, $M_r$ 
and the scale $\mu_0$.

We first formulate the matching condition for the scattering 
amplitude $T_{NR}$ introduced in Eq.~(\ref{eq:residue_TNR})   
and start with the observation that 
the normalization of the asymptotic states 
is different in the relativistic and in the non--relativistic
theory: in the relativistic case, one has
\eq\label{eq:alpha2R}
\langle{\bf k}|{\bf q}\rangle=(2\pi)^3\,2w({\bf p})\,\delta^3({\bf k}-{\bf q})
\quad\quad\mbox{[relativistic]}\, .
\en
From this we  conclude that, in order that the cross sections 
in the relativistic and in the non--relativistic 
theories are the same, the scattering matrix elements   
must satisfy 
\eq\label{eq:matching}
\!\!\!\!\!\!&&\!\!\!\!\!\!T({\bf p}_1, {\bf p}_2;{\bf q}_1,{\bf q}_2)=
\prod_{i=1}^2(2w({\bf p}_i))^{1/2}(2w({\bf q}_i))^{1/2}
T_{NR}({\bf p}_1,{\bf p}_2;{\bf q}_1,{\bf q}_2)\, ,
\nonumber\\
&&
\en
i.e.,  the non--relativistic 
scattering amplitude acquires an extra factor $(2w({\bf k}))^{-1/2}$ for 
each external leg. 
The equality Eq.~(\ref{eq:matching}), 
which goes under the name of {\it matching condition}, 
is understood to hold at small 3-momenta, order by order
in the expansion in ${\bf p}_i/M$, ${\bf q}_i/M$. This corresponds
to an expansion in the formal parameter $v$ introduced in
the previous subsection. We will provide illustrations below.

The matching condition for transition amplitudes with more than 4 legs can be
formulated in a similar fashion. However, in order to do so explicitly, one
must introduce the concept of connected $S$-matrix elements. As
we will not need these higher order transition amplitudes in the present
work, we do not go into more details here.

Finally, let us mention 
that the matching condition does not always
determine all couplings $g_i$ separately. This happens, for instance,
when one decides to retain a non--minimal 
set of operators in the non--relativistic
Lagrangian. The matching condition, which
is formulated for the scattering amplitudes, then determines only 
certain combinations of the non--relativistic couplings. The point is that 
exactly the same combinations then enter all observable quantities that can be 
calculated from the non--relativistic Lagrangian. Consequently, 
the non--relativistic effective theory leads to a
consistent treatment of the low--energy physics also in this case. Explicit
examples and further details can be found in Ref.~\cite{Antonelli}, 
chapter IIIc.

\subsection{Tree level}

As an illustration to the matching procedure, consider the scattering 
process $q_1+q_2\to p_1+p_2$ in the sector with $Q=0$. 
The relativistic scattering 
amplitude at tree level is  $\lambda_r/6$.
Using the matching condition Eq.~(\ref{eq:matching}), and expanding the 
non--relativistic amplitude in the CM frame 
(${\bf p}_1=-{\bf p}_2 ={\bf p}$, ${\bf q}_1=-{\bf q}_2={\bf q}$, 
$|{\bf p}|=|{\bf q}|$)
in powers of $v$, we get 
\eq\label{eq:TNR4}
\frac{T({\bf p}_1,{\bf p}_2;{\bf q}_1,{\bf q}_2)}{2w({\bf p})2w({\bf q})}=
\frac{\lambda_r}{6} \, 
\frac{1}{2w({\bf p})2w({\bf q})}=\frac{\lambda_r}{24 M^2}
-\frac{\lambda_r}{24 M^4}\,{\bf p}^2+O\biggl(\frac{1}{M^6}\biggr)\, .
\en 
On the other hand, the non-relativistic amplitude, evaluated from the 
Lagrangian (\ref{eq:LNR}) is given by
\eq\label{eq:LTNR4}
T_{NR}({\bf p}_1,{\bf p}_2;{\bf q}_1,{\bf q}_2)=
\frac{g_1}{M^2}-\frac{g_2{\bf p}^2}{M^4}-\frac{g_3({\bf p}-{\bf q})^2}{M^4}
+O({\bf p}^4)\, .
\en 
From these relations, we can read off the couplings $g_1,g_2,g_3$ 
at tree level, 
\eq\label{eq:g}
g_1=\frac{\lambda_r}{24}+O(\lambda_r^2)\, ,\quad
g_2=\frac{\lambda_r}{24}+O(\lambda_r^2)\, ,\quad
g_3=O(\lambda_r^2)\, .
\en 
The generalization of the tree--level 
matching procedure to terms containing
more field operators, or higher derivatives, is straightforward. Here, we only
mention that the
many--particle scattering amplitudes contain singular parts that 
stem from particles traveling near their mass shell between 
two interactions. These singular parts 
are exactly reproduced at low energies by the
tree graphs in the non--relativistic theory. The remaining part of the
relativistic amplitudes is a polynomial in the small external 3-momenta, and
can be reproduced by adjusting the couplings in  ${\cal L}_{NR}$.

\subsection{Loops}

At higher order in the low--energy expansion, loops also contribute.
 In the non--relativistic theory, they 
 reproduce the non--analytic behavior of the relativistic amplitudes at low
energy. The   polynomial parts can be fixed with the help
of the matching condition. 

\begin{figure}[t]
\begin{center}
\includegraphics[width=11.cm]{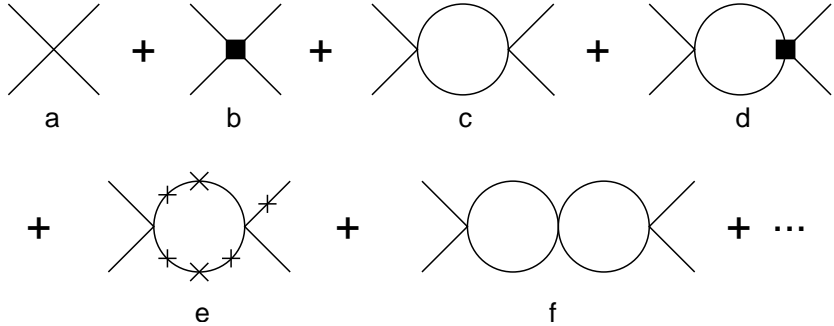}
\end{center}
\caption{Typical diagrams, contributing to the 2-particle elastic 
scattering amplitude in the non--relativistic theory. 
Filled boxes and crosses denote the derivative
vertices and the self--energy insertions, respectively.} 
\label{fig:bubbles}
\end{figure}

To illustrate the procedure, we again consider the two--particle amplitude.
 The structure of the perturbation series 
is particularly simple here, see Fig.~\ref{fig:bubbles}:
 at any order in the loop expansion, only
the chain of $S$-wave bubbles plus relativistic insertions 
in the internal and
external lines plus insertions of derivative 4-particle vertices contribute
to the scattering amplitude.  
The elementary building block to calculate a diagram with any number of
bubbles in the CM frame $P^\mu=p_1^\mu+p_2^\mu=(P^0,{\bf 0})$ is given by
\eq\label{eq:building_block}
J(P^0)&=&
\int\frac{d^Dl}{(2\pi)^Di}\,\frac{1}{M+\frac{{\bf l}^2}{2M}-P^0+l^0-i0}
\, \frac{1}{M+\frac{{\bf l}^2}{2M}-l^0-i0}
\nonumber\\[2mm]
&=&\frac{iM}{4\pi}\,(M(P^0-2M))^{1/2}\, ,\quad\quad 
{\rm at}~D\to 4\, ,~~P^0>2M\, .
\en 
The function $J(P^0)$ is analytic in the complex $P^0$ plane, cut along
the positive real axis for $P^0>2M$. The scattering amplitude 
is obtained by putting
$P^0=2w({\bf p})$, where ${\bf p}$ denotes the relative 3-momentum in the 
CM frame. Diagrams containing self--energy insertions, or derivative 
 couplings, are evaluated in a similar manner\footnote{Our
 theory is perturbative in dimensional regularization. 
We do not consider here the case when some of the non--relativistic 
couplings are ``unnaturally'' large, which would lead to the necessity
of partial re--summation of the perturbative series
(example: the ``pionless'' effective field theory in the two--nucleon 
sector~\cite{KSW1,KSW2}).}.
  
The perturbative expansion based on 
the Lagrangian Eq.~(\ref{eq:LNR}) generates a systematic low--energy expansion of
the scattering amplitudes. 
 The single loop $J(P^0)$ is proportional to the magnitude of the
small relative momentum $|{\bf p}|$.  Self--energy insertions and 
derivative vertices contribute additional powers of $|{\bf p}|$.
Hence, evaluating loop diagrams with vertices of higher dimension
yields contributions at higher order in the expansion in momenta. 
We conclude that, at any given order in $v^n$, only a finite number
of diagrams contributes.
For example, combining two non--derivative vertices through one loop
(Fig.~\ref{fig:bubbles}c) generates a contribution at  $O(v)$, and
 the contribution from the two--loop diagram  in Fig.~\ref{fig:bubbles}f 
is  $O(v^2)$, as is the derivative vertex 
displayed in Fig.~\ref{fig:bubbles}b.

From the above discussion one concludes that in the CM frame, the
non--relativistic scattering amplitude takes the form
\eq\label{eq:series}
 T_{NR}=M^{-2}\left(f_0+f_1\frac{|{\bf p}|}{M}+f_2\frac{|{\bf p}|^2}{M^2}
+f_3\frac{{\bf p}{\bf q}}{M^2}+\cdots\,\right) ,
\en
where the expansion coefficients $f_i$ are polynomials of finite order 
in the dimensionless couplings $g_i$.
 The matching condition Eq.~(\ref{eq:matching}) fixes these couplings
through $\lambda_r,M_r$ 
and the scale $\mu_0$ order by order in the perturbative expansion.

\begin{figure}[t]
\begin{center}
\includegraphics[width=11.cm]{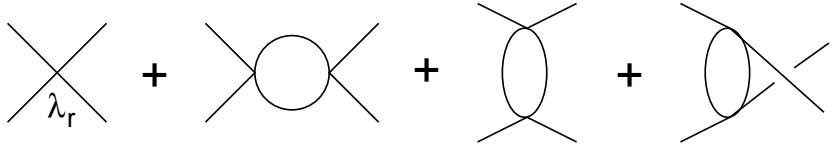}
\end{center}
\caption{One--loop scattering amplitude in the relativistic 
$(\phi^\dagger\phi)^2$ theory. Self-energy insertions are not displayed.}
\label{fig:lambda2}
\end{figure}

As one observes from Eq.~(\ref{eq:series}), the analytic structure of the 
amplitude is particularly simple. The only non--analytic piece of
the non--relativistic amplitude at low energy
is given by terms containing odd powers of $|{\bf p}|$, whereas the rest is
a polynomial in the external momenta. As mentioned above, 
only $s$-channel bubbles generate non-analytic contributions 
in the vicinity of the elastic threshold 
(graphs shown in Figs.~\ref{fig:bubbles}c,d,e,$\cdots$). 

Let us now verify that  the relativistic 
amplitude has the same non--analytic
structure at low energy as the non--relativistic one. 
 We consider the relativistic scattering amplitude at one loop
(see Fig.~\ref{fig:lambda2}), 
\eq\label{eq:lambda2}
T(s,t,u) = T_0 
+ \frac{\lambda_r^2}{72}\,(2\bar J(s)+2\bar J(t)+\bar J(u))+O(\lambda_r^3)\, ,
\en
where $T_0$ is a constant, 
$s=(p_1+p_2)^2,~t=(p_1-q_1)^2,~u=(p_1-q_2)^2$ 
are the usual Mandelstam variables, and where, for $x<0$,
\eq\label{Jbar}
\bar J(x)=
\frac{1}{16\pi^2}\,\biggl(\sigma_x\ln\frac{\sigma_x-1}{\sigma_x+1}
+2\biggr)\, ,
\quad
\sigma_x=\sqrt{1-\frac{4M^2}{x}}\,.
\en
The amplitude Eq.~(\ref{eq:lambda2}) has a more complicated analytic structure
than its non--relativistic counterpart Eq.~(\ref{eq:series}). For 
instance, its partial waves contain, in addition to a cut at $s>4M^2$,
 a left--hand cut at $s<0$. However, in the vicinity of the elastic
threshold, the contribution from these distant singularities can 
be approximated by a polynomial. Expanding the functions $\bar J(s)$,
$\bar J(t)$, $\bar J(u)$ near $s=4M^2$, $t=0$, $u=0$, we obtain
\eq\label{eq:Jbar_exp}
16\pi^2\bar J(s)&=&2+\frac{i\pi |{\bf p}|}{M}-\frac{2|{\bf p}|^2}{M^2}
+O(v^3)\, ,
\nonumber\\[2mm] 
16\pi^2\bar J(t)&=&-\frac{({\bf p}-{\bf q})^2}{6M^2}+O(v^4)
\, ,
\nonumber\\[2mm] 
16\pi^2\bar J(u)&=&-\frac{({\bf p}+{\bf q})^2}{6M^2}+O(v^4)
\, .
\en
Substituting this expansion into Eq.~(\ref{eq:lambda2}), one immediately
sees that the low--energy 
behavior of the non--relativistic amplitude Eq.~(\ref{eq:series}) is reproduced.
One may further check that
the non--analytic part $\propto \lambda_r^2|{\bf p}|$ 
in the expression for the amplitude, which is generated by 
 $\bar J(s)$, is automatically reproduced in the non--relativistic 
calculation with the correct coefficient, 
provided the matching at $O(\lambda_r)$ has been performed.

We end this subsection with a brief discussion of crossing and 
charge symmetry. In the relativistic theory, the scattering amplitude
obeys crossing symmetry. The crossed channels, along with 
$\phi^+\phi^-\to\phi^+\phi^-$, include also the reactions
$\phi^+\phi^+\to\phi^+\phi^+$ and $\phi^-\phi^-\to\phi^-\phi^-$.
 These reactions are described by a single analytic amplitude, 
derived from the relativistic Lagrangian Eq.~(\ref{eq:L4}).

In the non--relativistic case, 
crossing symmetry is apparently lost because, first of all,
 the non--re\-la\-ti\-vis\-tic 
expansion is performed in the vicinity of $s=4M^2$, $t=u=0$.
In difference to the relativistic case,
 an analytic continuation of the amplitude
from the vicinity of the $s$-channel threshold to the $t$- or $u$-channel 
thresholds
cannot be performed, because the distance between these 
two regions exceeds the radius of the convergence of the pertinent 
Taylor series
(see, e.g. Eq. (\ref{eq:Jbar_exp})). Furthermore, our non--relativistic 
Lagrangian describes only the sector with total 
charge zero, so that the scattering amplitudes for
 $\phi^+\phi^+\to\phi^+\phi^+$ and $\phi^-\phi^-\to\phi^-\phi^-$ vanish in
this theory (within the non--relativistic approach, this choice is a
consistent procedure). 
We therefore conclude that  crossing symmetry is not present
in the non--relativistic approach {\it ab initio}. At most, one may describe
the scattering amplitudes in all crossed channels, including all pertinent
terms in the non--relativistic  Lagrangian. The couplings, which are determined
from  matching to the relativistic theory,
 will then obey the restrictions imposed by  
crossing symmetry.
The same conclusion holds for charge invariance, which in general connects
 sectors of  different total charge.

\subsection{Relation to the effective range expansion}
\label{subsec:eff_range}

There is one property of the non--relativistic theories that makes the use
of this framework 
extremely effective in many areas of hadron physics. 
The property is related
to the expansion parameters in the non--relativistic 
perturbation series.

Following the line of reasoning outlined in the introduction, we consider
a generic hadronic scattering process at momenta which are 
much smaller
than any of hadron masses involved and any of the dynamically generated scales.
It is clear that any consistent EFT at such momenta should operate only
with observable characteristics ($S$-matrix parameters)
of this process (since the distances involved are already asymptotic). 
In other words, for e.g. the elastic two--particle scattering 
process, the expansion
 parameters in the non--relativistic EFT should be expressible through the 
parameters of the effective range expansion (scattering length, effective
range, shape parameters), rather than directly through the coupling 
constant $\lambda_r$, the 
running mass $M_r$, etc. It can be immediately seen that the 
non--relativistic EFT, which we have constructed here, passes this test:
at a given order $v^n$ only a finite number of the non--relativistic loops
contribute, and the coefficients $f_0,f_1,\cdots$ in Eq.~(\ref{eq:series})
can be  expressed through a finite number of 
effective--range parameters of the relativistic theory. For example, the 
coefficient of the diagram shown in Fig.~\ref{fig:bubbles}a is proportional
to the scattering length $a$ in the relativistic theory to all orders in 
$\lambda_r$ and not only at $O(\lambda_r)$, 
since all other diagrams Fig.~\ref{fig:bubbles} give vanishing 
contributions at threshold. Hence, the contributions Fig.~\ref{fig:bubbles}c
and Fig.~\ref{fig:bubbles}f are proportional to $a^2|{\bf p}|$ and 
$a^3{\bf p}^2$, respectively, so that the expansion in the scattering length 
$a$ and in the small momenta is correlated. Note that, 
in order to achieve
this convenient ordering of the various
 contributions, the regularization scheme -- which is used
to calculate loops -- should not contain a mass scale that destroys 
power--counting. We use dimensional regularization,
 because it has this property.

%%%%%%%%%%%%%%%%%%%%%%%%%%%%%%%%%%%%%

\setcounter{equation}{0}
\section{Non--relativistic effective theories: including photons}
\label{sec:including_gamma}

\subsection{The Lagrangian}

Photons are included in the theory through minimal coupling,

\eq\label{eq:L4A}
{\cal L}=D_\mu\phi(D^\mu\phi)^\dagger-
M_r^2\phi\phi^\dagger+\frac{\lambda_r}{4!}\,(\phi\phi^\dagger)^2
-\frac{1}{4}\,F_{\mu\nu}F^{\mu\nu}+{\rm counterterms}\, ,
\en
where $\phi$ is a charged scalar field as before,
$A_\mu$ stands for the photon field,
$D_\mu\phi=\partial_\mu\phi+ieA_\mu\phi$,
$F_{\mu\nu}=\partial_\mu A_\nu-\partial_\nu A_\mu$, and the gauge fixing 
term is not explicitly shown.
 The theory described by the Lagrangian
 Eq.~(\ref{eq:L4A}) contains two coupling constants $\lambda_r$ and $e$
and thus mimics the hadronic atom problem where an interplay of
electromagnetic and strong effects is present.
For convenience, we refer to the self-interaction 
of the scalar field as ``strong interactions''. 

The non--relativistic effective Lagrangian that describes this theory
 at low energy is constructed
along the same pattern as before. 
In the presence of photons, a few additional considerations 
should be taken into account.
First, one starts from a gauge--invariant  non--relativistic 
Lagrangian.
 Appropriate building blocks are covariant derivatives
of the non--relativistic fields, and the electric and magnetic
 fields ${\bf E}$ and ${\bf B}$. 
Second, the couplings are expressed in units of the {\it physical} mass $M$.
 The infinitely many terms in the
Lagrangian are then generated by expanding it in 
inverse powers of the mass. This amounts to an ordering according to the
number of space derivatives, and according to the number of electric and 
magnetic fields. Third,
 one may finally use different gauge fixing in 
the non--relativistic and in the 
relativistic theories. The Coulomb gauge is a convenient choice for the
non--relativistic theory.

The lowest--order terms in the non--relativistic theory are 
\eq\label{eq:LNRA}
{\cal L}_{NR}&=&-\frac{1}{4}\, F_{\mu\nu}F^{\mu\nu}
+\sum_{\pm} \Phi_\pm^\dagger\biggl(iD_t-M+\frac{{\bf D}^2}{2M} 
+\frac{{\bf D}^4}{8M^3}+\cdots
\nonumber\\[2mm]
&\mp& eh_1\frac{{\bf D}{\bf E}-{\bf E}{\bf D}}{6M^2}
+\cdots\biggr)\Phi_\pm
+\frac{g_1}{M^2}(\Phi_+^\dagger\Phi_-^\dagger)
(\Phi_+\Phi_-)+\cdots,
\en
where $\Phi_\pm$ denotes the non--relativistic field operator for 
charged particles, 
$D_t\Phi_\pm=\partial_t\Phi_\pm \mp ieA_0\Phi_\pm$,
${\bf D}\Phi_\pm=\nabla\Phi_\pm \pm ie{\bf A}\Phi_\pm$
are the covariant derivatives,
${\bf E}=-\nabla A_0-\dot{\bf A}$,
${\bf B}={\rm rot}\, {\bf A}$ denote the electric and magnetic fields, 
and $h_1,g_1$ are 
non--relativistic effective couplings\footnote{Unless stated otherwise, 
we use here and in the following the
  same symbols 
for the coupling constants, for the
  running and physical masses and for the running scale as in the previous 
section \ref{sec:nonrel}, where $e=0$ (both in the relativistic and 
non--relativistic theories). This avoids an
  unnecessary flooding of the text with symbols. Further, to the order
  considered in the following, we do not need to distinguish between the 
 bare and the
  renormalized charge, so we keep the symbol $e$ throughout. See also the
  discussion in  subsection \ref{subsec:vacuum} on vacuum polarization.}. 
For instance,  $h_1$ is 
 related to the electromagnetic charge radius of the particle, see below. 
The ellipses stand for higher--order derivative terms, as well as 
for non--minimal terms containing ${\bf E}$ and ${\bf B}$.

We add several comments.
 First,  the set 
of operators in the Lagrangian Eq.~(\ref{eq:LNRA}) is not minimal:
the term with $h_1$ could be eliminated in 
favor of  four--particle local terms 
 by using the equations of motion for
the Coulomb photon $A_0$. As a result of this mani\-pu\-lation, e.g. 
the operator
proportional to $g_1$ would receive a contribution. This means that the same
linear combination of  $h_1$ and $g_1$ enter
the expressions for the
two--particle scattering amplitude and  for the bound--state energies -- 
one does not need to know $h_1$ and $g_1$ separately.
We prefer to work with this
non--minimal set, because it renders the presentation more transparent 
and simplifies the comparison with  results available in the literature.
Second, calculating loops with ${\cal L}_{NR}$ generates 
ultraviolet (UV) and infrared (IR)
divergences.  Therefore, to obtain UV finite Green functions and $S$-matrix
elements  at $d=3$, counterterms should be introduced.
On the other hand, because the final NR amplitudes will be 
expressed in terms of the UV finite relativistic expressions, such a
renormalization is not needed: one simply performs all calculations at $d\neq
3$ whenever needed, then does the matching at $d>3$ in order to avoid infrared
divergences that show up in on--shell amplitudes, and sets $d=3$ at the end.
 This simplifies the calculation considerably, while the final result is 
the same. For a detailed discussion of this point, and for explicit examples,
 see, e.g., Refs.~\cite{Antonelli,Bern4}.

The LECs in the  the effective Lagrangian ${\cal L}_{NR}$ depend on the 
fine--structure constant. Indeed, they can be considered
as functions of $\lambda_r,e$, of the mass $M_r$ and of the scale $\mu_0$,
\eq\label{eq:1+alpha}
h_1=\bar h_1+\alpha h_1'+O(\alpha^2)\, ,\quad\quad
g_1=\bar g_1+\alpha g_1'+O(\alpha^2)\, ,
\en
such that
\bea
\frac{dh_1}{d\mu_0}=
\frac{dg_1}{d\mu_0}=0\,\,.
\eea
The barred quantities refer to the limit $\alpha=0$. 

Further note that, in general, the couplings of the non--relativistic
 Lagrangian are complex, because they contain contributions from 
 processes of particle creation and annihilation in the relativistic theory. 
The imaginary part arises,
if the threshold for such a process lies below the threshold for two
massive particles. In the case which is considered here, the imaginary
part of e.g. the coupling $g_1$ will be determined by the 
 $\phi^+\phi^-$-annihilation into  intermediate states with two or more
photons -- consequently, ${\rm Im}\,g_1=O(\alpha^2)$
(a generalized unitarity condition which holds in the presence of complex
couplings is briefly discussed in appendix~\ref{app:generalized}).
For this reason, the general 
non--relativistic Lagrangian is not a hermitian operator, 
and $T$-invariance implies (cf. with Eq.~(\ref{eq:T-hermit}))
\eq\label{eq:T}
T{\cal L}_{NR}^\dagger(x^0,{\bf x})T^\dagger={\cal L}_{NR}(-x^0,{\bf x})\, .
\en

\subsection{Perturbation theory}

\begin{figure}[t]
\begin{center}
\includegraphics[width=14cm]{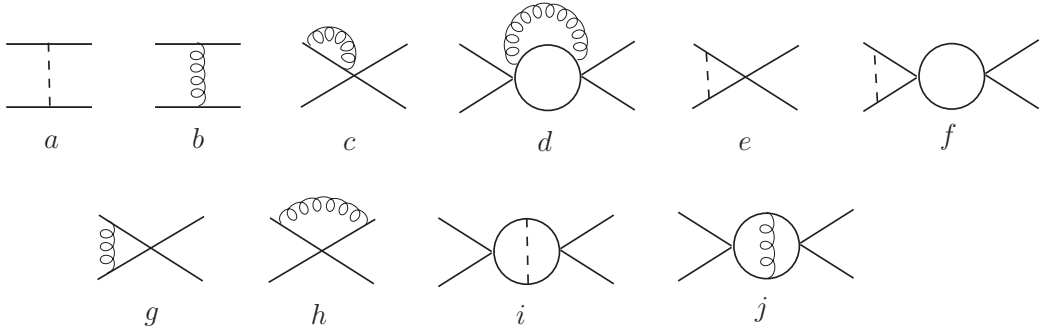}
\end{center}
\caption{One--photon corrections to the non--relativistic scattering amplitude
(representative set of diagrams).
Dashed and wiggly lines correspond to Coulomb and to transverse photons,
respectively. Diagrams containing non--minimal photon vertices, relativistic 
insertions and derivative couplings  are not shown.}
\label{fig:photons}
\end{figure}

To generate the perturbative expansion, one again splits the Lagrangian 
into a free and an
interacting part, 
${\cal L}_{NR}={\cal L}_{NR}^0+{\cal L}_{NR}^{\rm int}$ 
with ${\cal L}_{NR}^0=
-\frac{1}{4}\,F_{\mu\nu}F^{\mu\nu}+\sum_\pm
\Phi_\pm^\dagger(i\partial_t-M+\triangle/2M)\Phi_\pm$.
 In addition to the strong bubbles, 
relativistic insertions and  derivative vertices already 
discussed in the previous section, diagrams may now contain  photon lines.
The calculations in the Coulomb gauge are done as follows.
First, the non--dynamical field $A_0$ is removed from the Lagrangian
by using the EOM. This procedure generates the non--local operator
\eq
\triangle^{-1}=
-(2\pi)^{-3}\int d^3{\bf k}\,{\rm e}^{-i{\bf k}({\bf x}-{\bf y})}/{\bf k}^2\, 
\en
in the Lagrangian. In the following, for ease of understanding, 
we keep calling the pertinent diagrams as ``generated by
 the exchange of Coulomb photons''.
The propagator of  transverse photons in the Co\-u\-lomb gauge is given by
\eq\label{eq:Coulomb_gauge}
D^{ij}(k)&\!=\!&i\int dx\,{\rm e}^{ikx}\langle 0|TA^i(x)A^j(0)|0\rangle
=-\frac{1}{k^2}\,\biggl(\delta^{ij}-\frac{k^ik^j}{{\bf k}^2}\biggr)\, .
\en 
In the following, we restrict ourselves to the first nontrivial order in
 $\alpha$, and consider 2-particle
scattering in different channels. 
In Fig.~\ref{fig:photons},
 several examples of  diagrams with virtual photons are displayed.
These are: one--photon exchange in elastic channels
(Figs.~\ref{fig:photons}a,b), self--energy corrections in the external 
and internal
legs (Figs.~\ref{fig:photons}c,d), vertex corrections to the external legs
(Figs.~\ref{fig:photons}e,f,g,h), internal corrections 
due to the virtual photon exchange (Figs.~\ref{fig:photons}i,j). 
At higher orders in the momentum 
expansion, diagrams containing non--minimal photon couplings
occur, generated e.g. by  $h_1$. These are  not shown in 
Fig.~\ref{fig:photons}.

\subsection{Coulomb photons}

Here, we investigate diagrams containing Coulomb photons  
at order $\alpha$: one--photon exchange 
(Fig.~\ref{fig:photons}a), vertex correction, where the photon is attached
before or after all strong vertices
(Figs.~\ref{fig:photons}e,f), and an internal exchange between
two strong vertices (Fig.~\ref{fig:photons}i). 

The contribution from the graph Fig.~\ref{fig:photons}a to the 
scattering amplitude of oppositely charged particles is given by
\eq\label{eq:1gamma_C}
T_{NR}^{\ref{fig:photons}a}({\bf p},{\bf q})
=\frac{4\pi\alpha}{|{\bf p}-{\bf q}|^2}\, .
\en
This contribution is non--analytic at ${\bf p},{\bf q}\to 0$,
 and thus very different from the one encountered in 
Eq.~(\ref{eq:series}).

Next, we consider the vertex correction Fig.~\ref{fig:photons}e. 
Let $V_c({\bf p},P^0)$ denote the part of the
diagram which stands on the left or on the right of the first 
strong interaction vertex.
After integration over the zeroth component of the loop momentum, the
  integral to be evaluated is
\eq\label{eq:Vc_I}
V_c({\bf p},P^0)=e^2\int\frac{d^d{\bf l}}{(2\pi)^d}\frac{1}{{\bf l}^2}
\frac{1}{P^0-2M -\frac{({\bf p}-{\bf l})^2}{M}}\, .
\en
The contribution to the scattering amplitudes is obtained by
evaluating this expression at $P^0=2w({\bf p})$. The result 
(singular at $|{\bf p}|\to 0$) is
\eq\label{eq:Vc}
V_c({\bf p}, 2w({\bf p}))=-\frac{\pi\alpha M}{4 |{\bf p}|} 
- i\alpha\theta_c +O(|{\bf p}|,d-3)\, ,
\en
where
\eq\label{eq:tc}
\theta_c=\frac{M}{2|{\bf p}|}\mu^{d-3}\left\{\frac{1}{d-3}
-\frac{1}{2}\,[\Gamma'(1)+\ln 4\pi]
+\ln{\frac{2|{\bf p}|}{\mu}}\right\}
\en
denotes the (infrared--divergent) Coulomb phase\footnote{This phase is identical to the one
 in the relativistic
theory~\cite{Yennie} at this order.},
and $\mu$ denotes the scale of
dimensional regularization in the non--relativistic theory (not equal 
to $\mu_0$ in general).

Finally, we consider  the two--loop diagram Fig.~\ref{fig:photons}i.
Let $B_c(P^0)$ denote the part of the amplitude which corresponds to the
diagram with one photon exchange between two strong interaction vertices.  
Integrating over the zeroth components of
the loop momenta, it is given by
\eq\label{eq:Bc_I}
B_c(P^0)=\frac{e^2}{(2\pi)^{2d}}\int\frac{d^d{\bf l}_1}{P^0-2M 
-\frac{{\bf l}_1^2}{M}}\, 
\frac{1}{|{\bf l}_1-{\bf l}_2|^2}\frac{d^d{\bf l}_2}
{P^0-2M-\frac{{\bf l}_2^2}{M}}\, .
\en
Evaluating this expression at $P^0=2w({\bf p})$, we find a result which is
again singular at $|\bf{p}|\to 0$,
\eq\label{def_Lambda}
B_c(2w({\bf p}))&=&-\frac{\alpha M^2}{8\pi}
\left\{\Lambda(\mu) +2\ln{\frac{2|{\bf p}|}{\mu}}-1-i\pi\right\} 
+O(|{\bf p}|,d-3)\, ,
\nonumber\\[2mm]
\Lambda(\mu)&=&\mu^{2(d-3)}\left\{\frac{1}{d-3}-\ln{4\pi}
-\Gamma'(1)\right\}\, .
\en
Here, $\Lambda(\mu)$ contains an UV singularity.
 For the consistency of the method it is
important to note that the contributions from diagrams obtained 
by adding mass insertions and/or using vertices with derivative
couplings (not shown explicitly in Fig.~\ref{fig:photons}) 
are  suppressed by powers of  momenta with respect
to the leading terms $B_c$ and $V_c$. They are not
needed in the following.

\subsection{Transverse photons and the threshold expansion}
The exchange of one transverse photon (Fig.~\ref{fig:photons}b) 
contributes with
\eq\label{eq:1gamma_T}
T_{NR}^{\ref{fig:photons}b}({\bf p},{\bf q})=\frac{4\pi\alpha}
{|{\bf p}-{\bf q}|^2}\,
\frac{({\bf p}+{\bf q})^2}{4M^2}\, .
\en
This term is suppressed by two powers of
 momenta as compared to the Coulomb contribution Eq.~(\ref{eq:1gamma_C}). 
 This is due to the fact that the transverse 
photon vertex in the Lagrangian Eq.~(\ref{eq:LNRA}) contains a derivative, 
whereas the corresponding vertex for the Coulomb photon is point--like.
In other words, the ordering in the Lagrangian is preserved in the
contribution to the amplitude in this case: a term which is suppressed in the 
Lagrangian by powers of the  mass $M$ generates
 a term in the amplitude which is suppressed by powers of momenta.
This feature is called {\it power counting} in the following.

For the non--relativistic theory  to make sense, 
power--counting should also persist in loops. This is not, however, the case
if one uses  standard 
dimensional regularization in the loop calculations. 
To demonstrate this fact, we consider in detail the 
virtual photon contribution to the self--energy 
of the massive particle at one loop.
\begin{figure}[t]
\begin{center}
\includegraphics[width=8.cm]{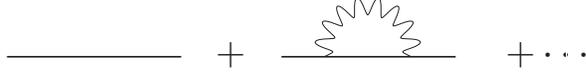}
\end{center}
\caption{NR propagator in the presence of transverse photons.}
\label{fig:photonloop}
\end{figure}
Summing up one--loop diagrams with a transverse photon to all orders 
(see Fig.~\ref{fig:photonloop}), one obtains
\eq\label{eq:self1}
D(p)=\frac{1}{\Omega}+\frac{1}{\Omega}\,\Sigma_\gamma(\Omega,{\bf p})\,
\frac{1}{\Omega}+\cdots=
\frac{1}{\Omega-\Sigma_\gamma(\Omega,{\bf p})}\, ,
\en
where $\Omega=M+\frac{{\bf p}^2}{2M}-p^0=O(v^2)$ and $\Sigma_\gamma(\Omega,{\bf p})$
denotes the one--loop diagram
\eq\label{eq:self--energy}
\Sigma_\gamma(\Omega,{\bf p})=\frac{e^2}{M^2}\int\frac{d^Dl}{(2\pi)^Di}\,\,
\frac{{\bf p}^2-({\bf p}{\bf l})^2/{\bf l}^2}
{-l^2(M+\frac{({\bf p}-{\bf l})^2}{2M}-p^0+l^0)}\, .
\en
 [Coulomb photons do not contribute to the self energy,
because the Coulomb photon propagator does not depend on the 
zeroth component $l^0$. Therefore,  one may close the
contour of integration over $l^0$  in that half--plane where the
propagator of the massive particle has no singularity.]

The self--energy is obviously of order ${\bf p}^2$,
\eq
\Sigma_\gamma(\Omega,{\bf p})=\frac{2e^2{\bf p}^2}{M}\,
\tilde\Sigma_\gamma(\Omega,{\bf p})\, .
\en
We combine the denominators by use of the Feynman--Schwinger trick
 and obtain at ${\bf p}=0$
\eq\label{eq:tilde_Sigma}
\tilde\Sigma_\gamma(\Omega,0)=-\frac{1}{12\pi^2}\ln\frac{\Omega}{2M}\,
\biggl(1-\frac{1}{\sqrt{1-\frac{2\Omega}{M}}}\biggr)
+\sum_{n=0}^\infty a_n\biggl(\frac{\Omega}{M}\biggr)^n\,\,, a_0\,\,\neq 0.
\en
The result consists of two parts: a non--analytic piece,
 which starts with a term
proportional to 
$\Omega\ln{\Omega}$, and a polynomial part, which starts with a constant.
The non--analytic part contributes all in all at order $v^4$ to the self--energy
and therefore respects power counting: its contribution is suppressed 
by $v^2$ with respect to the tree contribution $\Omega$ in the denominator 
of the last term in  Eq.~\ref{eq:self1}.
On the other hand, the polynomial part breaks power counting: it generates a
term which is of the same order as the tree level one.
Considering derivative vertices in Fig.~\ref{fig:photonloop},
 it is seen 
that the corresponding diagrams contribute at the same 
order in $v$, so that the low--energy effective theory 
is inconsistent in this sense.

It is easy to identify the reason for this failure.
First,  we note that the integration momenta 
$l^0,{\bf l}$  can vary in wildly different regions. 
For example, one or 
both of them can be of order of the hard scale $M$, be ``soft'' ($\sim v$)
or ``ultrasoft'' ($\sim v^2$). Consequently, this integral is given by
a sum of different terms, where each one corresponds to an integration 
in a different regime~(see, e.g., Ref.~\cite{Beneke}. For a detailed investigation
 of this statement for a specific one--loop diagram 
in the non--relativistic effective theory,
see also Ref.~\cite{Zemp}). Therefore, 
in order to construct a consistent low--energy theory,
one should modify the Feynman rules to get rid of the high--energy contribution.
In a most general way, this can be done by using the 
so--called ``threshold expansion''~\cite{Beneke,Pineda,Pineda_lamb,Pineda_ultrasoft,Pineda_QCD,Antonelli,Bern4,Manohar}\footnote{In 
case of $\pi N$ scattering at one loop in ChPT, the threshold expansion is
obviously 
equivalent to the infrared regularization introduced in Ref.~\cite{BLI}.}. 
A simple 
 algorithm for the calculation of Feynman diagrams that occur in 
the present work can be formulated as follows:
\begin{itemize}
\item[i)]
Perform contour integrals over the zeroth components of the momenta.
\item[ii)]
Assume all 3-momenta (external as well as the integration momenta) to be
much smaller than the hard scale $M$; expand the integrands in the small 
3-momenta.
\item[iii)]
Interchange the order of  integration and  expansion, then perform 
integrations in dimensional regularization.
\end{itemize}
After applying the threshold expansion, the integrand becomes 
a homogeneous function in small kinematic variables, 
and the counting rules are restored.
On the other hand, the high--energy contribution, which is polynomial at 
low energy, and which upsets power counting, is removed from the result.

Let us see how this procedure works in the present case. 
The contour integration in 
Eq.~(\ref{eq:self--energy}) gives
\eq\label{eq:SE_cont}
\Sigma_\gamma(\Omega,{\bf p})=\frac{e^2}{2M^2}\int\frac{d^dl}{(2\pi)^d}\,\,
\biggl({\bf p}^2-\frac{({\bf p}{\bf l})^2}{{\bf l}^2}\biggr)\,
\frac{1}{|{\bf l}|}\,\,
\frac{1}{\Omega+|{\bf l}|-\frac{{\bf p}{\bf l}}{M}
+\frac{{\bf l}^2}{2M}}\, .
\en
It is seen that ${\bf l}$ should scale like $\Omega\sim v^2$, otherwise
the expansion of the integrand leads to  no--scale integrals that vanish 
in  dimensional regularization. With such scaling assumed, 
the individual terms in the denominator of the last term in 
 Eq.~(\ref{eq:self--energy}) scale like $v^2,v^2,v^3,v^4$.
Therefore, this denominator is not a homogeneous function
in $v$, leading to a breakdown of the counting rules.
Applying the threshold expansion, we obtain (we put a hat
on threshold--expanded quantities)  
\eq\label{eq:hat-SE}
\hat\Sigma_\gamma(\Omega,{\bf p})&=&
\frac{e^2}{2M^2}\int\frac{d^d{\bf l}}{(2\pi)^d}\, 
\biggl({\bf p}^2-\frac{({\bf p}{\bf l})^2}{{\bf l}^2}\biggr)
\frac{1}{|{\bf l}|}
\biggl\{\frac{1}{\Omega+|{\bf l}|}
\nonumber\\[2mm]
&+&\biggl(\frac{{\bf p}{\bf l}}{M}
-\frac{{\bf l}^2}{2M}\biggr)\, \frac{1}{(\Omega+|{\bf l}|)^2}
+\cdots\biggr\}\fs
\en
Performing the remaining integration, we obtain
\eq\label{eq:hat-SE-1}
\hat\Sigma_\gamma(\Omega,{\bf p})&=&\frac{e^2}{2M^2}\,{\bf p}^2\,
\Omega^{d-2}\,\frac{\Gamma(d)\Gamma(2-d)}
{(4\pi)^{d/2}\Gamma(1+\frac{d}{2})}+O(M^{-3})
\nonumber\\[2mm]
&=&
\frac{e^2}{6\pi^2M^2}\,{\bf p}^2\,\Omega\,\,
\biggl\{\, L(\mu)+\ln\frac{2\Omega}{\mu}-\frac{1}{3}\biggr\}
+O\left(M^{-3},d-3\right)\, ,
\\[2mm]
\label{eq:Lmu}
L(\mu)&=&
\mu^{d-3}\biggl(\frac{1}{d-3}-\frac{1}{2}\,(\Gamma'(1)+\ln 4\pi+1)\biggr)\, .
\en
As expected, the non--analytic piece $\sim\Omega\ln\Omega$ stays 
unaffected. On the other hand,
 the part of the polynomial piece which  scales like $\sim v^2$
has disappeared from the result.
Since the threshold--expanded self--energy is proportional to $\Omega$,
the position of the particle pole is not affected by  radiative corrections.
It is furthermore seen that in
the vicinity of the mass shell, the propagator Eq.~(\ref{eq:self1}) is unaffected,
\eq\label{eq:Z-charged}
D(p)&\to&\frac{1}{\Omega}\, ,\quad \Omega\to 0\quad
(d>3)\, ,
\eea
as a result of which the external lines are not renormalized (wave function
renormalization constant $Z=1$.). Further, it is 
seen from Eq.~(\ref{eq:hat-SE-1}) that
the introduction of a counterterm rendering
the two--point function finite at $d=3$ would add a
contribution of order ${\bf p}^2$ to $Z$, which diverges as $d\to 3$.

We  investigate  a second example,
 and consider a diagram that describes the exchange of a
transverse photon between the initial pair of  charged particles
with  momenta $p_\pm^\mu=(p^0,\pm{\bf p})$,
 see Fig.~\ref{fig:photons}g. The contribution of this diagram is 
proportional to the integral
\eq\label{eq:Sakharov-transverse}
J_{+-\gamma}(|{\bf p}|)&=&-\frac{e^2}{M^2}\,\,
\int\frac{d^Dl}{(2\pi)^Di}\,\,\biggl({\bf p}^2
-\frac{({\bf p}{\bf l})^2}{{\bf l}^2}\biggr)\,
\frac{1}{(M+\frac{({\bf p}-{\bf l})^2}{2M}-p^0+l^0)} 
\nonumber\\[2mm]
&\times&
\frac{1}{l^2(M+\frac{({\bf p}-{\bf l})^2}{2M}-p^0-l^0)}\, .
\en
We put the external particles on the mass shell, 
$p^0=M+{\bf p}^2/(2M)+O({\bf p}^4)$, and perform the
threshold expansion in the integral. Note that with this procedure, the
integrands  should be expanded in the
$O({\bf p}^4)$ remainder of $p^0$.
The threshold--expanded integral 
can be rewritten in the following manner, 
\eq\label{eq:mpole-vertex}
\hat J_{+-\gamma}(|{\bf p}|)&=&\frac{e^2}{M}
\int\frac{d^d{\bf l}}{(2\pi)^d}\,
\frac{1}{{\bf l}^2}\,\biggl({\bf p}^2
-\frac{({\bf p}{\bf l})^2}{{\bf l}^2}\biggr)\,
\frac{1}{{\bf l}^2-2{\bf p}{\bf l}}+\cdots
\nonumber\\[2mm]
&=&\frac{e^2\,|{\bf p}|}{16M}
+\frac{ie^2\, |{\bf p}|}{8\pi M}\,\biggl( L(\mu)
+\ln\frac{2|{\bf p}|}{\mu}\biggr)+\cdots\, .
\en
The divergence at $d=3$ is an infrared one.
We see that  power-counting is at work also here: in comparison to the 
exchange of a Coulomb photon, this contribution is suppressed by ${\bf p}^2$. 
 This particular contribution to the scattering amplitude vanishes 
at threshold as well.

\subsection{Matching}
\label{subsec:matching}

The matching condition given in Eq.~(\ref{eq:matching}) is universal and
holds in the presence of photons as well (as photons are relativistic 
particles, the corresponding states
 have the same normalization as
in the relativistic theory). However, as we have explicitly seen,
the singularity structure of the amplitudes near threshold is different from
one given in Eq.~(\ref{eq:series}). Further, in order to determine all 
couplings in the non--relativistic Lagrangian
 Eq.~(\ref{eq:LNRA}) separately (e.g. the coupling constant $h_1$), it is convenient
to consider amplitudes with external photon legs as well.

We start with the coupling $h_1$  which 
is related to the charge radius of the scalar particle.
We closely follow the method described in Ref.~\cite{Kinoshita}.
We consider the transition amplitude ${\mathbold S}_{fi}$ of the 
charged particle
in  an external field $A^\mathrm{ext}_\mu$ 
and define the relativistic form factor 
$F(t)$ through the linear term in the
expansion in $A^\mathrm{ext}_\mu$,
\eq
\hspace*{-0.55cm}&&{\mathbold S}_{fi}=2w({\bf p})\,(2\pi)^3\delta^3({\bf p}-{\bf q})
+ie\tilde A^\mathrm{ext}_\mu(p-q)(p+q)^\mu F(t)+O(e^2)
\quad\mbox{[relativistic]}\, ,
\nonumber\\
\hspace*{-0.55cm}&&
\en
where $\tilde A^\mathrm{ext}_\mu(p-q)$ denotes the Fourier--transform of the
external field, and where $t=(p-q)^2$. At a small $t$, 
\eq
F(t)=1+\frac{1}{6}\,\langle r^2\rangle t +O(t^2)\, ,
\en
where $\langle r^2\rangle$ stands for the mean square radius in the limit
 $\alpha=0$.

The definition of the form factor in the non--relativistic case is 
 very similar,
\eq
\hspace*{-0.5cm}&&{\mathbold S}_{fi}
=(2\pi)^3\delta^3({\bf p}-{\bf q})
+ie\tilde A^\mathrm{ext}_\mu(p-q) F^\mu_{NR}(p,q)+O(e^2)
\quad\mbox{[non--relativistic]}\fs
\nonumber\\
\hspace*{-0.5cm}&&
\en
The matching condition Eq.~(\ref{eq:matching}) yields in our case
\eq\label{eq:form_match}
(p+q)^\mu F(t)
=(2w({\bf p})2w({\bf q}))^{1/2}F_{NR}^\mu(p,q)\, .
\en
The zeroth component of the 
non--relativistic form factor at the lowest order in $\alpha$
can be directly read off from the Lagrangian Eq.~(\ref{eq:LNRA}), 
considering the coefficient of the term linear in the field $A^0$,
\eq\label{eq:exp_NR}
F_{NR}^0(p,q)
=\biggl(1-\frac{h_1}{6M^2}\,({\bf p}-{\bf q})^2+O(v^4)\biggr)\,,
\en
and Eq.~(\ref{eq:form_match}) finally gives
\eq\label{h1_r2}
h_1=M^2\langle r^2\rangle+O(\alpha)\, .
\en
This relation is an example of matching to threshold
parameters, discussed in subsection~\ref{subsec:eff_range}. Namely,
the quantity $\langle r^2\rangle$ is the charge radius of the relativistic
scalar particle at $\alpha=0$ and to all orders in the strong coupling constant
$\lambda_r$. In order to perform matching 
in terms of $\lambda_r$, all what one has to do is to calculate 
$\langle r^2\rangle$ in the relativistic theory at a given order in $\lambda_r$
and substitute this result into Eq.~(\ref{h1_r2}). Additional calculations that
would invoke non--relativistic EFT are not needed.

We now perform the matching of the coupling $g_1$. For this, we consider
 the elastic 
scattering amplitude of two oppositely charged particles.
 Let us restrict to diagrams of order $\alpha$.
At this order, the relevant 
diagrams  can be  divided in two groups: those which can
be made disconnected by cutting one photon line, and those which cannot,
\eq\label{eq:1gamma+n}
T=T^{1\gamma}+\bar T\, ,\quad\quad
T_{NR}=T_{NR}^{1\gamma}+\bar T_{NR}\, .
\en
For the relativistic theory, this splitting is unambiguous. On the other hand,
 the non--relativistic Lagrangian is not unique -- 
e.g., depending on whether one keeps the coupling $h_1$ in the Lagrangian or
removes it using the equations of motion,
 the corresponding term in $T_{NR}^{1\gamma}$ will be absent. Here,
we stick to ${\cal L}_{NR}$ in  Eq.~(\ref{eq:LNRA}), which
generates the relativistic one--photon exchange amplitude to the relevant
order in the momentum expansion. 
The matching condition then holds separately for the 1-photon exchange
contribution, and for the remaining part 
of the amplitude.  For this reason, we
retain only the one--particle irreducible amplitudes $\bar T$ and $\bar
T_{NR}$ in the matching condition.

We now first discuss the singularity structure of the non--relativistic
scattering amplitude
at threshold, and  combine information about the strong sector from
Eq.~(\ref{eq:series}) with the known threshold behavior of the various 
diagrams with virtual photons, given in
Eqs. (\ref{eq:Vc}), (\ref{def_Lambda}), (\ref{eq:hat-SE-1}), 
(\ref{eq:Z-charged}) and
(\ref{eq:mpole-vertex}). It can be easily seen that at order
$\alpha$
\begin{itemize}
\item[i)]
diagrams with transverse photons do not contribute at threshold;
\item[ii)]
each strong loop gives rise to a suppression factor $|{\bf p}|$,
so that only diagrams with not more than one strong bubble contribute
at threshold;
\item[iii)]
in the expression for $\bar T_{NR}$, one may combine the strong diagram
shown in Fig.~\ref{fig:bubbles}a together with the Coulomb correction
to this diagram in Fig.~\ref{fig:photons}e (there is  second diagram of
this type, where the Coulomb photon is exchanged after the strong vertex).  
One may further check that
at order $\alpha$ the (infrared--divergent) Coulomb phase 
$1+2i\alpha\theta_c={\rm e}^{2i\alpha\theta_c}+O(\alpha^2)$
 can be factorized in the whole non--relativistic amplitude.
\end{itemize}
Using Eqs. (\ref{eq:Vc}), (\ref{def_Lambda}), (\ref{eq:Z-charged}),
(\ref{eq:mpole-vertex}), we find that the
threshold behavior Eq.~(\ref{eq:series})  is
 modified -- at order $\alpha$ -- by virtual photon
contributions in the following manner,
\eq\label{eq:nuclear}
{\rm e}^{-2i\alpha\theta_c}\bar T_{NR}
=\frac{A_1}{|{\bf p}|}\, 
+A_2\ln\frac{2|{\bf p}|}{M}+A_3 +O(|{\bf p}|)\, .
\en
A straightforward calculation of the coefficients
 gives
\eq\label{eq:A012}
A_1&=&\frac {\pi\alpha g_1 }{2M}\, ,\quad
A_2=-\frac{\alpha g_1^2}{4\pi M^2}\, ,\nonumber\\[2mm]
A_3&=&\frac{g_1}{M^2}\biggl\{1-\frac{\alpha g_1}{8\pi}\,
\biggl(\Lambda(\mu)+\ln\frac{M^2}{\mu^2}-1-2\pi i \biggr)\biggr\}\, .
\en
Suppose now that one calculates the relativistic amplitude at order $\alpha$,
and to any order in $\lambda_r$. In order to be consistent with the 
non--analytic behavior predicted by the non--relativistic theory,
the threshold behavior of the relativistic 1-particle irreducible amplitude 
must be given by
\eq\label{eq:curlyA}
{\rm e}^{-2i\alpha\theta_c}\bar T=\frac{B_1}{|{\bf p}|}
+B_2\ln\frac{2|{\bf p}|}{M}+{\cal T}+O(|{\bf p}|)\,  .
\en
The infrared--finite quantity ${\cal T}$ will be referred to hereafter  as the 
``relativistic threshold amplitude.'' It can be considered  
the  $O(\alpha)$ generalization of the standard definition of the scattering
amplitude, valid in the presence of real and virtual photons.
 [\underline {Remark:} While defining the threshold amplitude,
one has to discard in particular  terms which diverge logarithmically
at threshold. This procedure is ambiguous, and depends on the choice
of the scale at which the logarithm is set to vanish. 
In Eqs. (\ref{eq:nuclear})
and in (\ref{eq:curlyA}), it was chosen to be the reduced mass of the system.
We are free, however, to choose instead a different scale, as a result of
which the threshold amplitude will change accordingly.
On the other hand,  observable quantities, which are expressed through the
threshold amplitude (e.g. bound--state energies of hadronic atoms),
are independent of this scale. In the present model, this 
 can be verified order by order in the perturbative expansion.]

Finally, we   express the coupling $g_1$ in terms of ${\cal T}$,
\eq\label{eq:match_g2}
{\cal T}=4g_1\biggl\{1-\frac{\alpha g_1}{8\pi}\, 
\biggl(\Lambda(\mu)+\ln\frac{M^2}{\mu^2}-1-2\pi i \biggr)\biggr\}\, .
\en
This completes the matching of $g_1$ and of $h_1$. 

%%%%%%%%%%%%%%%%%%%%%%%%%%%%%%%%%%%%%%%%%%%%%%%%%%%%%%%%%%%%%%%%%%%%%%%%%
\setcounter{equation}{0}
\section{Bound states}
\label{sec:boundstates}

\subsection{Introductory remarks}

In the previous sections, we constructed 
a systematic non--relativistic field 
theory, which is equivalent to the underlying relativistic 
theory at small 3-momenta. The equivalence is achieved by performing a matching procedure 
order by order in the momentum expansion.  The 
non--relativistic approach does not provide new information about the 
low--energy behavior of the scattering  amplitudes, because this low--energy
behavior is the information that enters the matching condition.

The non--relativistic approach becomes useful in the 
 description of the shallow bound states in the theory.
To be specific, we consider bound states in the 
system described by the Lagrangian Eq.~(\ref{eq:L4A}). In this model, in the 
vicinity of the elastic threshold, there exists a tower
of nearly Coulombic bound states, whose energies 
are approximately given by 
\eq\label{eq:leading}
E_n=2M-\frac{M\alpha^2}{4n^2}\, ,\quad n=1,2\cdots.
\en
 Due to the combined effect of the strong and the 
residual electromagnetic
interaction, the energy levels are displaced and acquire a
finite width. Our aim is to find the pertinent corrections to the 
leading order formula Eq.~(\ref{eq:leading}). As already mentioned, the
 non--relativistic Lagrangian constructed in the previous section
 offers a simple,
 elegant and very efficient framework to solve the problem
 ~\cite{Caswell}. We illustrate in this section the procedure.

\subsection{Coulomb problem}

We start from the unperturbed solution which corresponds to 
a pure Coulomb potential.
The unperturbed situation is described by the Lagrangian
which is obtained from the original non--relativistic Lagrangian Eq.~(\ref{eq:LNRA})
by discarding everything but the minimal coupling of the Coulomb photons
to the non--relativistic massive particle,
\eq\label{eq:LC}
{\cal L}_{NR}=-\frac{1}{2}\, A^0\triangle A^0
+\sum_{\pm} \Phi_\pm^\dagger\biggl(iD_t-M+\frac{\triangle}{2M}\biggr)
\Phi_\pm\, .
\en
Eliminating the field $A^0$ through the EOM, the Hamiltonian becomes
\eq\label{eq:HC} 
&&{\bf H}_{\rm 0}+{\bf H}_{\rm C}=\int d^3{\bf x}\biggl\{\sum_{\pm}\Phi_\pm^\dagger
\biggl(M-\frac{\triangle}{2M}\biggr)\Phi_\pm 
\nonumber\\[2mm]
&-&\frac{e^2}{2}\,(\Phi_+^\dagger\Phi_+-\Phi_-^\dagger\Phi_-)\triangle^{-1}
(\Phi_+^\dagger\Phi_+-\Phi_-^\dagger\Phi_-)     \biggr\}\, .
\en
We  introduce  creation and annihilation operators,
\eq\label{eq:aa}
\Phi_\pm(0,{\bf x})=\int\frac{d^3{\bf p}}{(2\pi)^3}\,
{\rm e}^{i{\bf p}{\bf x}}a_\pm({\bf p})\, ,\quad\quad
[a_\pm({\bf p}),a_\pm^\dagger({\bf k})]=(2\pi)^3\delta^3({\bf p}-{\bf k})
\, ,
\en
and construct the bound state of two charged particles 
in Fock space,
\eq\label{eq:Psi_n}
&&|\Psi_{nlm},{\bf P}\rangle=
\int\frac{d^3{\bf k}}{(2\pi)^3}\,\Psi_{nlm}({\bf k})
|{\bf P},{\bf k}\rangle\, ,
\nonumber\\[2mm]
&&|{\bf P},{\bf k}\rangle
=a_+^\dagger(\frac{1}{2}\,{\bf P}+{\bf k})\,
a_-^\dagger(\frac{1}{2}\,{\bf P}-{\bf k})\,|0\rangle\, ,
\en
where $\Psi_{nlm}({\bf k})$ denotes the  Schr\"odinger 
wave function in  momentum space,  and $n,l,m$ stand for the 
principal quantum number, the angular momentum 
and its projection in the $z-$ direction, respectively. See appendix
\ref{app:notations} for further notation, in particular, 
 for an explicit expression of the wave functions $\Psi_{nlm}(\bf k)$.
The state vectors 
(\ref{eq:Psi_n}) satisfy 
\eq\label{eq:Schrodinger}
({\bf H}_{\rm 0}+{\bf H}_{\rm C})|\Psi_{nlm},{\bf P}\rangle=
\biggl(E_n+\frac{{\bf P}^2}{4M}\biggr)|\Psi_{nlm},{\bf P}\rangle\,  .
\en
To proceed,  we introduce the resolvent for the Coulomb problem,
\eq\label{eq:resolvent_C}
{\bf G}_C(z)=\frac{1}{z-{\bf H}_0-{\bf H}_C}={\bf G}_0(z)
+{\bf G}_0(z){\bf H}_C{\bf G}_C(z)\, ,
\en
 whose  matrix element between the states $|{\bf P},{\bf k}\rangle$ 
develops poles at $z=E_n$. To remove the CM momentum of the
matrix elements, we introduce the notation
\eq\label{eq:CM}
({\bf q}|{\bf r}(z)|{\bf p})=
\int \frac{d^3{\bf P}}{(2\pi)^3} \, 
\langle{\bf P},{\bf q}|{\bf R}(z)|{\bf 0},{\bf p}\rangle\, ,
\en
where ${\bf R}(z)$ is any operator in Fock space, and where 
${\bf r}(z)$ denotes
the pertinent operator in the CM system.
 The matrix element of  the resolvent ${\bf G}_C$ is related to 
Schwinger's Green function~\cite{Schwinger},
\eq\label{eq:Schwinger}
({\bf q}| {\bf g}_C(z)|{\bf p})&=& 
\frac{(2\pi)^3\delta^{3}({\bf q}-{\bf p})}
  {E-\frac{{\bf q}^2}{M}}-\frac{1}{E-\frac{{\bf q}^2}{M}}\,
  \frac{4\pi\alpha}{|{\bf q}-{\bf p}|^2}\,\frac{1}{E-\frac{{\bf p}^2}{M}}
  \nonumber\\[2mm]
  &-& \frac{1}{E-\frac{{\bf q}^2}{M}}\,\,
  4\pi\alpha\eta I(E;{\bf q},{\bf p})\,\,\frac{1}{E-\frac{{\bf p}^2}{M}}\, ,
\en
with
\eq\label{eq:Schwinger-I}
I(E;{\bf q},{\bf p})=
\int_0^1\frac{x^{-\eta} dx}
  {[({\bf q}-{\bf p})^2x+\eta^2/\alpha^2(1-x)^2(E-\frac{{\bf q}^2}{M})
    (E-\frac{{\bf p}^2}{M})]}\, ,
\en
where $\eta=\frac{1}{2}\,\alpha\,(-E/M)^{-1/2}$ and
$E=z-2M$.
This function develops poles at $\eta=1,2,\cdots$. The Coulomb wave functions
in momentum space 
can be read from the residues of these poles~\cite{Schwinger}.

\subsection{Feshbach formalism and the Rayleigh--Schr\"{o}dinger\\
perturbation theory}

We now construct a systematic perturbation theory to include
the remaining strong and electromagnetic interactions
which are contained in the Lagrangian Eq.~(\ref{eq:LNRA}). 
 It is convenient to apply
the so--called Feshbach formalism~\cite{Feshbach1,Feshbach2}.

The Hamiltonian of
the system takes the form
\eq\label{eq:Hfull}
{\bf H}={\bf H}_0+{\bf H}_C+{\bf V}\, ,
\en
where ${\bf V}$ stands for all interactions other that the static Coulomb 
potential. Consider the full resolvent
\eq\label{eq:Gz}
{\bf G}(z)=\frac{1}{z-{\bf H}}\, .
\en
It satisfies the equation
\eq
{\bf G}={\bf G}_C+{\bf G}_C \mathbold{ \tau} {\bf G}_C\, ,\quad\quad
\mathbold{\tau}={\bf V}+{\bf V}{\bf G}_C\mathbold{ \tau}\, .
\en
The spectral representation of this resolvent 
contains a sum over all states. In order to calculate
the energy shift of the $n,l$ level, it is convenient to remove the
corresponding unperturbed pole contribution $(z-E_n)^{-1}$ (in the CM frame) 
from ${\bf G}_C(z)$ by defining  
\eq\label{eq:GC_removed}
\bar{{\bf G}}_C^{nl}={\bf G}_C\left\{{\bf 1}-
\sum_m\int 
\frac{ d^3{\bf P}}{(2\pi)^3}\,|\Psi_{nlm},{\bf P}\rangle\langle\Psi_{nlm},{\bf P}|\right\} \, .
\en
We further introduce 
\eq\label{eq:tau-def}
\bar{\mathbold{\tau}}^{nl}={\bf V}
+{\bf V}\bar{\bf G}_C^{nl}\bar{\mathbold{\tau}}^{nl}\, ,
\en
and find for ${\bf G}$ the representation
\eq
{\bf G}=\bar{\bf G}_C^{nl}+\bar{\bf G}_C^{nl}
\bar{\mathbold{\tau}}^{nl}\bar{\bf G}_C^{nl}
+(1+\bar{\bf G}_C^{nl}\bar{\mathbold{\tau}}^{nl}){\bf \Pi}_{nl}
(1+\bar{\mathbold{\tau}}^{nl}\bar{\bf G}_C^{nl})\, ,
\en
where
\eq\label{eq:eqpoles}
{\bf \Pi}_{nl}=\sum_m\int\frac{d^3{\bf P}}{(2\pi)^3}\,
\frac{|\Psi_{nlm},{\bf P}\rangle\langle\Psi_{nlm},{\bf P}|}
{z-\frac{{\bf P}^2}{4M}-E_n-
(\Psi_{nl}|\bar{\mathbold{\tau}}^{nl}(z;{\bf P})|\Psi_{nl})}\, ,
\en
and
\eq\label{eq:eqtau}
(\Psi_{nl}|\bar{\mathbold{\tau}}^{nl}(z;{\bf P})|\Psi_{nl})=
\int \frac{d^3{\bf P'}}{(2\pi)^3}\,\langle\Psi_{nlm},{\bf P'}|
\bar{\mathbold{\tau}}^{nl}(z)|\Psi_{nlm},{\bf P}\rangle\, .
\en
Here we used the fact that the right--hand side does not 
depend on the magnetic quantum number $m$ --
 here and in the following,  we  therefore omit the subscript $m$, whenever 
no ambiguity can occur.
The singularity at $z=E_n$ is absent in the barred
quantities. Therefore, the
pertinent pole must occur through a zero in the denominator of the
expression Eq.~(\ref{eq:eqpoles}). In the CM frame, the relevant
 eigenvalue equation to be solved is
\eq\label{eq:key}
z_{nl}-E_n-
(\Psi_{nl}|\bar{\mathbold{\tau}}^{nl}(z)|\Psi_{nl}) =0\, ,
\en
where the  matrix element denotes the quantity on the left-hand
side of Eq. (\ref{eq:eqtau}), evaluated at ${\bf P}=0$.
  To ease notation, we often will omit the indexes in $z_{nl}$.

The  {\it master equation} (\ref{eq:key})  is a compact form of the
conventional Rayleigh--Schr\"{o}dinger perturbation theory, valid for
unstable systems as well.
 It  fixes the convergence domain of  perturbation theory:
the theory is applicable  as long as  the level shift 
 does not become comparable to the distance between the ground-state and
the first radial--excited Coulomb poles. Equation (\ref{eq:key}) is valid for
 a general potential -- containing e.g. the interaction with the
transverse photons -- since in the derivation, we did not use the explicit
form of the interaction Hamiltonian. 

Finally, the  level shifts and  the widths
 can be obtained from the solution of the master equation,
\eq\label{eq:obs}
z=E_n+\Delta E_{nl}\, ,
\en
where $\Delta E_{nl}$ is complex. 
As we shall see later, the poles of the resolvent ${\bf G}(z)$ 
occur on the second Riemann sheet in the complex $z$-plane,
in accordance with  analyticity requirements.

\subsection{Energy levels}

The  energy shifts admit a Taylor series expansion (up to logarithms),
\eq\label{eq:powers}
\Delta E_{nl}=a_{nl}\alpha^3+b_{nl}\alpha^4\ln\alpha
+c_{nl}\alpha^4+O(\alpha^5)\, .
\en
We now show how to calculate the coefficients $a_{nl},b_{nl}$ and ${c_{nl}}$.

We expand the matrix element in the master equation (\ref{eq:key})
 in a Taylor series in $(z-E_n)$ and obtain
\eq\label{eq:Taylor}
z-E_n=\frac{(\Psi_{nl}|\bar{\mathbold{\tau}}^{nl}(E_n)|\Psi_{nl})}
{1-\frac{d}{dE_n}\,(\Psi_{nl}|\bar{\mathbold{\tau}}^{nl}(E_n)|\Psi_{nl})}+\cdots\, .
\en
Because the numerator is a 
quantity of order $\alpha^3$, the second term in the
denominator starts to contribute at order $\alpha^5$ 
and can thus be dropped~\cite{Bern4}, and one finds
\eq\label{eq:Taylor1}
z-E_n=(\Psi_{nl}|\bar{\mathbold{\tau}}^{nl}(E_n)|\Psi_{nl})+O(\alpha^5)\, .
\en
It remains to evaluate the pertinent matrix elements of the
operator $\bar{\mathbold{\tau}}^{nl}(E_n)$.

We start with the determination of
the perturbation  ${\bf V}$ from 
the Lagrangian Eq.~(\ref{eq:LNRA}). 
 One first eliminates the non--dynamical 
field $A^0$ by using EOM. At the order of accuracy required by 
Eq.~(\ref{eq:powers}), it suffices to retain only following terms:
\eq\label{eq:V}
{\bf V}&=&{\bf H}_S+{\bf H}_R+e{\bf H}_\gamma+e^2{\bf H}_{fin}+\cdots\, ,
\nonumber\\[2mm]
{\bf H}_S&=&-\frac{g_1}{M^2} \int d^3{\bf x}\, 
\Phi^\dagger_+ \Phi^\dagger_- \Phi_+ \Phi_- \,  ,\nonumber\\[2mm] 
{\bf H}_R&=&-\sum_\pm\int d^3{\bf x}\,\Phi_\pm^\dagger\,
\frac{\triangle^2}{8M^3}\,\Phi_\pm\, ,
\nonumber\\[2mm]
e{\bf H}_\gamma &=& \sum_\pm\frac{\mp ie}{M}\,\int d^3{\bf x}\, {\bf A} \, 
\Phi_\pm^\dagger\, {\bf \nabla} \, \Phi_\pm \,  ,
\nonumber\\[2mm]
e^2{\bf H}_{fin}&=&-\frac{e^2h_1}{6M^2}\,\int d^3{\bf x}\,
(\Phi_+^\dagger\Phi_+-\Phi_-^\dagger\Phi_-)
(\Phi_+^\dagger\Phi_+-\Phi_-^\dagger\Phi_-)\, .
\en
Additional terms generated by the Lagrangian Eq.~(\ref{eq:LNRA}) do not
  contribute to the energy shift at 
$O(\alpha^4)$, and are hence omitted.
 Evaluating the energy shift is now straightforward.
Solving Eq. (\ref{eq:tau-def}) by iterations, at the order
of accuracy required we get
\eq\label{eq:tau_iter}
\bar{\mathbold{\tau}}^{nl}(z)={\bf H}_S+{\bf H}_S\bar{\bf G}_C^{nl}(z){\bf H}_S
+{\bf H}_R+e^2{\bf H}_{fin}+e^2{\bf H}_\gamma
\bar{\bf G}_C^{nl}(z){\bf H}_\gamma
+\cdots\, .
\en
Then, using Eq.~(\ref{eq:Taylor1}), we find
\eq\label{eq:DEn4}
\Delta E_{nl} &=&(\Psi_{nl}|{\bf H}_S+{\bf H}_S
\bar{\bf G}_C^{nl}(E_n){\bf H}_S
+{\bf H}_R+e^2{\bf H}_{fin}
\nonumber\\[2mm]
&+&e^2{\bf H}_\gamma\bar{\bf G}_C^{nl}(E_n){\bf H}_\gamma
|\Psi_{nl})+O(\alpha^5)\, .
\en
In order to evaluate this quantity, one inserts a complete set of 
states between the various operators -- only a finite number of 
terms is non--zero in each sum -- and calculates the integrals over the 
intermediate momenta.
It is convenient to further split the energy shift Eq.~(\ref{eq:DEn4})
 into  so--called ``strong''
and ``electromagnetic'' parts,
\eq\label{eq:DEn}
\Delta E_{nl}=\delta_{l0}\biggl(\Delta E_n^{\rm str}-\frac{i}{2}\,\Gamma_n\biggr)+
\Delta E^{\rm em}_{nl}+O(\alpha^5)\, \qquad\mbox{[scalar QED]}\fs
\en
\begin{sloppypar}
\noindent 
 The (real) electromagnetic shifts $\Delta E_{nl}^{\rm em}$ are evaluated below.
 The above equation should  be taken  as a
definition of the bracketed term on the right--hand side.
Since the strong Hamiltonian is local, the strong corrections
to the energy levels with  angular momentum $l\neq 0$
are suppressed by additional powers of $\alpha$. As a result of this, 
 they vanish at the accuracy
considered here, as is indicated by the Kronecker symbol in Eq.~(\ref{eq:DEn}).
\end{sloppypar}

The naming scheme for the various corrections
should not be understood literally.
For example, the ``electromagnetic'' contribution contains the coupling
$h_1$ which, according to Eq.~(\ref{h1_r2}), depends on the quantity
$\langle r^2\rangle$ which is a function of the ``strong'' coupling constant
$\lambda_r$. Vice versa, there are electromagnetic corrections
to the coupling $g_1$ which enters $\Delta E_n^\mathrm{str}$, 
 see Eq.~(\ref{eq:1+alpha}).

\begin{figure}[t]
\begin{center}
\includegraphics[width=9.cm]{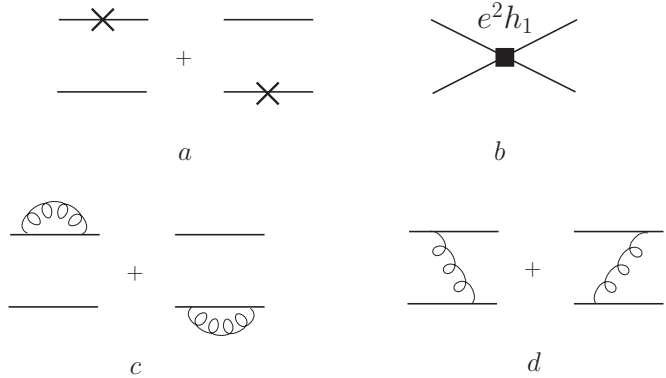}
\end{center}
\caption{Diagrams contributing to the electromagnetic energy shift: 
a) relativistic insertions; b) finite--size effect;
c) self--energy corrections due to transverse photons; d) transverse 
photon exchange. }
\label{fig:em_energy}
\end{figure}

Let us first consider the electromagnetic shift. At $O(\alpha^4)$, it has 
contributions from the relativistic insertions, finite--size effects, and
transverse photons (see Fig. \ref{fig:em_energy}), 
\eq\label{eq:int_em}
\Delta E^{\rm em}_{nl}&=&-\int\frac{d^3{\bf p}}{(2\pi)^3}\,
|\Psi_{nl}({\bf p})|^2
\frac{{\bf p}^4}{4M^3}+\frac{e^2h_1}{3M^2}|\tilde\Psi_{n0}(0)|^2\nnnl
&&-\int\frac{d^3{\bf p}}{(2\pi)^3}\,\frac{d^3{\bf q}}{(2\pi)^3}\,
\Psi_{nl}({\bf p})V_\gamma({\bf p},{\bf q})\Psi_{nl}({\bf q})\fs
\en
The contribution from the self--energy diagram Fig.~\ref{fig:em_energy}c is of order $\alpha^5$ and is therefore 
not displayed. The diagram with one transverse
photon exchange (Fig. \ref{fig:em_energy}d) gives
\eq\label{eq:Vgamma}
V_\gamma({\bf p},{\bf q})&=&\frac{e^2}{4M^2|{\bf p}-{\bf q}|}\,
\biggl(({\bf p}+{\bf q})^2-\frac{({\bf p}^2-{\bf q}^2)^2}
{|{\bf p}-{\bf q}|^2}\biggr)
\frac{1}{\frac{M\alpha^2}{4n^2}+\frac{{\bf p}^2}{2M}+\frac{{\bf q}^2}{2M} +|{\bf p}-{\bf q}|}
\nonumber\\[2mm]
&=&\frac{e^2}{4M^2|{\bf p}-{\bf q}|^2}\,
\biggl(({\bf p}+{\bf q})^2-\frac{({\bf p}^2-{\bf q}^2)^2}
{|{\bf p}-{\bf q}|^2}\biggr)+\cdots\, .
\en
The evaluation of  
the pertinent integrals in Eq.~(\ref{eq:int_em}) for arbitrary $n,l$
is made easier by eliminating
 the terms of the type ${\bf p}^2\Psi_{nl}({\bf p})$  
by use of the Schr\"odinger equation 
in momentum space~\cite{SchweizerHA}. We obtain
\eq\label{eq:Eem}
\Delta E^{\rm em}_{nl}=\alpha^4M\biggl(\frac{\delta_{l0}}{8n^3}
+\frac{11}{64n^4}-\frac{1}{2n^3(2l+1)}\biggr)
+\frac{\alpha^4M^3\langle r^2\rangle}{6n^3}\,\delta_{l0}\, .
\en
Note that in the calculations of the electromagnetic shift, we have
replaced ${\bf G}_C(z)$ by ${\bf G}_0(z)$. It can be verified by direct 
calculations, that this does not affect the result at $O(\alpha^4)$.
 In the limit of  point particles, the above result
coincides with the one from Ref.~\cite{Nandy}.
Next, we turn to the calculation of the strong shift. According to
Eq.~(\ref{eq:DEn4}), one has
\eq\label{eq:DEnstr}
\Delta E_n^{\rm str}-\frac{i}{2}\,\Gamma_n
=-\frac{\alpha^3M g_1}{8\pi n^3}
(1-\frac{g_1}{M^2}\langle \bar {\bf g}_C^{n0}(E_n)\rangle)+O(\alpha^5)\, ,
\en
 where
\eq\label{eq:bargc}
\!\!\!&&\!\!\!\!\!\!\!\langle \bar {\bf g}_C^{n0}(E_n)\rangle
=\!\!\int \frac{d^d{\bf p}}{(2\pi)^d}\,\frac{d^d{\bf q}}{(2\pi)^d}\,
({\bf p}|\bar {\bf g}_C^{n0}(E_n)|{\bf q})=\frac{\alpha M^2}{8\pi}
\biggl(\Lambda(\mu)+\ln\frac{M^2}{\mu^2}-1+s_n(\alpha)\biggr)\, ,
\nonumber\\[2mm]
\!\!\!&&\!\!\!\!\!\!\! s_n(\alpha)=2(\psi(n)-\psi(1)-\frac{1}{n}+\ln\alpha-\ln n)\, ,\quad\quad
\psi(x)=\Gamma'(x)/\Gamma(x)\, .
\en
\noindent
Finally, we substitute the value of the coupling constant
$g_1$ from Eq.~(\ref{eq:match_g2}) into the expression for the strong 
energy shift Eq.~(\ref{eq:DEnstr}). In this manner, one obtains an expression
which contains only quantities defined in the relativistic theory,
\eq\label{eq:NLOall}
\Delta E_n^{\rm str}-\frac{i}{2}\,\Gamma_n=-\frac{\alpha^3 M}{32\pi n^3}\,
{\cal T}\biggl(1-\frac{\alpha (s_n(\alpha)+2\pi i)}{32\pi}\,{\cal T}
\biggr)+O(\alpha^5)\, .
\en

In order to calculate the strong shift of a given level,
we take the real part of the this expression 
 and obtain
\eq\label{eq:NLO}
\Delta E_n^{\rm str}=-\frac{\alpha^3 M}{32\pi n^3}\,
{\rm Re}\,{\cal T}\biggl(1-\frac{\alpha s_n(\alpha)}{32\pi}\,{\rm Re}\,{\cal T}
\biggr)+O(\alpha^5)\, .
\en
Equations (\ref{eq:Eem}), (\ref{eq:NLOall}) and (\ref{eq:NLO}) 
are the main results of 
this section. They provide, at this order, a complete expression for the total
(electromagnetic and strong) shift of the energy level with a given
quantum numbers $n,l$. Furthermore, Eq.~(\ref{eq:NLO}) is the generalization
 of the DGBT formula for the strong shift of the energy level
of the bound state~\cite{Deser,Trueman,Uretsky,Tkebuchava} 
 to  next--to--leading order in $\alpha$.

\subsection{Decay into 2 photons}
\label{subsec:em_decays}

In the present theory, the ground state can only decay in two or more photons.
This is a hard process, with a mass gap of order of the heavy particle
mass $M$. 
Consequently, within the non--relativistic theory, the decay is described
through the imaginary parts of the coupling constants. At leading order,
this is the coupling $g_1$  in Eq.~(\ref{eq:DEnstr}), with
 ${\rm Im}\, g_1=O(\alpha^2)$. According to Eq.~(\ref{eq:Taylor}), we may then use  Eq.~(\ref{eq:DEnstr})
 to calculate this width up to and including terms of order $\alpha^5$.
 The result is
\eq\label{eq:2photon}
\Gamma_1^{2\gamma}=\frac{M\alpha^3 }{4\pi}\,
 {\rm Im}\, g_1+O(\alpha^6)\, .
\en
At the order of accuracy we are working, the imaginary part of $g_1$
is given by (see also appendix~\ref{app:generalized})
\eq
&&\hspace{-.5cm}{\rm Im}\, g_1=\frac{1}{16}\, \sum_{2\gamma}
(2\pi)^4\delta^4(p_1+p_2-q_\gamma-q'_\gamma)
|T^{\phi^+(p_1)\phi^-(p_2)\to 2\gamma}|^2\biggr|_{threshold}
+O(\alpha^3)\, .
\nonumber\\
&&
\en
Evaluating the matrix element at tree level, we find
\eq
{\rm Im}\, g_1=\pi\alpha^2+O(\alpha^2\lambda_r,\alpha^3)\, ,
\en
see also subsection \ref{subsec:twophotondecay}.
We finally arrive at the following result
for the two--photon decay width of the ground  state~\cite{pioniumgg}, 
\eq\label{eq:twophotonwidth}
\Gamma_1^{2\gamma}=\frac{M\alpha^5}{4}+O(\alpha^5\lambda_r,\alpha^6)\, .
\en
In the real world, additional channels 
 are open for decay (strong as well as electromagnetic).
 These will be treated in later sections.

\subsection{Vacuum polarization due to electrons}
\label{subsec:vacuum}

In this subsection we separately consider a particular correction to the hadronic
atom observables, which does not emerge in the model described by
the Lagrangian Eq.~(\ref{eq:L4A}), but is present in the case of real hadronic 
atoms -- namely, the correction due to vacuum polarization induced by an
electron loop.
In order to take this effect into account, one would have to use
the formulation of the low--energy effective theory of the Standard Model 
not in terms  of photons and
 hadrons only~\cite{Urech,KU,MMS,Rupertsberger}, but to consider 
the explicit inclusion of the leptonic sector of the Standard Model
as well~\cite{Neufeld}.
Further, although in the treatment of the hadronic atoms 
 weak interactions can be safely neglected,
one has still to take into account the presence of electrons, which couple 
to photons.
The reason for this is that the electron mass $m_e\approx 0.5\,{\rm MeV}$
is numerically of order or 
smaller than $\alpha M$ -- a scale which is ``resolved'' in 
the non--relativistic theory ($M$ stands now for a typical reduced mass
in the hadronic bound state). 
Examining, however, the possible contributions 
from the explicit electron degrees of freedom to the observables of hadronic
atoms, one may easily check that these all vanish at the next--to--leading order
in $\alpha$ with one exception. Modification of the static Coulomb potential
by an electron loop (vacuum polarization) gives rise to  corrections 
in the hadronic atom observables which are amplified by large numerical 
factors -- powers of $M/m_e$, which emerge from the calculation
of the matrix elements in Eq. (\ref{eq:key}). If one would count the quantity
$\alpha M/m_e$ at $O(1)$ in order to get rid of these large factors,
the vacuum polarization effect starts to contribute to the bound--state energy
at $O(\alpha^3)$, i.e. at the same order as the leading--order strong shift
which is calculated by using the DGBT formula. To the decay width, the 
contribution of the vacuum polarization effect leads to a $O(\alpha)$
 correction\footnote{For a systematic 
discussion of  vacuum polarization contributions within a potential 
model framework, see, e.g., Ref.~\cite{Blomqvist}.}.

\begin{figure}[t]
\begin{center}
\includegraphics[width=5.5cm]{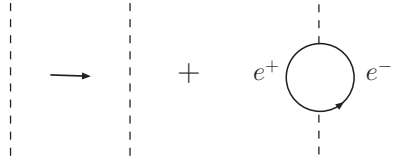}
\end{center}
\caption{Modification of the static Coulomb potential
  by an electron loop.}
\label{fig:vacpol}
\end{figure}

Since the effect from the presence of electrons
appears only at one place, it would be counterproductive to carry out a
full--fledged inclusion of the explicit electron degrees of freedom into
the framework.
More easily, the same goal may be achieved, using e.g. the methods
described in Ref.~\cite{Eiras} (see also Refs.~\cite{Bern3,vacpol1,vacpol2,vacpol3}). 
Here one first integrates 
out the electron field in the generating functional.
At the order of accuracy we are working, this amounts to a modification
of the kinetic Lagrangian $-\frac{1}{4}\,F_{\mu\nu}F^{\mu\nu}$ 
by the electron loop. Furthermore, at this order
one may safely neglect the corrections to
the transverse photon propagator, performing the static limit 
$k^2\to -{\bf k}^2$ in the photon self--energy part. Finally, the whole effect
reduces to the well--known modification of the static Coulomb potential in Eq.~(\ref{eq:HC}) 
 through the electron loop (see Fig. \ref{fig:vacpol}).
In momentum space, this modification reads
\eq\label{eq:vacpol}
-\frac{4\pi\alpha}{{\bf k}^2}&\rightarrow&
-\frac{4\pi\alpha}{{\bf k}^2}-\frac{4\alpha^2}{3}\,\int_{4m_e^2}^\infty
\frac{ds}{s+{\bf k}^2}\,\frac{1}{s}\,\biggl(1+\frac{2m_e^2}{s}\biggr)
\sqrt{1-\frac{4m_e^2}{s}}
\nonumber\\[2mm]
&\doteq& -\frac{4\pi\alpha}{{\bf k}^2}
+{\bf V}^{\rm vac}({\bf k})\, .
\en
[Here, one has removed the UV--divergence in the propagator
 by charge renormalization.] Considering 
the second term in Eq.~(\ref{eq:vacpol}) as a perturbation, one
may calculate the corrections to the bound--state observables in a standard 
manner. Namely, adding ${\bf V}^{\rm vac}$ to the Hamiltonian in 
Eq.~(\ref{eq:V}) and iterating, we arrive at the following correction
terms to the right--hand side of  Eq.~(\ref{eq:DEn4}),
\bea\label{eq:VP0}
&&\hspace{-1cm}
(\Psi_{nl}|{\bf V}^{\rm vac}+{\bf V}^{\rm vac}{\bf G}_C^{nl}(E_n){\bf H}_S
+{\bf H}_S{\bf G}_C^{nl}(E_n){\bf V}^{\rm vac}
+{\bf V}^{\rm vac}{\bf G}_C^{nl}(E_n){\bf V}^{\rm vac}+\cdots|\Psi_{nl})\, .
\nonumber\\&&\hspace{-1cm}
\eea
Starting with the first term, we note that the relevant matrix elements
\bea\label{eq:vacleading}
\Delta E_{nl}^\mathrm{vac}=
(\Psi_{nl}|{\bf V}^{\rm vac}|\Psi_{nl})
\eea
have been evaluated analytically
for any $n,l$ in  Ref.~\cite{Eiras}. For completeness, here we
reproduce the integral
 representation provided in Ref.~\cite[Eq.~(B.3)]{Eiras}
which is valid in the generic case of a bound state of two oppositely
charged particles with arbitrary masses,
\bea\label{eq:vacsecond}
\Delta E_{nl}^\mathrm{vac}&=&-\frac{\mu_c\alpha^3}{3\pi n^2}\sum_{k=0}^{n-l-1}
\left(\begin{array}{cc}n-l-1\\k\end{array}\right)\left(\begin{array}{cc}n+l\\n-l-k-1\end{array}\right)
\xi_n^{2(n-l-k-1)}\nnnl
&&\times \int_0^1\,dx\,\frac{x^{2l+2k+1}}{(\xi_n+x)^{2n}}\sqrt{1-x^2}\,(2+x^2)
\,;\,\quad \xi_n=\frac{nm_e}{\mu_c\alpha}\, .
\eea
Here $\mu_c$ denotes the reduced mass of the bound system, and $m_e$ is the
electron mass.

The second term in Eq.~(\ref{eq:VP0}) contains both
${\bf V}^{\rm vac}$ and ${\bf H}_S$. Because the strong Hamiltonian 
${\bf H}_S$ is local, the corrections in the states with $l\neq 0$ start at
higher order in $\alpha$ than in the $S$-states, see also the remark after
  Eq.~(\ref{eq:DEn}). We therefore stick to the corrections in $S$-states.
 It is convenient to introduce the quantity
\eq\label{eq:VP1}
\delta \Psi_{n0}({\bf p})=\int\frac{d^d{\bf k}}{(2\pi)^d}\,
\frac{d^d{\bf q}}{(2\pi)^d}\,({\bf p}|\bar{\bf g}_C^{n0}(E_n)|{\bf k})\,
{\bf V}^{\rm vac}({\bf k}-{\bf q})\Psi_{n0}({\bf q})\, ,
\en
which is called ``modification of the Coulomb wave function at
the origin, due to vacuum polarization.''
 The corrections are 
 then most easily expressed in terms of the  ratio
 \eq\label{eq:deltavp}
\delta^{\rm vac}_n\doteq 2
\delta \tilde\Psi_{n0}(0)/\tilde\Psi_{n0}(0)\, .
\en
 The next--to--leading correction to the energy shift and width emerges
through the modification of the value of the wave function at the origin
 (cf. with Eq.~(\ref{eq:DEnstr})),
\eq\label{eq:modify}
|\tilde\Psi_{n0}(0)|^2\to
|\tilde\Psi_{n0}(0)|^2(1+\delta_n^{\rm vac})\, .
\en
The quantity $\delta^{\rm vac}_n$
is explicitly evaluated  for $n=1$ in Ref.~\cite{Eiras}.
 At the present experimental accuracy, the correction
$\delta^{\rm vac}_n$ is not negligible  for \piH, where it 
amounts to approximately half a percent.

The notion of a
``modified Coulomb wave
function'' appears here for the sake of convenience only 
-- in this 
manner, one may easily parameterize the correction term and carry out a
comparison with other approaches. The Feshbach formalism does 
not refer to the exact wave function of the bound
system at all  -- this might even 
fail  to be a well--defined quantity in the 
case of  meta-stable bound states. All calculations
 in the Feshbach formalism
are performed in terms of the resolvent ${\bf G}(z)$, which is a well defined
quantity.

We expect that the remaining terms in Eq.~(\ref{eq:VP0}) are very small,
 and we discard them in the following.

\subsection{Energy shift and width: summary}
\label{subsec:scalarsummary}
Amazingly enough, the above results carry over nearly unchanged to the real
world of hadronic atoms, described in the framework of ChPT. For this reason, we
collect the results of this and the previous section for later reference.

The relevant {\it master equation} to be solved is
\eq\label{eq:summary1}
z-E_n-
(\Psi_{nl}|\bar{\mathbold{\tau}}^{nl}(z)|\Psi_{nl}) =0\, \,,
\en
where the matrix element on the right--hand side is
 the one in Eq.~(\ref{eq:eqtau}), evaluated at $\bf P=0$.
 We also introduced several energy shifts,
\bea\label{eq:summary2}
z&=&E_n+\Delta E_{nl}\,\,,\\
\label{eq:summary3}
\Delta E_{nl}&=&\Delta E^\mathrm{em}_{nl}+\Delta E^\mathrm{vac}_{nl}
+\delta_{l0}\left(\Delta
  E_n^\mathrm{str}-\frac{i}{2}\Gamma_n\right)+O(\alpha^5)\,.
\eea
The understanding of these relations is as follows. 
i) $\Delta E_{nl}$ is a complex
quantity, defined through Eq.~(\ref{eq:summary2}).
ii) The real shifts 
$\Delta E^\mathrm{em}_{nl}$ and $\Delta E^\mathrm{vac}_{nl}$ are defined in
Eqs.~(\ref{eq:Eem}) and (\ref{eq:vacleading}), 
respectively. iii) The remainder
$\Delta E_{nl}-\Delta E_{nl}^\mathrm{em}-\Delta E_{nl}^\mathrm{vac}$ is split
into the real components $\Delta E_n^\mathrm{str}$ and $\Gamma_n$,
which are related to the threshold amplitude in the underlying relativistic
theory. Including vacuum polarization effects, 
\eq
\Delta E_n^{\rm str}-\frac{i}{2}\,\Gamma_n=-\frac{\alpha^3 M}{32\pi n^3}\,
{\cal T}\biggl(1-\frac{\alpha (s_n(\alpha)+2\pi i)}{32\pi}\,{\cal T}
+\delta_n^\mathrm{vac}\biggr)+O(\alpha^5)\, ,
\en
 where ${\cal T}$ is the threshold amplitude defined in Eq.~(\ref{eq:curlyA}).

%%%%%%%%%%%%%%%%%%%%%%%%%%%%%%%%%%%%%%%%%%%%%%%%%%%%%%%%%%%%%%%%%%%%

\setcounter{equation}{0}
\section{On DGBT formulae in ChPT}
\label{sec:splitting}

In this section, we investigate in some detail the procedure 
to extract information on the hadronic scattering lengths 
from the energy spectrum of hadronic atoms. To start with, we
perform a
 {\em Gedankenexperiment}: 
let us assume that the fundamental
 relativistic Lagrangian is given by
Eq.~(\ref{eq:L4A}) (the effects of vacuum polarization do not play any role
 here and are ignored).
Assume further
that one measures the energy levels of the  two--body bound state and
extracts the real part of the threshold amplitude ${\rm Re}\,{\cal T}$
 via the expressions Eqs.~(\ref{eq:DEn}), (\ref{eq:Eem}) and (\ref{eq:NLO}).
 Can we purify ${\rm Re}\,{\cal T}$ from electromagnetic interactions 
 and  determine the ``purely strong'' 
threshold amplitude?
 The question is tailored to 
quantum--mechanical models, where one considers bound states formed by the
sum of Coulomb and short--range strong potentials, and where
the Coulomb potential may be easily switched on or off. After this
investigation, we  discuss
 the analogous question in the framework of ChPT. We follow the 
ideas outlined in Ref.~\cite{Scimemi}, see also 
 Refs.~\cite{Bijnens-Prades,Moussallam,Anant-M,DescG,Pineda_em,Gegelia_em}.

\subsection{Electromagnetic corrections in  scalar QED}
\label{subsec:mu1}

The amplitude ${\cal T}$ defined in 
Eq. (\ref{eq:curlyA}) depends on two coupling constants,
 $\lambda_r$ and $e$.
 We expand 
 ${\cal T}$ in powers of the fine--structure constant, 
\eq\label{eq:repr}
\mbox{Re}\,\, {\cal T}=\bar {\cal T}+\alpha {\cal T}^1+O(\alpha^2)\, .
\en
Here, $\bar{\cal T}$ denotes the ``purely strong'' threshold amplitude.
 Consider now the energy shift of the bound state, determined
by Eq. (\ref{eq:NLO}). At leading order we have the standard DGBT 
formula~\cite{Deser},
\eq\label{eq:inter_LO} 
\Delta E_n^{\rm str}&=&-\frac{\alpha^3 M}{32\pi n^3}\,
\bar {\cal T}\, +O(\alpha^4)\co
\en
which  may be used to measure  ${\bar{\cal T}}$,
\bea
\bar{\cal T}={\cal E}_n^{\rm str}+{ O}(\alpha)\,,\quad {\cal E}_n^{\rm str}=-\frac{32\pi n^3 \Delta E_n^{\rm str}}{\alpha^3M}.
\eea
 To include higher order corrections, we rewrite Eq. (\ref{eq:NLO}) as
\eq\label{eq:NLO1}
{\cal  E}_n^{\rm str}&=&
\bar{\cal T}\biggl(1-\frac{\alpha s_n(\alpha)}{32\pi}\,\bar{\cal T}
\biggr)+\alpha {\cal T}^1 + O(\alpha^2)\, .
\en
 This formula leads --
 provided that one can
 calculate ${\cal T}^1$  -- to a 
 refined determination of the 
threshold amplitude,
\bea\label{eq:accurate}
\bar{\cal T}={\cal E}_n^{\rm str}\left(1+\frac{\alpha s_n(\alpha)}{32\pi}\,{\cal E}_n^{\rm str}\right) 
-\alpha {\cal T}^1
+{ O}(\alpha^2)\,.
\eea
It turns out, however, that $\bar{\cal T}$ and ${\cal T}^1$ cannot be uniquely 
defined. Indeed, let us imagine a perturbative expansion
 of the  amplitude, and let us retain the terms up to and including one
 loop. The result will have the  structure
\eq\label{eq:Tlambda}
\mbox{Re}\,\, {\cal T}&=&\lambda_r 
+\frac{a_1}{2}\,\lambda_r^2 \ln\frac{M^2}{\mu_0^2}
+\frac{a_2}{2}\,\lambda_r \alpha \ln\frac{M^2}{\mu_0^2}
+b_1\lambda_r^2 +b_2\lambda_r \alpha \, + \cdots\,,
\en
where $a_i,b_i$ are pure numbers, and where the ellipsis stands for the
contributions beyond one loop. This amplitude is scale independent, because
 $\lambda_r$ satisfies the RG equation
\eq\label{eq:RG}
\mu_0\frac{d}{d\mu_0}\,\lambda_r&=&a_1\lambda_r^2+a_2\lambda_r\alpha
+\cdots .\quad\quad
\en
What is the meaning of the splitting in Eq.~(\ref{eq:repr})?
 The  amplitude $\bar {\cal T}$ is given by
\eq
\bar{\cal T}=\bar\lambda_r +\frac{a_1}{2}\,
\bar\lambda_r^2 \ln\frac{\bar M^2}{\mu_0^2}+b_1\bar \lambda_r^2 \, 
+\cdots\,,
\en
 where $\bar \lambda_r$ and $\bar M$ are the couplings and masses in 
pure $\phi^4$ theory, compare the footnote at the beginning of section 
\ref{sec:including_gamma}. The coupling constant  $\bar \lambda_r$ 
satisfies the RG equation  with $\alpha=0$, 
\eq\label{eq:RGbar}
\mu_0\frac{d}{d\mu_0}\,\bar\lambda_r=a_1\bar\lambda_r^2+O(\bar\lambda_r^3)\, .
\en
 The second term in the representation Eq.~(\ref{eq:repr}) is  then simply the
 difference ${\cal T}-\bar{\cal T}$. In order to actually calculate it, we need
to know the relation between $\lambda_r $ and $\bar\lambda_r $
 \cite{Scimemi}. For this, we may specify the scale $\mu_1$ 
at which the
 two couplings coincide. We denote the corresponding coupling by
 $\bar\lambda_r(\mu_0;\mu_1)$,
\eq
\bar\lambda_r(\mu_0;\mu_1)=\lambda_r(\mu_0)
\biggl(1-\alpha a_2\ln\frac{\mu_0}{\mu_1}
+\cdots\biggr)\,,
\en
where the arbitrary scale $\mu_1$ is referred to as the {\em matching scale}.
As $\lambda_r$ is independent of $\mu_1$, we have
\eq
\label{eq:mu1}
 \mu_1 \frac{d}{d\mu_1} \bar\lambda_r(\mu_0;\mu_1)=\alpha a_2
\bar\lambda_r +\cdots\,.
\en
We express the coupling $\lambda_r$ through $\bar\lambda_r$ 
and arrive at the representation Eq.~(\ref{eq:repr}), with
\eq\label{eq:a2b2}
{\cal T}^1=\bar\lambda_r(\mu_0;\mu_1)
 \biggl\{\frac{a_2}{2}\,\ln\frac{ M^2}{\mu_1^2}
+b_2\,\biggr\}\,+ \cdots\,.
\en
The difference between $M$ and $\bar M$  does not 
matter at the order of the perturbative 
expansion considered here.

We now come to the main point: because ${\cal T}^1$ depends on the scale $\mu_1$,
\eq
\mu_1\frac{d}{d\mu_1}\,{\cal T}^1&=&- a_2\bar\lambda_r+\cdots\co
\en
the amplitude $\bar{\cal T}$ in Eq.~\eqref{eq:accurate} becomes convention dependent 
as well [whereas  $\mbox{Re}{\cal T}=\bar{\cal T}+\alpha{\cal T}^1+\cdots$ is independent of $\mu_1$.]
The relation between the amplitudes evaluated from Eq.~\eqref{eq:accurate} 
with matching scales $\mu_1$
and $\mu_2$ is
\eq
\bar{\cal T}_{\mu_2}=\bar{\cal T}_{\mu_1}(1+\alpha 
a_2\ln{\frac{\mu_2}{\mu_1}})+\cdots\,.
\en
The ambiguity shows up at order $\alpha$ and does not affect
the interpretation of the ``purely strong'' amplitude at  leading order,
which is extracted by using Eq.~(\ref{eq:inter_LO}).

\subsection{Isospin breaking effects in ChPT}
\label{subsec:isospinbreaking}
 In the real world, the dynamics of hadronic atoms 
 can be analyzed  with ChPT. As this is a quantum field theory, 
 the issue of purification of the amplitude from electromagnetic contributions
 is affected with the same ambiguity as the toy example discussed above.
 Hadronic atoms are not an exception in this sense: whenever
 an attempt is made to split electromagnetic effects in ChPT, one is faced
 with this problem. We now explain how this can be handled.

 We start from the observation that the Lagrangian in ChPT contains, aside
 from the particle fields $\Phi$, a set
 of LECs, which are not determined by chiral 
symmetry alone. We consider
 ChPT in the hadronic sector, with photons included. Generically, one
 has
\eq
{\cal L}_\mathrm{ChPT}={\cal L}({\cal G}; {\cal K};{\cal M},e;\Phi)\,.
\en
We have classified the LECs in two groups: the ones in ${\cal G}$ ({\it
  strong} LECs), which
stand for those which survive at $e=0$, like the pion decay constant in the
chiral limit, the LECs $L_i$ at order $p^4$, etc.
The group ${\cal K}$ stands for the so called {\it electromagnetic }
 LECs that one has to introduce while incorporating 
electromagnetic interactions in the theory, like the pion mass difference in
the chiral limit, the $K_i$ introduced by Urech
 \cite{Urech}, etc. In addition, there are the quark masses collected in
 ${\cal M}$, and the electromagnetic coupling $e$.

Given the structure of this Lagrangian, Green
functions can be evaluated in a straightforward  manner.
The Lagrangian is so constructed that all UV divergences
cancel -- the result for any quantity is  independent of the renormalization
scale $\mu_0$ and can symbolically be written in the form
\eq
X&=&X({\cal G}^r;{\cal K}^r;{\cal M},e;\mu_0;p_i)\,,\nonumber\\
\frac{dX}{d\mu_0}&=&0.
\en
We have indicated the dependence on the
renormalized LECs ${\cal G}^r,{\cal K}^r$ 
and on  external momenta $p_i$. In particular, one can determine the algebraic
form of the masses $\pi^\pm,\pi^0, K^\pm,K^0,\eta,\ldots$. 
We now define the isospin symmetry
 limit in the following manner.
\begin{itemize}
\item
Set $e=0,m_u=m_d$. The quantities $\bar X$ 
so obtained depend on the renormalized parameters 
${\cal G}^r$, on the quark masses  and on the momenta.
\item
In the isospin symmetry limit so defined, 
physical masses are grouped into mass degenerate 
 (isospin) multiplets. Assign  numerical values to these. In
particular, for the pion, kaon and  proton mass, choose the physical 
values for $M_{\pi^+}, M_{K^+}, m_p$, and for the pion decay constant take $F_\pi=92.4$ MeV.
\item
Isospin breaking terms are defined to be the difference $X - \bar X$. These
depend on the full set of renormalized parameters 
${\cal  G}^r;e;{\cal K}^r$, and, in addition, on the physical masses, 
on quark mass ratios, on $F_\pi$ 
 and on the momenta.
\end{itemize}
To numerically calculate  the isospin breaking corrections, 
one uses measurements or estimates to assign
numerical values to the needed renormalized LECs, and on quark mass ratios, 
as a result of which
the isospin breaking terms can be calculated for any quantity.

We comment the procedure.
\begin{itemize}

\item
It is algebraically well defined and internally consistent. 
As far as we can judge, it agrees with
 calculations of isospin breaking corrections performed in recent years
 by many authors.

\item
The problem with the ambiguity in purifying quantities from electromagnetic 
interactions is hidden in the numerical values assigned to the LECs.
Here, we  assumed that all LECs are
known, to within reasonable error bars. 
In other words, fixing the values of all LECs unambiguously defines the
splitting in physical observables.
\item
The relation of the chiral Lagrangian 
to the underlying theory, QCD+QED, will not be discussed here 
[see Refs.~\cite{Scimemi,Bijnens-Prades,Moussallam,Anant-M,DescG,Pineda_em,Gegelia_em}
where certain aspects of the problem are considered]. 
It is however clear that
 the mentioned ambiguity will show up again once the 
matching of the chiral Lagrangian to QCD+QED is
performed. We take it that the uncertainties in the electromagnetic LECs
are chosen in such a manner that this ambiguity is taken care of
appropriately. 

\item
Assigning, in the isospin symmetry limit, physical values 
 to $F_\pi,M_{\pi^+}, M_{K^+}$ and to the proton mass overconstrains the 
available parameters in QCD ($\Lambda_{QCD}, \hat m,m_s$):
 fixing the first three quantities  allows one to calculate the proton mass,
and there is no reason why this mass should coincide with the physical value $m_p$.
 However, we expect the difference to be small, with a negligible effect
on the quantities considered here.

\item
The manner in which
electromagnetic corrections are treated in Ref.~\cite{Moussallam} differs from the
present prescription in at least 
one important aspect: in that treatment, the leading--order meson mass terms 
 that enter the strong
chiral Lagrangian become scale dependent and run with $\alpha$ 
 in the full theory. This is not the
case in the procedure advocated here.
\end{itemize}

We add a note concerning the bookkeeping of isospin breaking corrections.
Isospin breaking contributions  can be evaluated
 as a power series in $\alpha$ and 
$m_d- m_u$ (modulo logarithms).
It is useful to
define a small parameter $\delta$ as a bookkeeping device,
\eq\label{eq:delta}
\delta\sim\alpha,\,  m_d- m_u\, .
\en
A particular calculation is then carried out to a specific order in $\delta$.
We note that the above counting is  not the only possible choice. 
Indeed, in the treatment of 
pionium ~\cite{Bern1,Bern2,Bern4,SchweizerHA},
 the bookkeeping was performed 
 differently: $\delta\sim\alpha\,, (  m_d-  m_u)^2$. 
The reason for this is that,
both in the pion mass difference and in the $\pi\pi$ scattering amplitudes, the
terms linear in $  m_d-  m_u$ are absent -- 
 isospin breaking generated by the 
quark mass difference start at $O((  m_d-  m_u)^2)$.
On the other hand, if one still
wants to count $(  m_d-  m_u)$ like $\alpha$ in pionium, this  amounts
to discarding all strong isospin breaking corrections in the final result,
because the calculation was carried out at next--to--leading order in $\delta$.
 We have indeed found \cite{Bern2,Bern4} that numerically, these corrections are
tiny, and nothing changes in the answer if these are left out.

This concludes our review on the formulation of the general 
approach to hadronic atoms. We have used the lagrangian in Eq.~(\ref{eq:L4A})  to 
illustrate the main ideas,
which do depend neither on the details of the underlying 
interaction, nor on the choice of a  particular bound state.
In the remaining part  of this article we 
consider the application of this general theory
to several hadronic atoms, in the framework of chiral perturbation theory.

\bigskip
\setcounter{equation}{0}
\section{Hadronic atoms and scattering lengths:\\ 
General observations}\label{sec:atomsscattering}
\label{subsec:widthtable}

\begin{figure}[t]
\begin{center}
\includegraphics[width=9.4cm]{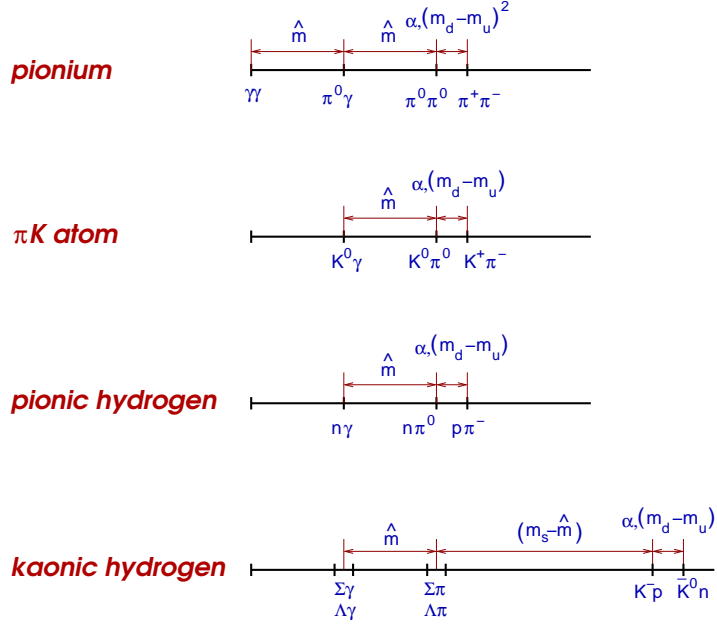}
\end{center}
\caption{Decay channels of various hadronic atoms. Different thresholds
are separated due to   electromagnetic/strong isospin breaking 
[$\alpha\neq 0,~m_u\neq m_d$],  and/or 
$SU(3)$ breaking [$m_s\neq \hat m$], and/or
 a non-zero value of the pion mass [$\hat m\neq 0$].
Multi-photon states are not explicitly indicated: e.g., the decay channel of 
pionium into $\pi^0\pi^0$ also implies decays into states with
any number of additional photons. The scales in the Figure 
 are chosen arbitrarily.}
\label{fig:widthplot}
\end{figure}

Before starting the investigation of the individual atoms listed in Table \ref{tab:atoms}, 
we find it useful to provide already here
 some general information on their spectra,
and on the relation of these to the pertinent threshold amplitudes.
We start with an investigation of the counting rules in the parameter $\delta$ introduced 
 at the end of the last section.

\subsection{Counting rules for the widths}
Strong interaction energy shifts  start at order $\delta^3$. On the other hand,
the counting rules 
 are slightly more complicated for the decay widths,
because  the bookkeeping
of the phase space factor (in powers of $\delta$) depends on the final state involved.

The hadronic atoms decay into  states whose thresholds lie below the 
bound--state energy. In the cases that we consider here, the separation 
of the thresholds is due to (one or several of) the following 
mechanisms (see Fig.~\ref{fig:widthplot}):
\begin{itemize}

\item[i)]
Level splitting in the atoms, which is  $O(\alpha^2)$;

\item[ii)]
Isospin breaking  [$\alpha\neq 0$ and/or $m_u\neq m_d$];

\item[iii)]
$SU(3)$ breaking  [$m_s\neq \hat m$];

\item[iv)]
Chiral symmetry breaking [$\hat m\neq 0$].
\end{itemize}

The phase space factor in the $S$-wave counts as $O(\delta^{1/2})$ for ii) and
as $O(1)$ for iii) and iv). Furthermore, up to order $\delta^5$, only the decay
width of the $S$-states is non--zero,
\eq
\mbox{Im}\,\Delta E_{nl}=-\frac{1}{2}\,\Gamma_n\delta_{l0}+O(\delta^5)\fs
\en
In the case of  particles with spin, $\Delta E_{nl}$ denotes the averaged 
energy shift, see appendix~\ref{app:notations}. Transition widths
between different energy levels also start at $O(\delta^5)$ (see,  
 e.g., Ref.~\cite{BetheSalpeter}).

Taking into account the fact that the square of the Coulomb wave function
at the origin counts at $\delta^3$, 
the decay widths into different final states
 are seen to obey the following counting rules:

\subsubsection*{Pionium}

In the $S$-wave,  pionium decays overwhelmingly into $\pi^0\pi^0$.
At leading order, the decay width counts at $O(\delta^{7/2})$. 
Next--to--leading order isospin breaking corrections to this partial decay
 are of order  $\delta^{9/2}$. The anomaly--induced decay into $\pi^0\gamma$
 cannot proceed in the $S$-wave, and the decay  into 
$\gamma\gamma$ starts at $O(\delta^5)$~\cite{pioniumgg}. At 
order $O(\delta^{9/2})$,
the multi--photon final states (e.g. $\pi^0\pi^0\gamma$) do not contribute 
either, see Ref.~\cite{SchweizerHA} for a detailed discussion.
 {\em At order $\delta^{9/2}$, only the decay into the $\pi^0\pi^0$
final state needs to be taken into account.}

\subsubsection*{$\pi K$ atom}

The decay width of the $\pi K$ atom into $\pi^0K^0$ starts at order $\delta^{7/2}$,
and the next--to--leading order corrections count as $O(\delta^{9/2})$.
Up to $O(\delta^5)$ there are no other contributions, see Ref.~\cite{SchweizerHA} for a detailed discussion.
{\em At order $\delta^{9/2}$, only the decay into the $\pi^0K^0$
final state needs to be taken into account.}

\subsubsection*{Pionic hydrogen}

In difference to  pionium and to the $\pi K$ atom, there are two decay channels
in \piH with comparable widths: the $n\pi^0$ and the $n\gamma$
states, with $\simeq 60\%$ and $\simeq 40\%$ decay probabilities, respectively.
The ratio of these two quantities gives the Panofsky ratio $P\simeq 1.546$~\cite{panofsky1,panofsky2}.
The decay width into $n\pi^0$ starts at $O(\delta^{7/2})$ and next--to--leading
isospin breaking corrections to it count as $O(\delta^{9/2})$. The
decay width into $n\gamma$ starts at $O(\delta^4)$ -- there are no corrections
to the leading order at the accuracy we are working. The Panofsky ratio
at leading order counts as $O(\delta^{-1/2})$. Again, there is no need
to consider isospin breaking corrections.
Further, at the accuracy we are working, the decays into  final states with
more than one photon can be neglected.
{\em At order $\delta^{9/2}$, only the decays into the $n\pi^0$ and $n\gamma$ 
final states need to be taken into account.}

\subsubsection*{Kaonic hydrogen}

Since $M_{\bar K^0}+m_n > M_{K^-}+ m_p$, the decays of  \kaH
proceed overwhelmingly into the strong channels $\pi\Lambda$ and $\pi\Sigma$.
The decay width at leading order counts as $O(\delta^3)$. The decay width
into $\Sigma\gamma$, $\Lambda\gamma$, as well as next--to--leading order 
isospin breaking corrections to the decay into strong channels
start at $O(\delta^4)$. Other decay channels are suppressed.

\bigskip

\noindent
\underline{Remark:} The couplings $\pi^+\pi^-\pi^0\gamma$
and $K^+\pi^-K^0\gamma$ in the relativistic theory are proportional to the
antisymmetric tensor $\varepsilon^{\mu\nu\alpha\beta}$.
For this reason, the pertinent transition amplitudes necessarily
contain at least one soft momentum and are suppressed as compared
 to the $\pi^-p\to n\gamma$ amplitude. There is therefore 
no need to introduce a Panofsky ratio for these channels at the order
 considered here.

\subsubsection*{Pionic deuterium}
Pionic hydrogen can decay  e.g. into $nn$. 
Therefore, the phase space factor is of order $\delta^0$, and the width 
is of the order $\delta^3$.

\subsubsection*{Kaonic deuterium}
Kaonic deuterium can decay e.g. into $\Sigma n \pi, \Lambda n\pi$.
 The corresponding phase space factor is of order $\delta^0$, 
and the width is of order $\delta^3$.

\subsection{Spectrum and scattering lengths}
We are concerned in this review to a large extent with the connection between the 
hadronic atom spectra and the underlying hadronic scattering
 amplitudes at threshold (scattering lengths). Four elastic reactions are involved: 
 $\pi\pi\to \pi\pi$, $\pi K\to \pi K$,
$\pi N\to \pi N$ and $\bar K N\to \bar KN$. In the isospin symmetry 
limit $\alpha =0, m_u=m_d$, each of them is described by   two independent isospin amplitudes. 
These are real in the first three reactions, and complex in the last reaction. The atoms considered
 fall into two distinct groups:
 
\underline{Group 1:}
To this group belong the first three atoms displayed in Figure~\ref{fig:widthplot}:
$A_{2\pi}, A_{\pi K}$, and \piH. At the order considered here, these atoms can decay 
into states that are separated in energy from the bound state through isospin breaking effects, 
which are calculable in ChPT (or which can be taken into account with the Panofsky 
ratio in case of \piH). For each of these three atoms, there are two real
 threshold amplitudes in the isospin symmetry limit. Therefore, it suffices to have two 
experimental numbers to pin them down: the strong energy level shift and width of the 
ground state.

\underline{Group 2:}
To this group belong \kaH, \kad and \pid. These atoms can decay into states that are separated 
in energy from the bound state also in the absence of isospin breaking, because the splitting  is 
due to $\hat m\neq 0$ and(or) $ \hat m\neq  m_s$. As a result of this, the widths are 
enhanced. We already 
 mentioned in subsection \ref{subsec:kaH} that the $\bar K N$ scattering lengths cannot be determined from 
the energy level shift and width of the ground state of \kaH even in principle, because one has
 to fix two complex numbers. As we will see later, it turns out that \kad 
provides the lacking information, see subsection \ref{subsec:kaondeuteron}. 
The case of \pid plays a special role here. From a  principle point of view, information on this compound 
is not needed:
 to measure the two real
 pion--nucleon scattering lengths,  knowledge of the strong  energy level shift and width 
of the ground state of \piH suffices.
  On the other hand, due to an insufficient knowledge of one of the LECs involved, 
information aside from \piH is presently needed. It is provided e.g. by the strong energy shift in \pid, which 
can be related to the pion--nucleon scattering lengths. On the other hand,  its width cannot be used for this purpose. 
The physical reason for this is the fact that \pid can decay e.g. into $nn$ final states, a reaction 
which cannot be related to the elastic pion--nucleon scattering amplitude.
The technical reason behind this is the fact that the ratio between the strong energy 
level shift and the width in \pid is of order $\delta^0$, while the imaginary part 
of the pion--nucleon amplitude is of order $\delta^{1/2}$: there can be no 
DGBT formula that relates the energy shift {\it and} width directly with the pion--nucleon 
scattering amplitude, because there would be a mismatch in counting  powers of $\delta$.

%%%%%%%%%%%%%%%%%%%%%%%%%%%%%%%%%%%%%%%%%%%%%%%%%%%%%%%%%%%%%%%%%%%%%%
\setcounter{equation}{0}
\section{Pionium}
\label{sec:pionium}

\subsection{DIRAC experiment at CERN}
\label{subsec:DIRAC}

As already mentioned,
the aim of the DIRAC collaboration at CERN is to measure 
the lifetime of the $\pi^+\pi^-$ atom in the ground state 
with  10\% precision~\cite{DIRAC,Adeva:2003up,Gomez,DIRAC_PRELIMINARY,Goldin:2005ce,DIRAC_RESULT,DIRAC:addendum}. This
allows one to determine the difference $|a_0 - a_2|$
of  $S$-wave $\pi\pi$ scattering 
lengths at  5\% accuracy. 
 Measurements 
of the lifetime of  $\pi^+ \pi^-$ atoms have also  been proposed 
 at J-PARC and at GSI~\cite{DIRAC1,DIRAC2}. For an earlier
 attempt to measure
 pionium production, see Ref.~\cite{Betker:1996dj}. 

Details of the set--up of the DIRAC experiment at the CERN Proton Synchrotron 
 can be found in Ref.~\cite{Adeva:2003up}. 
The underlying  idea is the following. 
High--energy proton--nucleus collisions  produce 
pairs of oppositely charged pions via strong decays of 
intermediate hadrons. 
Some of these pairs  form $\pi^+\pi^-$ atoms due to 
Coulomb final state interaction. 
 Once produced, the $\pi^+\pi^-$ atoms propagate with 
relativistic velocity.  Before they decay into
pairs of neutral pions, the atoms interact with the target atoms.
 This interaction excites/de-excites or 
breaks them up. The $\pi^+\pi^-$ pairs from the break--up 
exhibit specific kinematic features, 
which allow one to identify them experimentally. 
Excitation/de-excitation and break--up of the atom 
compete with its decay. Solving the transport 
equations  for excitation/de--excitation and 
break--up leads to a target--specific relation between 
break--up probability and lifetime, which is believed
to be known  at the 1\% level~\cite{DIRAC_RESULT}. 
Measuring the break--up 
probability then allows one to determine the lifetime 
of pionium~\cite{Nemenov:1984cq,Afanasev,Trautmann}.  

The first observation of  $\pi^+\pi^-$ atoms ~\cite{Afanasev:1993zp} has 
set a lower limit on 
its lifetime,
$\tau > 1.8 \times 10^{-15}$~s (90\%~CL).  
 Recently~\cite{DIRAC_RESULT}, 
the DIRAC collaboration reported a measurement 
 based on a large sample of data taken in 2001 with Ni 
targets, see Eq.~(\ref{eq:tau_pionium}).

 As pointed out Refs.~\cite{Ovsyannikov1,Ovsyannikov2}, 
 a measurement of the energy splitting between $2s$- and $2p$-states of  pionium 
allows one to determine the combination $2a_0 + a_2$ of scattering lengths,
see also section~\ref{sec:fundamental} of this report.
 Assuming $a_0 - a_2>0$, knowledge of the width and energy shift thus allows one
to determine separately $a_0$ and $a_2$. 

\subsection{Two--channel problem}

In contrast to scalar QED considered
in  previous sections, one is concerned here with two coupled channels
of non--relativistic particles, $\pi^+\pi^-$ and $\pi^0\pi^0$, which   are 
separated by a mass gap $M_\pi-M_{\pi^0}\ll M_\pi$, see Figure~\ref{fig:widthplot}. 
 As a result of this, the ground state of pionium dominantly decays 
into  $\pi^0\pi^0$, whereas its decay into $\gamma\gamma$ is suppressed by a factor $\sim 10^3$.

We investigate the decay process in the
non--relativistic framework developed in  previous sections.
 In particular, the master equation (\ref{eq:summary1}) and the decompositions Eqs.~(\ref{eq:summary2},
\ref{eq:summary3}) apply also here -- we simply need to 
adapt the potential $\bf V$ that occurs in the quantity $\bar{\bf \tau}^{nl}$, see  Eq.~(\ref{eq:tau-def}). 
This can be achieved by constructing the relevant Hamiltonian.
 We follow Refs.~\cite{Bern1,Bern2,Bern4,SchweizerHA}, which use a two--channel
formalism. An alternative method, based on  a one--channel framework, will be invoked
in the description of pionic hydrogen later in this report.

The Lagrangian of the system contains the non--relativistic charged and 
neutral pion fields $\pi_\pm(x),\pi_0(x)$ and the electromagnetic field
$A_\mu(x)$. At the accuracy we are working, it suffices to retain only the 
following terms (cf. Eq.~(\ref{eq:LNRA})),
\eq\label{eq:LNR_pipi}
{\cal L}_{NR}&\!=\!&-\frac{1}{4}\,F_{\mu\nu}F^{\mu\nu}+
\sum_\pm\pi_\pm^\dagger\biggl(iD_t-M_\pi+\frac{{\bf D}^2}{2M_\pi}
+\frac{{\bf D}^4}{8M_\pi^3}
+\cdots 
\nonumber\\[2mm]
&&\hspace{-1cm}\mp\,\, \frac{e}{6}\,\langle r_\pi^2\rangle
({\bf D}{\bf E}-{\bf E}{\bf D})+\cdots \biggr)\pi_\pm
+\pi_0^\dagger\biggl(i\partial_t-M_{\pi^0}+\frac{\triangle^2}{2M_{\pi^0}}
+\frac{\triangle^4}{8M_{\pi^0}^3}+\cdots\biggr)\pi_0
\nonumber\\[2mm]
&&\hspace{-1cm}+c_1\pi_+^\dagger\pi_-^\dagger\pi_+\pi_-
+c_2(\pi_+^\dagger\pi_-^\dagger\pi_0^2+\mbox{h.c.})
+c_3(\pi_0^\dagger)^2\pi_0^2+\cdots\, ,
\en
where $c_1,c_2,c_3$ denote  non--relativistic 
four--pion couplings, and $\langle r_\pi^2\rangle$ is the
charge radius of the pion.
 As is shown in Ref.~\cite{SchweizerHA}, the second--order
derivative term retained in Ref.~\cite{Bern1,Bern4} is in fact redundant and
can be eliminated by using the equations of motion. This procedure, which merely
amounts to choosing the coupling constant $c_4=0$ in Refs.~\cite{Bern1,Bern4},
does not affect any of observable quantities.

\begin{sloppypar}
Without performing any calculation, the electromagnetic energy shift in the
bound state with  angular momentum $l$ and  principal quantum
number $n$ up to  terms of order $\alpha^5$ 
can be directly read off from Eq.~(\ref{eq:Eem}). One finds~\cite{SchweizerHA}
\end{sloppypar}

\vspace*{-.3cm}

\eq\label{eq:Eem_pipi}
\Delta E^{\rm em}_{nl}=\alpha^4M_\pi\biggl(\frac{\delta_{l0}}{8n^3}
+\frac{11}{64n^4}-\frac{1}{2n^3(2l+1)}\biggr)
+\frac{\alpha^4M_\pi^3\langle r_\pi^2\rangle}{6n^3}\,\delta_{l0}
\, .
\en
\begin{sloppypar}
Further, the contribution due to vacuum polarization $\Delta E_{nl}^{\rm vac}$
 is defined by Eq.~(\ref{eq:vacleading}) and is
 calculated in Ref.~\cite{Eiras}, see Eq.~(\ref{eq:vacsecond}).
\end{sloppypar}

In order to evaluate the strong shift of a given level, we return
to Eq.~(\ref{eq:summary1}) and perform the calculations
by explicitly assuming $e=0$ everywhere, except in the 
static Coulomb interaction
(this is a perfectly legitimate 
procedure up to  terms of order $\delta^5$). Suppressing everything
but four--pion local interactions and the relativistic mass insertions, 
 one has
\eq\label{eq:tau-iterations}
\bar{\mathbold{\tau}}^{n0}(E_n)=
\bar {\bf V}+\bar {\bf V}\bar{\bf G}_C^{n0}\bar {\bf V}
+\bar {\bf V}\bar{\bf G}_C^{n0}\bar {\bf V}\bar{\bf G}_C^{n0}\bar {\bf V}
+\bar {\bf V}\bar{\bf G}_C^{n0}\bar {\bf V}\bar{\bf G}_C^{n0}\bar {\bf V}
\bar{\bf G}_C^{n0}\bar {\bf V}
+\cdots\, ,
\en
where $\bar{\bf V}={\bf H}_S+{\bf H}_R$, and
\eq\label{eq:HSHR}
{\bf H}_S&=&-\int d^3{\bf x}\,\,\biggl\{c_1\pi_+^\dagger\pi_-^\dagger\pi_+\pi_-
+c_2(\pi_+^\dagger\pi_-^\dagger\pi_0^2+\mbox{h.c.})
+c_3(\pi_0^\dagger)^2\pi_0^2\biggr\}\, ,
\nonumber\\[2mm]
{\bf H}_R&=&-\int d^3{\bf x}\,\,
\biggl\{\sum_\pm\pi_\pm^\dagger\, \frac{\triangle^4}{8M_\pi^3}\,\pi_\pm
+\pi_0^\dagger\,\frac{\triangle^4}{8M_{\pi^0}^3}\,\pi_0\biggr\}\fs
\en
Note that at the accuracy we are working, three interactions in 
Eq.~(\ref{eq:tau-iterations}) are sufficient (each iteration is suppressed by
 one power of $\delta$).
The general  expression for the strong energy shift was already worked out in Eq.~(\ref{eq:key}).
 Here, it remains  to  
 simply calculate the pertinent matrix elements. 

\begin{figure}[t]
\begin{center}
\includegraphics[width=8.cm]{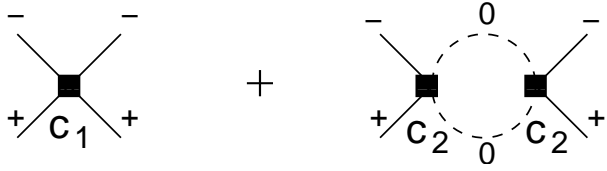}
\end{center}
\caption{Lowest--order contributions to the complex energy shift of 
the $\pi^+\pi^-$ atom, see Eq.~(\ref{eq:lowest}). The labels ``$+$'', 
``$-$'' and ``$0$'' (solid and dashed lines)
denote the charged and neutral pions, respectively.}
\label{fig:lowest}
\end{figure}

As an illustration, let us calculate the (complex)
level shift at  lowest order in $\delta$.
It is clear that one should obtain the DGBT
formula for the real and imaginary parts of the shift.
In order to demonstrate that this is indeed the case, 
it suffices to consider only those contributions
to the quantity $\bar{\mathbold{\tau}}^{n0}(z)$, which are shown
in Fig.~\ref{fig:lowest} (the tree diagram and the neutral pion loop). 
We obtain
\eq\label{eq:lowest}
(z-E_n)^{\rm str}=-|\tilde\Psi_{n0}(0)|^2 (c_1+2c_2^2 J_0(z))
+o(\delta^{7/2})\, ,
\en
where the quantity $J_0(z)$ is the loop integral 
defined in Eq.~(\ref{eq:building_block}), with
the mass $M$ replaced by $M_{\pi^0}$. This function has a branch point
at $z=2M_{\pi^0}$ and its imaginary part has the same sign as the imaginary
part of $z$ throughout the cut plane. Therefore, Eq.~(\ref{eq:lowest})
has no solution on the first Riemann sheet. On the other hand, if we
analytically continue $J_0(z)$ from the upper rim of the cut to the second
Riemann sheet, we find a zero at
\eq\label{eq:2Riemann}
{\rm Re}\, z&=&E_n-\frac{\alpha^3\mu_c^3}{\pi n^3}\,c_1+\cdots,\quad\quad
\nonumber\\[2mm]
{\rm Im}\, z&=&-\frac{\alpha^3\mu_c^3}{\pi n^3}\,
\frac{M_{\pi^0}\rho_n^{1/2}}{2\pi}\, c_2^2
+\cdots,
\en
where $\rho_n=M_{\pi^0}(E_n-2M_{\pi^0})$. Further,  matching
in the isospin limit yields
\eq\label{eq:m_iso}
3M_\pi^2c_1&=&4\pi(2a_0+a_2)+\cdots\, ,
\nonumber\\[2mm]
3M_\pi^2c_2&=&4\pi(a_2-a_0)+\cdots\, ,
\nonumber\\[2mm]
3M_\pi^2c_3&=&2\pi(a_0+2a_2)+\cdots\, ,
\en
where the ellipses stand for terms of order $\alpha$ and $(m_d-m_u)^2$.
Using  $|\tilde\Psi_{n0}(0)|^2=\alpha^3M_\pi^3/(8\pi n^3)$,
we finally conclude that i) Eq.~(\ref{eq:2Riemann})
reproduces the leading order expressions Eqs.~(\ref{eq:DeserGamma}) and
(\ref{eq:DE}),
 and  ii)  the bound--state pole
is indeed located on the second Riemann sheet.  
Note  that the real and imaginary parts of the energy shift
 are of order $\delta^3$ and $\delta^{7/2}$, respectively.

Calculations up to and including first non--leading order in $\delta$ are 
straightforward and proceed in  complete analogy to the case of 
scalar QED, considered earlier -- details may be found 
 in Refs.~\cite{Bern1,Bern2,Bern4,SchweizerHA}. The result is 
\eq\label{Gamma_E-updated1}
\Delta E_n^{\rm str} = -\frac{\alpha^3 M_\pi}{n^3} 
{\cal A}_c (1 + K'_n)+o(\delta^4) \, ,&&\,
\Gamma_n =\frac{2}{9 n^3}\,\alpha^3 p^\star_n {\cal A}_x^2(1+K_n) 
+o(\delta^{9/2})\, ,
\nonumber\\
&&
\en 
with
\eq\label{eq:KK} 
K_n&=&\frac{\Delta_\pi}{9M_\pi^2}\, 
(a_0+2a_2)^2 - \frac{\alpha}{3}\,s_n(\alpha)\,(2a_0+a_2)\, , 
\nonumber\\[2mm] 
K'_n&=&  - \frac{\alpha}{6}\,s_n(\alpha)\, 
(2a_0+a_2)\, , 
\nonumber\\[2mm] 
p^\star_n &=& \lambda^{1/2}(E_n^2,M_{\pi^0}^2,M_{\pi^0}^2)/(2E_n)\, 
,\quad \Delta_\pi=M_\pi^2-M_{\pi^0}^2\co
\en 
where  $s_n(\alpha)$ is given in Eq.~(\ref{eq:bargc}).
The quantities
${\cal A}_x$ and ${\cal A}_c$ coincide with the real parts of the relativistic 
threshold amplitudes in the channels
$\pi^+\pi^- \to \pi^0\pi^0$ and $\pi^+\pi^- \to \pi^+\pi^-$, respectively,
up to kinematic normalization factors. The precise relations are
 given in
appendix \ref{app:notations}. With the normalization chosen there, one has
\eq\label{AApr_exp}
{\cal A}_x = a_0 - a_2 + \delta{\cal A}_x + o(\delta) \, ,\quad 
{\cal A}_c = \frac{1}{6}\, (2 a_0 + a_2)  
+ \delta{\cal A}_c + o(\delta) \, ,
\en 
where $\delta{\cal A}_x$ and $\delta{\cal A}_c$ denote isospin breaking corrections
of order $\delta$. 
Finally, including vacuum polarization correction in the above expressions amounts to
$K_n\to K_n+\delta_n^{\rm vac}$ and $K_n'\to K_n'+\delta_n^{\rm vac}$, where 
$\delta_n^{\rm vac}$ is defined by Eq.~(\ref{eq:deltavp}).

For completeness, we also display the strong shift of
the $P$-wave states at leading order~\cite{SchweizerHA},
\bea\label{eq:pwave_pipi}
\Delta E^{\rm str}_{n1}=-\frac{(n^2-1)}{8n^5}\,\alpha^5 M_\pi^3 a_1^1\, ,
\en
where $a_1^1$ denotes the $P$-wave $\pi\pi$ scattering length.
The strong decay into $\pi^0\pi^0$ final states from  
$P$-wave states is forbidden by  $C$-invariance.

The equations (\ref{Gamma_E-updated1}), (\ref{eq:KK}), (\ref{AApr_exp}),
(\ref{eq:pwave_pipi})
 provide the framework for the numerical analysis of  pionium 
observables.

\subsection{DGBT formula   and  numerical analysis}
The relations
 Eqs.~(\ref{Gamma_E-updated1}), (\ref{eq:KK}) and (\ref{AApr_exp}) can be 
rewritten as follows,
\eq\label{Master_vac}
\hspace{-1cm}\Delta E^{\rm str}_n
  &=&-\frac{\alpha^3 M_\pi}{6n^3} \, (2a_0+a_2) \, 
\left(1 + \delta_n' \right)+o(\delta^4)\, ,\,
\delta_n' = \frac{6\, \delta{\cal A}_c}{2 a_0 + a_2} + K_n'\, , 
\nonumber\\[2mm]
\hspace{-1cm}\Gamma_n &=& \frac{2}{9n^3} \, \alpha^3 \, 
p_n^\star(a_0-a_2)^2 \left(1 + \delta_n\right)+o(\delta^{9/2}) \, ,\,
\delta_n = \frac{2\, \delta{\cal A}_x}{a_0 - a_2} + K_n\fs\nnnl
\en
At this stage one may invoke ChPT and evaluate the quantities
$\delta{\cal A}_x$ and $\delta{\cal A}_c$. 
 This procedure  leads to
a systematic calculation of the isospin breaking corrections.

We now demonstrate how the calculations at this 
concluding step are carried out. 
In general, the following representation for the quantities
$\delta{\cal A}_x$ and $\delta{\cal A}_c$ holds in ChPT,
\eq\label{chiral-A}
\delta{\cal A}_x= h_1(m_d-m_u)^2 + h_2\alpha, \quad
\delta{\cal A}_c= h_1'(m_d-m_u)^2 + h_2'\alpha,
\en
where $h_i,h_i'$ are  functions of the quark mass $\hat m$ and
of the RG--invariant scale of QCD. 
According to our definition of the isospin limit,
the parameter $\hat m$ is adjusted so that in this limit, the pion mass
is equal to the charged pion mass $M_\pi$.

Consider, for instance, the amplitude ${\cal A}_x$ at tree level in ChPT,
\eq\label{A-leading}
{\cal A}_x=\frac{3}{32\pi F^2}(4 M_\pi^2-M_{\pi^0}^2) 
+ O(p^4,e^2p^2)\, .
\en
In the isospin symmetry limit this expression reduces to
\eq
a_0-a_2=\frac{9M_\pi^2}{32\pi F^2} +O(p^4)\, .
\en
Comparing this with Eq.~(\ref{A-leading}), we find
\eq\label{eq:bulk}
{\cal A}_x = a_0-a_2 +\frac{3\Delta_\pi}{32 \pi F^2}
+O(p^4,e^2p^2)\, .
\en
 From this result, we may read off the coefficient  $h_2$ at
 leading order in the chiral expansion,
\eq
h_2= \frac{3 (M_\pi^2-M_{\pi^0}^2)}{32 \alpha\pi F^2}
+O(\hat{m})\, .
\en
[To be precise, the first term on the right--hand side of this equation
should be evaluated at $\alpha = 0$. To ease notation, we omit this
request here and in the following]. On the other hand, the above
calculation is not accurate enough to determine $h_1$  at leading
order, because for this purpose, the amplitude is needed at order $p^4$.
This procedure may obviously be carried out order by order in the chiral
expansion  -- all that is needed is the chiral expansion of the 
scattering amplitude at threshold, at $m_u\neq m_d$, $\alpha\neq 0$.
 As a result of this, the quantities $h_i$ are represented  as a
 power series in the quark mass $\hat{m}$ (up to logarithms). 

\begin{sloppypar}
The evaluation of  the amplitude for $\pi^+\pi^-\rightarrow\pi^0\pi^0$
was  carried out  at $O(p^4,e^2p^2)$ 
in Refs.~\cite{KU,Bern2,Bern4}, and a detailed 
numerical analysis of  the pionium decay width was performed
in Refs.~\cite{Bern2,Bern4}. This analysis results in
the following value for the ground-state correction,
\end{sloppypar}
\eq\label{eps-K}
\delta{\cal A}_x=(0.61\pm 0.16)\times 10^{-2}\, ,\quad\quad
K_1=(1.15\pm 0.03)\times 10^{-2}\, .
\en
Using input scattering lengths
from Refs.~\cite{Colangelo:2000jc,Colangelo:2001df}, we obtain 
 for the isospin breaking correction  $\delta_1$ in Eq.~(\ref{Master_vac}) and
 for the
 lifetime of the ground state,
\eq\label{tau}
\tau= \Gamma_{1}^{-1}=(2.9\pm 0.1)\times10^{-15}~{\mbox s}\, ,\quad\quad
\delta_1=(5.8\pm 1.2)\times 10^{-2} \, .
\en 
We note that $\delta_1$  amounts to a $6\%$  correction to the 
leading--order DGBT formula~\cite{Deser}.
 On the other hand, the preliminary result of DIRAC, displayed
in Eq.~(\ref{eq:tau_pionium}), corresponds to the following
value of the difference $a_0-a_2$,
\eq\label{a0a2_diff}
a_0  -  a_2  =  0.264~^{+0.033}_{-0.020} \,\, M_{\pi}^{-1} \,.
\en 
In Ref.~\cite{SchweizerHA}, the energy shift
of the ground state  of  pionium was worked out in an analogous manner
 (see also Ref.~\cite{Nehme}), with the result
\eq\label{correctionE0}
\Delta E_1^{\rm str}=(-3.8\pm 0.1)\, \mbox{eV}\co
\quad
\delta_1'= (6.2\pm 1.2)\times 10^{-2} \, .
\en
As was discussed in the previous sections,  
the width and strong energy shift 
are modified by vacuum polarization effects,
$\delta_n \rightarrow \delta_n + \delta_n^{\rm vac}$, 
$\delta_n' \rightarrow \delta_n' + \delta_n^{\rm vac}$,
where $\delta_n^{\rm vac}$ is defined in Eq.~(\ref{eq:deltavp}). 
For example, in the ground  state, the correction amounts to
$\delta_1^{\rm vac}=0.31 \cdot 10^{-2}$ \cite{Eiras}. 
We  conclude that it is safe to neglect, in this system, the  $\delta_n^{\rm vac}$
altogether:
the uncertainties in $\delta_n$ and $\delta_n'$ are larger than
$\delta^{\rm vac}_n$ itself. 

\begin{table}[t]
\begin{center}
\begin{tabular}{|l|r|r|r|r|}\hline
&&&&\\[-3.5mm]
& $\Delta E^{\rm em}_{nl}$ [eV]
& $\Delta E^{\rm vac}_{nl}$    [eV] 
& $\Delta E^{\rm str}_{nl}$ [eV]
& $10^{15}\tau_{nl}$ [s]\\[1mm] 
\hline
&&&&\\[-3.5mm]
 $n$=1, $l$=0&$-0.065$ 
 &$-0.942$&$-3.8\pm 0.1$ &$2.9\pm0.1$\\[1mm]
  $n$=2, $l$=0&$-0.012$ 
 &$-0.111$&$-0.47\pm 0.01$&$23.3\pm0.7$\\[1mm]
  $n$=2, $l$=1&$-0.004$&$-0.004$ 
 &$\simeq-1\cdot10^{-6}$&$\simeq1.2\cdot10^{4}$\\[1mm]
\hline
\end{tabular}
\end{center}
\caption{Numerical values for the energy shift and the lifetime 
of the $\pi^+\pi^-$ atom, taken from Ref.~\cite{SchweizerHA}.
The quantity
$\Delta E_{n0}^{\rm str}$ stands for $\Delta E_n^{\rm str}$, and
$\tau_{n0}=(\Gamma_n)^{-1}$. 
\label{tab:numerics_pi}}
 \medskip
\end{table}

Finally, in Table~\ref{tab:numerics_pi} we list
different contributions to the
 energy shift and the lifetime for the first few levels in  pionium \cite{SchweizerHA}, 
calculated by using
the input scattering lengths from 
Refs.~\cite{Colangelo:2000jc,Colangelo:2001df}.
The calculations for the $S$-wave states were carried out
at next--to--leading order in isospin
symmetry breaking. The bulk of the
uncertainty in these quantities is due to the
uncertainties in the pertinent scattering lengths. The lifetime of the
$2p$-state is calculated at leading order only and is determined
by the $2p-1s$ radiative transition~\cite{Ovsyannikov2}.   
Finally, the theoretical value for the $2p-2s$ energy splitting is given 
by~\cite{SchweizerHA}   
\eq
  \hspace{-.5cm}\Delta E^{2s-2p} = \Delta E_{2}^{\rm str}+
\Delta E_{20}^{\rm  em}-\Delta E_{21}^{\rm em}+
\Delta E_{20}^{\rm vac}-\Delta E_{21}^{\rm vac}
=-0.59\pm 0.01\,{\rm eV}.
\label{DeltaE2pi0}
\en
The uncertainty displayed is the one in $\Delta E_{2}^{\rm str}$ only. 
 To the accuracy we are working, we may neglect the strong shift in the 
$2p$ state, because it is of order $\alpha^5$.

\subsection{Two--photon decay of pionium}
\label{subsec:twophotondecay}

The  decay width of the pionium ground state in two photons is given by
\eq\label{eq:pionium-2gamma}  
\Gamma_1^{2\gamma}&=&\frac{M_\pi^5}{4}\alpha^5\, 
|A(4M_\pi^2,-M_\pi^2,-M_\pi^2)|^2 +o(\delta^5)\, ,
\en
where the invariant amplitude $A(s,t,u)$ for the process 
$\gamma\gamma\to\pi^+\pi^-$ is defined in Eq.~(2.4) of 
Ref.~\cite{Gasser:2006qa}. This amplitude was evaluated at NLO in chiral $SU(3)\times SU(3)$ by Bijnens and 
Cornet~\cite{bijnenscornet}, and to two loops in $SU(2)\times SU(2)$ in Refs.~\cite{buergi,Gasser:2006qa}. 
 Here, we use the $SU(2)\times SU(2)$ version displayed in
 Eq.~(5.1) of Ref.~\cite{Gasser:2006qa},
\eq\label{eq:pipigammagamma}
M_\pi^2\,A(4M_\pi^2,-M_\pi^2,-M_\pi^2)
=1+\frac{2M_\pi^2}{F_\pi^2}\,\biggl(\bar G_\pi(4M_\pi^2)
+\frac{\bar l_6-\bar l_5}{48\pi^2}\biggr)+O(M_\pi^4)\, ,
\en
where
\eq\label{eq:Gpi}
\bar G_\pi(s)&=&-\frac{1}{16\pi^2}\,\biggl(1+\frac{2M_\pi^2}{s}\int_0^1
\frac{dx}{x}\,\ln(1-\frac{s}{M_\pi^2}\,x(1-x))\biggr)\, ,
\nnnl
\bar G_\pi(4M_\pi^2)&=&\frac{\pi^2-4}{64 \pi^2}\fs
\en
The terms of order $M_\pi^4$ in Eq.~\eqref{eq:pipigammagamma} 
denote two--loop contributions, and  $\bar l_5,\bar l_6$ 
are $O(p^4)$ chiral LECs~\cite{GL_ChPT}. These are related to the 
pion polarizabilities ${\alpha}_{\pi}$ and
${\beta}_{\pi}$,
\eq
{\alpha}_{\pi}-{\beta}_{\pi}=
\frac{\alpha(\bar l_6-\bar l_5)}{24\pi^2F_\pi^2M_\pi}\,
+O(M_\pi)\fs
\en

 The two--photon decay
width of  pionium was evaluated already earlier~\cite{Uretsky,pioniumgg}.
Here, we comment on the  expression given by Hammer and Ng~\cite{pioniumgg}, 
who were the first to incorporate contributions from the pion polarizabilities.
Their expression does not contain the contribution from $\bar G_\pi$, which is, in the chiral expansion, of the same order as the one 
from the polarizabilities.
 Numerically, this difference amounts to a large  effect. Using 
$\bar l_6-\bar l_5=3.0\pm 0.3$, which corresponds to $(\alpha-\beta)_{\pi}
=6.0\cdot 10^{-4}\,\mbox{fm}^3$~\cite{Gasser:2006qa}, we obtain from Eq.~(\ref{eq:pipigammagamma}) 
\eq
M_\pi^2\,A(4M_\pi^2,-M_\pi^2,-M_\pi^2)=1+4.2\cdot 10^{-2}+2.9\cdot 10^{-2}
+O(M_\pi^4)\, ,
\en
i.e., the contribution from $\bar G_\pi(4M_\pi^2)$ is larger than the one from
the pion polarizabilities. Note also that in Ref.~\cite{pioniumgg} much larger
values of the polarizabilities 
$\alpha_{\pi}=-\beta_{\pi}=(6.8\pm 1.4\pm 1.2)\cdot 10^{-4}\,\mbox{fm}^3$ have been used,
tending to mask the absence of $\bar G_\pi(4M_\pi^2)$.

The reason why the contribution from $\bar G_\pi(s)$ was missed in
Ref.~\cite{pioniumgg} is the following. In that article,
the annihilation amplitude $\gamma\gamma\to\pi^+\pi^-$ is obtained from the
Compton amplitude through analytic continuation. Moreover, the authors 
expand the amplitude in small photon momenta and retain only terms which are  at most 
quadratic in this expansion.
 In the annihilation channel, however,
 the photon momenta are of the order of the pion mass. Therefore, the analytic
continuation of the {\em truncated} Compton amplitude can be justified 
if and only if the expansion of this amplitude in photon momenta has a 
convergence radius of the order of the heavy scale in ChPT. In this case, the 
coefficients of such an expansion are regular in the chiral limit. 
However, as can be seen from Eq.~(\ref{eq:Gpi}), this is not 
what happens: at the Compton threshold, the expansion
is carried out in powers of 
$s/M_\pi^2$ (with $s\to t$).
 The coefficients
of the higher--order terms in the expansion of $\bar G(s)$ 
become more and more singular in the chiral limit.  
 As a result of this, they contribute at the same order at $s=4M_\pi^2$
and must thus be kept.

\bigskip

We summarize this section with the observation 
that the $\pi^+\pi^-$ atom is now completely understood in the framework of 
QCD+QED,
 on a conceptual as well as on a quantitative level. 
For  bound--state observables, the theoretical predictions
are made  with  percent accuracy. If the experiments
are performed with the planned accuracy, a precise measurement of the $\pi\pi$
scattering lengths is indeed   feasible.

%%%%%%%%%%%%%%%%%%%%%%%%%%%%%%%%%%%%%%%%%%%%%%%%%%%%%%%%%%%%%%%%%%%%
\setcounter{equation}{0}
\section{$\pi K$ atom} 
\label{sec:piK}

As far as the description of the bound state within the non--relativistic
effective theory is concerned,
the $\pi K$ atom problem is completely analogous to the case of pionium 
considered in the previous section. For this reason, we shall display 
only the final results here. Details may be found in 
 Ref.~\cite{SchweizerHA}.

For definiteness, we consider the bound state of $K^+$ and $\pi^-$.
The strong energy shifts and widths of the $\pi K$ atom are given by
 expressions  similar to Eq.~(\ref{Gamma_E-updated1}),
\eq\label{eq:EG-piK}  
\Delta E^{\rm str}_{n} &=&
  -\frac{2\alpha^3\mu_c^2}{n^3}\,\mathcal{A}_c\left(1+K_n'\right)+o(\delta^4)\, ,
\nonumber\\[2mm]
  \Gamma_{n}&=&
  \frac{8\alpha^3\mu_c^2}{n^3}\,p^\star_n\mathcal{A}_x^2\left(1+K_n\right)
+o(\delta^{9/2})\, ,
\en
where ${\cal A}_x$ and ${\cal A}_c$ are now related to the relativistic 
threshold amplitudes
for the scattering processes $\pi^- K^+ \rightarrow \pi^0 K^0$ and 
$\pi^- K^+ \rightarrow \pi^- K^+$, respectively,
 see appendix~\ref{app:notations}. In particular,
\eq\label{eq:piK_iso}
&&\mathcal{A}_x=a^-_0+\delta{\cal A}_x+o{(\delta)}\, ,\quad\quad
\mathcal{A}_c=a_0^+ + a_0^- +\delta{\cal A}_c+o{(\delta)}\, ,
\nonumber\\[2mm]
&&a^+_0=\frac{a^{1/2}+2a^{3/2}}{3}\, ,\quad
a^-_0=\frac{a^{1/2}-a^{3/2}}{3}\, ,
\en
where  $a^{1/2},a^{3/2}$ are the $\pi K$ scattering lengths with  total
isospin $I=1/2$ and $I=3/2$, respectively, 
and $\delta{\cal A}_x,\delta{\cal A}_c$
 denote isospin breaking terms of order $\delta$.
 Furthermore,
\eq\label{eq:KK_piK}
 K_n &=& \frac{M_\pi\Delta_K+M_K\Delta_\pi}{\msigma}
(a^+_0)^2 - 2 \alpha\mu_c ({a_0^++a_0^-}) s_n(\alpha)\, ,
\nonumber\\[2mm]
K_n' &=& -\alpha{\mu_c} ({a_0^{+} + a_0^{-}})s_n(\alpha)\, ,
\en
where $\Delta_K=M_K^2-M_{K^0}^2$, $\msigma=M_\pi+M_K$,
$\mu_c$ is the reduced mass of the $\pi^-K^+$ system, and
\eq\label{eq:kin_piK}
E_n=\msigma-\frac{\alpha^2\mu_c}{2n^2}\, ,\quad\quad
p^\star_n = \lambda^{1/2}(E_n^2,M_{\pi^0}^2,M_{K^0}^2)/(2 E_n)\, .
\en

The equations (\ref{eq:EG-piK}) can be rewritten as
\eq\label{eq:corr_piK}
  \Delta E_n^{\rm str}&=&  - \frac{2\alpha^3\mu_c^2}{n^3} ({a_0^++a_0^-})
\left(1+\delta_n'\right)+o{(\delta^4)}, 
\quad\quad \delta_n' = \frac{\delta{\cal A}_c}{a_0^++a_0^-}+ K_n'\, ,
\nonumber\\[2mm]
  \Gamma_n &=& \frac{8\alpha^3\mu_c^2}{n^3}\, p_n^\star(a_0^-)^2\, 
 (1+\delta_n)+o(\delta^{9/2}), \quad \quad 
  \delta_n = \frac{2\,\delta{\cal A}_x}{a_0^-}+K_n\, .
\en
The electromagnetic shift is given by
\eq\label{DeltaEem}
\Delta E^{\rm em}_{nl}&=&\frac{\alpha^4\mu_c}{n^3}
\left(1-\frac{3\mu_c}{\msigma}\right)\left[\frac{3}{8n}
-\frac{1}{2l+1}\right]+\frac{4\alpha^4\mu_c^3\lambda}{n^3}\delta_{l0}
\nonumber\\[2mm]
&+&\frac{\alpha^4\mu_c^2}{\msigma}\left[\frac{1}{n^3}\delta_{l0}
+\frac{1}{n^4}-\frac{3}{n^3(2l+1)}\right]
\, ,
\en
where $\lambda=\frac{1}{6}\,(\langle r_\pi^2\rangle+\langle r_K^2\rangle)$. 
Vacuum polarization introduces a small change in 
Eq.~(\ref{eq:corr_piK}): $\delta_n\to \delta_n+\delta_n^{\rm vac}$,
 $\delta_n'\to \delta_n'+\delta_n^{\rm vac}$. As in  pionium, this 
correction is very small: for example, in the ground state it amounts to
$\delta_1^{\rm vac}=0.45\cdot 10^{-2}$~\cite{Eiras}. In the numerical
analysis given below this tiny contribution is neglected.

Finally, 
the strong shift in the $P$-wave state and  the $P$-wave decay width into 
$\pi^0 K^0$ at leading order are given by~\cite{SchweizerHA}
\eq\label{eq:Pwave}  
\Delta E^{\rm str}_{n1} &=&
  -\frac{2(n^2-1)}{n^5}\,\alpha^5\mu_c^4\left(a_1^++a_1^-\right)\, ,
\nonumber\\[2mm]
  \Gamma_{n1}^{\pi^0 K^0} &=& \frac{8(n^2-1)}{n^5}\, 
  \alpha^5\mu_c^4\,{(p_n^\star)}^3{(a_1^-)}^2,
\en
where $a_1^\pm$ denote $P$-wave $\pi K$ scattering lengths.

A comprehensive numerical analysis of the $\pi K$ atom observables
in ChPT has been carried out in Ref.~\cite{SchweizerHA}.
Below we give a short summary of the results of this analysis.
The isospin breaking corrections to the $\pi K$ threshold amplitudes 
Eq.~(\ref{eq:piK_iso}) have been worked out by several 
authors~\cite{Nehme:2001wa,Kubis:2001bx,Kubis:2001ij,Nehme:2001wf,Kubis_diss}
at $O(p^4,e^2p^2)$. 
Whereas the analytic expressions for $\delta{\cal A}_x$ and $\delta{\cal A}_c$ 
obtained in 
Refs.~\cite{Nehme:2001wa,Kubis:2001bx,Kubis:2001ij,Nehme:2001wf,Kubis_diss} 
are not identical,
the numerical values agree within the uncertainties. 
In the calculations of the
the $\pi K$ atom observables carried out
in Ref.~\cite{SchweizerHA}, the values 
\eq
\delta{\cal A}_x=(0.1\pm 0.1) \times 
10^{-2}M_\pi^{-1}\, ,\quad\quad \delta{\cal A}_c=(0.1\pm 0.3)\times 
10^{-2}M_\pi^{-1}
\en
were used.
Table~\ref{tab:delta} contains the final results of these calculations for
the isospin breaking corrections to the strong shift and width
of the $ns$ state with $n=1,2$.
\begin{table}[t]
\begin{center} 
\begin{tabular}{|c|c|c|c|}\hline
&&&\\[-3.5mm]
$10^2\delta_1$ & $10^2\delta_2$ & $10^2\delta_1'$ 
& $10^2\delta_2'$\\[1mm]
\hline
&&&\\[-3.5mm]
$4.0\pm2.2$&$3.8\pm2.2$ &$1.7\pm2.2$ &$1.5\pm2.2$ \\[1mm] 
\hline
\end{tabular}
\end{center}
\caption{Isospin breaking corrections in the DGBT--type formulae
for the $\pi K$ atom  in  $ns$ states, taken from Ref.~\cite{SchweizerHA}. See 
Eq.~(\ref{eq:corr_piK}) for the definition of $\delta_i$ and $\delta_i'$. \label{tab:delta}}
\end{table}

Next, we consider the calculation of
the lifetime of the $\pi K$ atom ground state, given the
input for the $S$-wave $\pi K$ scattering lengths. 
As discussed in section~\ref{sec:fundamental},  different values
for scattering lengths are available in the literature. In the analysis carried out in 
Ref.~\cite{SchweizerHA}, the results of $O(p^4)$ 
calculations in ChPT~\cite{BKM-Kpi,BKM-Kpi1}, as well as the solutions
of the Roy--Steiner equations~\cite{piKBuettiker} have been used as 
 an input. The prediction for the lifetime is\footnote{Schweizer \cite{SchweizerHA} does not quote an uncertainty
 for the lifetime which result from ChPT at order $p^4$. For completeness, we re-evaluated the
first line of Eq.~(\ref{eq:lifetime_piK}), using the value 
$a_0^-=0.079\pm 0.001~M_\pi^{-1}$, given in Ref.~\cite{SchweizerHA}.}
\eq\label{eq:lifetime_piK}
&&\tau=(4.8\pm 0.2)\cdot 10^{-15}~{\rm s}\hspace*{2.9cm}
\mbox{ChPT \,\,\,$O(p^4)$}\,,
\nonumber\\[2mm] 
&&\tau=(3.7\pm 0.4)\cdot 10^{-15}~{\rm s}\hspace*{1.5cm}
\mbox{Roy-Steiner equations}~\cite{piKBuettiker}\fs
\en
 Finally, 
in Table~\ref{tab:numericsK}
 we display various contributions to the energy shift and width of the 
low--lying levels of the $\pi K$ atom, taken from Schweizer \cite{SchweizerHA}. Moreover, because
 the strong shift in the  $P$-wave is small, one obtains for the $2s-2p$ level
splitting \cite{SchweizerHA}
\eq\label{DeltaE2}
  \Delta E^{2s-2p} = \Delta E_{2}^{\rm str}+\Delta E_{20}^{\rm em}-\Delta
  E_{21}^{\rm em}+\Delta E_{20}^{\rm vac}-\Delta E_{21}^{\rm vac}
=-1.4 \pm 0.1\,{\rm eV}\, .
\en

\begin{table}[t]
\begin{center}
\begin{tabular}{|l|r|r|r|r|}
\hline
&&&&\\[-3.5mm]
&$\Delta E^{\rm em}_{nl}$ [eV]
&$\Delta E^{\rm vac}_{nl}$ [eV]&
$\Delta E^{\rm str}_{nl}$ [eV]&
$10^{15}\tau_{nl}$ [s]\\[1mm]
\hline
&&&&\\[-3.5mm]
$n$=1, $l$=0&$-0.095$&$-2.56$&$-9.0\pm1.1$&$3.7\pm0.4$\\[1mm]
$n$=2, $l$=0&$-0.019$&$-0.29$&$-1.1\pm0.1$&$29.4\pm3.3$\\[1mm]
$n$=2, $l$=1&$-0.006$&$-0.02$&$\simeq -3\times 10^{-6}$
&$\simeq0.7\times 10^{4}$\\[1mm]
\hline
\end{tabular}
\end{center}
\caption{Numerical values for the energy shift and the lifetime 
of the $\pi K$ atom, taken from Ref.~\cite{SchweizerHA}.
The quantity
$\Delta E_{n0}^{\rm str}$ stands for $\Delta E_n^{\rm str}$, and
$\tau_{n0}=(\Gamma_n)^{-1}$. 
\label{tab:numericsK}}
\end{table}

We summarize that, as a  result of the investigation carried out in 
Ref.~\cite{SchweizerHA},  the 
next--to--leading order isospin breaking corrections to the observables 
of the $\pi K$ atom in the ground state are now known with a precision
that is comparable to the one in 
pionium. In our opinion, this is a completely satisfactory situation from the point of
view of the data analysis of  possible future experiment. We refer the reader back to 
 subsection \ref{subsec:pikatom} for  comments concerning the relevance 
of  measuring  $\pi K$  scattering lengths.

%%%%%%%%%%%%%%%%%%%%%%%%%%%%%%%%%%%%%%%%%%%%%%%%%%%%%%%%%%%%%%%%%%%%%%%%
\setcounter{equation}{0}
\section{Pionic hydrogen}
\label{sec:piH}

\subsection{Pionic hydrogen experiments at PSI} 

During the last decade, the Pionic Hydrogen Collaboration 
at PSI  performed high--precision 
measurements of strong interaction parameters of both, pionic 
hydrogen (\piH$\!$) and pionic 
deuterium 
(\pid$\!$)~\cite{Sigg:1995,Chatellard:1995,Sigg:1996qd,Chatellard:1997nw,Hauser:1998,Schroder:1999,Schroder:2001,R98,Gotta:2004rq,Gotta:2005,Gotta:2006}
 in order to extract independent information about 
the $S$-wave $\pi N$ scattering lengths. 
 In this section, we mainly discuss pionic hydrogen.

\begin{figure}[t]
\vspace*{.6cm}
\begin{center}
\includegraphics[width=7.cm,angle=0]{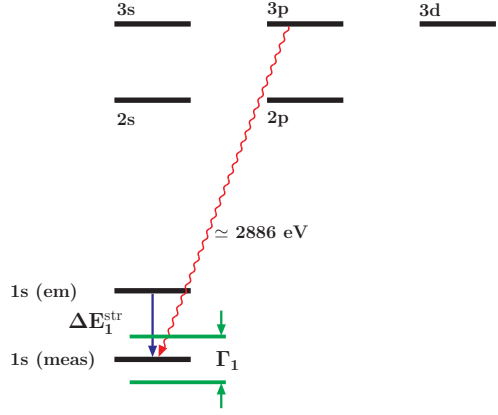}
\end{center}
\caption{The scheme of energy levels of \piH with 
the $3p-1s$ $X$-ray transition.
The displacement and the finite width of the ground state are explicitly
shown. }
\label{fig:piH_levels}
\end{figure}

This exotic atom is formed, when the 
kinetic energy of a negative pion is of the order of a few eV. 
The incoming pion is captured by the Coulomb field in highly excited 
states and a  de--excitation cascade starts, which proceeds through 
$X$-ray emission as well as through other (non--radiative) mechanisms. 
The atomic cascade ends in 
the ground state, which then decays, 
mainly in $n \gamma$ and $n \pi^0$ states.
The experimental setup uses the high--intensity low--energy pion beam 
$\pi E5$ at PSI.
The energy shifts are extracted from the 
measured $X$-ray lines, see Fig.~\ref{fig:piH_levels}.
In particular, the strong interaction shift in the ground state
  is obtained from
 the measured  mean transition energy $E_{3p-1s}$.

 The theoretical framework is analogous to the one summarized in 
subsection \ref{subsec:scalarsummary}, see also below.
  The quantity  $E_{3p-1s}$ is related
to the energy levels by
$E_{3p-1s}={\rm Re}\,(z_{31}-z_{10})$, where
\eq\label{eq:explicit}
z_{nl}=E_n+\Delta E_{nl},\quad
X_{nl}=\frac{1}{2(2l+1)}\,
\sum_{j=|l-\frac{1}{2}|}^{l+\frac{1}{2}}(2j+1)X_{nlj}\,\,;\,\,X=z,\Delta E\,.
\en
 Here, $E_n$ stands for the pure Coulomb energy,
$j=l\pm\frac{1}{2}$ denotes the total angular momentum of a given eigenstate,
 and $\Delta E_{nlj}$ is
the energy shift of a generic energy level $z_{nlj}$ 
of  \piH, labeled
by the quantum numbers $n,l,j$. 
Further, in analogy with Eqs.~(\ref{eq:summary1}),(\ref{eq:summary2}) and
 (\ref{eq:summary3}),
this energy shift is further split into an
 electromagnetic piece, a contribution from vacuum polarization and from the
strong shift,
\eq\label{eq:splitting}
\Delta E_{nlj}=\Delta E_{nlj}^{\rm em}+\Delta E_{nl}^{\rm vac}
+\delta_{l0}(\Delta E_{n}^{\rm str}-\frac{i}{2}\,\Gamma_{n})
+o(\delta^4)\, .
\en
The strong shift of the ground state is determined from
\eq \label{eq:strongshift}
\Delta E^{\rm str}_{1} = E_{3p-1s}^{\rm em}
+ E_{3p-1s}^{\rm vac}-  E_{3p-1s}\, ,
\en 
where $E_{3p-1s}^{\rm em}$ and $E_{3p-1s}^{\rm vac}$ are the 
(theoretical values for) the electromagnetic transition 
energy and the vacuum polarization contribution,
\eq\label{eq:strongshift1}
 E_{3p-1s}^{\rm em}=E_3-E_1+\Delta E_{31}^{\rm em}-\Delta E_{10}^{\rm em}\, ,
\quad
 E_{3p-1s}^{\rm vac}=\Delta E_{31}^{\rm vac}-\Delta E_{10}^{\rm vac}\fs
\en
Here, $\Delta E_{nl}^{\rm em}$ is defined through $\Delta E_{nlj}^{\rm em}$
according to Eq.~(\ref{eq:explicit}).
In addition, one makes  use of the fact  that the strong--interaction shift of the $3p$ state 
is suppressed by additional powers of $\alpha$ and is
therefore negligible (see, e.g., Ref.~\cite{Rasche:zg}).

The following values for the 
strong energy shift and width of the ground state have been reported back in 2001~\cite{Schroder:2001}, 
\eq\label{eq:schroder}
\Delta E_{1}^{\rm str}&=&-7.108\pm 0.013~\mbox{(stat)}
\pm 0.034~\mbox{(syst)}~{\rm eV}\, ,\nonumber\\[2mm]
\Gamma_{1}&=&0.868\pm 0.040~\mbox{(stat)}\pm 0.038~\mbox{(syst)}~{\rm eV}\, .
\en
One determines the 
$S$-wave $\pi N$ scattering lengths $a_{0+}^+$ and
$a_{0+}^-$ from
\eq\label{eq:Deser-piH}
\Delta E_{1}^{\rm str}&=&-2\alpha^3\mu_c^2(a_{0+}^++a_{0+}^-)(1+\delta'_1)+o(\delta^4)\, ,
\nonumber\\[2mm]
\Gamma_{1}&=&8\alpha^3\mu_c^2p_1^\star\biggl(1+\frac{1}{P}\biggr)[\,
a_{0+}^-(1+\delta_1)\, ]^2+o(\delta^{9/2})\, ,
\en
where $P$ denotes
the Panofsky ratio,
and $\mu_c$ is the reduced mass of the $\pi^- p$ system. Further,
$p_1^\star=\lambda^{1/2}(E_1^2,m_n^2,M_{\pi^0}^2)/(2E_1)$ is
the CM momentum of the $\pi^0 n$ pair after decay.

The quantities  $\delta'_1$ and $\delta_1$ 
stand for the isospin breaking corrections, which in ChPT are of order 
$\alpha$ or $m_d-m_u$. The
lowest--order DGBT formula has $\delta'_1=\delta_1=0$.
In the analysis of the data given in Ref.~\cite{Schroder:2001}, one has used
the values obtained in a potential scattering model~\cite{Sigg},
\eq\label{eq:Siggcorrections}
\delta'_1=(-2.1\pm 0.5)\cdot 10^{-2}\, ,\quad
\delta_1=(-1.3\pm 0.5)\cdot 10^{-2}\fs
\en
Recently, a new calculation of the isospin breaking corrections within
a potential model has been performed in Ref.~\cite{rascheetal}.
The pertinent scattering lengths still contain residual 
electromagnetic effects that cannot be removed with the potential 
model used in Ref.~\cite{rascheetal} (of course, the same is true for the analysis
of Ref.~\cite{Sigg}). We refer the reader to Ref.~\cite{rascheetal} for more
details. 

\begin{figure}[t]
\vspace*{.6cm}
\begin{center}
\includegraphics[width=12.cm,angle=0]{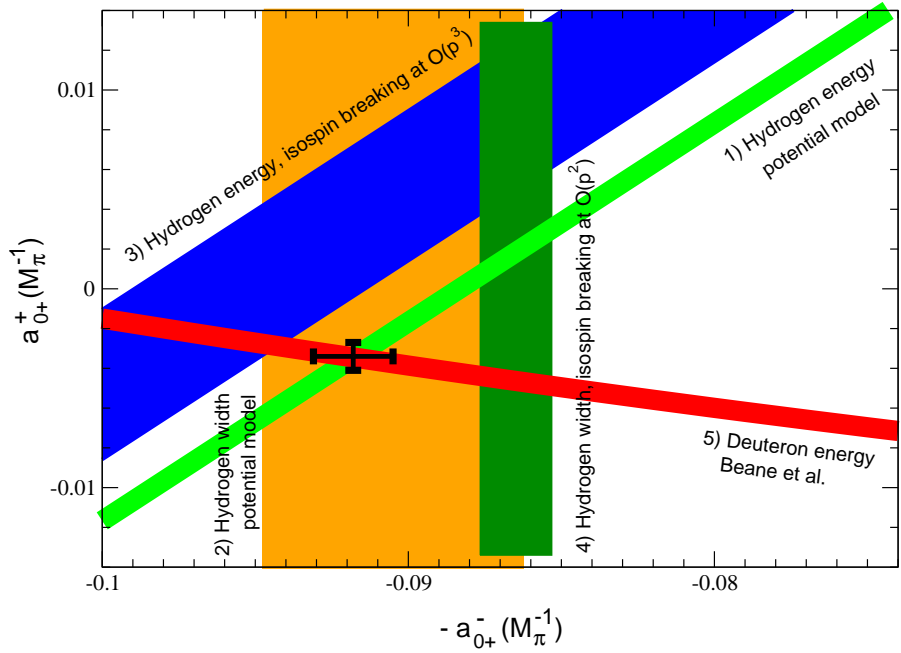}
\end{center}
\caption{Determination of the $\pi N$ $S$-wave scattering lengths from
the energy shift and width of \piH:
i) Old analysis (bands 1 and 2), which is using 
the data of Ref.~\cite{Schroder:2001} given in Eq.~(\ref{eq:schroder})
and applying the isospin breaking corrections, calculated in the potential 
model~\cite{Sigg}, see Eq.~(\ref{eq:Siggcorrections});
ii) New analysis (bands 3 and 4), which is using preliminary data 
of Ref.~\cite{Gotta:2006} given in Eq.~(\ref{eq:hyd_energywidth}),
 together with the isospin breaking corrections calculated in ChPT,
see Eqs.~(\ref{eq:Mojziscorrections}) and (\ref{eq:Zempcorrections}).
The scattering lengths evaluated \cite{Bernard2} 
from the \pid energy shift Eq.~(\ref{eq:deuteron}) 
 are shown in band 5. The cross indicates the solution Eq.~(\ref{eq:beaneetal}) for
 the scattering lengths in the old analysis.}
\label{fig:bands_new}
\end{figure}

The analysis of the experimental data proceeds as
follows. Equations~(\ref{eq:schroder}) and (\ref{eq:Deser-piH}) generate two
 constraints on the possible values of the scattering lengths
$a_{0+}^+$ and $a_{0+}^-$. They are displayed with the
two linear bands  1,2 in Fig.~\ref{fig:bands_new} (here, one assumes that $a_{0+}^-$ is positive).
 The widths of the bands 
 are determined by both, the
experimental error in $\Delta E_{1}^{\rm str}$ and in $\Gamma_{1}$, and by the
theoretical uncertainty in Eq.~(\ref{eq:Siggcorrections})\footnote{
We treat the uncertainties in the isospin breaking corrections Eq.~(\ref{eq:Siggcorrections}) 
as  systematic ones, combine them quadratically with the systematic uncertainty 
in the energy shift and width in Eq.~(\ref{eq:schroder}), and add to the result the statistical 
uncertainties in Eq.~(\ref{eq:schroder}) linearly.}.
In addition, in the same Figure we also plot
the band which emerges \cite{Bernard2} from the measurement of
the strong energy shift Eq.~\eqref{eq:deuteron} of  \pid$\!\!$, see comments in  subsection \ref{subsec:piHpiD}.

In the analysis carried out in Ref.~\cite{Bernard2}, 
all three bands happen to intersect in a very small domain, suggesting
a coherent overall picture. The $S$-wave $\pi N$ scattering lengths
can then be determined very well from the intersection domain of these
three bands:
\eq\label{eq:beaneetal}
\hspace*{-.3cm}&&a_{0+}^+=(-0.0034\pm 0.0007)M_\pi^{-1}\, ,\quad
a_{0+}^-=(0.0918\pm 0.0013) M_\pi^{-1}\, \quad\mbox{Ref.~\cite{Bernard2}}.
\nonumber\\
\hspace*{-.3cm}&&
\en
This picture, however, has undergone  substantial changes in the last few
 years, on the experimental as well as on the  theory side.
First, upgrading of the experiment has resulted
in an increased accuracy of the measured \piH
width, as well as in substantially decreasing its central value, 
compare Eqs.~(\ref{eq:schroder})   and  (\ref{eq:hyd_energywidth}).
 Second, as already stated in the introduction,
results like those given in Eq.~(\ref{eq:Siggcorrections}) cannot be trusted
{\it ab initio}: the potential model, which was used  to derive
these results, does not include the full content of isospin breaking in
QCD+QED (see also Ref.~\cite{rascheetal}). 
We will discuss this in a quantitative manner later in this section.
Here, we simply note that, if one applies
the isospin breaking corrections calculated in ChPT~\cite{Bern3,Mojzis,Zemp}
to the \piH energy and width given in Ref.~\cite{Gotta:2006} and 
Eq.~(\ref{eq:hyd_energywidth}), the central
values of the scattering lengths are shifted, see the bands 3 and 4 in the Figure.
There is now  no more  a common intersection area between the \piH and the 
\pid bands, hinting at an internal inconsistency of the theoretical 
methods  which were used to analyze the data.
The most radical proposal for circumventing the problem is to
exclude the \pid data from the
global analysis, thus avoiding the potentially largest source of the
systematic error which stems from a poor control of the
multiple--scattering series for the $\pi d$ scattering length.
It was argued that, if the planned 2\% accuracy for the width is achieved, 
it must be possible to
determine the $\pi N$ scattering lengths from \piH
alone~\cite{Gottaprecision}. We will investigate this possibility later in this section.

\subsection{Effective theory and counting rules}

The case of  \piH which is considered here, differs in
two aspect from  pionium and from the $\pi K$ atom: first and most importantly, it
decays with a $\simeq 40\%$ probability into the channel $n\gamma$,
where the CM momentum of the decay products is of order of the pion
mass. It is clear that, if one tries to include $n\gamma$
intermediate states explicitly, this will upset the whole
power--counting scheme of the non--relativistic approach, where $M_\pi$
plays the role of a hard scale~\cite{Zemp}. For this reason, we apply a method
already used in section~\ref{sec:boundstates} for the calculation of the 
two--photon decay 
width: we  construct a non--relativistic theory
in which the $n\gamma$ channel is ``integrated out'', and where its contribution
appears only through the low--energy effective couplings. 
One could then study \piH in the non--relativistic theory  
with two channels $\pi^-p$ and $\pi^0n$~\cite{Zemp}, which closely resembles
the pionium case. However, we choose a different strategy here. Namely,
in order to demonstrate the flexibility of the approach, 
we find it instructive to consider the inclusion of the $n\pi^0$ channel
into the list of ``shielded'' channels as well and work with a one--channel
theory. The final result must of course agree with the one obtained
in the two--channel setting~\cite{Zemp}.

The second generalization is related to the fact that
the nucleons are fermions. However,  the inclusion of 
particles with spin into the non--relativistic framework is nearly trivial
and proceeds straightforwardly.

In order to establish the Lagrangian of the system, one has to merely write 
down all possible operators with a minimal number of derivatives, which are
consistent with all symmetries (see also Refs.~\cite{Caswell,Kinoshita,Labelle_QED}). 
 The result is
\eq\label{eq:lagr_piN}
{\cal L}_{NR}
&=&
-\frac{1}{4}\,F_{\mu\nu}F^{\mu\nu}
+\psi^\dagger\biggl\{i{\cal D}_t-m_p+\frac{{\cal D}^2}{2m_p}
+\frac{{\cal D}^4}{8m_p^3}+\cdots
\nonumber\\[2mm]
&-&
c_p^F\,\frac{e{\mathbold{\sigma}}{\bf B}}{2m_p}
-c_p^D\,\frac{e({\cal D}{\bf E}-{\bf E}{\cal D})}{8m_p^2}
-c_p^S\,\frac{ie{\mathbold{\sigma}}({\cal D}\times{\bf E}-
{\bf E}\times{\cal D})}{8m_p^2}+\cdots\biggr\}\psi
\nonumber\\[2mm]
&+&
\sum_{\pm}\pi_\pm^\dagger\biggl\{ iD_t-M_\pi
+\frac{{\bf D}^2}{2M_\pi}+\frac{{\bf D}^4}{8M_\pi^3}+\cdots
\mp c^R\,\frac{e({\bf D}{\bf E}-{\bf E}{\bf D})}{6M_\pi^2}
+\cdots\biggr\}\pi_\pm
\nonumber\\[2mm]
&+&
g_1(\psi^\dagger\psi)(\pi_-^\dagger\pi_-)
+e_1\biggl\{(\psi^\dagger\stackrel{\longleftrightarrow}{{\cal D}^2}\psi)
(\pi_-^\dagger\pi_-)
+(\psi^\dagger\psi)(\pi_-^\dagger\stackrel{\longleftrightarrow}{{\bf D}^2}\pi_-)
\biggr\}
+\cdots\, ,
\nonumber\\
\en
where $\psi$ stands for the non--relativistic proton
field, ${\cal D}$ and ${\bf D}$ denote
the covariant derivatives acting on the proton and the charged pion, with 
$\psi^\dagger\stackrel{\longleftrightarrow}{{\cal D}^2}\psi=
\psi^\dagger({\cal D}^2\psi)+({\cal D}^2\psi^\dagger)\psi$ (similar for the pions)
and
$c_p^F,c_p^D,c_p^S,c^R,g_1,e_1$ are various non--relativistic couplings.
In the 4-particle sector, we discard the vertices that correspond to the
$P$-wave interactions, as well as spin--flip terms, since these do not
contribute at the accuracy we are working.

The effective theory based on the Lagrangian Eq.~(\ref{eq:lagr_piN}) enables
one to evaluate the energy shift and the decay width of  pionic
hydrogen perturbatively in the isospin breaking parameter 
$\delta\sim\alpha\sim(m_d-m_u)$.
The power counting rules in this parameter are as follows
(see also subsection~\ref{subsec:widthtable}). 
The leading--order strong energy shift in Eq.~(\ref{eq:Deser-piH}) 
is of order $\delta^3$, and the correction term $\delta'_1$ counts 
at $O(\delta)$. In the formula for the decay width,  
the CM momentum $p_1^\star$ corresponding to the decay 
into the $\pi^0n$ final state, counts at $O(\delta^{1/2})$. Further, since the
cross 
section for $\pi^-p\to n\gamma$ starts at $O(\alpha)$, one gets 
$1/P=O(\delta^{1/2})$. The correction term $\delta_1$ in 
Eq. ~(\ref{eq:Deser-piH}) counts again at $O(\delta)$.  
Our final aim is to evaluate 
the energy shift up to and including terms of order $\delta^4$, and 
the decay width up to and including terms of order 
$\delta^{9/2}$.
It can be checked that at this accuracy  
no other terms than already displayed in Eq.~(\ref{eq:lagr_piN}) are needed.

Next we  note that Eq.~(\ref{eq:lagr_piN}) now includes the effective--range
term with the coupling $e_1$. This term
 has been discarded previously for  scalar QED,
as well as for  pionium  and the $\pi K$ atom.
The reason for this is that, if the coupling $e_1$ stays finite in the limit
$\delta\to 0$, the effective--range term does not contribute at the accuracy we
are working. It can be however shown that in our case $e_1$ is singular 
as $\delta\to 0$. In general,
the counting of the effective couplings $g_1,e_1,\cdots$
in the parameter $\delta$ changes, when one integrates out the 
$\pi^0 n$ state. To illustrate this, we 
compare the threshold momenta
 for the $\gamma n$ and $\pi^0 n$ intermediate states in the CM frame, denoted
 by $p_\gamma^\star(0)$ and $p^\star(0)$ respectively (see appendix~\ref{app:unitarity}):
\eq
\pi^-p\to n\gamma &:\quad& 
p_\gamma^\star(0)=\frac{\lambda^{1/2}((m_p+M_\pi)^2,m_n^2,0)}{2(m_p+M_\pi)}=O(\delta^0)\, ,
\nonumber\\[2mm]
\pi^-p\to \pi^0n&:\quad& 
p^\star(0)=\frac{\lambda^{1/2}((m_p+M_\pi)^2,m_n^2,M_{\pi^0}^2)}{2(m_p+M_\pi)}
=O(\delta^{1/2})\, .
\en
As it will be demonstrated below,
in the presence of a shielded $\pi^0 n$ intermediate state,
 the dependence of the constants $g_1$ and $e_1$ on the parameter
$\delta$ becomes non--analytic. For example, unlike the previous cases,
  unitarity now gives
${\rm Im}\, g_1=O(\delta^{1/2})$
(see appendix~\ref{app:unitarity}). 
Moreover, we shall see that ${\rm Im}\, e_1=O(\delta^{-1/2})$ and, due
to this fact, the effective--range contribution survives 
at next--to--leading order in the decay width of pionic hydrogen. 
To summarize, more complicated power counting rules in the parameter $\delta$ 
is the price  one pays for using a one--channel formalism.

\subsection{Matching}
\label{subsec:matching_piN}

Let us next consider matching of  the parameters of the effective
non--re\-la\-ti\-vis\-tic Lagrangian Eq.~(\ref{eq:lagr_piN}) to
ChPT. The matching in the one--particle sector, which determines the
couplings $c_p^F$, $c_p^D$, $c_p^S$ and $c^R$, is completely
analogous to one carried out in subsection~\ref{subsec:matching}.
The matrix elements of the electromagnetic current between two proton / 
two charged
pion states at leading order in 
the coupling constant $e$
are given by the standard expressions
\eq\label{eq:form_rel1}
&&\langle p's'|J^{\rm em}_\mu(0)|ps\rangle
=e\bar u(p',s')\biggl\{ \gamma_\mu F_1(Q^2)+\frac{i}{2m_p}\,
\sigma_{\mu\nu}Q^\nu F_2(Q^2)\biggr\} u(p,s)\, ,
\nonumber\\[2mm]
&& \langle \pi^\pm(p')|J^{\rm em}_\mu(0)|\pi^\pm(p)\rangle
=\pm e(p'+p)_\mu F_\pi(Q^2)\, ,
\en
where $Q_\mu=(p'-p)_\mu$. At a small momenta, the
form factors can be expanded, yielding
\eq\label{eq:radii}
&&F_1(Q^2)=1+\frac{Q^2}{6}\,\langle r_p^2\rangle+O(Q^4)\, ,\quad\quad
F_2(Q^2)=\kappa_p+O(Q^2)\, ,\quad\quad
\nonumber\\[2mm]
&&F_\pi(Q^2)=1+\frac{Q^2}{6}\,\langle r_\pi^2\rangle+O(Q^4)\, .
\en
Here $F_1(Q^2),F_2(Q^2)$ and $F_\pi(Q^2)$ are the Dirac and Pauli form factors of
the proton and the electromagnetic form factor of the pion, respectively
(see appendix~\ref{app:notations}).
Further, $\langle r_p^2\rangle$ and $\kappa_p$ denote the
charge radius squared and the anomalous magnetic moment of the proton.
Expanding the form factors in powers of 3-momenta and
performing the matching enables one to read off the
values of the low--energy constants at
leading order in the parameter $\delta$~\cite{Bern3},
\eq\label{eq:cNR}
\begin{array}{l l l}
c_p^F=1+\kappa_p\, ,&\hspace*{1.cm}& c_p^D=1+2\kappa_p+\frac{4}{3}\, 
m_p^2\langle r_p^2\rangle\, ,\\[4mm]
c_p^S=1+2\kappa_p\, ,&& c^R=M_\pi^2\langle r_\pi^2\rangle\, .
\end{array}
\en
The four-particle couplings $g_1,e_1$ are determined through  matching
to the $\pi^-p$ elastic amplitude in the vicinity of  threshold, calculated in
ChPT. The procedure
is described in detail in the appendices~\ref{app:notations} and \ref{app:unitarity}.
 Below, we merely recall the notation. In total, we shall need the
pion--nucleon amplitudes in 4 different physical channels labeled 
``c/x/0/n'', respectively:
\begin{itemize}
\item[]
$p\pi^-\to p\pi^-~{ (c)}$ /
$p\pi^-\to n\pi^0~{ (x)}$ /
$n\pi^0\to n\pi^0~{ (0)}$ /
$n\pi^-\to n\pi^-~{ (n)}$ \fs
\end{itemize}
Channel ${n}$ shows up only in the \pid case. 
One calculates these amplitudes at $O(\alpha,(m_d-m_u))$ in ChPT and obtains
the threshold amplitudes  ${\cal T}_{c,x,0,n}$. 
 It is convenient to define the real threshold amplitudes
${\cal A}_{c,x,0,n}$, which are proportional to  
$\mbox{Re}\,{\cal T}_{c,x,0,n}$, see Eq.~(\ref{eq:appnorm}).
The normalization  is chosen so that, approaching the 
isospin limit, one has
\eq\label{eq:piN_ampl}
{\cal A}_c=a_{0+}^++a_{0+}^-+\delta{\cal A}_c+\cdots\, ,\quad \quad
{\cal A}_x=-a_{0+}^-+\delta{\cal A}_x+\cdots\, ,
\nonumber\\[2mm]
{\cal A}_0=a_{0+}^++\delta{\cal A}_0+\cdots\, ,\quad\quad
{\cal A}_n=a_{0+}^+-a_{0+}^-+\delta{\cal A}_n+\cdots\, ,
\en
where $\delta{\cal A}_c,\delta{\cal A}_x,\delta{\cal A}_n=O(\delta)$,
$\delta{\cal A}_0=O(\delta^{1/2})$ are
leading isospin breaking corrections and the ellipses stand for  
 higher--order terms. The corrections at order $\delta^{1/2}$ emerge due to
 the unitary cusp. 
At the order of accuracy we are working, the couplings
$g_1$ and $e_1$ are finally expressed in terms of ${\cal A}_c,{\cal A}_x$ and ${\cal A}_0$. 

\subsection{Bound state -- electromagnetic shift}

A generic solution of the unperturbed Schr\"odinger equation 
with pure Coulomb potential, which describes the bound state of spin $0$ 
(charged pion)
and spin $\frac{1}{2}$ (proton) particles
is characterized by the following set of quantum numbers: 
the principal quantum number $n=1,2,\cdots$,
the angular momentum $l=0,1,2,\cdots$, the total angular momentum
$j=|l-\frac{1}{2}|,l+\frac{1}{2}$, and its $z$-component  $m=-j,\cdots, j$.
The explicit expression for the wave function is given by
\eq\label{eq:Clebsch} 
&&|\Psi_{nljm}({\bf P})\rangle=\sum_\sigma\int\frac{d^3{\bf q}}{(2\pi)^3}\,
\langle jm|l(m-\sigma)\frac{1}{2}\sigma\rangle\,
\Psi_{nl(m-\sigma)}({\bf q})|{\bf P},{\bf q},\sigma\rangle\, ,
\nonumber\\[2mm]
&&
|{\bf P},{\bf q},\sigma\rangle=b^\dagger(\eta_1{\bf P}+{\bf q},\sigma)
a^\dagger(\eta_2{\bf P}-{\bf q})|0\rangle\, .
\en
Here, $\eta_1=m_p/\msigma$, $\eta_2=M_\pi/\msigma$, $\msigma=(m_p+M_\pi)$, $\mu_c=m_pM_\pi/\msigma$,
$a^\dagger,b^\dagger$ stand for the creation operators of the non--relativistic
pion and proton,
$\langle jm|l(m-\sigma)\frac{1}{2}\sigma\rangle$ are
the pertinent Clebsch-Gordan coefficients, and 
$\Psi_{nlm}({\bf q})$ denotes the Coulomb wave 
function in the momentum space. 

The energy shift is given by the counterpart of
 Eqs.~(\ref{eq:summary1},\,\ref{eq:summary2}),
\eq\label{eq:key_piN}
\Delta E_{nlj}=(\Psi_{nlj}|\bar{\mathbold{\tau}}^{nlj}(E_n)
|\Psi_{nlj})+o(\delta^4)\, ,
\en
where $E_n=\msigma-\frac{1}{2n^2}\,\mu_c\alpha^2$ is the Coulomb
energy for this state, and
 where we have again used the fact that the matrix element on the right--hand
side does not depend on $m$. The quantity
$\bar{\mathbold{\tau}}^{nlj}(E_n)$ is defined through the equation
\eq\label{eq:key_tau}
\bar{\mathbold{\tau}}^{nlj}(z)={\bf V}
+{\bf V}\bar {\bf G}_C^{nlj}(z)\bar{\mathbold{\tau}}^{nlj}(z)\, ,
\en
where, as before, ${\bf V}$ denotes the perturbation Hamiltonian (everything
except the static Coulomb interaction) and 
the pole--subtracted Coulomb resolvent is now given by 
 \eq\label{poleremoved}
\bar {\bf G}_C^{nlj}(z)={\bf G}_C(z)-\sum_m\int\frac{d^3{\bf P}}{(2\pi)^3}\,
\frac{|\Psi_{nljm}({\bf P})\rangle\langle\Psi_{nljm}({\bf P})|}
{z-E_n-\frac{{\bf P}^2}{2\msigma}}\, .
\en
According to Eq.~(\ref{eq:splitting}), the total energy shift 
$\Delta E_{nlj}$ is given by a sum of several
terms. The {\it electromagnetic shift} is~\cite{Bern3,Raha1}
 \eq\label{eq:splitting1}
\Delta E_{nlj}^{\rm em}&&\hspace{-7mm}= 
 -\frac{m_p^3+M_\pi^3}{8m_p^3M_\pi^3}\,\biggl(\frac{\alpha\mu_c}{n}\biggr)^4
\biggl\{\frac{4n}{l+\frac{1}{2}}-3\biggr\} 
 -\frac{\alpha^4\mu_c^3}{4m_pM_\pi n^4}\,
\biggl\{-4n\delta_{l0}-4+\frac{6n}{l+\frac{1}{2}}\biggr\} 
 \nonumber\\[2mm] 
&+&\frac{2\alpha^4\mu_c^3}{n^4}\,
 \biggl(\frac{c_p^F}{m_pM_\pi}+\frac{c_p^S}{2m_p^2}\biggr)\, 
\biggl\{\frac{n}{2l+1}-\frac{n}{2j+1}
 -\frac{n}{2}\,\delta_{l0}\biggr\} 
\nonumber\\[2mm]
 &+&\frac{4\alpha^4\mu_c^3}{n^3}\,\delta_{l0}\biggl(\frac{c_p^D}{8m_p^2} 
+\frac{c^R}{6M_\pi^2}\biggr)\fs
 \en 
\begin{sloppypar}
\noindent
 The above analytic calculation
 neatly reproduces the numerical results of Ref.~\cite{Sigg}
for the ground-state energy. 
 In order to facilitate the comparison,
we split the electromagnetic shift by introducing the same
naming scheme as in Ref.~\cite{Sigg} (see Table~\ref{tab:Sigg}),
\end{sloppypar}
 \eq
-\frac{1}{2}\,\mu_c\alpha^2+\Delta E_{10}^{\rm em}=E_{10}^{\rm KG}
 +E_{10}^{\rm fin}+E_{10}^{\rm rel}\, ,
\en
 where 
\eq
 E_{10}^{\rm KG}&=&-\frac{1}{2}\,\mu_c\alpha^2
\biggl(1+\frac{5}{4}\,\alpha^2\biggr)\, ,
 \nonumber\\[2mm]
 E_{10}^{\rm fin}&=&\frac{2}{3}\,\alpha^4\mu_c^3
 \biggl(\langle r_p^2\rangle+\langle r_\pi^2\rangle\biggr)\, ,
\nonumber\\[2mm]
 E_{10}^{\rm rel}&=&-\frac{1}{2}\,\mu_c\alpha^4
\biggl(\frac{\mu_c}{4\msigma}+\frac{m_p^2}{\msigma^2}-1 -\frac{2\kappa_p M_\pi^2}{\msigma^2}\biggr)\fs
 \en

\begin{table}[t]

\begin{center}

\begin{tabular}{ |l| l| r| r| }
\hline
&&&\\[-3.5mm]
Corrections & Notation & Ref.~\cite{Sigg} & Ref.~\cite{Bern3}     \\
&&&\\[-3.5mm]
\hline
&&&\\[-3.5mm]
Point Coulomb, KG equation              &$E_{10}^{\rm KG}$   & $-3235.156$ & $-3235.156$ \\
&&&\\[-3.5mm]
\hline
&&&\\[-3.5mm]
 Finite size effect (proton, pion)       &$E_{10}^{\rm fin}$  & $0.102$     & $0.100$     \\
&&&\\[-3.5mm]
  \hline
&&&\\[-3.5mm]
  Vacuum polarization, order $\alpha^2$   &$E_{10}^{\rm vac}$  & $-3.246$    & $-3.241$    \\
&&&\\[-3.5mm]
  \hline
&&&\\[-3.5mm]
  Relativistic recoil, proton spin and    &               &             &             \\
anomalous magnetic moment               &$E_{10}^{\rm rel}$  & $0.047$     & $0.047$     \\
 &&&\\[-3.5mm]
  \hline
&&&\\[-3.5mm]
  Vacuum polarization, order $\alpha^3$   &               & $-0.018$    &             \\
&&&\\[-3.5mm]
  \hline
&&&\\[-3.5mm]
  Vertex correction                       &               & $0.007$     &             \\[2.mm]
\hline
 \end{tabular}

\end{center}

\caption{
Contributions to the electromagnetic binding energy of the $\pi^-p$
atom ground state and the correction due to vacuum polarization, in eV. 
Vacuum 
polarization at order $\alpha^3$ and the vertex correction have not 
yet been  calculated within the non--relativistic approach.}
\label{tab:Sigg}
\end{table}

\noindent The comparison is displayed\footnote{We used  
$M_\pi=139.56995 \mbox{ MeV},m_p=938.27231\mbox{ MeV}$,
                     $m_e=0.51099907 \mbox{ MeV}$,
$\langle r_p^2 \rangle = (0.849 \mbox{ fm})^2$,
$\langle r_\pi^2\rangle=(0.657\mbox{ fm})^2$,
                     $\kappa_p=1.79284739$, $\alpha^{-1}=137.035989561$,
 see also Ref.~\cite{Sigg}. The above value for $\langle r_p^2 \rangle$ 
somewhat differs from the recent update that can be found in the
literature~\cite{emformfactor}. To make the comparison easy, we however do 
not change this value.}  in Table~\ref{tab:Sigg}.
 As is seen, the agreement is very good up to some higher--order
contributions, which have not been yet calculated in the effective field theory
framework. Further, we have checked that the analytic expressions for
$E_{10}^\mathrm{KG}$ and for $E_{10}^\mathrm{rel}$ agree with the 
analytic result given out in Ref.~\cite{austendeswart}.
 [These authors did not 
determine $E_{10}^\mathrm{fin}$].

We conclude this subsection the with a few remarks.
\begin{enumerate}
\item
Eq.~(\ref{eq:splitting1}) provides a compact analytic expression for the electromagnetic 
shifts at order $\alpha^4$, for any $n,l,j$.
\item
To include the contribution $\Delta E_{nl}^{\rm vac}$
 from vacuum polarization 
at order $\alpha^2$, one may use the
 integral representation worked out in Ref.~\cite{Eiras}. It is  reproduced  
for convenience in Eq.~(\ref{eq:vacsecond}).
\item
One may wish to include still vacuum polarization contributions from order $\alpha^3$, and
 vertex corrections, as displayed in Table \ref{tab:Sigg}.  Note, however, 
that the uncertainty generated by the LEC $f_1$ that occurs in the strong energy shift 
is an order of magnitude larger than vacuum polarization at order $\alpha^3$. 
Indeed, one has [see Eq.~(\ref{eq:delta-p2}) below]
\bea
\Delta E_1^\mathrm{str}|_\mathrm{f_1}&=&\frac{4\alpha^4\mu_c^3}{M_\pi}f_1 \fs
\eea
For $f_1=-1$ GeV$^{-1}$, this amount to a shift of $-0.15$ eV.
An uncertainty in $f_1$ of the order of $-100$ MeV$^{-1}$ induces therefore a 
shift in $\Delta E_1^\mathrm{str}$ which is of the same size as the 
contribution from vacuum polarization at order $\alpha^3$. We see no way to pin 
down $f_1$ to this precision in the foreseeable future.
\item
For these reasons, the electromagnetic shifts Eqs.~(\ref{eq:splitting1}) and (\ref{eq:vacsecond})
provide a  convenient representation, of sufficient accuracy, 
that allows one to easily incorporate changes in the values of the 
pertinent parameters $\langle r^2\rangle_\pi,\cdots,$ whenever needed.
 Together with Eqs.~(\ref{eq:splitting}), (\ref{eq:strongshift}) and (\ref{eq:strongshift1}), 
 they allow one to translate measured energy shifts into strong shifts in an easy manner.
\end{enumerate}

\subsection{DBGT formula for \piH}
Finally, at the accuracy we are working, the strong shifts and widths
 are (cf. with Eq.~(\ref{eq:DEnstr}))
  \eq\label{eq:DE_piH}
\Delta E_n^{\rm str}-\frac{i}{2}\,\Gamma_n
  =-\frac{\alpha^3\mu_c^3}{\pi n^3}\,
(g_1+4\gamma_n^2e_1-g_1^2\langle \bar {\bf g}_C^{n0}(E_n)\rangle)+\cdots\, ,
 \en
where $\gamma_n=\gamma/n$,   and
\eq\label{eq:bar_piH}
  \langle \bar {\bf g}_C^{n0}(E_n)\rangle
=\frac{\alpha \mu_c^2}{2\pi}
  \biggl(\Lambda(\mu)+\ln\frac{4\mu_c^2}{\mu^2}-1+s_n(\alpha)\biggr)\, .
\en
  Moreover, in order to arrive at Eq.~(\ref{eq:DE_piH}), 
we have used that
  \eq
 \int\frac{d^d{\bf p}}{(2\pi)^d}\,{\bf p}^2\Psi_{n0}({\bf p})=
  -\gamma_n^2\tilde \Psi_{n0}(0)\, .
\en
 To arrive at this result, we used the Schr\"odinger equation in momentum
space for the Coulomb wave function $\Psi_{n0}({\bf p})$,  and took into
 account that no--scale integrals vanish in dimensional
 regularization. 

  Expressing now the couplings $g_1$ and $e_1$ through the threshold scattering
amplitudes (see appendix~\ref{app:unitarity} for details), we arrive at our
  final result
\eq\label{eq:NLO-piN}
  &&\Delta E_{n}^{\rm str}=-\frac{2\alpha^3\mu_c^2}{n^3}\,
{\cal A}_c\,(1+K'_n)+o(\delta^4)\, ,
 \nonumber\\[2mm]
&&\Gamma_{n}=\frac{8\alpha^3\mu_c^2p_n^\star}{n^3}\,\biggl(1+\frac{1}{P}\biggr)\,
  {\cal A}_x^2\,(1+K_n)
+o(\delta^{9/2})\, ,
  \en
where  $p_n^\star$ stands for the relative 
  3-momentum of the $n\pi^0$ pair after the decay of the Coulombic
  bound state,
\eq
p_n^\star=\frac{\lambda^{1/2}(E_n^2,m_n^2,M_{\pi^0}^2)}{2E_n}\, .
\en

 Finally, the
correction terms at this order are given by
  \eq\label{eq:KK1}
K'_n&=&-\alpha\mu_c\,(a_{0+}^++a_{0+}^-)s_n(\alpha)+\delta_n^{\rm vac}\, ,
  \nonumber\\[2mm]
K_n&=&-2\alpha\mu_c\,(a_{0+}^++a_{0+}^-) s_n(\alpha)
  +2\mu_c\Delta m (a_{0+}^+)^2+\delta_n^{\rm vac}\, ,
\en
where $s_n(\alpha)$ is given in Eq.~(\ref{eq:bargc}).
  For $n=1$, this result coincides with the one given in Ref.~\cite{Zemp}.

\subsection{Isospin breaking corrections and data analysis}

\subsubsection*{Chiral expansion of the threshold amplitude}

 The threshold amplitudes which enter Eq.~(\ref{eq:NLO-piN}) contain 
itself isospin breaking corrections, that we identify in the manner 
displayed in Eq.~(\ref{eq:piN_ampl}).
The correction to the bound state energy shift in the notation
  Eq.~(\ref{eq:Deser-piH})
 is then given by
\eq\label{eq:delta-r}
 \delta'_n=\frac{\delta{\cal A}_c}{a_{0+}^++a_{0+}^-}+K'_n\fs
\en
 The correction ${\delta\cal A}_c$  can be 
calculated systematically, order by order in ChPT.
 Further, $K'_n$ can be evaluated by
  using the values of the scattering lengths from e.g. Ref.~\cite{Schroder:2001}
 and the vacuum polarization correction  from Ref.~\cite{Eiras}.
 For example, for the ground state, this gives
 $K'_1=0.66\cdot 10^{-2}+0.48\cdot 10^{-2}=1.14\cdot 10^{-2}$,
where $\delta_1^{\rm vac}=0.48\cdot 10^{-2}$. For the chiral expansion 
of the isospin breaking part in the threshold amplitudes, we write
\eq\label{eq:isobr-piN}
\delta{\cal A}_c=\delta{\cal A}_{c,2}+\delta{\cal A}_{c,3}+O(p^4)\, .
\en
The first two terms have been evaluated by
  Refs.~\cite{Fettes1,Fettes2,Bern3,Mojzis} (Ref.~\cite{Fettes2} does not 
provide a complete analytic expression of the scattering amplitude). 
At order $p^2$, 
the result is
\eq\label{eq:delta-p2}
\delta{\cal A}_{c,2}=\frac{1}{4\pi(1+M_\pi/m_p)}\,\biggl(
\frac{4\Delta_\pi}{F_\pi^2}\,c_1-\frac{e^2}{2}\,(4f_1+f_2)\biggr)\, .
\en
For $\delta{\cal A}_{c,3}$, see Ref.~\cite{Mojzis}.
 Unlike in the case of pionium, this correction
 contains  nontrivial strong and electromagnetic LECs
$c_1,f_1,f_2$~\cite{FettesI,Muller} already at lowest order. This
 introduces a sizable systematic error in the theoretical
prediction of the corresponding contributions to $\delta_n'$.
In order to proceed further,
 we have first to specify the values of these LECs. 

\subsubsection*{Strong LECs: $c_i$}

There are different options for fixing the values of the LECs in the 
strong sector. In the context of the present problem, it is consistent to 
determine these constants from the threshold data on $\pi N$ scattering,
similar to Refs.~\cite{BLII,Mojzis,Bernard:2007zu}. Below, we  closely follow
the method of Ref.~\cite{BLII}, using the same conventions and notations.
We denote the isospin symmetric spin--nonflip $\pi N$
amplitude by $D^+(q^2,t)$, where
$q^2=\lambda(s,m_p^2,M_\pi^2)/4s$ is the square of the CM momentum and 
$s,t,u$ are the usual Mandelstam variables.  In the vicinity
of threshold the quantity $\mbox{Re}\,D^+(q^2,t)$ can be expanded in 
Taylor series,
\eq
\mbox{Re}\,D^+(q^2,t)=D^+_{00}+D^+_{10}q^2+D^+_{01}t+\cdots\, .
\en
The amplitudes $D^+_{ij}$ contain the LECs $c_1,c_2,c_3$ [out of 
which we need $c_1$]. On the other hand,
these amplitudes can be expressed in terms of the threshold parameters 
(see, e.g., Ref.~\cite{BLII}), 
\eq
D^+_{00}&=&4\pi(1+\eta)\, a_{0+}^+\, ,
\nonumber\\[2mm]
D^+_{10}&=&4\pi(1+\eta)\,\biggl\{\frac{a_{0+}^+}
{2\eta m_p^2}+b_{0+}^++a_{1-}^++2a_{1+}^+\biggr\}\, ,
\nonumber\\[2mm]
D^+_{01}&=&2\pi\,\biggl\{\frac{a_{0+}^+}
{4 m_p^2}+a_{1-}^++a_{1+}^+(2+3\eta)\biggr\}\,,
\en
where $\eta=M_\pi/m_p$ and $a_{l\pm}^+$, $b_{l\pm}^+$ stand for the scattering
lengths and for the effective ranges.
 Consequently, given an algebraic expression for the amplitude $D^+(q^2,t)$,
one may solve for the constants $c_1,c_2,c_3$ in terms of the experimentally 
measured threshold parameters and other strong LECs, whose contribution enters
$D^+(q^2,t)$ first at order $p^4$ and is small. Moreover, using $D^+(q^2,t)$
calculated at $O(p^2),O(p^3),O(p^4),\cdots$, one may extract the couplings
$c_i$ from data at the same order -- the differences between the numerical
values corresponds to the residual quark mass dependence in these
constants\footnote{We thank T. Becher and H. Leutwyler for providing us with 
the  explicit analytic expression for the amplitude 
$D^+(q^2,t)$ up to and including the  fourth order terms.}
(note that, while doing this, we treat $D^+_{00}$,  $D^+_{10}$ and $D^+_{01}$
as a fixed input and do not expand in the variable $\eta$).

In Table~\ref{tab:ci} we summarize the central values of $c_1,c_2,c_3$ at
different chiral orders. As an experimental input, we have used Koch's
values for the threshold parameters~\cite{Koch2}. Other LECs which contribute
to $D^+(q^2,t)$ at order $p^4$ are: $O(p^4)$ 
 LEC $\bar l_3$ from the pion sector and the
fourth--order pion--nucleon LECs $e_1\cdots e_6$~\cite{BLII}.
In the calculations we use $\bar l_3=2.9$ and set
the finite part of the constants $e_i$ to 0 at the scale $\mu=m_p$.
 We have
checked that the dependence of $c_1,c_2,c_3$ on these LECs is rather weak, 
so even a large uncertainty here does not affect the final result. 
From this Table one readily sees that, 
going from $O(p^2)$ to $O(p^3)$ leads to substantial quark mass 
 effects in all $c_i$, see also Ref.~\cite{BLI}. On the other hand,
 $c_1$ and $c_3$  become stable already at $O(p^3)$,
 suggesting that the procedure is  convergent.

\begin{table}[t]
\begin{center}
\def\arraystretch{1.2}
\begin{tabular}{|c|c|c|c|}
\hline
LECs & $O(p^2)$ & $O(p^3)$ & $O(p^4)$ \\
\hline
$c_1$ &-0.9&-1.2&-1.2\\
\hline
$c_2$ &2.6&4.0&2.6\\
\hline
$c_3$ &-4.4&-6.1&-6.1\\
\hline
\end{tabular}
\end{center}
\caption{The central values of the LECs $c_1,c_2,c_3$, extracted from the 
experimental data at different chiral orders 
(in units of $\mbox{GeV}^{-1}$).}\label{tab:ci}
\end{table}

Table~\ref{tab:ci} contains only central values. The estimate of
the uncertainties is more subtle. The quoted experimental uncertainties 
on the threshold parameters are rather
small. However, since in actual calculations one is truncating amplitudes at 
a given order in ChPT, the uncertainty in the LECs is set to reflect the
uncertainty due to the higher orders as well. Bearing this in mind, we assign
the following error estimate to our central value of $c_1$:
\eq\label{eq:c1p3}
c_1=-1.2\pm 0.3~\mbox{GeV}^{-1}\, \quad\quad\mbox{[order $p^4$]}\fs
\en
At order $O(p^2)$ we assign an asymmetric error:
\eq\label{eq:c1_p2}
c_1=-0.9^{+0.2}_{-0.5}~\mbox{GeV}^{-1}\quad\quad\mbox{[order $p^2$]}\, .
\en
\begin{sloppypar}
\noindent
Here we take the same uncertainty as in the recent analysis of
LECs of
 the $O(p^2)$ Lagrangian~\cite{Meissner:2005ba}. In this paper, the author
 compiles various
determinations of the couplings known in the literature, including the
information about the sub-threshold $\pi N$ amplitudes and the 
$NN$ 
potential~\cite{BLI,Buettiker:1999ap,FettesI,Bernard:1996gq,Epelbaum:2003gr,Rentmeester:2003mf,Entem:2003ft}. 
The resulting values of $c_1$ range from $-0.7~{\rm GeV}^{-1}$ to 
$-1.4~{\rm GeV}^{-1}$, which is in  reasonable agreement with our finding
(the sign of the uncertainty for $c_1$ in Ref.~\cite{Meissner:2005ba}  must 
be reversed). 
On the other hand, our error bars
are generous enough to include the shift of the central
value when going from $O(p^2)$ to $O(p^3)$. We therefore expect that the 
result remains stable with respect to  higher--order quark 
mass effects.
\end{sloppypar}

Finally, we note that our former result
$c_1=-(0.93\pm 0.07)~{\rm GeV}^{-1}$ at $O(p^2)$ \cite{Mojzis}
includes the uncertainty of the
 scattering lengths  only. Even though that result is compatible 
with the above one within the error bars, we prefer to use a
 more conservative error estimate here.

\subsubsection*{Electromagnetic LECs: $f_1$ and $f_2$}

Next, we turn to the
electromagnetic constants $f_1,f_2$. 
The quantity $f_1$ occurs in the 
 chiral expansion of the nucleon mass and  in elastic 
pion--nucleon scattering $\pi^\pm p(n)\rightarrow \pi^\pm p(n)$.
The electromagnetic part of the proton--neutron mass difference
 is given by the constant $f_2$
at leading order in the chiral expansion,
\eq
-e^2F^2f_2=(m_p-m_n)^{\rm em}\, .
\en
Here we disagree with  Ref.~\cite[Eq.~(12)]{Steininger}
by a factor of 2. Numerically, we use $(m_p-m_n)^{\rm em}
=(0.76\pm0.3)$ MeV~\cite{physrep}, or
\eq\label{f2protonneutron}
f_2&=&-(0.97\pm 0.38)\ {\mbox{GeV}}^{-1}\, .
\en
We are  left with the determination of $f_1$.
The sum $m_p+m_n$ contains the combination $e^2(f_1+f_3)$ - 
 the constants $f_1$ and $f_3$  can thus not be disentangled from 
information on the nucleon masses. 
We may consider $m_p+m_n$  as a quantity that 
fixes $f_3$, once $f_1$ is known. Therefore, elastic pion--nucleon
scattering is  the only realistically accessible  source of information on
 $f_1$. In principle, one may consider combinations of amplitudes 
that vanish in the isospin symmetry limit, and determine $f_1$ from those.
The combination 
\eq\label{f1determination}
X=T^{\pi^+p\rightarrow \pi^+p}+T^{\pi^-p\rightarrow\pi^-p}-2T^{\pi^0p\rightarrow\pi^0p}
\en
has this property. The tree graphs 
 of  $X$ start at order $p^2$ and contain $f_1$ - that one 
may try to determine hence from here. Of course, one is 
faced with a problem of accuracy: in order to determine $X$, 
one needs to consider the difference of two large numbers, quite aside from 
the fact that the cross section $\pi^0p\rightarrow\pi^0p $ is not known 
experimentally\footnote{\label{foot:bernstein}For a proposal to measure 
the elastic scattering of neutral pions, see, e.g., Ref.~\cite{Bernstein:1998ip}.}. 
 It remains also to be seen whether a combination of experimental  
data and lattice calculations can provide a reliable estimate of $f_1$.

In the absence of precise experimental information, 
we can i) rely on order--of--magnitude estimates, 
or ii) consider model calculations.
As to order--of--magnitude estimates, we follow Fettes and 
Mei\ss ner~\cite{Fettes:2000vm} and write
\eq\label{f1ordermeissner}
F^2e^2|f_1|\simeq \frac{\alpha}{2\pi}m_p\, ,\nonumber
\en
or
\eq\label{f1estimate}
|f_1|\simeq 1.4\ {\rm {GeV}}^{-1}\, ,
\en
 because $f_1$ is due to a genuine  photon loop at the quark level 
(we divide by $2\pi$ rather than by 4$\pi$~\cite{Fettes:2000vm} to 
be on the conservative side).
This estimate also confirms the expectation~\cite{Bern3} 
 that $|f_1|$ has the same size as $|f_2|$, see Eq.~(\ref{f2protonneutron}).

In Refs.~\cite{Faessler1,Faessler2}, $O(p^2)$ LECs have been estimated in 
the framework of a quark model, with the result (in our notation)
\eq\label{c1faessler}
c_1&=&-1.2\ {\rm GeV}^{-1}\, , 
\nonumber\\[2mm]
(f_1,f_2,f_3)&=&(-2.3\pm 0.2,-1.0\pm 0.1,2.1\pm 0.2)\, {\rm GeV}^{-1}\, .
\en
For a recent attempt to calculate $f_1$ by using a method related
to  resonance saturation, see Ref.~\cite{Ericson-Ivanov}.

\subsubsection*{Isospin breaking correction $\delta_1,\delta_1'$ 
and data analysis}

Using the values of LECs at $O(p^2)$, which are given in Eqs.~(\ref{eq:c1_p2}),
(\ref{f2protonneutron}) and (\ref{f1estimate}), we find that
\eq\label{eq:corrp2}
{\delta'_1}=(-4^{+3}_{-4})\cdot 10^{-2}\hspace*{3.cm}\mbox{[order $p^2$]}\, ,
\en
where the central value corresponds to $f_1=0$. We add the uncertainties quadratically.
The bulk of the uncertainty  is due to
the one in $f_1$ and $c_1$.

As mentioned above,
the calculation of the isospin breaking correction at $O(p^3)$ in ChPT
has been carried out in Ref.~\cite{Mojzis}, see also Ref.~\cite{Fettes2}.
 Using the estimate Eq.~(\ref{eq:c1p3}) for $c_1$, these calculations lead to
\eq\label{eq:Mojziscorrections}
{\delta'_1}=(-9.0\pm 3.5)\cdot 10^{-2}\hspace*{3.cm}\mbox{[order $p^3$]}\, .
\en
It turns out that, albeit the loop contributions at $O(p^3)$ are sizable
due to the presence of chiral logarithms, 
the contribution of LECs at this order is quite small. The uncertainty
is dominated
by the uncertainty in $f_1$ and, to a certain extent, by the one in $c_1$. 
We expect that the contributions of the 
higher--order LECs, which are multiplied by additional powers of $M_\pi$,
will be even more suppressed and do not alter significantly the uncertainty
displayed.
At this order,  isospin breaking
in the $\pi^-p$ elastic scattering amplitude is still of purely electromagnetic
origin: the terms with $(m_d-m_u)$ appear first
 at $O(p^4)$.

The isospin breaking corrections in the charge--exchange and to the $\pi^-n\to\pi^-n$
elastic threshold amplitudes at $O(p^2)$ are~\cite{Zemp,Raha3} 
\eq\label{eq:delta-p2_xn}
\delta{\cal A}_x&=&\frac{1}{16\pi(1+M_\pi/m_p)}\,\biggl(
\frac{g_A^2\Delta_\pi}{m_pF_\pi^2}+2e^2f_2\biggr)+O(p^3)\, ,
\nonumber\\[2mm]
\delta{\cal A}_n&=&\frac{1}{4\pi(1+M_\pi/m_p)}\,\biggl(
\frac{4\Delta_\pi}{F_\pi^2}\,c_1-\frac{e^2}{2}\,(4f_1-f_2)\biggr)+O(p^3)\, .
\en
One may use the above expression to evaluate the correction term to the 
decay width in Eq.~(\ref{eq:Deser-piH}). At this order, one 
has~\cite{Zemp}\footnote{Note that in Ref.~\cite{Zemp}, $O(p^2)$
  expressions for the scattering lengths were used in numerical 
calculations, instead of their experimental
  values. If we use the $a_{0+}^+$ and $a_{0+}^-$ from
  e.g. Ref.~\cite{Schroder:2001}, the result changes insignificantly to
$\delta_1=(0.7\pm 0.2)\cdot 10^{-2}~\mbox{[at order $p^2$]}$.}
\eq\label{eq:Zempcorrections}
\delta_1=(0.6\pm 0.2)\cdot 10^{-2}\hspace*{3.cm}\mbox{[order $p^2$]}\, ,
\en
where we take $g_A=1.27$, and where
the error comes from the uncertainty in $f_2$.
Note that, unlike the amplitude ${\cal A}_c$, the
isospin breaking part of the charge--exchange 
amplitude at $O(p^2)$ does not contain the quantities $f_1$ and $c_1$. 
 In
this channel, one is, therefore, able to determine 
the isospin breaking corrections to a 
much better accuracy. Still,  achieving
an accuracy that is comparable to the present experimental 
precision is not possible
without the evaluation of the corrections of order $O(p^3)$ (at least).
This task should  urgently be addressed.

The results given in Eqs.~(\ref{eq:Mojziscorrections}) and 
(\ref{eq:Zempcorrections}) should be contrasted with the ones obtained in 
the framework of potential models, see Eq.~(\ref{eq:Siggcorrections}).
As anticipated, these two sets are rather different, 
because  potential model do not, in general, fully reflect QCD+QED.

We now come back to the discussion of Fig.~\ref{fig:bands_new}, which was 
presented in the beginning of this section.
It is seen that presently, the theoretical uncertainty in
 the scattering length $a_{0+}^+$ 
 is much larger 
than assumed in the potential model approach. For this reason, 
a precise determination of $a_{0+}^+$ 
from  hydrogen data alone  is not possible, unless one finds 
a way to pin down $f_1$ more precisely.
On the other hand, as we will show later in this article, 
a simultaneous analysis of  \piH and 
\pid data may allow one to 
pin down the  scattering lengths in a more reliable manner.

%%%%%%%%%%%%%%%%%%%%%%%%%%%%%%%%%%%%%%%%%%%%%%%%%%%%%%%%%%%%%%%%%%%%%%
\setcounter{equation}{0}
\section{Kaonic hydrogen}
\label{sec:KH}

\subsection{The kaonic hydrogen and kaonic deuterium experiments at DA$\Phi$NE}

An unique source of negative kaons, which provides important 
conditions for the study of the low--energy kaon--nucleon interaction,
has been made available by the DA$\Phi$NE electron--positron collider 
in Frascati.
 The DEAR experiment at DA$\Phi$NE and its successor SIDDHARTA 
aim at a precision measurement of the strong interaction shifts 
and widths of \kaH and \kad. The final aim is  
to extract the antikaon--nucleon scattering lengths from the measured 
characteristics of these 
atoms~\cite{Baldini,Bianco,Beer,Zmeskal,Lucherini:2007ki,SIDDHARTANEW}.

In the DEAR experiment the low--momentum negative kaons, 
produced in the decay of the $\phi$-mesons at DA$\Phi$NE, leave the 
thin--wall beam pipe, are degraded in energy to a few MeV, enter 
a gaseous target through a thin window and are finally stopped 
in the gas. The stopped kaons are captured in an outer orbit of 
the gaseous atoms, thus forming the exotic kaonic atoms. The kaons 
cascade down and some of them will reach the ground state emitting 
$X$-rays. 

As mentioned before, 
recent results of the DEAR collaboration~\cite{SIDDHARTANEW} for 
the shift and the width of the kaon hydrogen in its $1s$ ground state 
(see Eq.~(\ref{eq:dear}))
considerably improve the accuracy of the earlier
KpX experiment at KEK~\cite{KEK} and confirm the repulsive
character of the $K^- p$ scattering at threshold. 
It should be also pointed out that the DEAR central result
deviates from all 
earlier experiments~\cite{Davies,Izycki,Bird,KEK}, see also
Fig.~\ref{fig:K-Hydrogen} below. Below we shall discuss
the implications of these beautiful new 
measurements for establishing the precise
values of the $\bar KN$ scattering lengths $a_0$ and $a_1$.
The discussion is  based on Ref.~\cite{Raha1}.

\subsection{DGBT--type formulae for  kaonic hydrogen}

The non--relativistic theory of \kaH~\cite{Raha1} 
is almost a carbon copy
of the \piH case. It is, on the other hand,
 amusing that the existing small differences at the end result 
 in a strikingly different picture
(this, in our opinion, is  another demonstration of the 
power and flexibility of the NRQFT approach). 
 We do not present many explicit formulae, because they often look identical
in the \piH and \kaH case -- we rather concentrate
 on those properties of these two systems which are not the same.
For example, there is no need to carry out again the
calculation of the electromagnetic shift -- it is given 
by the corresponding expression derived in the pionic
hydrogen case Eq.~(\ref{eq:splitting1}) after obvious replacements $M_\pi\to M_K$
and $\langle r_\pi^2\rangle\to\langle r_K^2\rangle$.

We start to list those characteristic features of the $\bar KN$ system which 
cause a difference to the case of pionic hydrogen.
\begin{itemize}
\item[i)]
The only states that are degenerate in mass with the $K^-p$ state in the
isospin limit $\delta\to 0$, are the states $K^-p+m\gamma$, 
$\bar K^0n+m'\gamma$, with
$m,m'=0,1,\cdots$. It is convenient to use a framework which 
explicitly ``resolves'' all these states in the
non--relativistic theory.
On the contrary, the effect of other intermediate states, whose mass is not
degenerate with that of the $K^-p$ state in the isospin limit, 
could be included in the couplings
of the non--relativistic effective Lagrangian.

\item[ii)]

Breaking of  $SU(3)$ symmetry, which is proportional to the quark
mass difference $m_s-\hat m$, is much larger than  isospin breaking
 effects. For 
this reason, we count $m_s-\hat m$ as 
$O(1)$ in $\delta$.
Because there are open strong channels below $K^-p$ threshold 
-- e.g. $\pi\Sigma,~\pi^0\Lambda$, the real and 
imaginary parts of strong kaon--nucleon couplings count as $O(1)$.
 This leads to a completely different power counting as compared to  
the \piH case (see also subsection~\ref{subsec:widthtable}):
 the strong shift and width are quantities of order $\delta^3$.

In the non--relativistic theory, 
where the neutral ($\bar K^0n$) channel is explicitly resolved, the couplings
are analytic in $\delta$ (there are no corrections that go like
 $\sqrt{\delta}$, cf. with the discussion in section~\ref{sec:piH} and in 
appendix~\ref{app:unitarity}). 
The expansion of a typical 4-particle coupling constant
(similar as in Eq.~(\ref{eq:lagr_piN})) is given by
\eq
g_i=g_i^{(0)}+\alpha g_i^{(1)}+(m_d-m_u)g_i^{(2)}+O(\delta^2)\, ,
\en
where
$g_i^{(0)},g_i^{(1)},g_i^{(2)},\cdots$ are functions of the strange
quark mass $m_s$, along with $\hat m=\frac{1}{2}\,(m_u+m_d)$.

\item[iii)]
The Panofsky ratio  Eq.~(\ref{eq:Panofsky}) in \piH 
is of order $\delta^{-1/2}$.
 Numerically, the branching ratio into the
$n\gamma$ channel amounts  to a contribution of 
$\simeq 40~\%$ in the total decay width. 
In contrast to this, in the case of \kaH, this branching ratio 
counts as
$O(\delta^{-1})$. The measured branching ratios into the leading
$\Lambda\gamma$, $\Sigma\gamma$
 channels are of the order of a per mille~\cite{Whitehouse} (the theoretical
description of this quantity by using chiral Lagrangians~\cite{Ivanov3,Oset_Lee} 
gives a result which is consistent with the experiment
by order of magnitude). Consequently, the perturbative
treatment of the effects due to these channels is justified.
At the numerical precision one is working, and bearing in mind present
 experimental and theoretical uncertainties, they may even be
neglected altogether.

\item[iv)]
The $\bar K^0n$ intermediate state lies in the vicinity
 the $K^-p$ threshold. Due to 
this fact, the well--known unitary cusp emerges in the
$K^-p$ elastic
scattering amplitude (see, e.g., Refs.~\cite{Dalitz1,Dalitz2,Deloff} and 
appendix~\ref{app:unitarity})
which is the source of huge isospin breaking
corrections
(note that the
cusp effect is also the dominant isospin breaking effect in some other
low--energy processes, e.g. in neutral pion photo--production
off nucleons~\cite{cusp1_o,cusp2_o} as well as $K\to 3\pi$ decays~\cite{Cabibbo1,Cabibbo2,Prades,cusp1,cusp2,Batley:2005ax}.).

\item[v)]
The $\Lambda(1405)$ never appears explicitly in this approach, that is
correlated with counting $(m_s-m_u)$ as a hard scale. 
The influence of the resonance
is only indirect and results in a large  $\bar KN$ 
threshold scattering amplitude, which then leads to a significant
increase of the isospin breaking corrections.

\end{itemize}

\begin{figure}[t]
\begin{center}
\includegraphics[width=10.cm]{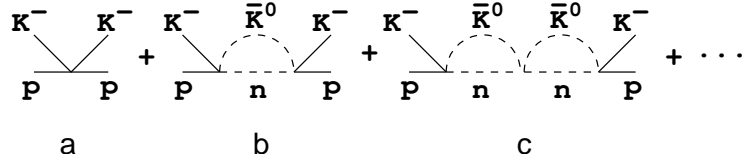}
\end{center}
\caption{$\bar K^0n$ bubbles in the threshold amplitude of elastic $K^-p$
scattering. Note that the bubbles with $K^-p$ in the intermediate state vanish
at threshold.}
\label{fig:KN-bubbles}
\end{figure}

Due to the presence of the  $\bar K^0n$ states,
 the relative weight of the isospin breaking correction in the
$ K^-p$ threshold amplitude is dramatically increased. Consider, for 
instance, the real part of this amplitude, which contains any number
of the $s$-channel bubbles with $K^-p$ and $\bar K^0n$ intermediate states.
Further, the contribution of the former vanishes at threshold, and
only the neutral loops contribute to the threshold amplitude 
(see Fig.~\ref{fig:KN-bubbles}).

Consider now the series of the bubble diagrams shown in this Figure. 
Since the $\bar K^0n$ state lies higher than $K^-p$ state, the
contribution from a single $\bar K^0n$ bubble, which is shown in 
Fig.~\ref{fig:KN-bubbles}b, at threshold is real and proportional
to $q_0=(2\mu_0(m_n+M_{\bar K^0}-m_p-M_K))^{1/2}=O(\sqrt{\delta})$,
where $\mu_0$ is the reduced mass of the $\bar K^0n$ pair. 
Thus, the isospin breaking correction to the threshold amplitude,
which stems from the diagram Fig.~\ref{fig:KN-bubbles}b, starts at 
$O(\sqrt{\delta})$ (note that $q_0=0$ in the isospin limit). 
Moreover, 
since both real and imaginary parts of the $\bar KN$
couplings $g_i$, which appear in the vertices, count as $O(1)$,
the contribution of the diagram  in Fig.~\ref{fig:KN-bubbles}b to the real part
of the threshold amplitude counts always as $O(\sqrt{\delta})$, irrespective
of the fact whether the neutral loop is real or imaginary (in the latter case,
the unitary cusp is below threshold and influences the real part of the 
amplitude through coupling to the inelastic channels).

For comparison, let us consider 
the counterpart of the diagram  Fig.~\ref{fig:KN-bubbles}b in the $\pi N$ 
case, in the non--relativistic theory where the $\pi^0 n$ state is 
explicitly resolved.
It is purely imaginary at threshold, since $m_n+M_{\pi^0}<m_p+M_\pi$.
Furthermore, the imaginary part of the non--derivative $\pi N$ couplings counts
as $O(\delta)$ since there are no open strong inelastic channels in this case
(the channel $\gamma n$ contributes at $O(\delta)$).
For this reason, there is no contribution from this diagram to the real part
of the threshold amplitude at $O(\sqrt{\delta})$. Only the product of two
neutral bubbles, which is shown in Fig.~\ref{fig:KN-bubbles}c can contribute
to the real part, but this contribution starts at $O(\delta)$.
To summarize, the isospin breaking
corrections in the $\bar KN$ and $\pi N$ amplitudes emerge at 
$O(\sqrt{\delta})$
and $O(\delta)$, respectively -- no wonder that the former are much larger in
magnitude than the latter.

The crucial observation which enables one to find the way out is the following:
albeit the isospin breaking corrections which are non--analytic in the
parameter $\delta$  are large, these can be expressed solely in terms of
those scattering lengths which one tries to extract from the experiment. 
Consequently, the presence of these large corrections does not affect the
accuracy of the extraction procedure.
On the other hand, the isospin breaking corrections at $O(\delta)$ cannot be
expressed it terms of the scattering lengths alone -- these in particular
contain the LECs from the isospin breaking  sector of the ChPT Lagrangian.
 One expects, however,
that this error, introduced by the $O(\delta)$ terms, should be
 of the order of magnitude of a few percent, as 
in the pionic hydrogen.

To illustrate the above statement, let us consider the bubble sum shown in
Fig.~\ref{fig:KN-bubbles} in the non--relativistic effective theory, 
using only non--derivative 4-particle vertices.
The corresponding approximate non--relativistic $K^-p\to K^-p$ scattering
amplitude, which we denote as ${\cal T}_{c,NR}^{(0)}$, is given by the
following expression,
\eq\label{eq:Kbubbles}
{\cal T}_{c,NR}^{(0)}=\frac{2\pi}{\mu_c}\, 
\frac{\frac{1}{2}\,(a_0+a_1)+q_0a_0a_1}{1+\frac{q_0}{2}\,(a_0+a_1)}
\doteq \frac{2\pi}{\mu_c}\, a_c\, ,
\en
where $\mu_c$ is the reduced mass of the $K^-p$ system, and where we have used
the matching in the absence of isospin breaking to relate various 
non--derivative $\bar KN$ couplings to the pertinent combinations of two
scattering lengths $a_0$ and $a_1$ (the error introduced is of order $\delta$).
Note also that for our
purposes, it would be enough to include the contributions from the first two
diagrams only. We shall however sum up the whole series shown in 
Fig.~\ref{fig:KN-bubbles} -- the difference which we introduce 
is again of order $\delta$.
The crucial observation is that, albeit the isospin breaking corrections
in ${\cal T}_{c,NR}^{(0)}$ are of order $\sqrt{\delta}$, the difference
${\cal T}_{c,NR}-{\cal T}_{c,NR}^{(0)}=O(\delta)$, 
where ${\cal T}_{c,NR}$ denotes the full non--relativistic amplitude.
In other words, the leading--order isospin breaking
effect due to the neutral loop has been explicitly included
in the quantity ${\cal T}_{c,NR}^{(0)}$.

Finally, using the matching condition, one may define the relativistic
amplitude, corresponding to the relativistic bubble sum given by 
Eq.~(\ref{eq:Kbubbles}),
\eq
{\mathcal T}_c^{(0)}=2m_p\,2M_K\,{\mathcal T}_{c,NR}^{(0)}=
8\pi\,(m_p+M_K)\,a_c\, ,
\en
which has the same property as its non--relativistic counterpart, namely, the
difference ${\cal T}_c-{\cal T}_c^{(0)}=O(\delta)$, where ${\cal T}_c$
is the relativistic elastic amplitude for $K^-p\to K^-p$. In general, one has
\eq\label{sc_lnew}
{\mathcal T}_c={\mathcal T}_c^{(0)}
+\frac{i\alpha\mu_c^2}{4M_Km_p}\,({\mathcal T}_c^{(0)})^2
+\delta{\mathcal T}_c+o(\delta)\, ,
\en
which is nothing but the definition of the correction term 
$\delta{\mathcal T}_c$ (the second term in the above definition with
the imaginary coefficient has been added for convenience.
It starts at order $\delta$.). Note also, that Eq.~(\ref{sc_lnew})
is the generalization of the relation
\eq\label{sc_lnew_bad}
{\mathcal T}_c=8\pi\,(m_p+M_K)\,\frac{1}{2}\, (a_0+a_1)+O(\sqrt{\delta})\, .
\en
The main message here is:
in contrast to Eq.~(\ref{sc_lnew_bad}), the quantity $\delta{\mathcal T}_c$
in Eq.~(\ref{sc_lnew}) is of order $\delta$
and therefore can be assumed to be not too large.
Note also that Eq.~(\ref{eq:Kbubbles}) exactly coincides with 
the amplitude introduced
in Refs.~\cite{Dalitz1,Dalitz2}. In our approach we however in addition 
demonstrate that this 
modification accounts for {\em all} potentially large $O(\sqrt{\delta})$
corrections to the threshold amplitude.

The results of our findings can be finally summarized an a modified
DGBT--type formula for generic $ns$ energy levels of  \kaH \hspace{-1mm},
\eq\label{eq:final}
\hspace*{-.4cm}&&\Delta E_{n}^{\rm str}-\frac{i}{2}\,\Gamma_{n}
=-\frac{\alpha^3\mu_c^3}{4\pi m_pM_Kn^3}
({\mathcal T}_c^{(0)}+\delta{\mathcal T}_c)\biggl\{1
-\frac{\alpha\mu_c^2s_n(\alpha)}{8\pi m_pM_K}{\mathcal T}_c^{(0)}
+\cdots\biggr\}+o(\delta^4) ,
\nonumber\\
\hspace*{-.4cm}&&
\en
where $\delta{\cal T}_c=O(\delta)$ and ellipses stand for (tiny) vacuum 
polarization contribution, which we do not display explicitly.

One expects that the equation (\ref{eq:final}) is much better suited 
for the analysis of the 
DEAR experimental data than the original DGBT formula. In this equation,
potentially large (parametrically enhanced) isospin breaking corrections
at $O(\sqrt{\delta})$ and\footnote{The corrections at $O(\delta\ln\delta)$,
which are referred to as ``Coulomb corrections'' hereafter, emerge from the
second term in curly brackets in Eq.~(\ref{eq:final}).} $O(\delta\ln\delta)$
 are explicitly separated from 
the rest, which is analytic in $\delta$ and is assumed to be small. 
Indeed, in Ref.~\cite{Raha1} an 
estimate of this $O(\delta)$ term has been carried out at tree level
in the $SU(3)$ version of ChPT. The calculations result in the effect at a 
percent level that supports the above conjecture. In the numerical analysis
of the DEAR data that follow, we shall always set $\delta{\mathcal T}_c=0$.

Finally, 
it is interesting to note that Eq.~(\ref{eq:Kbubbles}) enables one 
to independently test the
limits of applicability of the method for a given values of the scattering 
lengths {\it a posteriori}. Namely, in order to be consistent,
the term of order of $\delta$ in the expansion of the above
amplitude should not exceed a few percent, and the following terms must be 
negligible (see Table \ref{tab:KN}). This is, however, not the case
for all input values of scattering lengths available in the 
literature, see Refs.~\cite{Raha1,Raha4} for a detailed discussion.
\begin{table}[t]
\begin{center}
\begin{tabular}{|l|l|l|l|}
\hline
& Ref.~\cite{MO} & Ref.~\cite{Martin} \\
& $a_0=-1.31+1.24i$ & $a_0=-1.70+0.68i$ \\
& $a_1=0.26+0.66i$  & $a_1=0.37+0.60i$  \\
\hline
$a_{c,0}$ &$-0.52+ 0.95i$ & $-0.66+ 0.64i$ \\
$a_{c,1}$ &$-0.68+ 1.09i$
& $-0.98+ 0.66i$ \\
$a_{c,2}$ &$-0.67+ 1.15i$
& $-1.04+ 0.73i$ \\
$a_{c,3}$ &$-0.65+ 1.16i$
& $-1.04+ 0.75i$ \\
\hline
$a_{c,\infty}\doteq a_c$ &$-0.65+
1.15i$  & $-1.03+ 0.76i$ \\
\hline
\end{tabular}
\end{center}

\caption{Expansion of the $K^-p\to K^-p$ scattering length $a_c$
in powers of $q_0$ (bubble approximation, see Eq.~(\ref{eq:Kbubbles})). 
The index $n$ in $a_{c,n}$ corresponds to the $n$-th 
iteration. The results are given in fm.}
\label{tab:KN}
\end{table}

\renewcommand{\arraystretch}{1.6}
\begin{table}[t]
\begin{center}
\begin{tabular}{|l|l|l|}
\hline
Source & $a_0$ & $a_1$\\
\hline
Mei\ss ner and Oller \cite{MO} & $-1.31 + i 1.24 $ & 
$0.26 + i 0.66  $ \\
\hline
Borasoy, Ni\ss ler and Weise, fit ``$u$'' \cite{Nissler2} & $-1.48 + i 0.86 $ &
 $0.57 + i 0.83 $ \\
\hline
Oller, Prades and Verbeni, fit ``$A_4^+$'' \cite{Verbeni1} & $-1.23 + i 0.45 $ &
 $0.98 + i 0.35 $ \\
\hline
Martin \cite{Martin} & $-1.70 + i 0.68 $ &
$0.37 + i 0.60 $ \\
\hline
Borasoy, Mei\ss ner and Ni\ss ler,
fit ``full'' \cite{BMN} & $-1.64 + i 0.75 $ &
 $-0.06+i 0.57 $ \\
\hline
\end{tabular}
\end{center}
\caption{$\bar KN$ scattering lengths $a_0$ and $a_1$ (in fm) from 
the literature. 
These scattering lengths are used as an input in the calculations of the
kaon-deuteron scattering length.}
\label{tab:input}
\end{table}

\subsection{Analysis of the DEAR data}
\begin{figure}[t]
\begin{center}
\includegraphics[width=11.cm]{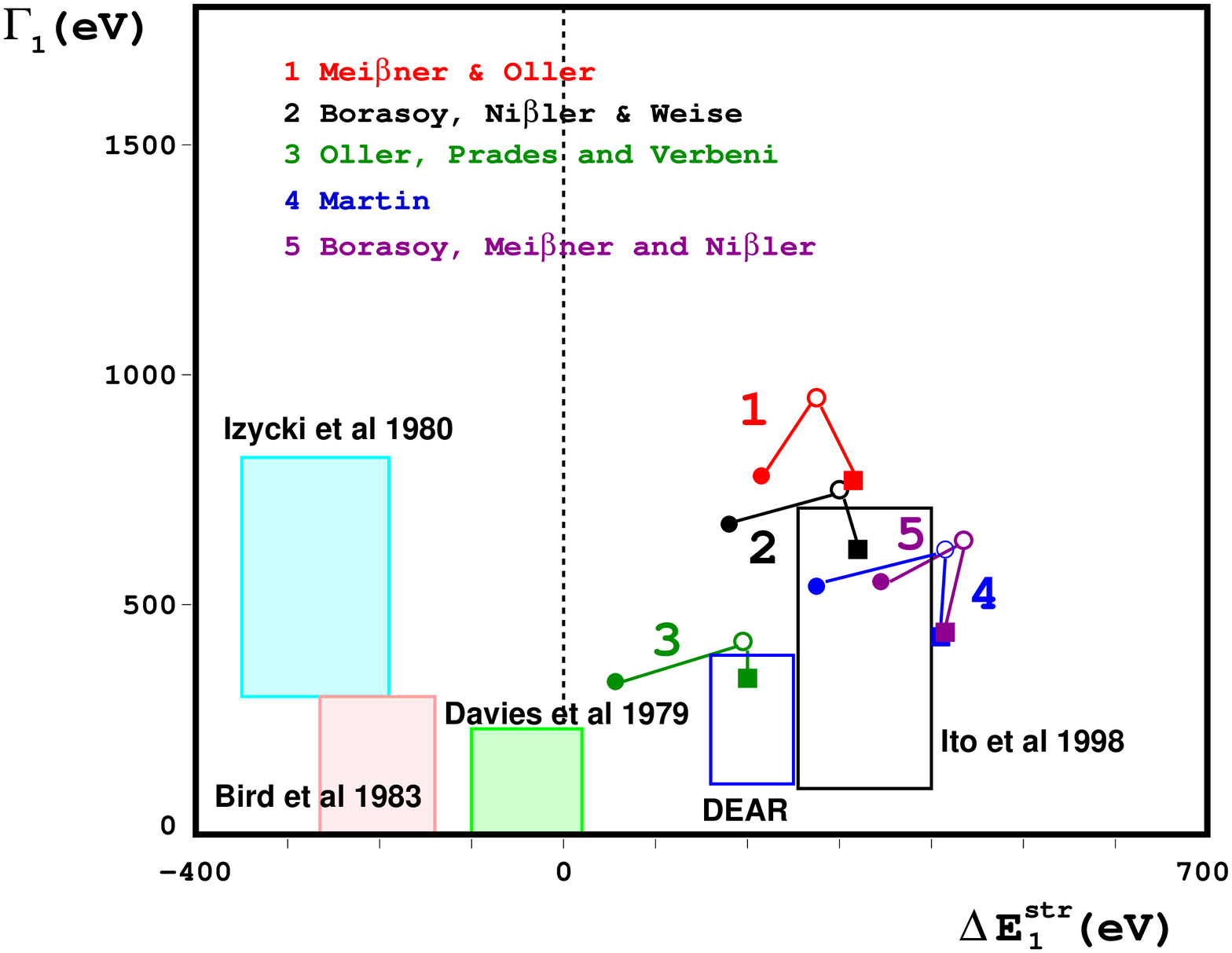}
\end{center}
\caption{Isospin breaking corrections in the energy shift and width of kaonic
hydrogen: different input for scattering lengths $a_0$ and $a_1$ from 
Refs.~\cite{MO,Nissler2,Verbeni1,Martin,BMN}, see Table~\ref{tab:input}.
Filled dots correspond to the DGBT formula (no isospin breaking), empty dots
include the effect of the unitary cusp and filled squares, in addition, 
take the Coulomb corrections into account.}
\label{fig:K-Hydrogen}
\end{figure}

The Figure~\ref{fig:K-Hydrogen} summarizes the analysis of the DEAR data
with the help of Eq.~(\ref{eq:final}).
Here we display the predicted value of the energy shift and width in the
ground state, using different scattering lengths as an input, see
Table~\ref{tab:input}. These predictions
are compared with the DEAR measurement, as well as earlier experimental data.
The comparison enables one to immediately conclude that the scattering data,
to which the parameters of Refs.~\cite{MO,Nissler2,Verbeni1,BMN} are fitted,
are, in general, not consistent with the DEAR data~\cite{Raha1}. 
This conclusion is valid for all input scattering lengths 
shown in Fig.~\ref{fig:K-Hydrogen}, except those from Ref.~\cite{Verbeni1}.
For a further discussion on this issue, we refer to 
Refs.~\cite{Nissler3,Verbeni2}.

The comparison allows one to conclude that:
\begin{itemize}
\item[i)] The corrections due to the unitary cusp, which start 
at $O(\sqrt{\delta})$, are indeed huge.
Even at the present accuracy level, it is absolutely necessary to take them
into account.
\item[ii)]
The Coulomb corrections that are amplified by $\ln\alpha$,
are also quite sizable. For example, choosing
scattering lengths from Refs.~\cite{MO,Martin}, we obtain that
the real part of the
correction term in the ground state amounts up to $9\%$ and $15\%$, 
respectively. 
\item[iii)]
The key point here is that one does not indeed need to know
the numerical values of these large corrections very accurately.
Since both the unitary and Coulomb corrections 
depend on the scattering lengths $a_0,a_1$ only, in the numerical analysis,
which aims to extract exactly these scattering lengths from the data,
it suffices to know the algebraic form of this dependence.
\end{itemize}

\begin{figure}[t]
\begin{center}
\vspace*{.7cm}
\includegraphics[width=9.cm]{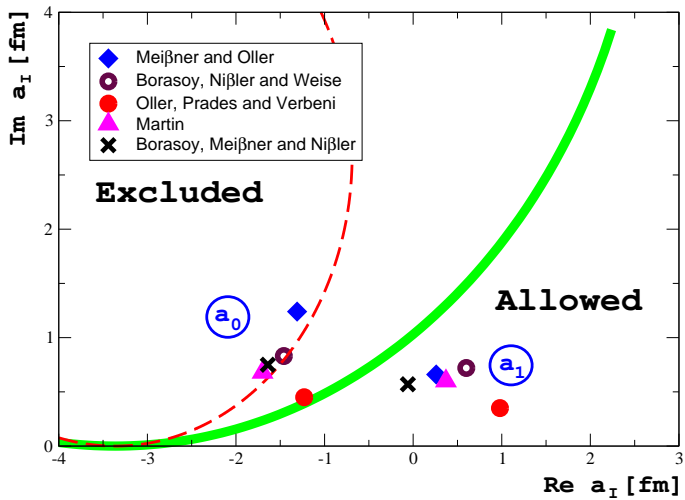}
\end{center}
\caption{Restrictions set by the DEAR data on the values of the scattering 
lengths $a_0$ and $a_1$ (thick solid line). For comparison, we give the scattering lengths from Table~\ref{tab:input}. 
The dashed line corresponds to the restrictions obtained by using 
KpX data instead of DEAR data.}
\label{fig:circle}
\end{figure}

One may further refine the analysis of \kaH data, fully
including the restrictions from unitarity~\cite{Raha4}. 
 We first rewrite Eq.~(\ref{eq:Kbubbles}),
\eq\label{eq:circle}
a_0+a_1+\frac{2q_0}{1-q_0a_c}\,\, a_0a_1-\frac{2a_c}{1-q_0a_c}=0\, .
\en
The quantity $a_c$ is fixed from experiment, by using
 Eqs.~(\ref{eq:Kbubbles}), (\ref{eq:final}), 
and neglecting corrections of order $\delta$. We decompose the 
scattering lengths into their real and imaginary parts,
\bea
a_I=x_I+i y_I\,,\, I=0,1\, ; \, a_c=u+i v\fs
\eea
Solving Eq.~(\ref{eq:circle}) for $a_0$ and requiring that 
its imaginary part is positive leads to a quadratic constraints 
on $x_1,y_1$: in the ($x_1, y_1$) -- plane, $a_1$ is expelled from a 
circle determined by $u,v$ and $q_0$,  and similarly for $1\leftrightarrow 2$.
 Explicitly,
\bea
&&(x_I+\frac{1}{q_0})^2+(y_I-r)^2\geq r^2\,;\, I=0,1\co\nonumber\\
&&r=\left[(1+q_0 u)^2+q_0^2v^2\right](4vq_0^2)^{-1}\fs
\eea
 Part of this circle is shown
in Fig.~\ref{fig:circle} (note that, bearing in mind the preliminary 
character of the DEAR data~\cite{SIDDHARTANEW}, we use only central values
in order to illustrate the construction of
the plot. We do not provide a full error analysis). 
In order to be consistent with the DEAR data, both $a_0$ and $a_1$ should be
on the outside of this universal DEAR circle\footnote{Note that an analysis of
the \kaH data alone does not allow one to determine both scattering
lengths $a_0$ and $a_1$ separately, even at leading order in $\delta$. The 
reason for this lies in the fact that both real and imaginary parts of
$a_I,~I=0,1$ count at $O(1)$, due to the presence of the open channels 
below threshold.}.
For comparison, on the same Figure
we also indicate (much milder) restrictions, which arise, when KpX data
are used instead of DEAR data. The values of
$a_0$ and $a_1$, plotted in this Figure, are again taken
from Table~\ref{tab:input}. As before, we see that
in most of the approaches it is rather
problematic to get a value for $a_0$ which is compatible with DEAR
(the price to pay for this is the presence of a very narrow $I=1$ pole
in the scattering amplitudes close to $K^-p$ threshold for the solutions from
Ref.~\cite{Verbeni1}).
This kind of analysis may prove useful in the near future,
when the accuracy of
the DEAR is increased that might stir  efforts on the theoretical side,
aimed at a systematic quantitative description of the $\bar KN$ 
interactions within  unitarized ChPT.

In conclusion we note that, in our opinion,
the present experimental data are still 
not precise enough to clearly distinguish
between different low--energy approaches to $\bar K N$ interactions which
are based on the unitarization of ChPT. 
In the future, however, DEAR/SIDDHARTA can
pose very stringent constraints on theoretical models, provided the 
measurements are carried out with the announced accuracy.

%%%%%%%%%%%%%%%%%%%%%%%%%%%%%%%%%%%%%%%%%%%%%%%%%%%%%%%%%%%%%%%%%%%%%%%%%%
\setcounter{equation}{0}
\section{Pionic and kaonic deuterium}
\label{sec:deuteron}

\subsection{Introductory remarks and DGBT formulae}

Pionic (kaonic)  deuterium is a hadronic atom, 
 made up from  $\pi^-$ ($K^-$) and
the deuteron, which in  turn is  a shallow bound state of the proton
and the neutron, held together by strong interactions. 
Constructing a theory of these bound states constitutes one
step further in the sophistication of the NRQFT approach.

The characteristic momentum for the deuteron is
$\bar\gamma=\sqrt{m_N\varepsilon_d}\simeq 46~{\rm MeV}$, with
 $\varepsilon_d=2.22~{\rm MeV}$ its  binding energy.
 This quantity is still three times 
smaller than the pion mass, but much
larger than the typical momenta in hadronic atoms $\simeq 1~{\rm MeV}$.
Consequently, in a very good approximation one may consider the pionic 
deuterium
as a 2-body bound state and apply the machinery that has been developed in
sections~\ref{sec:nonrel}, \ref{sec:including_gamma} and
\ref{sec:boundstates}. This program has been carried out in 
Refs.~\cite{Irgaziev,Raha2} for \pid\!\!. It can be generalized to the
case of  \kad without any change -- the interested reader may
consult these article for details. The analysis of 
pionic (kaonic) deuterium data directly determines the
pion--deuteron (kaon--deuteron) threshold amplitude. For illustration, we
give the expression for the \pid energy shift at next--to--leading
order in isospin breaking~\cite{Raha2},
\eq\label{eq:determine_pid}
\Delta E_1^{\rm str,d}-\frac{i}{2}\, \Gamma_1^{\rm d}=-2\alpha^3\mu_d^2\,
 {\cal A}_{\pi d}\,
\biggl\{1-2\alpha\mu_d\, {\cal A}_{\pi d}\,(\ln\alpha-1)+\cdots\biggr\}\, ,
\en
where $\mu_d$ denotes the reduced mass of the $\pi^-d$ system and
(cf. with Eq.~(\ref{sc_lnew})), and 
\eq\label{eq:d-scattl}
{\cal A}_{\pi d}={\cal A}_c-2\pi i\alpha\mu_d ({\cal A}_c)^2
=a_{\pi d}+\Delta{\cal A}_{\pi d}\, .
\en
Here, ${\cal A}_c$ denotes the pion--deuteron threshold amplitude, defined in 
Eq.~(\ref{eq:appdeuteron}), $a_{\pi d}$ is the pion--deuteron
scattering length in the isospin limit 
and $\Delta{\cal A}_{\pi d}$ stands for the 
isospin breaking correction.
The energy shift of  \kad at next--to--leading order is
also given by Eq.~(\ref{eq:determine_pid}), after an obvious 
replacement ${\cal A}_{\pi d}\to {\cal A}_{\bar Kd}$ and using an 
appropriate reduced mass.
However, the case of \kad inherently differs from  
\pid in one important aspect.
Namely, the estimated ratio of the binding energy to the decay width in 
the ground state of \kad amounts up only to $\simeq 8.6$ 
(see Ref.~\cite{Raha4}) and is much larger in the
pionic deuterium. This value can be still considered as 
large enough~\cite{Raha4} to justify
using the machinery based on the Rayleigh--Schr\"odinger perturbation theory 
(see section~\ref{sec:boundstates}), but the corrections at 
next--to--next--to--leading order might be not completely negligible, when the
accuracy of SIDDHARTA data is close to the planned 
one~\cite{Lucherini:2007ki,SIDDHARTANEW}. 
We therefore conclude that it could be interesting to perform 
-- at some point in the future -- an
estimate of these corrections within the non--relativistic EFT in a manner
described in section~\ref{sec:boundstates}.

At the next step of the investigation 
one has to ``resolve'' the scattering length (threshold amplitude) on the 
composite object (deuteron) in terms of the underlying hadron dynamics,
since for us this scattering length is primarily interesting
as an additional source of information about the pion--nucleon 
(kaon--nucleon) scattering lengths. This is the most difficult 
part of the problem.

Let us first ignore the isospin breaking corrections altogether and start
with \pid. In lowest order in $\bar\gamma/M_\pi$ there exists
an universal relation, which relates the pion--deuteron threshold amplitude
to the threshold parameters of the pion--nucleon
interactions,
\eq\label{eq:d_universal}
a_{\pi d}=\frac{1+\eta}{1+\eta/2}\,(a_{\pi p}+a_{\pi n})
+O(\bar\gamma/M_\pi)\, ,\quad\quad \eta=M_\pi/m_N\, ,
\en
where $a_{\pi p}$ and $a_{\pi n}$ denote pertinent linear
combinations of the $S$-wave $\pi N$ scattering lengths, 
see, e.g., Ref.~\cite{ThomasLandau}.
The relation for the $\bar Kd$ scattering length is completely similar. 
However the correction term
in Eq.~(\ref{eq:d_universal}), albeit down by the small factor 
$\bar\gamma/M_\pi\simeq 1/3$,
turns out to be even larger numerically than the first term. 
The reason for this
in the case of  \pid is that the first term 
is chirally suppressed. A similar situation emerges also
in the case of \kad, however for a different reason.
Here, the kaon--nucleon scattering lengths
are so large that the multiple scattering expansion
seems not to converge anymore (see, e.g.,  
 Ref.~\cite{Gal-conf} for a nice discussion of this issue).
We conclude that, in order to extract useful information from 
 experiments on  pionic (kaonic) deuterium, an accurate evaluation of the correction
term in Eq.~(\ref{eq:d_universal}) is necessary.

\begin{sloppypar}
Existing calculations of  pion--deuteron and kaon--deuteron scattering 
lengths have been carried out in different settings. The description of
low--energy meson scattering on the deuteron is one of the central topics
of the potential scattering theory and has been thoroughly investigated
during  decades. It would be absolutely impossible to cover all this 
very interesting work here, or 
even to provide a fairly complete bibliography.
We cite here only few sources~\cite{Ericson-Weise,Afnan:1974ye,Mizutani:1977xw,ThomasLandau,Kolybasov:1972bn,Deloff:2001zp,Baru,Ericson:2000md,Hetherington,Torres:1986mr,Deloff:1999gc,Barrett:1999cw,Bahaoui:2003xb,Baru:1997xf,Baru:1996pd,Tarasov:2000yi,Deloff:2003ns}, 
which we have consulted on the subject. 
\end{sloppypar}

In recent years,  calculations based on  low--energy effective theories
of QCD in the two--nucleon sector have started to appear. These calculations
enable one to extract (in principle) the pion--nucleon and kaon--nucleon 
scattering lengths in QCD directly from the experimental data, 
without any additional model--dependent input. 
Below, we give a very condensed review of these calculations.

\subsection{Pion--deuteron  scattering}
\label{subsec:pid}

Recently, there has been a considerable activity~\cite{Bernard1,Bernard2,Hanhart_Ko,Hanhart:2007ym,Valderrama:2006np,Platter:2006pt,Baru:2004kw,Hanhartnew} 
in the study of  
$\pi d$ scattering on the basis of the low--energy effective
field theory of QCD with non--perturbative pions, where  multi--pion 
exchanges are included 
into the $NN$ potential that is later iterated to all orders. 
The approach was first
formulated in Weinberg's seminal paper~\cite{Weinberg-deuteron} and has
recently reached a level of sophistication that allows one to perform
  systematic precision calculations in two--nucleon as well as in many--nucleon
systems. For a recent review of this approach, see
Ref.~\cite{Epelbaum:2005pn}. Alternatively, the calculations of the $\pi d$
scattering length have been also 
performed in the EFT with heavy/perturbative
pions~\cite{Borasoy:2003gf,Beane:2002aw,Raha2}.

It is instructive to briefly compare calculations carried out in the two
different settings. The couplings of the EFT with heavy pions include
 the threshold parameters of the $\pi N$ scattering (scattering
length, effective range, etc). Thus, the perturbative expansion in this EFT,
which is carried in powers of $\bar\gamma/M_\pi$, 
produces the multiple--scattering expansion exactly in a form one is looking 
for. Unfortunately, individual terms in the
 multiple--scattering series get strongly 
scale--dependent after the renormalization. The scale dependence can be
only canceled by a (large) three--body contribution, coming from the 6--particle 
LECs. And, since the value of these LECs is completely unknown, this
leads to a large uncertainty in the multiple--scattering 
series~\cite{Borasoy:2003gf,Beane:2002aw,Raha2}.

On the other hand, it has been 
shown~\cite{Bernard1,Bernard2,Nogga,Valderrama:2006np,Platter:2006pt} that
the scale dependence in the EFT
with perturbative pions is rather mild and therefore the pertinent
three--body LECs need not be large. 
This conjecture is independently supported by using dimensional 
estimate and resonance saturation for these LECs~\cite{Bernard2,Raha2}.
Physically, the strong scale dependence in the heavy pion EFT
(large LECs) reflects the increased uncertainty caused by working with a 
smaller energy cutoff, than for the EFT with non--perturbative pions. 
The price to pay for
suppressing the size of the three--body force is however that  
the expansion parameters in the latter approach are initially quark masses
and not the $\pi N$ scattering lengths.

\begin{figure}[t]
\begin{center}
\vspace*{.7cm}
\includegraphics[width=9.cm]{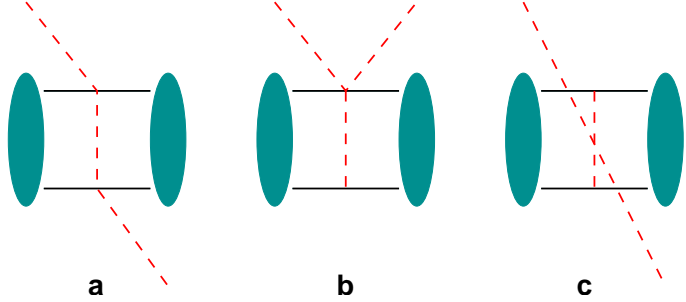}
\end{center}
\caption{Illustration of the modified power counting in the theory with
non--perturbative pions. The shaded blobs denote the wave function of the 
deuteron, and the solid 
and dashed lines correspond to the nucleons and pions, respectively.
Diagrams b) and c), albeit being of the same order 
in ChPT, are strongly suppressed as compared to the diagram a),
see Ref.~\cite{Bernard2}.}
\label{fig:figbernard}
\end{figure}

The conversion of the chiral expansion for the $\pi d$ scattering length
into the form of the multiple--scattering series can be 
achieved on the basis of the following 
heuristic observation (see Ref.~\cite{Bernard2}): it turns out that those
contributions in the chiral expansion, which do not have a counterpart
in multiple--scattering series (these are 
the diagrams that correspond to virtual 
annihilation and creation of pions, see e.g. Fig.~\ref{fig:figbernard}b,c),
are numerically suppressed with respect to the diagrams with no mass gap
(Fig.~\ref{fig:figbernard}a), although both emerge at the same chiral order.
In Ref.~\cite{Bernard2} this property was formalized by
introducing the so--called ``modified power--counting.''
Now, re--grouping the contributions in accordance to this new
counting, in Ref.~\cite{Bernard2} it has been shown 
that up--to--and--including fourth order 
the pion--deuteron scattering length can be
expressed in terms of the pion--nucleon scattering lengths (calculated 
at $O(p^3)$ in ChPT) and a remainder (the boost, or Fermi--motion correction), 
which turns out to be numerically not very large~\cite{Bernard2} albeit 
strongly scale--dependent.
Moreover, unknown LECs emerge first at fifth order in modified counting and
are expected to give a small contribution.

If one neglects the nucleon recoil  (static approximation), 
the final expression for
the pion--deuteron scattering length can be rewritten in a remarkably simple 
form~\cite{Bernard2},
\eq\label{eq:bernard}
&&\hspace{-1.5cm} {\rm Re}\, a_{\pi d}=2\,\frac{1+\eta}{1+\eta/2}\,a_{0+}^+
+2\,\frac{(1+\eta)^2}{1+\eta/2}\,\left((a_{0+}^+)^2-2(a_{0+}^-)^2\right)\,
\frac{1}{2\pi^2}\,
\left\langle\frac{1}{{\bf q}^2}\right\rangle_{\rm wf}
\nonumber\\[2mm]
&&\hspace{-1.5cm}+\,2\,\frac{(1+\eta)^3}{1+\eta/2}\,\left((a_{0+}^+)^3
-2(a_{0+}^-)^2(a_{0+}^+-a_{0+}^-)\right)
\,\frac{1}{4\pi}\,
\left\langle\frac{1}{|{\bf q}|}\right\rangle_{\rm wf}
\!+a_{\rm boost}+\cdots\, ,
\en
where $\langle\cdots\rangle_{\rm wf}$ stands for various wave function 
averages
(we remind the reader that the above expression is obtained
under the assumption of  exact isospin symmetry).
Note that Eq.~(\ref{eq:bernard}) has 
indeed the form of usual multiple--scattering series, known from EFT with 
heavy pions~\cite{Beane:2002aw,Raha2} (or the potential scattering theory).
The difference however is that the wave functions, which are used to calculate
the averages, are the wave functions in the EFT with 
non--perturbative pions. In the calculations in Ref.~\cite{Bernard2}
the NLO wave functions with the cutoff mass in the interval
$\Lambda=(500\cdots 600)~{\rm MeV}$  \cite{Epelbaum:1999dj}
have been used, yielding 
$\left\langle 1 / {\bf q}^2 \right\rangle_{\rm wf} = (12.3\pm 0.3)M_\pi\,$
and $\left\langle 1 /|{\bf q}|\right\rangle_{\rm wf}
= (7.2\pm 1.0) M_\pi^2\,$. 
The boost correction in Eq.~(\ref{eq:bernard}) is
$a_{\rm boost}=(0.00369\cdots 0.00511)M_\pi^{-1}$ (the strong scale dependence
of the boost correction is related to neglecting $\Delta$-resonance 
contribution, see below).

Further, the first term in the series is
chirally suppressed and is anomalously small. The contribution of the second 
term (which is quadratic in the scattering lengths) is large, but higher--order
contributions become again smaller, so that the series are likely to converge.
The measured value of the pion--deuteron scattering length is~\cite{Hauser:1998}
\eq
{\rm Re}\,{\cal A}_{\pi d}^{\rm exp}=-(0.0261\pm 0.0005)M_\pi^{-1}\, .
\en
From Eqs.~(\ref{eq:d-scattl}) and (\ref{eq:bernard}) one obtains a relation
 between $a_{0+}^+$ and $a_{0+}^-$, provided that the isospin-breaking part
$\Delta {\cal A}_{\pi d}$ in Eq.~(\ref{eq:d-scattl}) is dropped.
This relation is shown in Fig.~\ref{fig:bands_new}, see the band named
``Deuteron, Beane et al.''
 In the same Figure we plot the bands which correspond to the $a_{0+}^+$ and
$a_{0+}^-$, determined from the \piH energy shift and width~\cite{Gotta:2006}.
  The combined analysis of the \piH and the \pid data finally yields
the result displayed in Eq.~(\ref{eq:beaneetal}),
 see Ref.~\cite{Bernard2}.

\begin{sloppypar}
We would like to mention that there were additional approximations made in 
Eq.~(\ref{eq:bernard}). In particular, as already stated, 
 the static approximation for the pion
 propagator was used, resulting in averages 
$\langle 1/|{\bf q}|^n\rangle_{\rm wf}$. The possibility of lifting this approximation
 has been considered in Refs.~\cite{Faldt,Baru:2004kw,Raha2}, where it has been 
demonstrated that different corrections to the static limit largely cancel
 each other, so that Eq.~(\ref{eq:bernard}) indeed describes the exact result
quite accurately. What is important, corrections to the static limit can be
calculated perturbatively, because  the multiple--scattering series converges
in the pion--deuteron case. Further, higher--order contributions (most notably,
the so--called dispersive contribution) have been calculated recently
and shown to be rather small~\cite{Hanhart_Ko}
(for another estimate of the size of this correction 
in a different setting, see, e.g., Ref.~\cite{Doring:2004kt}).
The boost corrections were recently re--calculated taking into account 
the explicit $\Delta$-resonance~\cite{Hanhartnew}. It has been in particular
shown that the inclusion of the $\Delta$-resonance 
removes the large scale--dependence of the boost correction.
However, since our main aim here was to demonstrate the general
framework for the
numerical analysis of the \pid data, we have refrained
 from including
the results of the ongoing work into our final plot shown in
Fig.~\ref{fig:bands_new}.
Note also that the error bars in Fig.~\ref{fig:bands_new}
reflect the uncertainty in the coefficients of Eq.~(\ref{eq:bernard}), but not
these additional approximations or the higher--order terms.
In other words, 
we believe that there is still some room left for systematically
improving the numerical precision in Eq.~(\ref{eq:bernard}) in the framework
with non--perturbative pions.
\end{sloppypar}

\subsection{Isospin breaking in the $\pi d$ scattering length}

As already mentioned in section~\ref{sec:piH}, the 
pion--deuteron scattering length,
given by Eq.~(\ref{eq:bernard}), is not consistent with new data
on pionic hydrogen. Namely, 
from Fig.~\ref{fig:bands_new} one immediately observes
that the intersection area for new \piH bands moves far
away from the \pid band.
However, the correction
terms, which were mentioned in the previous subsection, are too small to
 be responsible for the large shift of the \pid
band, which is needed to reconcile the new \piH and \pid data.
Consequently, new mechanisms should be sought that could be a possible source
of large contributions to Eq.~(\ref{eq:bernard}). It should be also understood
that the only  loophole left in the multiple--scattering 
series Eq.~(\ref{eq:bernard}) is assuming  isospin symmetry. It is 
therefore interesting to check  whether  isospin breaking corrections to 
Eq.~(\ref{eq:bernard}) can bridge the gap with the new hydrogen 
data~\cite{Raha3}.

Already in 1977, Weinberg has pointed out~\cite{Weinberg:1977hb}
that the isospin breaking corrections to certain pion--nucleon scattering
amplitudes could become large, if the leading iso\-spin--sym\-met\-ric
contributions to these amplitudes are chirally suppressed. Unfortunately,
Weinberg's statement refers to the scattering processes with neutral pions
that makes it difficult to verify with present experimental 
techniques [see, however, footnote \ref{foot:bernstein}]. 

It turns out, however, that this large isospin breaking
correction emerges in the quantity ${\cal A}_{\pi d}$ as well, since the 
leading--order term
is proportional to the isospin even pion--nucleon scattering length $a_{0+}^+$
and is thus very small (see Eq.~(\ref{eq:bernard})). 
Quite surprisingly, such (a rather obvious) phenomenon
has not been explored so far until very recently~\cite{Raha3}. 
Studies of  isospin breaking in the 
$\pi d$ system 
(see, e.g., Refs.~\cite{Deloff:2001zp,Doring:2004kt,Rockmore,Hanhart:2007ym})
 include effects coming from the
virtual photons at low energy and/or the particle mass differences 
in the loops. 
Numerically these effects, which emerge at higher orders in ChPT, 
indeed turn out to be moderate. 
However, as it is well known (see section~\ref{sec:piH}), 
 isospin breaking in ChPT at leading order emerges through the direct 
quark--photon coupling encoded in the
LECs $f_1,f_2$ as well as due to the explicit quark mass dependence of 
the pion--nucleon amplitudes, which have not been taken into account in
these investigations.

Quite obviously, the leading--order isospin breaking correction to the
pion--deuteron scattering length is given by 
(cf. with Eq.~(\ref{eq:d_universal}))
\eq\label{eq:d_LO}
 {\rm Re} \,\Delta {\cal A}_{\pi d}=\frac{1+\eta}{1+\eta/2}\,
(\delta{\cal A}_c+\delta{\cal A}_n)+O(p^3)\, ,
\en
where the quantities $\delta{\cal A}_c$ and $\delta{\cal A}_n$ are defined
by Eqs.~(\ref{eq:delta-p2}) and (\ref{eq:delta-p2_xn}), respectively.

\begin{sloppypar}
In the numerical calculations the same parameters as
in section~\ref{sec:piH} are used. Namely, since the isospin breaking 
correction for ${\cal A}_{\pi d}$ is evaluated at $O(p^2)$ only, 
for consistency we use the value
$c_1=-0.9^{+0.2}_{-0.5}~{\rm GeV}^{-1}$ at order $p^2$, 
see Eq.~(\ref{eq:c1_p2}). For the same reason, here we have applied $O(p^2)$ 
isospin breaking corrections everywhere, although for the energy shift
$O(p^3)$ result is also known (see section~\ref{sec:piH}).
\end{sloppypar}

 \begin{figure}
\begin{center}
 \includegraphics[width=12.cm]{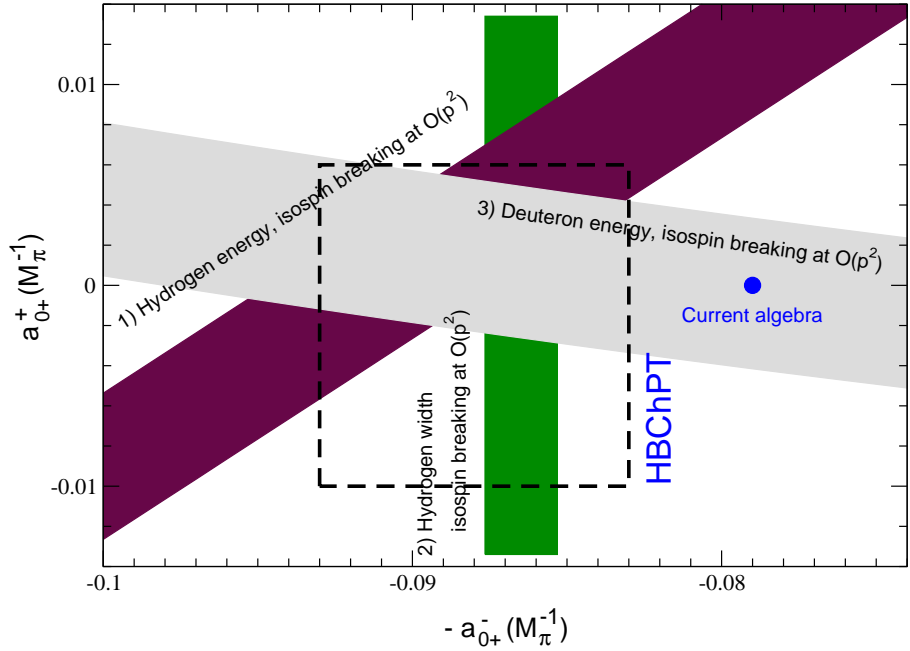}
\end{center}
  \caption{Determination of the $\pi N$ $S$-wave scattering lengths
$a_{0+}^+$ and $a_{0+}^-$ from the combined analysis of the experimental 
data on the \piH
energy shift and width, as well as
 the \pid energy shift.
Isospin breaking corrections have been
 evaluated in the EFT framework: 
see Eqs.~(\ref{eq:corrp2}), (\ref{eq:Zempcorrections}) and (\ref{eq:d_LO}) 
for bands 1), 2) and 3), respectively.
 The filled circle denotes the current algebra prediction~\cite{Weinberg:1966kf}, 
and the dashed box corresponds to the $O(p^3)$ calculation in Heavy Baryon
ChPT~\cite{FettesI}.  See main text for more comments. 
\label{fig:plot}}
 \end{figure}

 \begin{figure}
\begin{center}
 \includegraphics[width=12.cm]{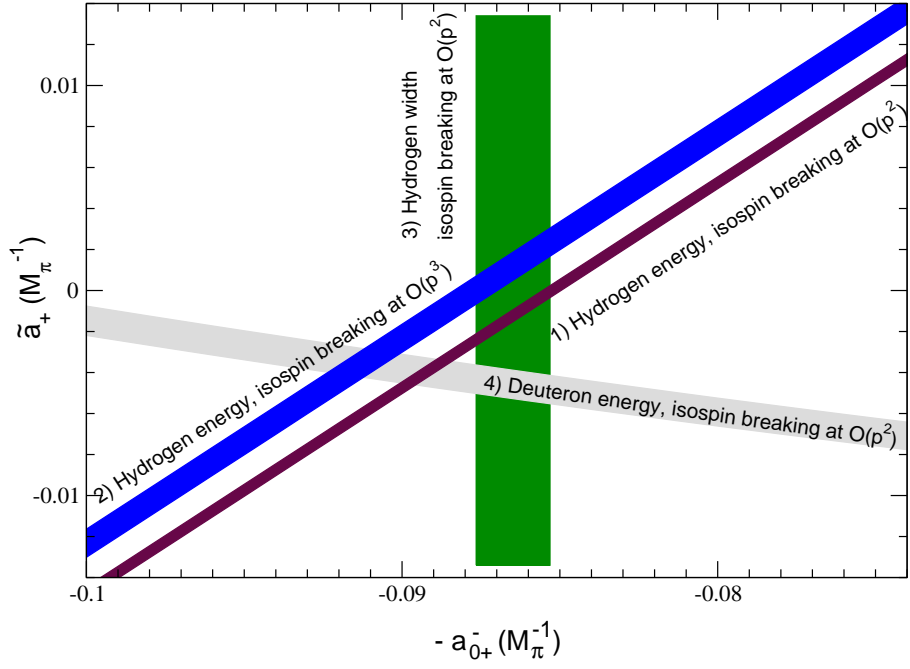}
\end{center}
  \caption{The quantity $\tilde a_+$ defined in Eq.~(\ref{eq:bplus}),
 plotted against $a_{0+}^-$, see
    Ref.~\cite{barumenu2007}. It is seen that, with the isospin breaking
    corrections evaluated at $O(p^2)$, the three bands 1), 3) and 4) 
have no common intercept~\cite{barumenu2007}. 
However, one also concludes that the corrections from higher 
orders are larger than the
    uncertainty at $O(p^2)$ coming from the LECs alone. The distance between
    the  bands 1) and 2), corresponding to the same \piH energy shift and to the
    isospin-breaking corrections evaluated at $O(p^2)$ and at $O(p^3)$,
    respectively, may serve as a rough estimate of the higher-order terms. }
\label{fig:bplus}
 \end{figure}

At the leading order, the isospin breaking correction to the $\pi d$
scattering length is independent
on the deuteron structure and is extremely large
\begin{equation}\label{eq:110}
{\rm Re}\,\Delta {\cal A}_{\pi d} 
= -(0.0110^{+0.0058}_{-0.0081}) \, M_\pi^{-1}~,
\end{equation}
or,
${\rm Re}\,\Delta {\cal A}_{\pi d}/ {\rm Re}\,a_{\pi d}^{\rm exp}=0.42$ 
(central values). Plotting again the bands in the $(a_{0+}^+,a_{0+}^-)$-plane
(see Fig.~\ref{fig:plot}),
 it is  seen that
the correction in Eq.~(\ref{eq:110}) moves the de\-u\-te\-ron band in 
 in the right direction: the isospin breaking 
corrections amount for the bulk of the apparent discrepancy between the experimental
data on  \piH and \pid in Figure~\ref{fig:bands_new}.

It is useful to comment here on the peculiar algebraic structure 
of these isospin breaking terms, which allows one set up a more convenient framework for the 
analysis of  \piH and \pid data.
 First, we note that the widths of the two bands 1) and 3) in Figure~\ref{fig:plot} are mainly 
due to the uncertainties in the LECs $c_1,f_1$ that occur in $\delta{\cal A}_{c,n}$. 
On the other hand, as    pointed out in Refs.~\cite{Raha3,rus:trento,barumenu2007},
 these two LECs occur in the same combination in the energy shifts of \piH and \pid at order $p^2$ in 
isospin breaking. Introducing the quantity~\cite{barumenu2007}
\bea\label{eq:bplus}
\tilde a_+=a_{0+}^++\frac{1}{4\pi(1+\eta)}\,\biggl(
\frac{4\Delta_\pi}{F_\pi^2}\,c_1-2e^2 f_1\biggr)\,,
\eea
the expressions for the energy shifts in \piH and \pid then contain 
only $\tilde a_+$ and $ a_{0+}^-$ (up to small contributions from the scattering length $a_{0+}^+$). 
One may then display the available information in 
the ($\tilde a_+$,$a_{0+}^-$)--plane~\cite{barumenu2007}, 
see Figure~\ref{fig:bplus}, from where it is seen that the data on \piH and \pid do not lead to
a common intersection\footnote{This seems to be in contradiction to the common 
intersection of the bands 1) and 3) in Fig.~\ref{fig:plot}. However, that intersection 
corresponds to a different choice of the LECs in the \piH and \pid energy 
shifts, and signals an apparent consistency only~\cite{baruprivate}.}.

Taking into account higher order terms in the multiple scattering 
theory of \pid improves the situation~\cite{barumenu2007}.
 On the other hand, up to now,  isospin breaking effects have been taken into account
only at $O(p^2)$ in \pid$\!\!$, albeit
 certain contributions are already calculated~\cite{Rockmore,Hanhart_Ko}.
 We illustrate in Figure~\ref{fig:bplus}  that these higher order terms do indeed matter:
 the blue band displays the band from the energy shift in \piH$\!\!$, provided that 
terms at order $p^3$ in isospin breaking are taken into account (that band corresponds to the
 blue band in Figure~\ref{fig:bands_new}, translated here to the variable $\tilde a_+$).
 We conclude that a similar 
investigation for \pid is urgently needed. A more complete analysis may then lead to
an improved determination  of the scattering length $a_{0+}^-$ and of the quantity $\tilde a_+$, because
there is an additional consistency constraint between \piH and \pid data.  
 On the other hand, the problem with 
the measurement of $a_{0+}^+$ persists: unless one finds means to pin down $f_1$ 
more reliably, the uncertainty in $a_{0+}^+$, which must ultimately be determined from $\tilde a_+$, 
 is of the order of $a_{0+}^+$ itself.

For completeness,  we now display results
of  independent analyses of the \piH data
 performed by other groups.

\begin{itemize}

\item
In Ref.~\cite{Sainio-new} an analysis of the
\piH data has been performed and the  GMO sum rule 
has been used for a precise determination of the
$\pi NN$ coupling constant. The energy shift and width of  \piH
were taken from Ref.~\cite{Gotta:2005} and Ref.~\cite{Schroder:2001}, respectively, and
the isospin breaking corrections
\eq\label{eq:Mojzisold}
\delta_1'=(-7.2\pm 2.9)\cdot 10^{-2}\quad\cite{Mojzis}\, ,\quad
\delta_1=(0.6\pm 0.2)\cdot 10^{-2}\quad\cite{Zemp}
\en
have been applied. The final results for the scattering lengths are
\eq
\hspace{-.9cm}a_{0+}^++a_{0+}^-=0.0933\pm 0.0029\,M_\pi^{-1},  \quad
a_{0+}^-=0.0888\pm 0.0040\,M_\pi^{-1}\fs
\en

\item
In the second work of 
Ref.~\cite{Gotta:2006}, the isospin breaking corrections from 
Eq.~(\ref{eq:Mojzisold}) have been applied to the recently measured values 
of the energy shift and width given in Eq.~(\ref{eq:hyd_energywidth}). The
result is 
\eq
a_{0+}^+=0.0069\pm 0.0031\,M_\pi^{-1}\, ,\quad\quad
a_{0+}^-=0.0864\pm 0.0012\,M_\pi^{-1}\, .
\en
\end{itemize}

\subsection{Kaon-deuteron scattering}
\label{subsec:kaondeuteron}
It is expected that the forthcoming measurement of the (complex) energy shift of \kad
by the SIDDHARTA collaboration at \dafne~\cite{Lucherini:2007ki,SIDDHARTANEW} enables one to extract -- in combination with \kaH data --
independently both $S$-wave $\bar KN$ scattering 
lengths $a_0$ and $a_1$. We show in this subsection that this indeed appears to be a feasible task.

We start with the kaon--deuteron threshold amplitude
 ${\cal A}_{\bar Kd}$, which is extracted
from the analysis of \kad data. This amplitude
 must be ``unfolded'' in
terms of  elementary $\bar KN$ scattering lengths,  via  multiple--scattering
theory. Because the 
multiple--scattering series for the kaon--deuteron scattering length is likely 
not to converge at all,  a re--summation is needed. A detailed investigation of the problem
within an EFT framework has been carried
out recently in Refs.~\cite{Kamalov:2000iy,Raha4}, see also Ref.~\cite{Chand-Dalitz} 
for a treatment of the same
  problem in  potential scattering theory.
The multiple--scattering
series  has been
re--summed in Refs.~\cite{Kamalov:2000iy,Raha4} to all orders, assuming that the nucleons are static (this is also
referred to as Fixed Centers Approximation (FCA)), and neglecting
derivative $\bar KN$ interactions. The resulting expression  is given 
by
\eq\label{eq:final-Kamalov}
\biggl(1+\frac{M_K}{M_d}\biggr){\cal A}_{\bar Kd} 
=\int_0^\infty dr\,(u^2(r)+w^2(r))\,
\hat {\cal A}_{\bar Kd}(r)\, ,
\en
where $u(r)$ and $w(r)$ denote the usual 
$S$- and $D$-wave components of the deuteron
wave function, which are normalized via the condition 
$\int_0^\infty dr\,(u^2(r)+w^2(r))=1$.
The NLO wave functions in the theory with non--perturbative 
pions~\cite{Epelbaum:2005pn,Evgeny-private} have been used in the 
calculations~\cite{Raha4}. Furthermore,
\eq\label{eq:ratio-Kamalov}
\hspace{-1cm}\hat {\cal A}_{\bar Kd}(r)=\frac{\tilde {\cal A}_c+\tilde {\cal A}_n
+(2\tilde {\cal A}_c\tilde {\cal A}_n-{\cal B}_x^2)/r-2{\cal B}_x^2
\tilde {\cal A}_n/r^2}
{1-\tilde {\cal A}_c\tilde {\cal A}_n/r^2+{\cal B}_x^2
\tilde {\cal A}_n/r^3}+\delta \hat {\cal A}_{\bar Kd}\,,
\en
with ${\cal B}_x^2=\tilde {\cal A}_x^2/(1+\tilde {\cal A}_0/r)$, and
\eq\label{eq:ratio-Kamalov1}
\biggl(1+\frac{M_K}{m_p}\biggr){\cal A}_{c,n,x,0}=\tilde {\cal A}_{c,n,x,0}\, ,
\en
\begin{sloppypar}
\noindent
where ${\cal A}_{c,n,x,0}$ denote the threshold scattering ampli\-tu\-des
for $K^-p\to K^-p$,  $K^-n\to K^-n$, $K^-p\to \bar K^0n$ and 
$\bar K^0n\to \bar K^0n$, respectively, see appendix~\ref{app:notations}. 
Retaining only the leading isospin breaking effects which are due to the 
unitary cusps, these amplitudes can be expressed through the scattering lengths
$a_0$ and $a_1$ (cf. with Eq.~(\ref{eq:Kbubbles})),
\eq\label{eq:cusps}
{\cal A}_c&=&\frac{\frac{1}{2}\,(a_0+a_1)+q_0a_0a_1}{1+\frac{q_0}{2}\,(a_0+a_1)}+\cdots\, ,
\quad
{\cal A}_n\!=\!a_1+\cdots\, ,
\nonumber\\
{\cal A}_x&=&\frac{\frac{1}{2}\,(a_0-a_1)}{1-\frac{iq_c}{2}\,(a_0+a_1)}+\cdots\, ,
\quad
{\cal A}_0\!=\!\frac{\frac{1}{2}\,(a_0+a_1)-iq_ca_0a_1}{1-\frac{iq_c}{2}\,(a_0+a_1)}+\cdots\, ,
\nonumber\\
&&
\en
where
\eq
&&q_c=\sqrt{2\mu_c\Delta}\, ,\quad
q_0=\sqrt{2\mu_0\Delta}\, ,\quad \Delta=m_n+M_{\bar K^0}-m_p-M_K\, ,
\nonumber\\[2mm]
&&\mu_c=\frac{m_pM_K}{m_p+M_K}\, ,\quad
\mu_0=\frac{m_nM_{\bar K^0}}{m_n+M_{\bar K^0}}\, .
\en
The isospin limit corresponds to $q_c=q_0=0$~\footnote{
Here, we identify
${\cal A}_{c,n,x,0}$ in Eqs.~(\ref{eq:ratio-Kamalov}) and 
(\ref{eq:ratio-Kamalov1})
with the $\bar K N$ amplitudes evaluated
at the pertinent physical thresholds. In the language of non--relativistic EFT,
 this amounts to neglecting  diagrams which describe 
 rescattering of $\bar K$ on the same nucleon. 
Including the rescattering contributions leads to the
replacement of ${\cal A}_{x,0}$ by the amplitudes evaluated at $s=(M_K+m_p)^2$
instead of $s=(M_{\bar K^0}+m_n)^2$. In addition, in the static approximation
the mass difference $\Delta$ is
neglected in the neutral kaon propagator  
(for comparison, 
see also Refs.~\cite{Chand-Dalitz,Bahaoui:2003xb}, where the problem has been
addressed within the framework of a potential model).
Numerically, the effect may not be negligible. 
However, following Ref.~\cite{Raha4},  we do not take it into account at this point.}. 
Finally, the quantity $\delta \hat {\cal A}_{\bar Kd}$ contains
the 6-particle non--derivative LEC, which describes  
interaction of the kaon with two nucleons. A dimensional estimate
yields a few percent systematic uncertainty
due to the presence of this LEC~\cite{Raha4}.
In the numerical results, which are displayed
below, this term is neglected completely.

\underline{Remark:} We comment on an important difference between the kaon-deuteron
  and pion-deuteron scattering in this respect. 
In the case of pion-deuteron scattering, a formula 
similar to  Eq.~(\ref{eq:ratio-Kamalov}) can be derived,
see, e.g., Ref.~\cite{Deloff:2001zp}. The $\pi N$ threshold
amplitudes  that enter the expression are real, up to small isospin breaking
effects. On the other hand, due e.g. to the process $\pi d \to nn$,
  the imaginary part of the corresponding correction
 $\delta\hat { \cal
  A}_{\pi d}$ does not vanish in the isospin symmetry limit. It is, therefore,
 expected to dominate the 
contributions from the first part, as a result of which
 the imaginary part of the 
pion--deuteron scattering amplitude at threshold cannot provide 
information on the $\pi N$ 
scattering lengths. In contrast to this, the imaginary parts of the
 $\bar K N$ amplitudes in Eq.~(\ref{eq:ratio-Kamalov}) are of order one.
As just mentioned, it is expected that 
 $\delta\hat{\cal A}_{\bar Kd}$ is
small as compared to the first term, and is neglected here.
 This explains why kaonic
deuterium does provide information on the imaginary part of the 
scattering lengths, while pionic deuterium does not.
 For  an investigation of this issue 
 in the framework of a potential model, see
Ref.~\cite{Bahaoui:2003xb}.) \underline{End of remark}

\end{sloppypar}

We would like to emphasize that, in order to  reduce the 
systematic error on the calculated kaon--deuteron scattering length,
it would be very important to
extend the calculations beyond the lowest--order 
formula Eq.~(\ref{eq:final-Kamalov}) by using the 
non--relativistic Lagrangian me\-thod of Ref.~\cite{Raha4}.
In particular, it would be interesting to estimate the effect of lifting FCA, as
well as to evaluate the contribution of the $D$-waves (derivative interactions).
The comparison with the Faddeev approach indicates that these corrections
may not be negligible (see, e.g., Ref.~\cite{Bahaoui:2003xb,Gal-conf}).

The main goal of the DEAR/SIDDHARTA experiment is to determine
individual $\bar KN$ scattering lengths $a_0$ and $a_1$ from the analysis
of combined \kaH and \kad data. Since the experimental
results on the \kad are still absent,
in Ref.~\cite{Raha4}
such an analysis has been carried out, using synthetic input \kad
data. The procedure can be schematically described as follows. From 
Eq.~(\ref{eq:circle}) one determines one of the scattering lengths, say $a_1$.
Substituting this expression into Eqs.~(\ref{eq:final-Kamalov}),
(\ref{eq:ratio-Kamalov}) and (\ref{eq:cusps}), one arrives at a 
non--linear equation for determining $a_0$ with a given input value
of ${\cal A}_{\bar Kd}$. 

\begin{figure}[t]
\begin{center}
\vspace*{.7cm}
\includegraphics[width=8.cm]{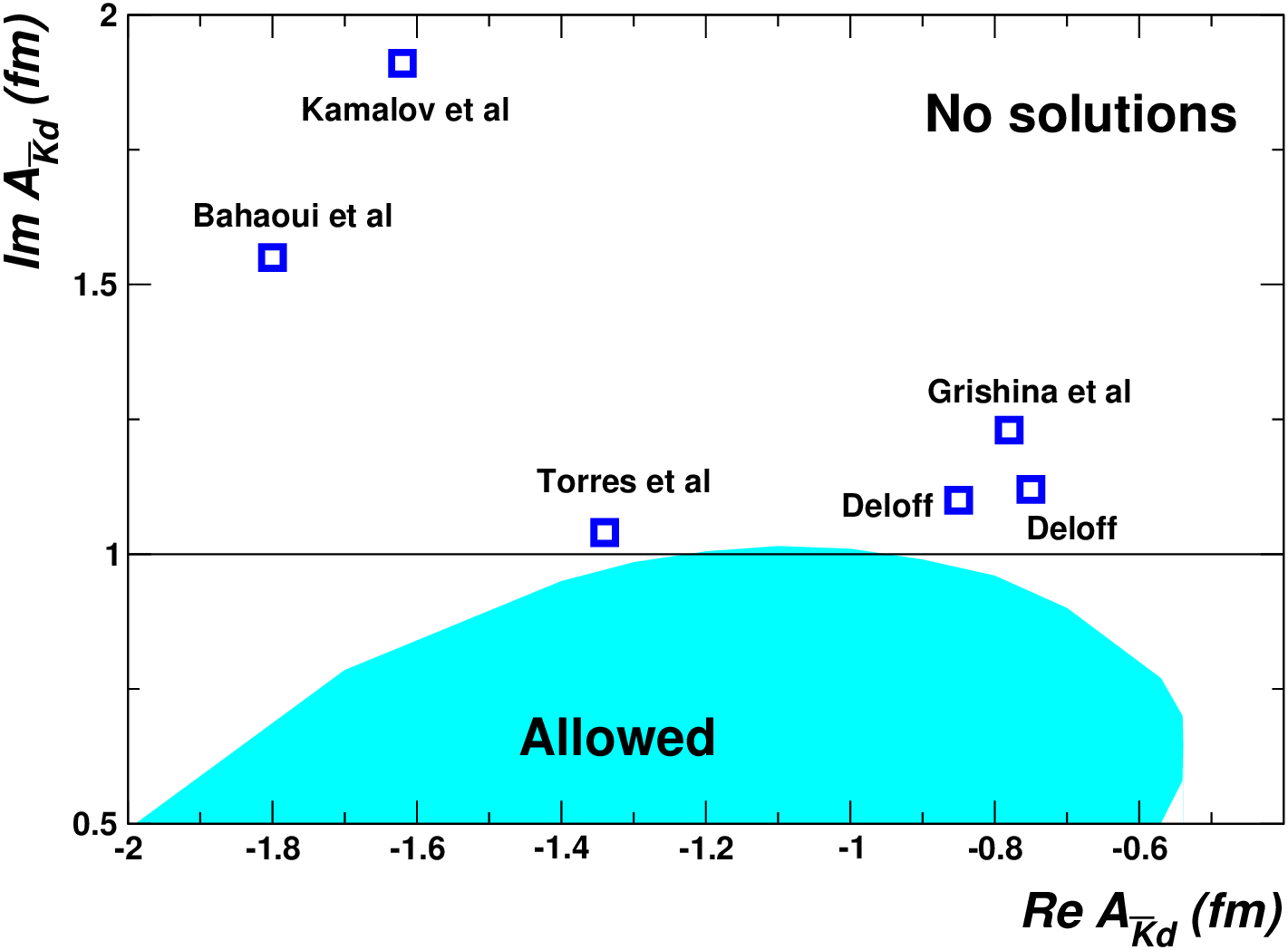}
\end{center}
\caption{The region in the $({\rm Re}\,{\cal A}_{\bar Kd},  
{\rm Im}\,{\cal A}_{\bar Kd})$--plane
where  solutions for $a_0$ and $a_1$ do exist. NLO wave 
functions~\cite{Epelbaum:2005pn,Evgeny-private} with the cutoff
parameter $\Lambda=600~{\rm MeV}$ have been used in the calculations 
(for $\Lambda=450~{\rm MeV}$, the result changes 
insignificantly). For comparison, we also show the results of
various calculations of 
${\cal A}_{\bar  Kd}$~\cite{Torres:1986mr,Deloff:1999gc,Grishina,Bahaoui:2003xb}
 (squares). 
As we see, none of the calculated scattering lengths is located in the shaded
 area.
The figure is the same as in Ref.~\cite{Raha4}.
}
\label{fig:area}
\end{figure}

The numerical solution of the above equation, carried out in Ref.~\cite{Raha4},
leads to an interesting conclusion. It turns out that unitarity and input DEAR
data for \kaH impose severe constraints on the possible input
values of ${\cal A}_{\bar Kd}$, for which the solutions for 
the $\bar KN$ scattering lengths with $\mbox{Im}\, a_I\geq 0$ do exist. 
The allowed region in the
 $({\rm Re}\,{\cal A}_{\bar Kd},{\rm Im}\,{\cal A}_{\bar Kd})$--plane
is shown Fig.~\ref{fig:area}. It also turns out that
the allowed values for ${\cal A}_{\bar Kd}$ qualitatively
agree with the value of the $\bar K^0d$ scattering length extracted from
the $pp\to d\bar K^0K^+$ reaction~\cite{Sibirtsev}.
 Finally,  we note that the region where solutions do exist
is much larger in the case of the KpX input~\cite{KEK}, than for the DEAR
input of kaonic hydrogen~\cite{Raha4}.

 The message of the investigation carried out in Ref.~\cite{Raha4}
 is very clear:  the 
combined analysis of DEAR/SIDDHARTA
data on \kaH and \kad 
is more restrictive than one would {\it a priori} expect. 
Moreover, if the corrections to the lowest--order approximate result
(going beyond FCA and including  derivative couplings)
 are moderate, they will not change the qualitative picture
shown in Fig.~\ref{fig:area}.  On the other hand, they
 constitute the largest potential source of 
theoretical uncertainty at present. 

We conclude with the expectation that the combined analysis of the forthcoming 
high--precision data from DEAR/SID\-DHAR\-TA collaboration on \kaH 
and de\-u\-te\-ri\-um will enable one to perform a stringent test of the 
framework used to describe low--energy kaon--deuteron scattering,
as well as to extract the values of $a_0$ and $a_1$ with  reasonable accuracy.
On the other hand, a considerable amount of theoretical work -- related to a systematic
calculation of higher--order corrections --
is still to be carried out before this goal is reached.

%%%%%%%%%%%%%%%%%%%%%%%%%%%%%%%%%%%%%%%%%%%%%%%%%%%%%%%%%%%%%%%%%%%%%%
\setcounter{equation}{0}
\section{Potential scattering theory}
\label{sec:potential}

In this section, we provide a condensed version of our view 
of the relation between EFT
and potential model calculations.

Many calculations of various hadronic atom 
properties were carried out within the framework of  potential
scattering theory since the seminal work of Deser et al.~\cite{Deser}.
 Potential model calculations (see, e.g., Refs.~\cite{rascheetal,Scheck,Deloff:2003ns,Trueman,Partensky,Lambert,Rasche:dz,Rasche:mp,Pilkuhn:rf,Rasche:zg,Moor:ye,Gashi:1997ck,Gashi:2001wv,Rasche:2001talk,Kaufmann:sf,Sigg,Ericson:2002km,Ericson:2003ju,Ericson:1981hs,Ericson-Weise,Afnan:1974ye,Mizutani:1977xw,ThomasLandau,Deloff:2001zp,Baru,Ericson:2000md,Hetherington,Torres:1986mr,Deloff:1999gc,Barrett:1999cw,Bahaoui:2003xb,Gashi:2001vt,Suebka:2004zi,Amirkhanov:1998ir,Kulpa:1996ry})
were basically the only theoretical setting to address the issue until rather
recently, when  methods of (effective) QFT came in use to tackle the
problem. Already the very first example investigated in this new framework
 -- the lifetime of the ground state of pionium --
 revealed a qualitative difference of the two approaches:
 A major contribution of  the
 so--called isospin breaking corrections to the energy--level shift
 turned out to be of opposite sign in the two approaches 
(cf. Ref.~\cite{Gashi:1997ck} and Refs.~\cite{Sazdjian1,Dubna3}).
As mentioned in section \ref{sec:piH}, similar
 discrepancies were later found  in the calculation of the energy shift
 in \piH (Ref.~\cite{Sigg} and Refs.~\cite{Bern3,Mojzis}), 
 see also the recent work Ref.~\cite{rascheetal}. 
These isospin breaking corrections are of the order of 
the envisaged experimental accuracy,
 and it is therefore  important to reveal the reason for the difference. 
 This section is
 devoted to a clarification of this point.

We find it most instructive to illustrate the two approaches in the case of
the strong energy shift of \piH. 
We display three expressions for the energy shift in the ground state,
in chronological order:
\bea
\Delta E_1^\mathrm{str}&=&\left\{\begin{array}{lcr}
N\,\, \mbox{Re}\,\, {\cal T}&&(a)\\[2mm]
N\,\,\mbox{Re}\,\, \hat {\cal T}\,\biggl\{1
+\beta_1\frac{\mbox{Re}\,\mbox{$\hat {\cal T}$}}{\mbox{$r_B$}}
+O(r_B^{-2})\biggr\}
&&(b)\\[3mm]
N\,\,\mbox{Re}\,{\cal T}_c\,\biggl\{1-\frac{\mbox{$\mu_c^2\alpha $}}{\mbox{$4\pi m_pM_\pi$}}\,
(\ln{\alpha}-1) \,\mbox{Re}\,{\cal T}_c\biggr\}
+o(\delta^4)&&(c)
\end{array}
\right.\nonumber\\[3mm]
&&N=-\frac{\alpha^3\mu_c^3}{4\pi m_p M_\pi}\, .\nonumber
\eea
The amplitudes ${\cal T},\hat {\cal T}$ and ${\cal T}_c$ stand for the elastic $\pi^-
p\to \pi^- p$ amplitude at threshold, in specific settings.
The formulae $(a)\,,\,(b)$ 
and (c) were derived by Deser et al.~\cite{Deser}, 
by Trueman~\cite{Trueman} (properly adapted here),
 and by Lyubovitskij and Rusetsky~\cite{Bern3}, 
respectively. The first two
are potential model ones, whereas the last one uses the EFT framework
advocated in this report\footnote{As mentioned, there were very many potential
  model calculations over the last four decades, improving $(b)$ in several
  respects. However, the basic difference between these approaches and EFT
  remained untouched. For this reason, we stick to 
the evaluation by Trueman for simplicity.}.
We now discuss
these expressions  in turn, and start with (a). Here,
${\cal T}$ is the elastic scattering amplitude  in the
absence of Coulomb interaction. 
 Additional terms
on the right hand side of this relation were not considered in Ref.~\cite{Deser}.
 The second
relation corresponds to an expansion in powers of 
$\hat{\cal T}/r_B$, where $\hat{\cal T}$ is
 evaluated  in the presence of the Coulomb
interaction. The coefficient $\beta_{1}$ depends on the dimensionless ratio
$r_0/r_B$, where $r_0 (r_B) $ stands for the effective range 
(Bohr radius). In order to compare
these two formulae, one needs to know the relation between the amplitudes
${\cal T}$
and $\hat {\cal T}$. In Ref.~\cite{Trueman}, the relation 
\cite{chewandgoldberger}
\bea\label{eq:cg}
\mbox{Re}\,\hat {\cal T}=\mbox{Re}\,{\cal T}\,\biggl\{1-
\frac{\mu_c^2\alpha\ln{\alpha}}{4\pi m_pM_\pi}\,
\,\mbox{Re}\,{\cal T}+O(\alpha)\biggr\}
\eea
was invoked, which indicates that (a) is a first order
approximation of (b)\,, in the sense of the expansion performed in 
(b).

We now come to the relation displayed in (c), worked out in Ref.~\cite{Bern3},
see also section \ref{sec:piH}. Optically, the relation is similar 
to $(a), (b)$. However, 
as is worked out at length in this report, $(c)$ is based
on a very general framework: quantum field theory. This framework
allows one to include in a systematic manner strong and 
electromagnetic interactions. The amplitude ${\cal T}_c$ is of the form
\bea
\mbox{Re}\,{\cal T}_c=8\pi(m_p+M_\pi)\,(a_{0+}^++a_{0+}^-)+\alpha F_1+(m_d-m_u) F_2\, + o(\delta)\,,
\eea
where the first term on the right hand side is a particular 
combination of $\pi N$ scattering lengths, and where the coefficients
$F_{1,2}$ depend on the underlying theory. In this report, this is taken to be
 QCD+QED,
 and the  corrections encoded by $F_{1,2}$ stand for the
so--called short--distance electromagnetic contributions to the scattering
amplitude. The quantities $F_{1,2}$ can be calculated 
in ChPT, in an expansion in the quark masses $m_u=m_d$, in terms of
a well--defined set of LECs.

It is at this stage that the EFT framework and the potential model
calculations differ:
EFT allow one to provide a direct contact
between the measured energy shifts and the calculated scattering 
lengths in QCD. 
For more details on the relation between the EFT and potential model
framework, we refer the interested reader to Ref.~\cite{Lipartia:2001zh}. 
In this reference, it is proved that the relation between the strong energy
shift in the form $(c)$ also holds in potential scattering, a property that
goes under the name {\it universality}. 
 Universality
clearly shows that it is not the use of the potential framework that causes 
discrepancies with the EFT approach, but the choice of the short--range
potential, which must include isospin breaking effects properly.

In our opinion, for the reasons just outlined, 
potential model calculations 
are superseded by EFT methods in the case of 
simple systems like $A_{2\pi}$, $A_{\pi K}$, \piH and \kaH.
On the other hand,  in more complicated systems like
\pid and \kad, the results of 
calculations performed within potential models may  still be useful as a hint
about the expected magnitude of different contributions. However, in order to
avoid an uncontrollable systematic error, one would have to address the same
calculations within the framework of low--energy effective theories of QCD 
as well.

%%%%%%%%%%%%%%%%%%%%%%%%%%%%%%%%%%%%%%%%%%%%%%%%%%%%%%%%%%%%%%%%%%%%%%%
\setcounter{equation}{0}
\section{Summary and outlook}
\label{sec:concl}

In recent years, a general   theory
of hadronic atoms has been developed, as is outlined in this review. It relies in an essential manner on 
effective field theories: ChPT, which describes
the interactions of hadrons and photons at low energy, and 
 non--relativistic effective QFT,
 that are used to describe
hadronic bound states. It turns out that the 
treatment of any hadronic bound system
 is pretty universal within this theory (i.e., independent of any details
characterizing the constituents) and can be carried out with a surprising
ease. 
The approach is {\em systematic:} the terms that are neglected 
count at higher orders in a power--counting scheme 
as compared to the ones that are retained.
 Another important  feature of the present approach
is the fact  that the needed effective field theories  are built
on  top of QCD+QED by successively integrating out various 
high--energy scales. Consequently, calculating observables 
 of a hadronic bound state within this approach, one ends up with a
quantity that is evaluated in the fundamental theory of strong and 
electromagnetic interactions.

The theory of hadronic atoms presented here merges  several fields of 
theoretical physics. As already pointed out, it heavily relies on ChPT
and on  non--relativistic effective field theories.
Presently, the attention is shifted to  
hadronic atoms containing the deuteron or other (light as well as heavy) 
nuclei. As a result of this, methods of  few--body physics 
 and in--medium properties of hadrons are needed in addition.

We believe that the conceptual problems of a
theory of hadronic atoms 
have now been  clarified to a large extent.
 In this review, we have described several specific applications.
 There is, however, still  room left for future
investigations on the subject. We have collected in Table \ref{tab:status} 
our view of 
the status of the {\it theoretical description} of the various hadronic 
compounds. The first column in the Table displays the system under
 consideration, and the second one contains
the status of the relation between the energy spectra and the pertinent
scattering amplitudes. The third column displays information about the relation
between the hadronic amplitudes and the various scattering lengths, whereas
the last one refers to keywords related to the underlying physics.

The main points displayed in the Table are the following.

\begin{table}
\noindent  
\renewcommand{\arraystretch}{1.1}
\begin{tabular}{|l|l|l|l|}
\hline
&&&\\[-3mm]
&$(\Delta E,\Gamma)\to{\cal T}$&${\cal T}\to a$ & Underlying physics \\[1mm]
\hline
&&&\\[-3mm]
$A_{2\pi}$&perfect&perfect&$\cdot\,\,\pi\pi$ scatt. lengths\\
&&&$\cdot\,\,$large/small quark\\[-0mm]
 &&&\,\, condensate\\
[1mm]\hline
 $A_{\pi K}$&perfect&perfect&$\cdot\,\,\pi K$ scatt. lengths\\[-0mm]
&&&$\cdot\,\,$large/small $SU(3)$\\[-0mm]
  &&& \,\,\,quark condensate\\
[1mm]\hline
  &&&\\[-3mm]
\piH &perfect&$\cdot\,\,$large uncertainty (LECs)&$\cdot\,\,\pi N$ scatt. lengths\\
 &&$\cdot\,\,$to be done:&$\cdot\,\,\pi N$ $\sigma$-term\\
 &&\,\,\,$O(p^4)$ in energy shift &$\cdot\,\,$ pion--nucleon \\
 &&\,\,\,$O(p^3), O(p^4)$ in the width&\,\,\,\,\,coupling constant\\
 [1mm]\hline
 &&&\\[-3mm]
 \kaH &reasonably&$\cdot \,\,O(\sqrt{\delta}),O(\delta\ln\delta)$: done& $\cdot\,\,\bar K N$
 \,\,\,scatt. lengths\\
 &good&$\cdot\,\,O(\delta)$ (tree level): done&$\cdot\,\,$unitarized ChPT\\
 &&$\cdot\,\,O(\delta)$ (loops): unrealistic &\\
 [1mm]\hline
 &&&\\[-3mm]
 \pid&good&{\it isospin conserving sector:}&$\cdot\,\,\pi N$ scatt. lengths\\[-0mm]
&&in good shape [still& $\cdot\,\,$EFT in  $2N$ sector\\[-0mm]
 &&in progress]&$\cdot\,\,$3-body calculations\\[-0mm]
&&{\it isospin breaking sector:}&\,\,\,in EFT\\[-0mm]
 &&leading order: done (huge)&$\cdot\,\,$fixing LECs\\[-0mm]
&&higher orders: partly done/&\\[-0mm]
 &&in progress&\\
[1mm]\hline
 &&&\\[-3mm]
\kad&satisfactory&{\it isospin conserving sector:}&$\cdot\,\,\bar K N$ scatt. lengths \\[-0mm]
&&nonder. coupling,&$\cdot\,\,$ unitarized ChPT\\[-0mm]
 &&stat. approx.: done&$\cdot\,\,$EFT in the $2N$ sector\\[-0mm]
&&deriv. coupling: to be done&$\cdot\,\,$3-body calculations\\[-0mm]
 &&{\it isospin breaking sector:}&\,\,\,in EFT\\
&&lead. order: to be completed &\\[-0mm]
 &&higher orders: unrealistic& \\
[1mm]\hline
\end{tabular}

\caption{Status of the theoretical description of various hadronic
  compounds, ordered according to increasing complexity. The first column
  displays the system, the second the relation between the spectrum and the
  scattering amplitude ${\cal T}$, the third column concerns the relation between the
  scattering amplitude ${\cal T}$ and the scattering lengths, and the last column displays
  issues of the underlying physics.}\label{tab:status}
\end{table}

\begin{itemize}

\item[i)]
A precise calculation of  isospin breaking corrections in  pionic 
hydrogen energy shift and width should be performed. 
This implies calculations done solely
in ChPT, in analogy with Ref.~\cite{Mojzis}. In particular, it would be 
useful to evaluate the charge--exchange amplitude at $O(p^3)$ (for the width)
 and, possibly,
at $O(p^4)$ for both $\pi^-p$ elastic and
charge--exchange amplitudes (energy shift and width).

\item[ii)]
Using the results of these calculations, the analysis of the \piH data
should be done anew. In particular, it would be  
very interesting to update the value
of $a_{0+}^+$ and of the pion-nucleon $\sigma$-term.

 \item[iii)]
In  $\pi K$ scattering, 
the structure of higher order ChPT contributions~\cite{piKbijnens} remains to be 
understood in view of the low--energy theorem Eq.~\eqref{eq:roessl}.
An issue still to be investigated is the relation
between the large/small $SU(3)$ quark condensate scenario, and the pertinent
 scattering lengths. 

\item[iv)]
 A substantial progress in the precise quantitative description
 of  pion-de\-u\-te\-ron scattering in the chiral EFT 
would be extremely desirable. This implies systematic calculations
of various higher--order correction terms, including both, isospin conserving 
 [Eq.~(\ref{eq:bernard})]
and isospin breaking [Eq.~(\ref{eq:d_LO})] contributions, see e.g. 
Ref.~\cite{Hanhart_Ko}.

\item[v)]
\begin{sloppypar}
We believe that \kaH and \kad,
which will be investigated by the SIDDHARTA experiment at 
 \dafne
\cite{Lucherini:2007ki,SIDDHARTANEW},
 represent
the most challenging theoretical task at pre\-sent. Many
issues should be addressed in this context. For example,
one has to gain a deeper insight into the incompatibility of the DEAR and 
scattering data. Is it possible to fit all data by using  unitarized ChPT?
What kind of  additional experimental input could help  to critically
constrain the parameters of  a fit?  
\end{sloppypar}

\item[vi)]
Still in connection with the SIDDHARTA experiment, it would be a major
breakthrough to present a systematic EFT calculation of the kaon--deuteron 
scattering length in terms of the threshold parameters of the $\bar KN$ 
interaction beyond the approximations used in Refs.~\cite{Kamalov:2000iy,Raha4}. 

\end{itemize}
On the {\it experimental side}, very precise data on \piH are 
underway~\cite{Gotta:2006}.
 The DIRAC collaboration has improved on the uncertainty of the 
lifetime measurement of pionium,
 see Ref.~\cite{tauscher_kaon2007}, and an update of the 
central value of $a_0-a_2$ is anxiously awaited. A successful measurement of 
individual scattering lengths via excited states, 
and $\pi K$ scattering lengths, would be extremely welcome.
We are looking forward to precise data on \kaH and \kad 
from SIDDHARTA. In conclusion, there are exciting times ahead on 
the experimental side.

\bigskip

{\it Acknowledgments:}
We thank W. Weise for inviting us to write this report. 
We  are grateful for discussions and/or comments on the manuscript to
V.~Baru,
B.~Borasoy, 
C.~Curceanu-Petrascu, 
E.~Epelbaum,
T.~Ericson,
A.~Gal,
D.~Gotta, 
C.~Guaraldo,
 H.-W.~Hammer,
Ch.~Hanhart,
A.N.~Ivanov,
M.A.~Ivanov,
S.~Kar\-shen\-bo\-im,
A.~Kudryavtsev,
H.~Leutwyler,
V.~E. Mar\-kus\-hin,
J.~Marton,
E.~Matsinos, 
Ulf-G.~Me\-i\ss ner,
L.~Nemenov,
G.~Rasche,
J.~Schacher,
L.~Simons,
D.~Trautmann. 
Partial financial support from the EU Integrated Infrastructure
Initiative Hadron Physics Project (contract number RII3--CT--2004--506078)
and DFG (SFB/TR 16, ``Subnuclear Structure of Matter'') 
and from the Swiss National Science Foundation 
is gratefully acknowledged. 
This research is also part of the project supported 
by the DFG under contracts FA67/31--1, GRK683, President grant of Russia 
``Scientific Schools'' No. 5103.2006.2, 
and by the EU Contract No. MRTN--CT--2006--035482
\lq\lq FLAVIAnet''.
 One of us (J.G.) is grateful to the Alexander von Humboldt--Stiftung and to 
the Helmholtz--Gemeinschaft for the award of  a prize 
 that allowed him to stay at the HISKP at the University of Bonn, 
where part of this work was performed. 
He also thanks the HISKP for the warm hospitality during these stays.

\clearpage

\renewcommand{\thefigure}{\thesection.\arabic{figure}}
\renewcommand{\thetable}{\thesection.\arabic{table}}
\renewcommand{\theequation}{\thesection.\arabic{equation}}

%%%%%%%%%%%%%%%%%%%%%%%%%%%%%%%%%%%%%%%%%%%%%%%%%%%%%%%%%%%%%%%%%%%%%%%%%%%%%%%%%
\appendix

%%%%%%%%%%%%%%%%%%%%%%%%%%%%%%%%%%%%%%%%%%%%%%%%%%%%%%%%%%%%%%%%%%%%%%%%%%%%%%
\setcounter{equation}{0}
\setcounter{figure}{0}
\setcounter{table}{0}

\section{Notation}
\label{app:notations}
In this appendix, we collect  some of the notation used.

\subsection{General}

The following  Coulombic bound states are considered:
\begin{itemize}
\item[]
Scalar QED
/
$\pi^+\pi^-$
/
$\pi^\mp K^\pm$
/
$\pi^- p$
/
$K^- p$
/
$\pi^- d$
/
$K^- d$\, .
\end{itemize}
In all these cases, at least one of the particles has  spin zero. We always 
attach the label 1 to the particle with  non--zero spin (if present). The masses
of the particles are denoted by $m_1$ and $m_2$. We further define
\eq
\msigma=m_1+m_2\, ,\quad\quad\mu_c=\frac{m_1m_2}{\msigma}\, ,\quad\quad
\eta_{1,2}=\frac{m_{1,2}}{\msigma}\, ,
\en
where $\msigma$ and $\mu_c$ denote the full mass and the reduced mass of a 
system of two particles, respectively. 

\subsection{Coulombic bound states}

For  Coulomb bound states, we use
\eq\label{eq:Acoulomb}
E_n&=&\msigma-\frac{\mu_c\alpha^2}{2n^2}\,\, ,\quad\quad
\gamma_n=\frac{\gamma}{n}\, ,\quad \gamma=\alpha\mu_c\, ,\quad
r_B=\gamma^{-1}\, ,\nnnl
\alpha^{-1}&=&137.036,
\en
where $E_n$ stand for the unperturbed Coulomb energies and $r_B$ is the Bohr
radius for a given state.
Further, in the case of the two spin-0 particles,
the bound--state  obeys the Schr\"odinger equation
\eq
({\bf H}_0+{\bf H}_C)|\Psi_{nlm}({\bf P})\rangle
=\biggl( E_n+\frac{{\bf P}^2}{2\msigma}\biggr)|\Psi_{nlm}({\bf P})\rangle,
\en
with the wave function given by
\eq
&&|\Psi_{nlm}({\bf P})\rangle=
\int\frac{d^3{\bf q}}{(2\pi)^3}\,\Psi_{nlm}({\bf q})
\,|{\bf P},{\bf q}\rangle\, ,
\nonumber\\[2mm]
&& |{\bf P},{\bf q}\rangle
=a_1^\dagger(\eta_1{\bf P}+{\bf q})a_2^\dagger(\eta_2{\bf P}-{\bf q})|0\rangle\fs
\en
The same formulae in case of a particle with spin is
\eq
({\bf H}_0+{\bf H}_C)|\Psi_{nljm}({\bf P})\rangle
=\biggl( E_n+\frac{{\bf P}^2}{2\msigma}\biggr)|\Psi_{nljm}({\bf P})\rangle,
\en
and
\eq
&&|\Psi_{nljm}({\bf P})\rangle=\sum_{\sigma}
\int\frac{d^3{\bf q}}{(2\pi)^3}\,
\langle jm|l(m-\sigma) s\sigma\rangle \,
\Psi_{nl(m-\sigma)}({\bf q})\,|{\bf P},{\bf q},\sigma\rangle\, ,
\nonumber\\[2mm]
&& |{\bf P},{\bf q},\sigma\rangle
=b_1^\dagger(\eta_1{\bf P}+{\bf q},\sigma)a_2^\dagger(\eta_2{\bf P}-{\bf q})|0\rangle\, .
\en
In the above formulae, $a_{1,2}^\dagger$ and $b_1^\dagger$ denote creation 
operators of particles without and with spin.

The wave functions $\Psi_{nlm}({\bf q})$ are the Fourier 
transform of standard Coulomb wave functions $\tilde\Psi_{nlm}({\bf x})$
 in coordinate space,
\eq
\Psi_{nlm}({\bf q})=\int d^3{\bf x}\,
{\rm e}^{-i{\bf q}{\bf x}}\tilde \Psi_{nlm}({\bf x})\, .
\en
The normalization is
\eq\label{eq:normwave}
\int d^3{\bf x}\,|\tilde\Psi_{nlm}({\bf x})|^2=1\, ,\quad\quad
\int \frac{d^3{\bf q}}{(2\pi)^3}\,|\Psi_{nlm}({\bf q})|^2=1\, .
\en 
Explicitly \cite{coulombwave},
\bea
\Psi_{nlm}({\bf q})&=&
\frac{N_{nlm}\,|{\bf q}|^l}{[{\bf q^2}+\gamma_n^2]^{l+2}}\,
C_{\,n-l-1}^{\,l+1}
\left(\frac{{\bf q^2}-\gamma_n^2}{{\bf q^2}+\gamma_n^2}\right)
Y_l^m(\theta,\varphi)\,,
\eea
where $C_n^m(x)$ are Gegenbauer polynomials, with generating function
\bea
(1-2xs+s^2)^{-m}=\sum_{n=0}^\infty C_n^m (x) s^n\,,
\eea
and $N_{nlm}$ are normalization constants, chosen in conformity with
 Eq.~(\ref{eq:normwave}). From this representation, 
\bea
\Psi_{nlm}({\bf q})\sim\left\{\begin{array}{ll}
|{\bf q}|^l,&|{\bf q}| \to 0\\
|{\bf q}|^{-4-l},&|{\bf q}|\to \infty
\end{array}
\right. \fs
\eea
The ground-state  wave function is 
\eq
\Psi_{100}({\bf q})=\frac{(64\pi\gamma^5)^{1/2}}{({\bf q}^2+\gamma^2)^2}\fs
\en

Due to  rotational symmetry, the corrections to the energy levels, 
which are calculated with the use of the above wave functions,
do not depend on the index $m$.
Therefore, we use everywhere $m=0$ and introduce the shorthand 
notation
\eq
\Psi_{nl}({\bf q})=\Psi_{nl0}({\bf q})\, ,\quad\quad
\tilde \Psi_{nl}({\bf x})=\tilde \Psi_{nl0}({\bf x})\, .
\en
We often need the wave function at the origin,
\eq
|\tilde\Psi_{n0}(0)|^2=\frac{\alpha^3\mu_c^3}{\pi n^3}\, .
\en
In the bound-state calculations we use the notation
\eq
s_n(\alpha)=2(\psi(n)-\psi(1)-\frac{1}{n}+\ln\alpha-\ln n)\, ,\quad\quad
\psi(x)=\Gamma'(x)/\Gamma(x)\, .
\en

\subsection{Master equation}

When only the Coulomb interactions are present, the resolvent $(z-\mathbold{H})^{-1}$
 is meromorphic in the complex $z$-plane, cut along the positive real axis above the elastic threshold.
The poles are located at $z=E_n$. If other interactions are turned on,
the bound--state poles are shifted from their pure Coulomb values into
the complex plane. The degeneracy of the energy eigenvalues is
lifted as well. The (complex) energy shift is given by 
\eq\label{eq:Amaster}
\Delta E_{nlj}=z_{nlj}-E_n=(\Psi_{nlj}|\bar{\mathbold{\tau}}^{nlj}(E_n)|\Psi_{nlj})
+o(\delta^4)\, ,
\en
where the matrix element is calculated between unperturbed Coulomb wave 
functions and does not depend on the quantum number $m$. Further,
$z_{nlj}$ denotes the new position of the pole, which was shifted
from its original position at $E_n$ and the pole--subtracted amplitude
$\bar{\mathbold{\tau}}^{nlj}(z)$ is defined by Eq.~(\ref{eq:key_tau}). 
In order to simplify notations, we often do
not attach explicit indexes $nlj$ to the pole position in the text.
In the case of two spin-0 particles, the index $j$ must be suppressed. 
In the main text, we refer to equation \eqref{eq:Amaster} as {\it master equation}.
\subsection{Energy shift of the atom}

The energy shift is further split into the electromagnetic and strong
pieces, as well as the term corresponding to the vacuum polarization
(whenever electrons are present) 
\eq\label{eq:appsplitting}
\Delta E_{nlj}=\Delta E_{nlj}^{\rm em}+\Delta E_{nl}^{\rm vac}+
\delta_{l0}\biggl(\Delta E_n^{\rm str}-\frac{i}{2}\,\Gamma_n\biggr)
+o(\delta^4)\, .
\en
(If both
particles have spin 0, the above formulae are modified by merely discarding
the index $j$).
The first term in this expression is given by Eq.~(\ref{eq:Eem_pipi}),
 Eq.~(\ref{DeltaEem}) or Eq.~(\ref{eq:splitting1}), for 
the case of $\pi^+\pi^-$, $\pi^\mp K^\pm$ and $\pi^-p$ atoms, respectively.
The second term, which corresponds to the vacuum polarization contribution,
in the case of the spin-0 particles is given by 
Eq.~(\ref{eq:vacleading}) and its generalization to the case of particles with
spin is straightforward. Namely, this contribution does not depend
on the total momentum $j$. Further,
at the order in isospin breaking parameter $\delta$ we are working, only the
$S$-wave strong shift should be taken into account.
The strong shift in higher partial waves is suppressed by additional powers of
$\delta$ and does not contribute at this order.

Further, note that if the splitting between the energy levels with the same 
orbital momentum $l$
is tiny, it is convenient to define the averaged level energies 
\eq\label{eq:naming}
\Delta E_{nl}={\cal N}_\ell^{-1}\sum_{j=|l-s|}^{l+s}(2j+1)\,
\Delta E_{nlj}
\, ,\quad\quad 
{\cal N}_\ell=\sum_{j=|l-s|}^{l+s}(2j+1)\, .
\en
When one speaks, e.g., about the energy of $1s$ or $3p$ levels in  
\piH, this average is meant.

\subsection{Threshold amplitude}

Let us now consider the definition of the threshold amplitude. 
The scattering amplitude for the process $1+2\to 3+4$ is given by
(the trivial term without interaction is omitted)
\eq\label{eq:app1}
\langle p_1\sigma',p_2;out|q_1\sigma,q_2;in\rangle=i(2\pi)^4\delta^4(p_1+p_2-q_1-q_2)
T_{\sigma'\sigma}(p_1,p_2;q_1,q_2)\, .
\en
The above amplitude describes the scattering in the channel with oppositely
charged particles, as well as in other channels (e.g. in the ``neutral''
channel). The Condon--Shortley--de Swart phase convention is used.
If there are no particles with spin, the indexes $\sigma',\sigma$
 should be dropped.

\begin{figure}[t]
\vspace*{.6cm}
\begin{center}
\includegraphics[width=10.cm,angle=0]{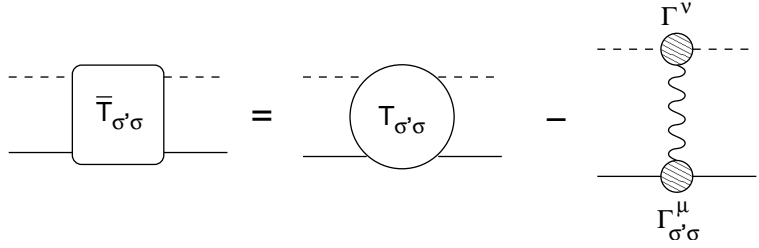}
\end{center}
\caption{Subtraction of the one--photon exchange diagram, see
  Eq.~(\ref{eq:subtraction}).}
\label{fig:onegamma}
\end{figure}

At the order we are working, it suffices to evaluate the scattering amplitude
 Eq.~(\ref{eq:app1}) at order $\alpha$. Starting from Eq.~(\ref{eq:app1}), one 
arrives at the threshold amplitude as a result of the following procedure.
In a first step, the one--photon exchange piece is subtracted in the elastic
 scattering amplitude of two oppositely charged particles,
\eq\label{eq:subtraction}
\bar T_{\sigma'\sigma}(p_1,p_2;q_1,q_2)=T_{\sigma'\sigma}(p_1,p_2;q_1,q_2)
+\Gamma^\mu_{\sigma'\sigma}(p_1,q_1)\,\frac{e^2g_{\mu\nu}}{t}\,
\Gamma^\nu(p_2,q_2)\, ,
\en
where $t=(p_1-q_1)^2=(p_2-q_2)^2$,  and $\Gamma^\mu_{\sigma'\sigma}(p_1,q_1)$ and
$\Gamma^\nu(p_2,q_2)$ denote the electromagnetic current matrix elements of  particles
1 and 2, respectively (if the particle 1 has no spin, the 
$\Gamma^\mu_{\sigma'\sigma}(p_1,q_1)$ is replaced by $\Gamma^\mu(p_1,q_1)$).
This procedure is demonstrated in Fig.~\ref{fig:onegamma}.
In the inelastic channels, there is no one--photon exchange diagram and
$\bar T_{\sigma'\sigma}= T_{\sigma'\sigma}$.
Moreover, it is not needed to remove the one--photon piece
in the elastic scattering amplitude with at least one neutral 
particle as well, because  the pertinent contribution to the
spin--nonflip amplitude is not singular at threshold.

We now list the decomposition of the matrix elements 
$\Gamma^\mu_{\sigma'\sigma}(p_1,q_1)$.
\begin{itemize}

\item[{\bf i)}] {\bf Spin 0 (pion, kaon)}
\eq
\Gamma^\mu(p,q)=(p+q)^\mu F(Q^2)\, , \quad Q=p-q\,; \quad F(0)=1\, .
\en

\item[{\bf ii)}] {\bf Spin $\mathbold {\frac{1}{2}}$ \bf (proton)}
\eq
\Gamma^\mu_{\sigma'\sigma}(p,q)=\bar u(p,\sigma')
\biggl(\gamma^\mu F_1(Q^2)+i\sigma^{\mu\nu}Q_\nu\frac{F_2(Q^2)}{2m_p}\biggr)u(q,\sigma)\, ,
\en
\eq
F_1(0)=1,\quad F_2(0)=\kappa_p\, ,
\en
where $F_1$ and $F_2$ denote the Dirac and Pauli form factors, respectively, 
and $\kappa_p$ is the anomalous magnetic moment

\item[{\bf iii)}] {\bf Spin--1 (deuteron)}

The decomposition is given,   e.g., in Refs.~\cite{Zuilhof,Gross}.

\end{itemize}

In the next step, we define the spin--nonflip amplitude by averaging over spins
\eq\label{eq:spinaverage}
\bar T(p_1,p_2;q_1,q_2)=\frac{1}{2s+1}\sum_\sigma
\bar T_{\sigma\sigma}(p_1,p_2;q_1,q_2)\, .
\en
At the final step, we go to the two particle CM frame, defined by
${\bf p}_1=-{\bf p}_2={\bf p}$ and
${\bf q}_1=-{\bf q}_2={\bf q}$,
 and remove the Coulomb phase from the 
(dimensionally regularized) spin-nonflip amplitude $\bar T$
\eq\label{eq:app2}
{\rm e}^{-in\alpha\theta_c} \bar T({\bf p},{\bf q})
=\frac{B_1}{|{\bf p}|}+B_2\ln\frac{|{\bf p}|}{\mu_c}+{\cal T}+O(|{\bf p}|)\, ,
\en
where $n=2$ for the scattering of two oppositely charged particles into
two charged particles (example: $\pi^+\pi^-\to\pi^+\pi^-$),
$n=1$ for the transition of two charged particles into a pair of 
neutral particles or vice versa (example: $\pi^+\pi^-\to\pi^0\pi^0$) and
$n=0$ for the scattering of two neutral particles into two neutral particles
(example: $\pi^0\pi^0\to\pi^0\pi^0$).
The (infrared--divergent) Coulomb phase is given by
\eq
\theta_c=\frac{\mu_c}{|{\bf p}|}\,\mu^{d-3}
\biggl(\frac{1}{d-3}-\frac{1}{2}\,(\Gamma'(1)+\ln 4\pi)
+\ln\frac{2|{\bf p}|}{\mu}\biggr)\, ,
\en
where $\mu$ stands for the scale of dimensional regularization.

The equation~(\ref{eq:app2}) defines the relativistic 
threshold amplitude ${\cal T}$ in all cases of interest.

\subsection{Real part of the threshold amplitudes}

 In particular cases, it turns out convenient to introduce the threshold amplitude
${\cal A}$, which coincides with ${\cal T}$ (or the real part thereof)
up to a normalization constant. Below we give the normalization for all
amplitudes needed (for a general scattering process 
$1+2\to\bar 1+\bar 2$ the notation ${\cal N}^{-1}=8\pi(m_1+m_2)$ is used, where
$m_1$ and $m_2$ stand for the mass of the particles 1 and 2, respectively).
The ellipses stand for the isospin breaking corrections.

\subsubsection*{$-$ Pionium}

\noindent {\it Scattering channels:}\\[2mm]  $\pi^+\pi^-\to\pi^+\pi^-~{(c)}$ / 
$\pi^+\pi^-\to\pi^0\pi^0~{(x)}$
\eq
{\cal A}_c&=&\frac{1}{32\pi}\,{\rm Re}\,{\cal T}_c=\frac{1}{6}\,(2a_0+a_2)+\cdots\, ,
\nonumber\\[2mm]
{\cal A}_x&=&-\frac{3}{32\pi}\,{\rm Re}\,{\cal T}_x=(a_0-a_2)+\cdots\, .
\en
The scattering lengths $a_0,a_2$ are denoted by $a_0^0,a_0^2$ in Ref.~\cite{GL_ChPT}.
\subsubsection*{$-$ $\mathbold {\pi K}$ atom} 
\noindent {\it Scattering channels:}\\[2mm]  $K^+\pi^-\to K^+\pi^-~{ (c)}$ / 
$K^+\pi^-\to\bar K^0\pi^0~{ (x)}$
\eq
{\cal A}_c&=&{\cal N}\,{\rm Re}\,{\cal T}_c=\frac{1}{3}\,(2a^{1/2}+a^{3/2})+\cdots\, ,
\nonumber\\[2mm]
{\cal A}_x&=&-{\cal N}\,{\rm Re}\,{\cal T}_x/\sqrt{2}
=\frac{1}{3}\,(a^{1/2}-a^{3/2})+\cdots\, .
\en
The scattering lengths are normalized as in Ref.~\cite{BKM-Kpi1}.

\subsubsection*{$-$ Pionic hydrogen}
\noindent {\it Scattering channels:}\\[2mm] 
$p\pi^-\to p\pi^-~{ (c)}$ /
$p\pi^-\to n\pi^0~{ (x)}$ /
$n\pi^0\to n\pi^0~{ (0)}$ /
$n\pi^-\to n\pi^-~{ (n)}$ 
\eq\label{eq:appnorm}
&&{\cal A}_{c,0,n}={\cal N}\,{\rm Re}\,{\cal T}_{c,0,n}\, ,\quad\quad
{\cal A}_x={\cal N}\,{\rm Re}\,{\cal T}_x/\sqrt{2}\, ,
\nonumber\\[2mm]
&&{\cal A}_c=a_{0+}^++a_{0+}^-+\cdots\, ,\quad\quad
{\cal A}_0=a_{0+}^++\cdots\, ,
\nonumber\\[2mm]
&&{\cal A}_n=a_{0+}^+-a_{0+}^-+\cdots\, ,\quad\quad
{\cal A}_x=-a_{0+}^-+\cdots\, .
\en
The scattering lengths are normalized as in Refs.~\cite{Hoehler,BLII}.
\subsubsection*{$-$ Kaonic hydrogen}

\noindent {\it Scattering channels:}\\[2mm] 
$pK^-\to pK^-~{ (c)}$ /
$pK^-\to n\bar K^0~{ (x)}$ /
$n\bar K^0\to n\bar K^0~{ (0)}$ /
$nK^-\to nK^-~{ (n)}$ 
\eq
&&{\cal A}_{c,0,n}={\cal N}\,{\cal T}_{c,0,n}\, ,\quad\quad
{\cal A}_x={\cal N}\,{\cal T}_x\, ,
\nonumber\\[2mm]
&&{\cal A}_c=\frac{1}{2}\,(a_0+a_1)+\cdots\, ,\quad\quad
{\cal A}_0=\frac{1}{2}\,(a_0+a_1)+\cdots\, ,
\nonumber\\[2mm]
&&{\cal A}_n=a_1+\cdots\, ,\quad\quad
{\cal A}_x=\frac{1}{2}\,(a_0-a_1)+\cdots\, .
\en
The scattering lengths are normalized as in Ref.~\cite{OsetRamos}.
\subsubsection*{$-$ Pionic deuterium}
\noindent {\it Scattering channels:}\\[2mm] 
$\pi^-d\to\pi^-d~{ (c)}$
\eq\label{eq:appdeuteron}
{\cal A}_c&=&{\cal N}\,{\cal T}_c=a_{\pi d}+\cdots\, .
\en

\subsubsection*{$-$ Kaonic deuterium}
\noindent {\it Scattering channels:}\\[2mm] 
$K^-d\to K^-d~{ (c)}$
\eq
{\cal A}_c&=&{\cal N}\,{\cal T}_c=a_{\bar Kd}+\cdots\, .
\en

\subsection{Two--particle phase space}

In the text, we use the symbols
\eq
\hspace{-.5cm} p_n^\star=\frac{\lambda^{1/2}(E_n^2, m_1^2, m_2^2)}{2E_n}\, ,
\quad\lambda(x,y,z)=x^2+y^2+z^2-2xy -2xz-2yz\, ,
\en
with $E_n$ given in Eq.~\eqref{eq:Acoulomb}.
\subsection{Panofsky ratio}
The Panofsky ratio is defined by
\eq\label{eq:Panofsky}
P=\frac{\sigma(\pi^-p\to\pi^0 n)}{\sigma(\pi^-p\to \gamma n)}\, ,
\en
evaluated at the $\pi^-p$ threshold. Its value is $P=1.546\pm0.009$ 
\cite{panofsky1} ($P=1.546\pm0.010$ \cite{panofsky2}).
 At the accuracy we are working, one 
does not need to consider electromagnetic corrections to the  cross 
sections. $P$ is a quantity of  order $\delta^{-1/2}$.

%%%%%%%%%%%%%%%%%%%%%%%%%%%%%%%%%%%%%%%%%%%%%%%%%%%%%%%%%%%%%%%%%%%%%%%
\setcounter{equation}{0}
\setcounter{figure}{0}
\setcounter{table}{0}
\section{Generalized unitarity}
\label{app:generalized}

Let ${\bf H}$ be a (non--hermitian) Hamiltonian.
We further write ${\bf H}={\bf H}_{0R}+{\bf H}_I$, where ${\bf H}_{0R}$
denotes  one--particle Hamiltonian, which includes all relativistic
corrections. Acting on the free two--particle states, it gives
\eq
{\bf H}_{0R}|{\bf q}_1,{\bf q}_2\rangle=(q_1^0+q_2^0)
|{\bf q}_1,{\bf q}_2\rangle\, ,
\en
where $q_i^0=\sqrt{M_i^2+{\bf q}_i^2},~i=1,2$. Next, we define the free
resolvent
\eq
{\bf G}_{0R}(z^\pm)&=&(z-{\bf H}_{0R}\pm i0)^{-1}
\en
and the $T$-operator
\eq
{\bf T}_{NR}(z)&=&{\bf H}_I+{\bf H}_I{\bf G}_{0R}(z^+){\bf T}_{NR}(z)=
{\bf T}_{NR}(z){\bf G}_{0R}(z^+){\bf H}_I+{\bf H}_I\, ,
\nonumber\\[2mm]
{\bf T}_{NR}^\dagger(z)&=&{\bf H}_I^\dagger+{\bf H}_I^\dagger{\bf G}_{0R}(z^-){\bf T}_{NR}^\dagger(z)=
{\bf T}_{NR}^\dagger(z){\bf G}_{0R}(z^-){\bf H}_I^\dagger+{\bf H}_I^\dagger\fs
\en
The relation to the $T$-matrix elements, introduced in 
Eq.~(\ref{eq:residue_TNR}) is given by
\eq
\langle {\bf p}_1,{\bf p}_2|{\bf T}_{NR}(z)|{\bf q}_1,{\bf q}_2\rangle
&=&-(2\pi)^3\delta^3({\bf p}_1+{\bf p}_2-{\bf q}_1-{\bf q}_2)\,
T_{NR}({\bf p}_1,{\bf p}_2;{\bf q}_1,{\bf q}_2)\, ,
\nonumber\\[2mm]
z&=&p_1^0+p_2^0=q_1^0+q_2^0\, .
\en
Express now the Hamiltonian through the $T$-operator
\eq
{\bf H}_I&=&{\bf T}_{NR}(z)\biggl(1+{\bf G}_{0R}(z^+){\bf T}_{NR}(z)\biggr)^{-1}
\, ,
\nonumber\\[2mm]
{\bf H}_I^\dagger&=&\biggl(1+{\bf T}_{NR}^\dagger(z){\bf G}_{0R}(z^-)\biggr)^{-1}
{\bf T}_{NR}^\dagger(z)\, .
\en
Subtracting these two equations and using the identity
\eq
{\bf G}_{0R}(z^+)-{\bf G}_{0R}(z^-)=-2\pi i\delta(z-{\bf H}_{0R})\, ,
\en
we finally arrive at the generalized unitarity relation for the $T$-operator.
\eq\label{eq:gen}
&&{\bf T}_{NR}(z)-{\bf T}_{NR}^\dagger(z)
=-2\pi i\,{\bf T}_{NR}^\dagger(z)\,\delta(z-{\bf H}_{0R})\,{\bf T}_{NR}(z)
\nonumber\\[2mm]
&+&\biggl(1+{\bf T}_{NR}^\dagger(z){\bf G}_{0R}(z^-)\biggr)
\biggl({\bf H}_I-{\bf H}_I^\dagger\biggr)\biggl(1+{\bf G}_{0R}(z^+){\bf T}_{NR}(z)\biggr)\, .
\en
On the other hand, let us consider the non-relativistic 
multichannel scattering with
hermitian Lagrangian. Unitarity condition in this case takes the form
\eq\label{eq:gen_i}
{\bf T}_{ii}(z)-{\bf T}_{ii}^\dagger(z)
=-2\pi i\,\sum_k
{\bf T}_{ik}^\dagger(z)\,\delta(z-{\bf H}_{0R})\,{\bf T}_{ki}(z)\, ,
\en
where ${\bf T}_{ii}(z)={\bf T}_{NR}(z)$ describes the scattering in the
elastic channel. From comparison of the above two equations it becomes clear
that the term in Eq.~(\ref{eq:gen}), which contains
${\bf H}_I-{\bf H}_I^\dagger$, describes the ``flow of the 
probability'' into the shielded channels $i\neq k$.
These channels are therefore encoded in the 
imaginary  parts of the couplings in the Lagrangian Eq.~(\ref{eq:LNR}).

%%%%%%%%%%%%%%%%%%%%%%%%%%%%%%%%%%%%%%%%%%%%%%%%%%%%%%%%%%%%%%%%%%%%%%%%%
\setcounter{equation}{0}
\setcounter{figure}{0}
\setcounter{table}{0}

\section{Matching and unitarity}
\label{app:unitarity}

Here we  express of the coupling constants
$g_1$ and $e_1$ -- introduced in subsection~\ref{subsec:matching_piN} --
 through the threshold scattering amplitudes.
 The couplings $g_1,e_1$ enter the expression
for the energy shift and width of  \piH at next--to--leading
order, see Eq.~(\ref{eq:DE_piH}). Taking the real and imaginary parts of
this equation, we get
\eq\label{eq:DEGamma}
\Delta E_n^{\rm str}&=&-\frac{\alpha^3\mu_c^3}{\pi n^3}\,
\mbox{Re}\,
(g_1+4\gamma_n^2e_1-g_1^2\langle \bar {\bf g}_C^{n0}(E_n)\rangle)+o(\delta^4)\, ,
\nonumber\\[2mm]
\Gamma_n&=&\frac{2\alpha^3\mu_c^3}{\pi n^3}\,
\mbox{Im}\,
(g_1+4\gamma_n^2e_1-g_1^2\langle \bar {\bf g}_C^{n0}(E_n)\rangle)+o(\delta^{9/2})\fs
\en
For the real part, the procedure is 
straightforward since, as it can be checked, $\mbox{Re}\,e_1=O(1)$ that 
means that the contribution,
 proportional to $\mbox{Re}\,e_1$, does not emerge at the accuracy we
are working. In is now clear that the answer -- after 
adjusting the normalization accordingly -- is given by Eq.~(\ref{eq:match_g2}),
\eq
\mbox{Re}\,g_1=\frac{\mbox{Re}\,{\cal T}_c}{4m_pM_\pi}\,
\biggl\{1+\frac{\alpha\mu_c^2}{2\pi}\,
\biggl(\Lambda(\mu)+\ln\frac{4\mu_c^2}{\mu^2}-1\biggr)\,
\frac{\mbox{Re}\,{\cal T}_c}{4m_pM_\pi}\biggr\}+o(\delta)\, .
\en
Matching of the imaginary parts
of $g_1$ and $e_1$ is more complicated, 
because these break the naive counting rules in $\delta$.
A most straightforward way to address this problem is to use unitarity.
The key observation is that, although the counting rules are modified,
the modification at this order comes only from one term, namely 
the contribution of the $\pi^0n$ intermediate state, which can be 
easily singled out.  

An additional (technical) complication, 
which arises here, is related to the fact
that we have to consider the unitarity condition at threshold in the presence
of photons and to subtract the infrared--singular pieces in accordance to the
definition of the threshold amplitude. 
Most easily, this problem can be circumvented by using the trick described in 
Ref. \cite{Zemp}: one analytically
continues the unitarity condition for the negative values of the 
CM momentum squared ${\bf p}^2$ and approaches the threshold from below.
We shall explain below, how this can be achieved.

\begin{figure}[t]
\vspace*{.6cm}
\begin{center}
\includegraphics[width=12.cm,angle=0]{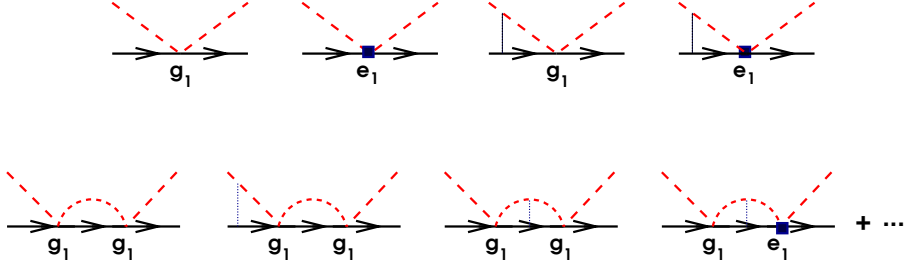}
\end{center}
\caption{Representative diagrams that describe the behavior 
of the non--relativistic spin--nonflip 
$\pi^-p\to\pi^-p$ amplitude near threshold, 
see Eq.~(\ref{eq:1+2Vc}). The vertex correction corresponds to $V_c$ and the
bubbles with and without Coulomb photons give $J_c$ and $B_c$, respectively.
The ellipses stand for diagrams with two and more pion bubbles. These do
not contribute to the matching condition at the accuracy we are working.
The vertex corrections in the final state are not shown.}
\label{fig:appendixplot}
\end{figure}

At the first step, it is necessary to establish the singularity structure
of the physical amplitudes, when the threshold is approached from below
(note that previously we have defined the threshold amplitudes through the
limiting procedure, when the threshold is approached from above). To this end
it will be again useful to invoke the non--relativistic effective theory, which
gives an analytic representation of the amplitude in the vicinity of the
threshold in terms of a finite number of simple loop integrals.
In the following, we discard the transverse photons as well as
non--minimal couplings of the Coulomb photons completely. The reason is
that these do not affect the matching of  $\mbox{Im}\,g_1$ at 
$O(\delta^{3/2})$ and $\mbox{Im}\,e_1$ at $O(\delta^{-1/2})$, which suffices 
at the accuracy we are working.

We start with the non--relativistic amplitude for the 
$\pi^-p$ elastic scattering near threshold and remove the
one--photon exchange contribution. In this manner, one obtains
 the counterpart of the amplitude
$\bar T_{NR}$, defined in the scalar case by Eq.~(\ref{eq:1gamma+n}).
In our case, the scattering amplitude contains the spin--nonflip as well as
spin--flip parts. In the CM frame it is given by 
\eq
\bar T_{\sigma'\sigma}({\bf p},{\bf q})=\delta_{\sigma'\sigma}
\bar F ({\bf p},{\bf q})+i\,\mbox{$\mathbold{\sigma}$}_{\sigma'\sigma}\cdot 
[{\bf q}\times{\bf p}]\,\bar G({\bf p},{\bf q})\, ,
\en
where ${\bf p}$ and ${\bf q}$ denote the CM 3-momenta for the
outgoing/incoming pion--nucleon pair. 
For determining the coupling constants $g_1,e_1$ it suffices to consider the
spin--nonflip part of the amplitude and to project out the $S$-wave component 
\eq
\bar F ({\bf p},{\bf q})=\frac{1}{2}\sum_\sigma
\bar T_{\sigma\sigma}({\bf p},{\bf q})\, ,\quad\quad
\bar F_0({\bf p}^2)=\frac{1}{2}\,\int_{-1}^1d\cos\theta\, 
\bar F ({\bf p},{\bf q})\, ,
\en
where $\theta$ is the angle between the vectors ${\bf p}$ and ${\bf q}$.

Discarding all but Coulomb photons,
one is left with only a few diagrams that determine the behavior
of $\bar F_0({\bf p}^2)$ at very small momenta, see 
Fig.~\ref{fig:appendixplot}. At order $\alpha$ one obtains
\eq\label{eq:1+2Vc}
\bar F_0({\bf p}^2)&=&(1-2V_c)\biggl(
g_1-4{\bf p}^2e_1+(g_1-4{\bf p}^2e_1)^2(J_c+B_c)+\cdots\biggr)\, .
\en
Note that, in order to keep the expressions as transparent as possible, we
did not further expand Eq.~(\ref{eq:1+2Vc}) in $\delta$ and ${\bf p}^2$,
retaining some higher--order terms as well.
The loop integrals, entering  Eq.~(\ref{eq:1+2Vc}), are given by
\eq\label{eq:loops_ex}
V_c&=&-\frac{\alpha\mu_c}{(-{\bf p}^2)^{1/2}}\,
\mu^{d-3}\biggl(\frac{1}{d-3}-\frac{1}{2}\,(\Gamma'(1)+\ln 4\pi)
+\frac{1}{2}\,\ln\frac{-4{\bf p}^2}{\mu^2}\biggr)
\nonumber\\[2mm]
&+&O(|{\bf p}|,d-3)\, ,
\nonumber\\[2mm]
J_c&=&-\frac{\mu_c}{2\pi}\,(-{\bf p}^2)^{1/2}+O(|{\bf p}|^3,d-3)\, ,
\nonumber\\[2mm]
B_c&=&\bar B_c
-\frac{\alpha\mu_c^2}{2\pi}\,\ln\frac{-{\bf p}^2}{\mu_c^2}+O(|{\bf p}|,d-3)
\, ,
\nonumber\\[2mm]
\bar B_c&=&-\frac{\alpha\mu_c^2}{2\pi}\,\biggl(\Lambda(\mu)-1
+\ln\frac{4\mu_c^2}{\mu^2}\biggr)\, ,
\en
where ${\bf p}^2<0$. Further, the quantity $J_c$ corresponds
to the single loop with the proton and the charged pion, no photons.
The quantities $V_c$ and $B_c$ are similar to the quantities defined in
section~\ref{sec:including_gamma} and correspond to the vertex correction and
to the internal exchange of the Coulomb photon, respectively. 

In Eq.~(\ref{eq:1+2Vc}) we now multiply both sides
 by $(1+2V_c)$ (remember that the quantity $V_c$ is real 
below threshold) and calculate the imaginary part. Using explicit expressions
for the loop integrals, we see that the following is valid below threshold,
\eq\label{eq:p2-piN}
(1+2V_c)\,\mbox{Im}\,\bar F_0({\bf p}^2)&=&
\biggl(h_0+h_0'\ln\frac{-{\bf p}^2}{\mu_c^2}\biggr)
+|{\bf p}|\biggl(h_1+h_1'\ln\frac{-{\bf p}^2}{\mu_c^2}\biggr)
\nonumber\\[2mm]
&+&|{\bf p}|^2\biggl(h_2+h_2'\ln\frac{-{\bf p}^2}{\mu_c^2}\biggr)
+o({\bf p}^2)\, ,
\en
where the coefficients $h_i,h_i'$ can be expressed in terms of $g_1, e_1$.
We shall further assume that $\mbox{Im}\, g_1=O(\delta^{1/2})$,
$\mbox{Im}\, e_1=O(\delta^{-1/2})$ (these assumptions will be verified
{\it a posteriori}). 
To determine the constants $g_1,e_1$ from the matching, we need 
following relations, which can be established from Eqs.~(\ref{eq:1+2Vc}),
(\ref{eq:loops_ex}) and (\ref{eq:p2-piN}),
\eq
h_0=\mbox{Im}\,(g_1+g_1^2\bar B_c)+o(\delta^{3/2})\, ,
\quad\quad h_2=-4\,\mbox{Im}\, e_1+o(\delta^{-1/2})\, .
\en
Let us now use the matching condition, which relates the imaginary part of
the quantity 
$\bar F_0({\bf p^2})$ to the (one--photon exchange--removed) 
relativistic spin--nonflip amplitude, defined in
accordance with Eq.~(\ref{eq:spinaverage}). We denote the latter quantity as
$T_c$, where the subscript ``c'' corresponds to the elastic channel 
$\pi^-p\to\pi^-p$. In the CM frame the matching condition reads
\eq\label{eq:subs}
(1+2V_c)\,\mbox{Im}\,\bar F_0({\bf p}^2)=\frac{1}{2}\,\int_{-1}^{1}d\cos\theta\,
\frac{(1+2V_c)}{2w_\pi({\bf p})\,2w_p({\bf p})}\,\mbox{Im}\,
\bar T_c({\bf  p},{\bf q})\, ,
\en
where $w_\pi({\bf p})$ and $w_p({\bf p})$ are the relativistic energies
of the charged pion and the proton in the CM frame, respectively.
Further, one may replace $\bar T_c$ by the full spin--nonflip 
amplitude $T_c$, since the one--photon exchange piece is real.

We further use unitarity in the relativistic theory to evaluate
the imaginary part of $T_c$ near threshold. 
Since the threshold is approached from below, the elastic contribution
from the $\pi^-p$ intermediate state, as well as  $\pi^-p$ plus any number
of photons do not contribute. Then, at the order of accuracy 
we are working, only the contributions from $n\pi^0$ and 
$n\gamma$ intermediate states should be retained. 
Substituting the result in Eq.~(\ref{eq:subs}), we finally get
\eq\label{eq:2terms}
(1+2V_c)\,\mbox{Im}\,\bar F_0({\bf p}^2)=p^\star({\bf p}^2)H({\bf p}^2)+
p_\gamma^\star({\bf p}^2)H_\gamma({\bf p}^2)+\cdots\, ,
\en
with
\eq
p^\star({\bf p}^2)&=&\frac{\lambda^{1/2}(s({\bf p}^2),m_n^2,M_{\pi^0}^2)}
{2\sqrt{s({\bf p}^2)}}\, ,\quad\quad
p_\gamma^\star({\bf p}^2)=\frac{\lambda^{1/2}(s({\bf p}^2),m_n^2,0)}
{2\sqrt{s({\bf p}^2)}}\, ,
\nonumber\\[2mm]
s({\bf p}^2)&=&(w_p({\bf p})+w_\pi({\bf p}))^2\,,
\en
and $H({\bf p}^2)$,  $H_\gamma({\bf p}^2)$ denote the pertinent angular 
integrals containing the scattering amplitudes into various intermediate 
states. We do not display the explicit expressions for $H({\bf p}^2)$ and  
$H_\gamma({\bf p}^2)$ here. The threshold behavior of $F_0({\bf p}^2)$
is determined by
\eq
p^\star({\bf p}^2)&\!\!=\!\!&p^\star(0)+(p^\star(0))'{\bf p}^2+O({\bf p}^4)\, ,\quad
p_\gamma^\star({\bf p}^2)=p_\gamma^\star(0)+O({\bf p}^2)\, ,
\nonumber\\[2mm]
H({\bf p}^2)&\!\!=\!\!&\biggl(k_0+k_0'\ln\frac{-{\bf p}^2}{\mu_c^2}\biggr)
             +O(|{\bf  p}|)\, ,\quad
H_\gamma({\bf p}^2)=H_\gamma(0)+O(|{\bf  p}|)\, .
\en
Note that at this order in $\delta$ 
there are no logarithms in $H_\gamma({\bf p}^2)$. Various coefficients count as
\eq
p^\star(0)&=&O(\delta^{1/2})\, ,\quad
(p^\star(0))'=O(\delta^{-1/2})\, ,\quad
p_\gamma^\star(0)=O(1)\, ,
\nonumber\\[2mm]
k_0&=&O(1)\, ,\quad
H_\gamma(0)=O(\delta) \, .
\en
The main property of the above representation is that the expansion 
coefficients in $H({\bf p}^2)$, $H_\gamma({\bf p}^2)$ are not enhanced in 
$\delta$-counting to the order we are working. At this order,
$\sim\delta^{-1/2}$ behavior in $e_1$  comes solely from the
 derivative  $(p^\star(0))'$. The matching therefore gives
\eq
\mbox{Im}\,(g_1+g_1^2\bar B_c)&=&p^\star(0)k_0+p_\gamma^\star(0)H_\gamma(0)
+o(\delta^{3/2})\, ,\quad
\nonumber\\[2mm]
-4\,\mbox{Im}\, e_1&=&(p^\star(0))'k_0+o(\delta^{-1/2})\, .
\en
From the above expression one may readily verify the power counting
$\mbox{Im}\,g_1=O(\delta^{1/2})$ and $\mbox{Im}\,e_1=O(\delta^{-1/2})$.
Further, for the particular combination of the low--energy constants
$g_1$ and $e_1$, which enters the expression for the decay
width Eq.~(\ref{eq:DEGamma}), we get
\eq
\mbox{Im}\,(g_1+g_1^2\bar B_c+4\gamma_n^2e_1)&=&(p^\star(0)-(p^\star(0))'\gamma_n^2)
k_0+p_\gamma^\star(0)H_\gamma(0)+o(\delta^{3/2})
\nonumber\\[2mm]
&=&\biggl(1+\frac{1}{P}\biggr)p^\star(-\gamma_n^2)k_0+o(\delta^{3/2})\, .
\en
In the above
expression $P$ denotes the Panofsky ratio which, at the order of accuracy we
are working, is
\eq
P=\frac{\sigma(\pi^-p\to\pi^0n)}
{\sigma(\pi^-p\to\gamma n)}\biggr|_{\mbox{thr}}
=\frac{p^\star(0)k_0}{p_\gamma^\star(0)H_\gamma(0)}+O(\delta^{1/2})\, .
\en
Further, note that the net effect
of the effective--range term at this order is to merely shift the
argument of the phase space factor to the correct bound--state value
$p_n^\star=p^\star(-\gamma_n^2)$, since
$\sqrt{s(-\gamma_n^2)}=E_n+O(\delta^4)$.

In order to evaluate the width at the required accuracy, it
suffices to determine the coefficient $k_0$. To this end, we recall that
the quantity $H({\bf p}^2)$ is defined as an angular integral over 
$|T_x|^2=\bigl\{\mbox{Re}\,T_x\bigr\}^2+\bigl\{\mbox{Im}\,T_x\bigr\}^2$, where $T_x({\bf p},{\bf q})$ denotes
the scattering amplitude for the process 
$\pi^-p\to\pi^0n$. Thus, at lowest order, $k_0$ must be proportional to
$\bigl\{\mbox{Re}\,{\cal T}_x\bigr\}^2$. Further, 
the imaginary part of the amplitude gives the
correction at $O(\delta)$ to this result. To find this correction,
one may invoke unitarity once again and make sure that at this order
only $\pi^0n$ intermediate state contributes. 
We skip all details here and display the final result,
\eq
\hspace*{-.4cm}&&\hspace*{-.4cm}
k_0=\frac{1}{32\pi m_pM_\pi(m_p+M_\pi)}\,
\biggl\{\mbox{Re}\,{\cal T}_x\biggr\}^2
\biggl\{1+\biggl(\frac{p^\star(0)}{8\pi(m_p+M_\pi)}\, 
\mbox{Re}\,{\cal T}_0\biggr)^2\biggr\}+o(\delta)\, .
\nonumber\\
\hspace*{-.4cm}&&\hspace*{-.4cm}
\en
From this equation we get
\eq\label{eq:Im-T222}
\hspace*{-.4cm}&&\hspace*{-.4cm}
\mbox{Im}\,(g_1+g_1^2\bar B_c+4\gamma_n^2e_1)
=\frac{4\pi p^\star(-\gamma_n^2)}{\mu_c}\,\biggl(1+\frac{1}{P}\biggr)\,
{\cal A}_x^2\,\biggl\{1+\bigl(p^\star(0){\cal A}_0\bigr)^2\biggr\}
+o(\delta^{3/2})\,  ,
\nonumber\\
\hspace*{-.4cm}&&\hspace*{-.4cm}
\en
where we have used Eq.~(\ref{eq:appnorm})
in order to rewrite the above expression in terms of the real amplitudes
${\cal A}_c$, ${\cal A}_x$ and ${\cal A}_0$.
Finally, using
Eqs.~(\ref{eq:Im-T222}), (\ref{eq:DEGamma}) and (\ref{eq:KK1}), 
we arrive at the
complete expression for the  decay width,  displayed
in Eq.~(\ref{eq:NLO-piN}).

\underline{Remark:} Although the modification of the counting rules, which was
considered in this appendix, seems a
bit complicated at a first glance, it is in fact just a remnant of the unitary
cusp effect. This effect has nothing to do with photons and can be established
 in the purely strong non--relativistic theory. Consider, for example, the
two--channel model like the one used for  pionium (without transverse photons). 
The coupling constants $c_1,c_2,c_3$ in this model obey the usual counting rules. 
Construct now an effective one--channel Hamiltonian, integrating out the
neutral channel and expressing new effective couplings in terms of
$c_1,c_2,c_3$ and the neutral bubble (see, e.g., Refs.~\cite{Bern1,Bern4}). The
reader is invited to check himself that exactly the above modified counting rules
for the new effective couplings emerge in the one--channel theory.
\underline{End of remark.}

\ed